# RELATIVISTIC QUANTUM FIELD THEORY OF HIGH–SPIN MATTER FIELDS: A PRAGMATIC APPROACH FOR HADRONIC PHYSICS

A Dissertation

by

DHARAM VIR AHLUWALIA

Submitted to the Office of Graduate Studies of
Texas A&M University
in partial fulfillment of the requirements for the degree of

DOCTOR OF PHILOSOPHY

August 1991

Major Subject: Physics

# RELATIVISTIC QUANTUM FIELD THEORY OF HIGH–SPIN MATTER FIELDS: A PRAGMATIC APPROACH FOR HADRONIC PHYSICS

A Dissertation

by

DHARAM VIR AHLUWALIA

Approved as to style and content by:

| | |
|---|---|
| David J. Ernst (Chair of Committee) | Richard L. Arnowitt (Member) |
| Che-Ming Ko (Member) | Roger A. Smith (Member) |
| Joseph B. Natowitz (Member) | Richard L. Arnowitt (Department Head) |

August 1991

# ABSTRACT

Relativistic Quantum Field Theory of High–Spin
Matter Fields: A Pragmatic Approach for Hadronic Physics.
(August 1991)
Dharam Vir Ahluwalia, B. Sc.(Hons.); M. Sc., University of Delhi, India
M. A., SUNY at Buffalo; M. S., Texas A&M University
Chair of Advisory Committee: Dr. David J. Ernst

A consistent phenomenology of the interaction of particles of arbitrary spin requires covariant spinors, field operators, propagators and model interactions. Guided by an approach originally proposed by Weinberg, we construct from group theoretical arguments the $(j,0)\oplus(0,j)$ covariant spinors and the field operators for a massive particles. Specific examples are worked out in the familiar language of the Bjorken and Drell text for the case of the $(1,0)\oplus(0,1)$, $(3/2,0)\oplus(0,3/2)$ and $(2,0)\oplus(0,2)$ matter fields. The $m\to 0$ limit of the covariant spinors is shown to have the expected structure. The algebra of the $\gamma^{\mu\nu}$ matrices associated with the $(1,0)\oplus(0,1)$ matter fields is presented, and the conserved current derived. The procedure readily extends to higher spins. The causality problem associated with the $j\geq 1$ wave equations is discussed in detail and a systematic procedure to construct causal propagators is provided. As an example a spin two wave equation satisfied by the $(2,0)\oplus(0,2)$, covariant spinors is found to support not only ten correct and causal solutions, but also thirty physically unacceptable acausal solutions. However, we demonstrate how to construct the Feynman propagator for the higher spin particles directly from the spinors and thus avoid the shortcomings of the wave equation in building a phenomenology. The same exercise is repeated for the $(1,0)\oplus(0,1)$ and $(3/2,0)\oplus(0,3/2)$ matter fields, and the same conclusions obtained.

A well–known set of linear equations for massless free particles of arbitrary spin is found to have acausal solutions. On the other hand, the $m\to 0$ limit of the wave equations satisfied by $(j,0)\oplus(0,j)$ covariant spinors are free from all kinematical acausality. This paradoxical situation is resolved and corrected through the introduction of a constraining principle.

The appendix reviews and presents in a unified framework classic works of Schwinger, Weinberg and Wigner regarding the elements of canonical quantum field theory, thus establishing the logical context of our work.

.

*In fond memories of two men*

*my father*
*B. S. Ahluwalia (1933-1977), a man I have missed most*

*and*

*my uncle*
*P. R. Rajgopal (     -1990), a man India will miss most.*

# ACKNOWLEDGEMENTS

Dave Ernst, the chair of my committee, is and has been a splendid collaborator, friend and advisor. I do not know what words to use to thank him, nor do I know how to say, given our western culture, that I am fond of him as a man. As such, I tell him: thank you, in a soft voice of affection. Then there are other teachers and mentors, and some friends, to whom all I must say that not only that they have added to my academic education, they have also served as symbols of personal strength, excellence and integrity. It is impossible to name them all, but here is a partial list (in alphabetic order): Dick Arnowitt, George Kattawar, Che-Ming Ko, Eckhard Krotschek, Wayne Saslow and Roger Smith. To Roger Smith I must add a further note of thank you for being always there in my personal and academic life. Without him my graduate school experience would have been less colorful and would have lacked certain human warmth. It is Roger who showed me how, in appropriate context, the distance between zero and one can be infinitely greater than distance from one to infinity. For George Kattawar, I register my thanks for his incessant encouragement and support. I thank Che-Ming Ko for a wonderful introductory course on Quantum Mechanics, and his unlimited accessibility. Dick Arnowitt is a physicist's physicist, a teacher from whom I learned many things in physics through his courses of unsurpassed excellence. I did not get to know Eckhard Krotschek and Wayne Saslow as much as I would have liked, but I have enjoyed their friendly presence in Liquid Physics Seminars and their offices. I also thank Joseph Natowitz and Paul Nelson for agreeing to advise me on academic matters by being on my committee. Besides these I must register my sincere thanks to Tom Ainsworth (Texas A&M), Allan Widom (Northeastern), Mikolaj Sawicki (Texas A&M), John Reading (Texas A&M), Bhanu Das (Oxford), Madhu Menon (Kentucky), Nathan Isgur (CEBAF), Allan Picklesimer (Los Alamos), Peter Herczeg (Los Alamos), Roy Thaler (Los Alamos), Bunny Clark (Ohio State), Peter Tandy (Kent State), Khin Maung Maung (Hampton/CEBAF), Michael Frank (Kent), Jim Vary (Iowa State), Ta-You Wu (Academia Sinica), Mendel Sachs (SUNY at Buffalo), Alex Vilenkin (Tufts), Eugene Barasch (Texas A&M), Chris

Pope (Texas A&M), Ron Bryan (Texas A&M), Michael Duff (Texas A&M), Ergin Sezgin (Texas A&M), Lewis Ford (Texas A&M), Glenn Agnolet (Texas A&M), Chia-Ren Hu (Texas A&M), Russ Huson (Texas A&M), Peter McIntyre (Texas A&M), Bob Webb (Texas A&M), Don Naugle (Texas A&M) and Marco Jaric (Texas A&M) for their encouragement in my work and friendship and hospitality. A special thanks are due to Steven Weinberg who from time to time took time to answer my correspondence seeking advice on matters of physics. For Warren Buck (Hampton/CEBAF) I must register my affectionate thanks for his charm, enthusiasm, and hospitality at HUGS at CEBAF 1990.

On a more personal side, away from the corridors of physics, I must thank Regina Fuchs-Ahluwalia for being a wonderful friend, a wife, and supporter of my work. Even as I write, at this late hour, she tolerates and understands the demands of physics, and the sacrifices it entails, of which she has become almost unknowingly a partner. Jugnu, Vikram and Shanti, our three wonderful sons, have added pleasure and warmth not only to my life but also to the corridors of physics. I thank them with affection and wish them better than the best, and hope one day they will find a life-long-passion of their own, and not complain that being sons of a graduate student they could not afford many luxuries their playmates took for granted.

I have also been extremely fortunate to have had a few friends whose warmth always keeps me enveloped wherever I am. They are: Barbara König, K. S. Balaji, Dave Fisher, Roxanne Amico, Edward Fuchs Jr, Eunice Castillo, Yagoda Ristowski-Najdowski, Mauro Parisi, Edna Menchaca–Johnson, Margeret Murry and Gerhard Baumann. They are the contents of my dreams and poetry. Without them, this thesis and everything else would lose significance and fire.

To these I must add, in thanks, my colleagues and friends at the graduate school: Jim Smith, Charlie Albert, Hong-Wei He, Chahriar Assad, Ahmed Mahmoud, Kirk Fuller, Dave Ring, Shen Shawn, Shinichi Urano, Frank Sogandares, Bob Bell, Jason Chen, Janice Epstein, Xu-Shan Fang, Fan Liu, David and Rhonda Batchelder, Jizhu Wu, John Burciaga, Mark Collins, and Jim Smith. Each, in their own way, added to my studies and life. I thank them all.

Special thanks are due to Christoph Burgard who read parts of this work, and discovered a few self cancelling mistakes of minus signs and factors of two. Iris Wellner, whom I only met a few months ago, and has now returned to Germany, added much fun during some of the most difficult periods of this work. Frank Schnorrenberg and Christina Maimarides need a mention for their kind friendship. I record my gratitude to Margrit and Greg Moores for their warm and supportive friendship, and introduction to glühwein.

For my detour to the outer boundaries of the academic world at the Jet Propulsion Laboratory I must thank Morteza Khorami, Faiza Lansing, Lynn Gresham, Rich Santiago, Alan Nikora and all those who serve excellence with integrity, at times risking their personal well being.

For my detour to film studies at SUNY at Buffalo I must thank Gerry O'Grady, Paul Sharits, Mauro Parisi, Tony Conrad, "Vieka", John Hand and many other friends with whom I had many a wonderful evenings and nights in dark and light. I must also thank my friends at Buffalo: Bob Wise and Alexander and Thersa Gella for their friendship and caring.

Finally, but not in the least, I thank my brothers: Karan and Bobby; my sisters: Sneh, Sarita and Mamta; my uncle: Shamsher Singh; and my mother: Bimla. Their contribution to my academic and personal life is a matter of much affection and pride for me. I salute them in gratitude. No one could wish to have been born in a a more affectionate family. I thank Miki, Carmen and Vicki Ernst for making our family a part of theirs, and accepting my calls at all hours of day and night.

There are many who I must thank but fail to recall their contributions to my work, at this hour, hours beyond midnight. I seek their forgiveness.

# TABLE OF CONTENTS

Page

Page

Page

# LIST OF TABLES

# 1. INTRODUCTION

## 1.1 Short Historical Review

Eight years after the publication of his relativistic wave equation [1] for spin one half particles in 1928, Dirac [2] proposed high–spin equations with constraints. In this 1936 paper he wrote: "... it is desirable to have the equations ready for a possible future discovery of an elementary particle with a spin greater than a half, or for approximate application to composite particles." Today, this speculation regarding the composite particles is remarkably confirmed. Neglecting the "non established" resonances, the following high–spin composite particles [3] have been experimentally observed: baryonic resonances with $1/2 \leq j \leq 11/2$, and mesonic resonances with $0 \leq j \leq 4$. The most common approach [4] to a relativistic field theory of particles with $j \geq 1$ is that of Proca, Rarita and Schwinger. This formalism [4, 5] owes its logical history to the 1936 paper of Dirac [2], the works of Fierz and Pauli [6,7,8] and depends for its present form [4] on a classic paper of Bargmann and Wigner [9]. For our purposes we note that this formalism consists of high–spin wave equations along with constraint equations. There is, however, a well recognised problem with this formalism. Corben and Schwinger [10], Johnson and Sudarshan [11], and Kobayashi and Shamaly [12] have respectively established for spin $1,\ 3/2,$ and 2 that electromagnetic coupling cannot be consistently introduced via the standard replacement of derivatives in the Lagrangians with gauge covariant derivatives: $\partial_\mu \rightarrow \partial_\mu + iqA_\mu$. Kobayashi and Takahashi [13,14] have recently argued that this pathology is generic to all equations of the Rarita–Schwinger type and has its origin in the existence of the associated constraint equations. In addition, this commonly used formalism becomes increasingly more complicated for $j \geq 3/2$.

Parallel to the development of the Rarita–Schwinger formalism, there exist studies of Duffin [15], Kemmer [16], Harish-Chandara [17,18] and Bhabha [19] which propose high–spin equations of the type:

---

This dissertation follows the style of *Annals of Physics.*

$$(i\,\beta^{\mu}\,\partial_{\mu} - m)\,\psi(x) = 0. \tag{1.1}$$

These equations lie outside the applicability of the Kobayashi–Takahashi theorem [13,14] and, therefore, it is not yet completely clear if one can consistently introduce the electromagnetic coupling into these equations through the minimal substitution. A significant part of the work on the high–spin wave equations [20], done after the well known paper of Johnson and Sudarshan [11], is concentrated in this direction and generally deals with the reducible representations of the Lorentz group $\sim [(SU(2)_{R} \otimes SU(2)_{L}]$.

It is within this historical context that we now discuss some of the needs of the nuclear physics community and our contribution to nuclear phenomenon involving high–spin mesons and baryons.

## 1.2 Motivation, Objectives and Achievements

At the present time several new accelerators are planned or are under construction: CEBAF, PILAC at LAMPF, and KAON. These nuclear physics facilities will be able to explore high–spin hadronic physics in greater detail than has been previously possible. Thus, although there remain fundamental difficulties with the quantum field theory of high–spin particles, a need for an internally consistent phenomenology of these particles is required. Such a phenomenology, at a minimum, must provide:

1.) Fully covariant spinors for any spin.

2.) Relativistic wave equations for any spin, and the associated conserved current.

3.) Complete understanding of the possible acausality, and a systematic procedure to construct propagators for any spin which propagate only the causal (i.e. physical) solutions.

4.) A general method for constructing model couplings.

Such a theory will allow us to do systematic calculations for the type of experiments one envisages at the above mentioned accelerators. Not only is such a work important from a pragmatic point of view, it is bound to provide a deeper understanding of the underlying structure of relativistic quantum fields.

In this work, following the work of Weinberg [21], we provide a general and explicit procedure [22] for constructing irreducible $(j,0)\oplus(0,j)$ covariant spinors and wave equations. The covariant spinors are obtained via construction of a general boost for any $j$ rather than as solutions of a specific wave equation. This procedure reproduces the standard Dirac spinors for $j = 1/2$. The relativistic covariant spinors for $j = 1,\ 3/2$ and 2 are explicitly constructed. For $j = 1$ we show the connection between the $(1,0)\oplus(0,1)$ covariant spinors and the standard Proca $A^\mu$ and $F^{\mu\nu}$. This is accomplished through the introduction of a procedure called "spinorial summation." We obtain $2(2j{+}1)$ covariant spinors for spin $j$, thus incorporating spinorial as well as particle–antiparticle degrees of freedom in a very natural fashion. The wave equations are derived so as to obtain conserved currents and phenomenological interaction Lagrangian densities. These wave equations are again obtained from a set of coupled equations valid for any spin, and have no associated constraints. As a result they lie outside the applicability of the Kobayashi–Takahashi theorem already cited. Again, for $j = 1/2$, these coupled equation reproduce the standard Dirac equation for spin one half particles. For $j = 1$, the $m \rightarrow 0$ limit yields the standard source free Maxwell equations of electromagnetism. The connection between the $(1,0)\oplus(0,1)$ covariant spinors and $\vec{E}$ and $\vec{B}$ fields is explicitly exhibited.

The $(j \geq 1,0)\oplus(0,j \geq 1)$ relativistic wave equations involve second and higher–order spacetime derivatives. A new technique is developed to study the origin of acausality in the high–spin wave equations. We show that [23] equations developed here do indeed have solutions which are acausal in character, for $j \geq 1$. However, the necessary propagators are constructed from the vacuum expectation values of the appropriate time ordered product of field operators. These field operators contain only the physically acceptable causal solutions of the wave equations and hence are free from any unphysical characteristics contained in the green functions associated with the relativistic wave equations.

In addition the quasi–relativistic and extreme relativistic limits of the $(j,0)\oplus(0,j)$ covariant spinors and equations are seen to provide some interesting insights into the nature of high–spin fields.

The remaining outstanding problem is a technique for generating the general forms for the phenomenological couplings. We take the conventional approach of modeling the interactions via the construction of the possible Poincaré scalars. The composite character of the hadrons is incorporated by including phenomenological form factors into the interaction. Some explicit models of such interaction Lagrangian densities are provided.

In addition to the facts that the formalism developed here is valid for any spin and is able to address the question of causality in a rather elegant fashion, it is also particularly suited for modern computer oriented numerical and symbolic manipulation technology.

Not only is the formalism developed here specifically designed to address theoretical issues arising from CEBAF, PILAC at LAMPF and KAON, the formalism is equally useful in studies of heavy ion collisions at RHIC where high spin particles such as the $f_2(1720)$ can be produced. Towards this end we have obtained the S–Matrix elements needed for cross section calculations of two photon mediated production of scalar and pseudoscalar particles. Extension of these results to the production of arbitrary spin particles now requires interaction Lagrangian densities which couple the gauge vector potential $A^\mu(x)$ to arbitrary spin. The problem of constructing this coupling is under investigation at present.

While the main text of this work is contained in seven chapters which follow, the appendix reviews some of the standard elements of the quantum field theory in the light of our work thus making the spin–dependence of the arguments involved explicit. The appendix presents some original arguments and derivations but is essentially based on the classic papers of Schwinger [24–32], Weinberg [21, 33–36] and Dyson [37], and presents the connection between these classic works in a unified fashion. Finally we note that since a beginning student in the nuclear physics community may not be familiar with some of the group theoretical nomenclature, we have incorporated appropriate definitions in the text. This should make this work immediately accessible to as large a research community as possible.

# 2. POINCARÉ TRANSFORMATIONS IN (1,3) SPACETIME AND THEIR EFFECT ON PHYSICAL STATES

## 2.1 Logical Structure of Quantum Field Theory

We take the view that the essential flat–spacetime kinematical and dynamical structure of quantum field theory, apart from some principle generating masses for some of the gauge bosons, is determined by the combined requirements of relativistic and gauge invariances. The relativistic invariance is defined as the invariance of the *form* of the laws of nature for all inertial observers. This invariance of the form, rather than a specific physical quantity, is sometimes called "Poincaré covariance" or "gauge covariance" depending upon the transformation under consideration[1]. The demand of Poincaré invariance determines what *matter fields* exist in nature. It is found that the wave equations, or the Lagrangian densities from which they follow, for the "free" matter fields are *not* invariant under local phase transformations of the form

$$\begin{pmatrix} \psi'_1(x) \\ \vdots \\ \psi'_n(x) \end{pmatrix} = \exp\Big[i\, g \sum_i \alpha_i(x)\, \Lambda_i\Big] \begin{pmatrix} \psi_1(x) \\ \vdots \\ \psi_n(x) \end{pmatrix}, \tag{2.1}$$

with $\Lambda = n \times n$ norm preserving $SU(n)$ matrices. The simplest such transformation, with $n = 1$, is the local $U(1)$ transformation

$$\psi'(x) = \exp\big[i\, g\, \alpha(x)\big]\, \psi(x). \tag{2.2}$$

Demanding invariance of the equations of motion[2] under local $U(1)$ transformation (2.2) naturally introduces a vector potential $A^\mu(x)$ which one identifies with

1 Since in the literature this fine distinction has almost disappeared we would often succumb not to explicitly distinguish between "covariance" and "invariance." However, almost invariably, the context should provide the needed distinction.

2 And hence the invariance of the Lagrangian density. The reason for this rather "inverted" emphasis lies in the fact that we will obtain equations of motion without reference to the Lagrangian formalism.

the electromagnetic interaction. The electromagnetic interaction is introduced by replacing

$$\partial^\mu \rightarrow D^\mu \equiv \partial^\mu \, + \, i\, q\, A^\mu, \tag{2.3}$$

in the kinematical equations of motion. The resulting dynamical equations are then invariant under Poincaré as well as local $U(1)$ gauge transformation (2.2) provided as

$$\psi'(x) \rightarrow \psi'(x) = \exp\big[i\, g\, \alpha(x)\big]\; \psi(x), \tag{2.4}$$

we simultaneously let

$$A^\mu(x) \rightarrow A'^\mu(x) = A^\mu(x) - \partial^\mu \alpha(x). \tag{2.5}$$

Why some $SU(n)$ invariances are physically realised and not others, is an unanswered question and perhaps points towards a yet undiscovered constraining principle of nature. Similarly why some representations of the Poincaré group are physically realised, and not others, is not yet known. It is quite possible that these unknown constraining principles contain in them solution of the yet unsolved problems of quantum field theory: such as a quantum field theory incorporating gravitational interaction.

This chapter is devoted to a systematic study of the relativistic invariance which all laws of nature, at least locally, are expected to respect on empirical grounds. It should perhaps be noted explicitly that the idea of "gauge invariance" is secondary to that of "Poincaré covariance" – for its very definition (in the quantum field theory) one needs the notion of matter fields first; and matter fields arise as finite dimensional representations of the Lorentz group.

So we begin, *ab initio*, with the fundamentals and arrive at the various results claimed in Sec. (1.2) in a logical and self contained fashion.

## 2.2 Poincaré Transformations

If two physical events occur at $x^\mu = (t, \vec{x})$ and $x^\mu + dx^\mu = (t + dt, \vec{x} + d\vec{x})$ then the observed constancy of the speed of light, for all inertial observers, requires that the interval

$$ds^2 = dt^2 - (d\vec{x})^2 \equiv \eta_{\mu\nu} dx^\mu dx^\nu \tag{2.6}$$

be invariant. The set of *linear* and *continuous* spacetime transformations which preserve $ds^2$ are as follows:

<u>Three Rotations</u> about each of the $(x, y, z)$–axes. The transformation matrices relating $x'^\mu$ with $x^\mu$, $x'^\mu = R^\mu{}_\nu x^\nu$, are given by

$$[R^\mu{}_\nu(\theta_x)] = \begin{pmatrix} 1 & 0 & 0 & 0 \\ 0 & 1 & 0 & 0 \\ 0 & 0 & \cos(\theta_x) & -\sin(\theta_x) \\ 0 & 0 & \sin(\theta_x) & \cos(\theta_x) \end{pmatrix}, \tag{2.7}$$

$$[R^\mu{}_\nu(\theta_y)] = \begin{pmatrix} 1 & 0 & 0 & 0 \\ 0 & \cos(\theta_y) & 0 & \sin(\theta_y) \\ 0 & 0 & 1 & 0 \\ 0 & -\sin(\theta_y) & 0 & \cos(\theta_y) \end{pmatrix}, \tag{2.8}$$

$$[R^\mu{}_\nu(\theta_z)] = \begin{pmatrix} 1 & 0 & 0 & 0 \\ 0 & \cos(\theta_z) & -\sin(\theta_z) & 0 \\ 0 & \sin(\theta_z) & \cos(\theta_z) & 0 \\ 0 & 0 & 0 & 1 \end{pmatrix}. \tag{2.9}$$

$[R^\mu{}_\nu(\theta_i)]$ represents a rotation by $\theta_i$ about the $ith$–axis. The rows and columns are labelled in the order $0, 1, 2, 3$.

Three Lorentz Boosts along each of the $(x, y, z)$–axes. The boost matrix for a boost along the positive direction of the unprimed $x$-axis, by velocity[3] $v$, is given by

$$[B^{\mu}{}_{\nu}(\varphi_x)] = \begin{pmatrix} \cosh(\varphi_x) & \sinh(\varphi_x) & 0 & 0 \\ \sinh(\varphi_x) & \cosh(\varphi_x) & 0 & 0 \\ 0 & 0 & 1 & 0 \\ 0 & 0 & 0 & 1 \end{pmatrix}, \tag{2.10}$$

with $x'^{\mu} = B^{\mu}{}_{\nu}(\varphi_x)x^{\nu}$. Similarly

$$[B^{\mu}{}_{\nu}(\varphi_y)] = \begin{pmatrix} \cosh(\varphi_y) & 0 & \sinh(\varphi_y) & 0 \\ 0 & 1 & 0 & 0 \\ \sinh(\varphi_y) & 0 & \cosh(\varphi_y) & 0 \\ 0 & 0 & 0 & 1 \end{pmatrix}, \tag{2.11}$$

$$[B^{\mu}{}_{\nu}(\varphi_z)] = \begin{pmatrix} \cosh(\varphi_z) & 0 & 0 & \sinh(\varphi_z) \\ 0 & 1 & 0 & 0 \\ 0 & 0 & 1 & 0 \\ \sinh(\varphi_z) & 0 & 0 & \cosh(\varphi_z) \end{pmatrix}, \tag{2.12}$$

where

$$\cosh(\varphi) = \gamma = \frac{1}{\sqrt{1 - v^2}}, \tag{2.13}$$

$$\sinh(\varphi) = v\gamma. \tag{2.14}$$

Four Translations

$$x'^{\mu} = x^{\mu} + a^{\mu}, \tag{2.15}$$

with $a_{\mu}$ as real constant displacements.

3 This is the velocity which a particle at rest in the unprimed frame acquires when seen from the primed frame.

*Parenthetic observations:* The time-order of physical events is preserved if $ds^2 \geq 0$. As a result, for any two physical events for which $ds^2 \geq 0$ the possibility of a cause and effect relationship exists. In quantum mechanics, of course, appropriate thoughts need to be given to the fundamental uncertainty in the measurement of $ds^2$ itself. This aspect of the subject, however, lies outside the boundaries of our immediate interest. In addition it should be noted explicitly that the Lorentz boosts become singular for the massless particles for which $v = c = 1$. In a closed system completely composed of massless particles, the measuring devices and the observer all move at relative speeds of unity. No measuring devices, such as clocks or rods, exist which can be at *rest* in any frame. Consequently the meaning of "physical measurement" seems to require a new definition for such a system. Poincaré covariance seems to acquire meaning only if we introduce some massive particles into our system, even if we do so only hypothetically. These massive particles can then be used to construct measuring devices, which can be at rest in frames occupied by observers made of massive particles. These observers can then study the laws governing the massless as well as massive particles. It would of course be interesting to formulate a theory, and experiments, without recourse to the existence of massive particles. To construct a theory of the early universe these epistemological questions must be confronted.

The above set of transformations can be summarised by

$$x'^\mu = \Lambda^\mu{}_\nu x^\nu + a^\mu, \tag{2.16}$$

with the constraint (required to preserve $ds^2$)

$$\Lambda^\mu{}_\rho \Lambda^\nu{}_\sigma \eta_{\mu\nu} = \eta_{\rho\sigma}. \tag{2.17}$$

The metric $\eta_{\mu\nu}$ is

$$[\eta_{\mu\nu}] = \begin{pmatrix} 1 & 0 & 0 & 0 \\ 0 & -1 & 0 & 0 \\ 0 & 0 & -1 & 0 \\ 0 & 0 & 0 & -1 \end{pmatrix}. \tag{2.18}$$

It is readily verified that

$$\Lambda_\mu{}^\rho \Lambda_\nu{}^\sigma = \delta^\rho{}_\sigma, \tag{2.19}$$

$$(\Lambda^{-1})^\mu{}_\nu = \Lambda_\nu{}^\mu. \tag{2.20}$$

Using these well known results it is immediately established that Poincaré transformations form a group, with

1. Multiplication law

$$\{\overline{\Lambda}, \overline{a}\}\{\Lambda, a\} = \{\overline{\Lambda}\Lambda, \overline{\Lambda}a + \overline{a}\}. \tag{2.21}$$

2. Inverse element

$$\{\Lambda, a\}^{-1} = \{\Lambda^{-1}, -\Lambda^{-1}a\}. \tag{2.22}$$

3. Identity element

$$\{I, 0\}. \tag{2.23}$$

In (2.23) above $I$ is a $4 \times 4$ identity matrix and 0 is a "zero" vector[4].

Using the group multiplication law (2.21), we find that the commutator of the two group elements is

$$\big[\{\Lambda_1, a_1\}, \{\Lambda_2, a_2\}\big] = \big\{(\Lambda_1\Lambda_2 - \Lambda_2\Lambda_1), (\Lambda_1 a_2 - \Lambda_2 a_1) + (a_1 - a_2)\big\}.$$

Consequently the $ds^2$ preserving continuous spacetime transformation form a non-Abelian group. It is called the "Poincaré group."

4 A "zero" vector $a^\mu \equiv (a^0 = 0, \vec{a} = \vec{0})$ is to be distinguished from a "null" vector for which only $ds^2 = 0$ is required

## 2.3 Generators of Poincaré Transformations and Associated Lie Algebra

For infinitesimal transformations, Eq. (2.16) can be written as

$$x'^{\mu} = (\delta^{\mu}{}_{\nu} + \lambda^{\mu}{}_{\nu})x^{\nu} + a^{\mu}, \tag{2.24}$$

where $\lambda^{\mu}{}_{\nu}$ and $a^{\mu}$ are infinitesimal constants. For various transformations the *nonvanishing* $\lambda^{\mu\nu} = \lambda^{\mu}{}_{\epsilon}\eta^{\epsilon\nu}$ are summarised in Table I

TABLE I

*Nonvanishing* $\lambda^{\mu\nu} = \lambda^{\mu}{}_{\epsilon}\eta^{\epsilon\nu}$. Note we only tabulate the nonvanishing $\lambda^{\mu\nu}$, as such, for example: $\lambda^{\mu\neq 2,\nu\neq 3} = -\lambda^{\nu\neq 3,\mu\neq 2} = 0$ for a rotation about the x–axis. Similar comments apply for other transformations.

| Rotation about: | | | Boost along: | | |
|---|---|---|---|---|---|
| x–axis | y–axis | z–axis | x–axis | y–axis | z–axis |
| $\lambda^{23} = -\lambda^{32}$ $= \theta_x$ | $\lambda^{31} = -\lambda^{13}$ $= \theta_y$ | $\lambda^{12} = -\lambda^{21}$ $= \theta_z$ | $\lambda^{10} = -\lambda^{01}$ $= \varphi_x$ | $\lambda^{20} = -\lambda^{02}$ $= \varphi_y$ | $\lambda^{30} = -\lambda^{03}$ $= \varphi_z$ |

Given Table I, we define ten linearly independent hermitian operators $X_\alpha$, called "generators" of the Poincaré transformations, corresponding to a parameter $\lambda^\alpha$ $[\lambda^1 = \theta_x, \ldots; \lambda^4 = \varphi_x, \ldots; \lambda^7 = a_0, \ldots]$

$$X_\alpha \equiv i\frac{\partial x'^{\,\mu}}{\partial \lambda^\alpha}\bigg|_{\lambda=0}\frac{\partial}{\partial x^\mu} \qquad (\alpha = 1, \ldots\ldots, 10). \tag{2.25}$$

Corresponding to the three rotations given by equations (2.7)– (2.9) we obtain the following three <u>generators of rotations</u>

$$L_x \equiv -X_{\theta_x} = -i\left(y\frac{\partial}{\partial z} - z\frac{\partial}{\partial y}\right), \tag{2.26}$$

$$L_y \equiv -X_{\theta_y} = -i\left(z\frac{\partial}{\partial x} - x\frac{\partial}{\partial z}\right), \tag{2.27}$$

$$L_z \equiv -X_{\theta_z} = -i\left(x\frac{\partial}{\partial y} - y\frac{\partial}{\partial x}\right). \tag{2.28}$$

The three boosts given by equations (2.10)– (2.12) yield the generators of Lorentz boosts

$$K_x \equiv X_{\varphi_x} = i\left(t\frac{\partial}{\partial x} + x\frac{\partial}{\partial t}\right), \tag{2.29}$$

$$K_y \equiv X_{\varphi_y} = i\left(t\frac{\partial}{\partial y} + y\frac{\partial}{\partial t}\right), \tag{2.30}$$

$$K_z \equiv X_{\varphi_z} = i\left(t\frac{\partial}{\partial z} + z\frac{\partial}{\partial t}\right), \tag{2.31}$$

Finally the translations given by Eq. (2.15) are produced by the four[5] generators of translations

$$P_\mu \equiv X_{a^\mu} = i\frac{\partial}{\partial x^\mu}. \tag{2.32}$$

It should be explicitly noted that the rotations, boosts and the translations under consideration here are globally constant.

---

5 At present there is nothing quantum mechanical about these generators. Subsequently, we will identify $\hbar\vec{L}$ and $-\hbar\vec{P}$ as the orbital angular momentum and linear momentum operators respectively in the $|x\rangle$ basis. Note: the linear momentum $\vec{P}$ is the spacial part of $P^\mu = \eta^{\mu\nu}P_\nu$.

If we introduce

$$L_{12} = L_z = -L_{21}, \quad L_{31} = L_y = -L_{13}, \quad L_{23} = L_x = -L_{32}, \quad L_{ij} = \epsilon^{ijk} L_k, \quad (2.33)$$

$$L_{i0} = -L_{0i} = -K_i, \qquad (i = 1, 2, 3), \quad (2.34)$$

then the effect of the infinitesimal transformations (2.24) can be summarised by the expression

$$x'^{\sigma} = \left[ 1 + \frac{i}{2} \lambda^{\mu\nu} L_{\mu\nu} - i a^{\mu} P_{\mu} \right] x^{\sigma}. \quad (2.35)$$

For instance the effect of a infinitesimal rotation about the z–axis is

$$\begin{aligned} t' &= \left[ 1 + \frac{i}{2} \left( \lambda^{12} L_{12} + \lambda^{21} L_{21} \right) \right] \; t = \left[ 1 + i \lambda^{12} L_{12} \right] \; t \\ &= t + i\theta_z \left[ -i \left( x \frac{\partial}{\partial y} - y \frac{\partial}{\partial x} \right) \right] \; t = t, \end{aligned} \quad (2.36)$$

$$x' = \left[ 1 + i \lambda^{12} L_{12} \right] x = x + i\theta_z \left[ -i \left( x \frac{\partial}{\partial y} - y \frac{\partial}{\partial x} \right) \right] \; x = x - \theta_z y, \quad (2.37)$$

$$y' = \left[ 1 + i \lambda^{12} L_{12} \right] y = y + \theta_z x, \quad (2.38)$$

$$z' = \left[ 1 + i \lambda^{12} L_{12} \right] z = z. \quad (2.39)$$

The effect of a finite rotation (again about the z–axis.), say on x, is

$$x' = \lim_{N\to\infty} \left(1 + i\frac{\theta_z}{N} L_z\right)^N x = [\exp(i\theta_z L_z)]\ x, \tag{2.40}$$

Consequently, the finite transformations (2.16), generated by any *one*[6] of the Poincaré generators, have the form

$$x'^{\sigma} = \left[\exp\left(\frac{i}{2}\lambda^{\mu\nu}L_{\mu\nu} - ia^{\mu}P_{\mu}\right)\right] x^{\sigma}. \tag{2.41}$$

Before we embark on the next logical question, we pause to collect the commutation relations between the various generators of the Poincaré transformations. The commutation relations read

$$[L_{\mu\nu}, L_{\rho\sigma}] = i(\eta_{\nu\rho}L_{\mu\sigma} - \eta_{\mu\rho}L_{\nu\sigma} + \eta_{\mu\sigma}L_{\nu\rho} - \eta_{\nu\sigma}L_{\mu\rho}), \tag{2.42}$$

$$[P_{\mu}, L_{\rho\sigma}] = i(\eta_{\mu\rho}P_{\sigma} - \eta_{\mu\sigma}P_{\rho}), \tag{2.43}$$

$$[P_{\mu}, P_{\nu}] = 0. \tag{2.44}$$

*Definitions:* A group formed by *continuous* transformations is called a Lie group. The set of commutators $\{\ [X_\alpha, X_\beta]\ \}$ associated with the generators are said to constitute the *algebra* associated with the Lie group. If all the commutators associated with the generators commute the group is said to be *Abelian.* If at least some of the commutators are non-zero, the group is called *non–Abelian.* As such, in the literature one often refers to commutators (2.42)–(2.44) as the Lie algebra associated with the Poincaré group.

6 We emphasise "any *one*", because generators corresponding to two different Poincaré transformations do not commute in general.

## 2.4 Poincaré Transformations and Quantum Mechanical States

Recalling the definition of Poincaré covariance:

> The form of the underlying laws of Nature, as determined by an inertial observer, which determine the nature of any observable phenomenon remain *unchanged* under a Poincaré transformation,

we ask the question:

> What constraints does the requirement of Poincaré covariance impose on quantum mechanical states and physical observables?

To study this question let us *experimentally* prepare a system[7] in a state $|state\rangle$. Let the *same* system now be observed by another inertial observer characterised by $\{\Lambda, a\}$. Denote the state as observed by this new observer by $|state\rangle'$. In order that $|state\rangle$ and $|state\rangle'$ be physically acceptable states, they *must* transform as

$$|state\rangle' = U(\{\Lambda, a\})\ |state\rangle, \tag{2.45}$$

where $U(\{\Lambda, a\})$ is an unitary operator constrained to satisfy:

$$U(\{\overline{\Lambda}, \overline{a}\})U(\{\Lambda, a\}) = U(\{\overline{\Lambda}\Lambda, \overline{\Lambda}a + \overline{a}\}). \tag{2.46}$$

This constraint that $U(\{\Lambda, a\})$ furnish a representation of the Poincaré group is required in order that a Poincaré transformation $\{\Lambda, a\}$ followed by $\{\overline{\Lambda}, \overline{a}\}$ has the same effect as the Poincaré transformation $\{\overline{\Lambda}, \overline{a}\}\{\Lambda, a\}$. Strictly speaking (2.46) is true for infinitesimal transformations. The finite Poincaré transformations which are constructed by successive application of infinitesimal transformations will occasionally have a minus sign on the *r.h.s* of (2.46). The representation is then said to be a *representation up to a sign.* This situation will arise when considering half–integral (spinor) representations. In such situations spinor fields

7 It will be seen in the Appendix that single particle free states of a massive particle are specified by specifying the four momentum, spin, and its projection along an observer chosen axis. With a similar specification, involving "helicity," for massless particles must be so combined as to yield observables which are even functions of spinor fields. For further comments on this point the reader is referred to Sec. 2.12 of Ref. [41].

The linearity of the unitary operator $U(\{\Lambda, a\})$ implies that for an infinitesimal Poincaré transformation

$$\{\Lambda, a\} = \{1 + \lambda, \epsilon\}, \tag{2.47}$$

$U(\{\Lambda, a\})$ has the form

$$U(\{\Lambda, a\}) = 1 - \frac{i}{2}\lambda^{\mu\nu}\Omega_{\mu\nu} + i\epsilon^{\mu}\Omega_{\mu}. \tag{2.48}$$

The factors of $-\frac{i}{2}$ and $i$ are purely for reasons of historical convention, and $\Omega_{\mu\nu}$ and $\Omega_\mu$ are yet undetermined generators. Determining the effect of a Poincaré transformation $\{\Lambda, a\}$ on physical states, therefore, involves explicit determination of these unknown generators. For determining the generators of Poincaré transformations we began with the known transformations, and formally determined the generators which induce those transformations. However, if we are given the generators of the Poincaré transformations, we could from them construct the transformation which the spacetime coordinates undergo under a Poincaré transformation. We could even go further, and claim that we could have obtained the generators as *one of the* representations of the algebra satisfied by these generators, if the Lie algebra associated with the Poincaré group was given *a priori.*

Taking this philosophical point of view we now proceed to find the algebra satisfied by $\Omega_{\mu\nu}$ and $\Omega_\mu$. Towards this end we follow Weinberg [34] and consider the transformation *of* $U(\{1+\lambda, \epsilon\})$ *by* an arbitrary $U(\{\Lambda, a\})$

$$\begin{aligned} &U(\{\Lambda, a\})\; U(\{1+\lambda, \epsilon\})\; U^{-1}(\{\Lambda, a\}) \\ &\qquad\qquad [Using\ (2.46)\ and\ U^{-1}(\{\Lambda, a\}) = U(\{\Lambda, a\}^{-1})] \\ &\quad = U(\{\Lambda(1+\lambda), \Lambda\epsilon + a\})\; U(\{\Lambda^{-1}, -\Lambda^{-1}a\}) \\ &\qquad\qquad [Using\ (2.46),\ again] \\ &\quad = U(\{1 + \Lambda\lambda\Lambda^{-1}, -a - \Lambda\lambda\Lambda^{-1}a + \Lambda\epsilon + a\}), \\ &\quad = U(\{1 + \Lambda\lambda\Lambda^{-1}, -\Lambda\lambda\Lambda^{-1} + \Lambda\epsilon\}), \end{aligned} \tag{2.49}$$

Using (2.48) in the *r.h.s* of the above expression we obtain

$$\begin{aligned} &U(\{\Lambda, a\})\ U(\{1+\lambda,\epsilon\})\ U^{-1}(\{\lambda, a\}) \\ &\quad = 1 - \frac{i}{2}(\Lambda\lambda\Lambda^{-1})^{\mu\nu}\Omega_{\mu\nu} + i(-\Lambda\lambda\Lambda^{-1}a + \Lambda\epsilon)^{\mu}\Omega_{\mu} \\ &\quad = 1 - \frac{i}{2}\Lambda^{\mu}{}_{\rho}\lambda^{\rho\sigma}(\Lambda^{-1})_{\sigma}{}^{\nu}\Omega_{\mu\nu} + i\left[-\Lambda^{\mu}{}_{\rho}\lambda^{\rho\sigma}(\Lambda^{-1})_{\sigma}{}^{\nu}a_{\nu}\Omega_{\mu} + \Lambda^{\mu}{}_{\rho}\epsilon^{\rho}\Omega_{\mu}\right]. \end{aligned} \tag{2.50}$$

But $(\Lambda^{-1})_{\sigma}{}^{\nu} = \Lambda^{\nu}{}_{\sigma}$ according to (2.20) which relates the inverse of $\Lambda$, $\Lambda^{-1}$, to $\Lambda$. Therefore

$$\begin{aligned} &U(\{\Lambda, a\})\ U(\{1+\lambda,\epsilon\})\ U^{-1}(\{\lambda, a\}) \\ &\qquad = 1 - \frac{i}{2}\Lambda^{\mu}{}_{\rho}\lambda^{\rho\sigma}\Lambda^{\nu}{}_{\sigma}\Omega_{\mu\nu} - i\Lambda^{\mu}{}_{\rho}\lambda^{\rho\sigma}\Lambda^{\nu}{}_{\sigma}a_{\nu}\Omega_{\mu} + i\Lambda^{\mu}{}_{\rho}\epsilon^{\rho}\Omega_{\mu}. \end{aligned} \tag{2.51}$$

On the *l.h.s* of the above expression substitute for $U(\{1+\lambda,\epsilon\})$ from (2.48) to get

$$\begin{aligned} &U(\{\Lambda, a\})\ U(\{1+\lambda,\epsilon\})\ U^{-1}(\{\lambda, a\}) \\ &= U(\{\Lambda, a\})\ \left[1 - \frac{i}{2}\lambda^{\mu\nu}\Omega_{\mu\nu} + i\epsilon^{\mu}\Omega_{\mu}\right]\ U^{-1}(\{\lambda, a\}) \\ &= 1 - \frac{i}{2}\lambda^{\mu\nu}\ U(\{\Lambda, a\})\ \Omega_{\mu\nu}\ U^{-1}(\{\Lambda, a\}) + i\epsilon^{\mu}\ U(\{\Lambda, a\})\ \Omega_{\mu}\ U^{-1}(\{\Lambda, a\}). \end{aligned} \tag{2.52}$$

Going back to (2.51), interchange $\mu \leftrightarrow \rho$ *and* $\nu \leftrightarrow \sigma$ on the *r.h.s* to obtain

$$\begin{aligned} &U(\{\Lambda, a\})\ U(\{1+\lambda,\epsilon\})\ U^{-1}(\{\lambda, a\}) \\ &= 1 - \frac{i}{2}\Lambda^{\rho}{}_{\mu}\lambda^{\mu\nu}\Lambda^{\sigma}{}_{\nu}\Omega_{\sigma\rho} - i\Lambda^{\rho}{}_{\mu}\lambda^{\mu\nu}\Lambda^{\sigma}{}_{\nu}a_{\sigma}\Omega_{\rho} + i\Lambda^{\rho}{}_{\mu}\epsilon^{\mu}\Omega_{\rho} \\ &= 1 - \frac{i}{2}\lambda^{\mu\nu}\Lambda^{\rho}{}_{\mu}\Lambda^{\sigma}{}_{\nu}\Omega_{\rho\sigma} \\ &\quad - i\lambda^{\mu\nu}\Lambda^{\rho}{}_{\mu}\Lambda^{\sigma}{}_{\nu}\left[\frac{1}{2}(a_{\sigma}\Omega_{\rho} - a_{\rho}\Omega_{\sigma}) + \frac{1}{2}(a_{\sigma}\Omega_{\rho} + a_{\rho}\Omega_{\sigma})\right] + i\epsilon^{\mu}\Lambda^{\rho}{}_{\mu}\Omega_{\rho}, \end{aligned} \tag{2.53}$$

where $a_{\sigma}\Omega_{\rho}$ is broken into its antisymmetric and symmetric parts. Now note that if $Q_{\sigma\rho}$ is symmetric then $\Lambda^{\rho}{}_{\mu}\Lambda^{\sigma}{}_{\nu}Q_{\sigma\rho}$ is symmetric in the indices $\mu, \nu$. The proof goes as follows. $\Lambda^{\rho}{}_{\mu}\Lambda^{\sigma}{}_{\nu}Q_{\sigma\rho} =$ [by symmetry of $Q$] $\Lambda^{\rho}{}_{\mu}\Lambda^{\sigma}{}_{\nu}Q_{\rho\sigma} =$ [renaming the summed indices] $\Lambda^\sigma{}_\mu \Lambda^\rho{}_\nu Q_{\sigma\rho}$ = [by rearranging] $\Lambda^\rho{}_\nu \Lambda^\sigma{}_\mu Q_{\sigma\rho}$. Exploiting the antisymmetry of $\lambda^{\mu\nu}$ we get

$$\begin{aligned} &U(\{\Lambda, a\})\; U(\{1+\lambda, \epsilon\})\; U^{-1}(\{\lambda, a\}) \\ &\qquad = 1 - \frac{i}{2}\lambda^{\mu\nu}\Lambda^\rho{}_\mu \Lambda^\sigma{}_\nu(\Omega_{\sigma\rho} + a_\sigma \Omega_\rho - a_\rho \Omega_\sigma) + i\epsilon^\mu \Lambda^\rho{}_\mu \Omega_\rho, \end{aligned} \tag{2.54}$$

Comparison of (2.54) and (2.52) yields the *Poincaré transformation properties of* $\Omega_{\mu\nu}$ *and* $\Omega_\mu$

$$U(\{\Lambda, a\})\; \Omega_{\mu\nu}\; U^{-1}(\{\Lambda, a\}) = \Lambda^\rho{}_\mu \Lambda^\sigma{}_\nu(\Omega_{\sigma\rho} + a_\sigma \Omega_\rho - a_\rho \Omega_\sigma), \tag{2.55}$$

and

$$U(\{\Lambda, a\})\; \Omega_\mu\; U^{-1}(\{\Lambda, a\}) = \Lambda^\rho{}_\mu \Omega_\rho. \tag{2.56}$$

It is immediately observed, from the absence of $a^\mu$ in the *r.h.s* of (2.56), that while $\Omega_\mu$ *is* translationally invariant $\Omega_{\mu\nu}$ is *not.*

The algebra satisfied by these generators can now be obtained by setting $\{\Lambda, a\} = \{1+\lambda, \epsilon\}$, where $\lambda$ and $\epsilon$ are a *new* set of infinitesimals, in the Poincaré transformation properties of $\Omega_{\mu\nu}$ and $\Omega_\mu$ given by (2.55) and (2.56). To implement this calculation we need to know the expansion of $U^{-1}(\{\Lambda, a\})$ to order $\mathcal{O}(\lambda, \epsilon)$

$$\begin{aligned} U^{-1}(\{\Lambda, a\}) &= U(\{(1+\lambda), \epsilon\}^{-1}) \\ &= U(\{(1+\lambda)^{-1}, -(1+\lambda)^{-1}\epsilon\}) \\ &= U(\{1-\lambda, -(1-\lambda)\epsilon\}) \\ &= 1 - \frac{i}{2}(-\lambda)^{\gamma\delta}\Omega_{\gamma\delta} + i[-(1-\lambda)\epsilon]^\eta \Omega_\eta \\ &= 1 + \frac{i}{2}\lambda^{\gamma\delta}\Omega_{\gamma\delta} - i\epsilon^\eta \Omega_\eta. \end{aligned} \tag{2.57}$$

We first consider (2.55), which gives the Poincaré transformation property of $\Omega_{\mu\nu}$, and substitute for $U(\{\Lambda, a\})$ for an infinitesimal transformation given by

(2.48), and use (2.57) for $U^{-1}(\{\Lambda, a\})$, in the *l.h.s (*of (2.55))

$$[1-\frac{i}{2}\lambda^{\alpha\beta}\Omega_{\alpha\beta}+i\epsilon^{\kappa}\Omega_{\kappa}]\Omega_{\mu\nu}[1+\frac{i}{2}\lambda^{\gamma\delta}\Omega_{\gamma\delta}-i\epsilon^{\eta}\Omega_{\eta}] \\ =(1+\lambda)^{\rho}{}_{\mu}(1+\lambda)^{\sigma}{}_{\nu}(\Omega_{\rho\sigma}+\epsilon_{\sigma}\Omega_{\rho}-\epsilon_{\rho}\Omega_{\sigma}). \tag{2.58}$$

To order $\mathcal{O}(\lambda, \epsilon)$,

$$L.H.S\ of\ (2.58)=\Omega_{\mu\nu}+\frac{i}{2}\lambda^{\rho\sigma}\left(\Omega_{\mu\nu}\Omega_{\rho\sigma}-\Omega_{\rho\sigma}\Omega_{\mu\nu}\right)+i\epsilon^{\theta}\left(\Omega_{\theta}\Omega_{\mu\nu}-\Omega_{\mu\nu}\Omega_{\theta}\right). \tag{2.59}$$

Similarly to order $\mathcal{O}(\lambda, \epsilon)$,

$$\begin{aligned} &R.H.S\ of\ (2.58) \\ &=\Omega_{\mu\nu}+\lambda^{\rho\sigma}\left(\eta_{\mu\sigma}J_{\rho\nu}-\eta_{\nu\rho}\Omega_{\mu\sigma}\right)+\epsilon^{\theta}\left(\eta_{\nu\theta}\Omega_{\mu}-\eta_{\mu\theta}\Omega_{\nu}\right) \\ &[\text{where have freely raised and lowered indices and used } \lambda^{\sigma\rho}=-\lambda^{\rho\sigma}] \\ &=\Omega_{\mu\nu}+\lambda^{\rho\sigma}\left[\frac{1}{2}\left(\eta_{\mu\sigma}\Omega_{\rho\nu}-\eta_{\mu\rho}\Omega_{\sigma\nu}\right)+\frac{1}{2}\left(\eta_{\mu\sigma}\Omega_{\rho\nu}+\eta_{\mu\rho}\Omega_{\sigma\nu}\right)\right] \\ &\qquad -\lambda^{\rho\sigma}\left[\frac{1}{2}\left(\eta_{\nu\rho}\Omega_{\mu\sigma}-\eta_{\nu\sigma}\Omega_{\mu\rho}\right)+\frac{1}{2}\left(\eta_{\nu\rho}\Omega_{\mu\sigma}+\eta_{\nu\sigma}\Omega_{\mu\rho}\right)\right] \\ &\qquad +\epsilon^{\theta}\left[\eta_{\nu\theta}\Omega_{\mu}-\eta_{\mu\theta}\Omega_{\nu}\right] \\ &=\Omega_{\mu\nu}+\frac{1}{2}\lambda^{\rho\sigma}\left[\eta_{\mu\sigma}\Omega_{\rho\nu}-\eta_{\mu\rho}\Omega_{\sigma\nu}-\eta_{\nu\rho}\Omega_{\mu\sigma}+\eta_{\nu\sigma}\Omega_{\mu\rho}\right] \\ &+\epsilon^{\theta}\left[\eta_{\nu\theta}\Omega_{\mu}-\eta_{\mu\theta}\Omega_{\nu}\right], \end{aligned} \tag{2.60}$$

where we have again used the antisymmetry of $\lambda^{\rho\sigma}$.

By comparison of (2.59) and (2.60) we arrive at the following results [we have rearranged certain terms and made replacements like $\Omega_{\sigma\nu} \to -\Omega_{\nu\sigma}$. Note such replacements are allowed without loss of generality because of the antisymmetry of $\lambda^{\mu\nu}$ (see Table I) which allows only the antisymmetric part of $\Omega_{\mu\nu}$ to contribute to $U(\{\Lambda, a\})$. ]

$$[\Omega_{\mu\nu}, \Omega_{\rho\sigma}]=i(\eta_{\nu\rho}\Omega_{\mu\sigma}-\eta_{\mu\rho}\Omega_{\nu\sigma}+\eta_{\mu\sigma}\Omega_{\nu\rho}-\eta_{\nu\sigma}\Omega_{\mu\rho}), \tag{2.61}$$

and

$$[\Omega_{\theta}, \Omega_{\mu\nu}]=i(\eta_{\theta\mu}\Omega_{\nu}-\eta_{\nu\theta}\Omega_{\mu}). \tag{2.62}$$

When the same exercise is carried out with (2.56), we reproduce (2.62) and complete the Lie algebra associated with the generators of $U(\{\Lambda, a\})$ by obtaining

$$[\Omega_\mu, \Omega_\nu] = 0. \tag{2.63}$$

It is immediately observed that the algebra associated with the generators of the continuous spacetime transformations $\{\Lambda, a\}$, given by (2.42)–(2.44), is identical with the algebra (2.61)–(2.63) associated with the generators of the Lie group formed by $U(\{\Lambda, a\})$. As a result we make the following customary identifications

$$\Omega_{\mu\nu} \equiv J_{\mu\nu}, \quad \Omega_\mu \equiv P_\mu. \tag{2.64}$$

Consequently

$$[J_{\mu\nu}, J_{\rho\sigma}] = i(\eta_{\nu\rho} J_{\mu\sigma} - \eta_{\mu\rho} J_{\nu\sigma} + \eta_{\mu\sigma} J_{\nu\rho} - \eta_{\nu\sigma} J_{\mu\rho}), \tag{2.65}$$

$$[P_\mu, J_{\rho\sigma}] = i(\eta_{\mu\rho} P_\sigma - \eta_{\mu\sigma} P_\rho), \tag{2.66}$$

$$[P_\mu, P_\nu] = 0. \tag{2.67}$$

The same commutation relations are displayed in a more visually accessible form in Table II.

TABLE II

Lie algebra associated with the generators of the *infinite* dimensional representations of the Poincaré group [Generator in the first vertical column , Generator in the first horizontal row]=Entry at the intersection. For example: $[J_x, P_y] = iP_z$.

| | $K_x$ | $K_y$ | $K_z$ | $J_x$ | $J_y$ | $J_z$ | $P_0$ | $P_x$ | $P_y$ | $P_z$ |
|---|---|---|---|---|---|---|---|---|---|---|
| $K_x$ | 0 | $-iJ_z$ | $-iJ_y$ | 0 | $iK_z$ | $-iK_y$ | $-iP_x$ | $-iP_0$ | 0 | 0 |
| $K_y$ | $iJ_z$ | 0 | $-iJ_x$ | $-iK_z$ | 0 | $iK_x$ | $-iP_y$ | 0 | $-iP_0$ | 0 |
| $K_z$ | $iJ_y$ | $iJ_x$ | 0 | $iK_y$ | $-iK_x$ | 0 | $-iP_z$ | 0 | 0 | $-iP_0$ |
| $J_x$ | 0 | $iK_z$ | $-iK_y$ | 0 | $iJ_z$ | $-iJ_y$ | 0 | 0 | $iP_z$ | $-iP_y$ |
| $J_y$ | $-iK_z$ | 0 | $iK_x$ | $-iJ_z$ | 0 | $iJ_x$ | 0 | $-iP_z$ | 0 | $iP_x$ |
| $J_z$ | $iK_y$ | $-iK_x$ | 0 | $iJ_y$ | $-iJ_x$ | 0 | 0 | $iP_y$ | $-iP_x$ | 0 |
| $P_0$ | $iP_x$ | $iP_y$ | $iP_z$ | 0 | 0 | 0 | 0 | 0 | 0 | 0 |
| $P_x$ | $iP_0$ | 0 | 0 | 0 | $iP_z$ | $-iP_y$ | 0 | 0 | 0 | 0 |
| $P_y$ | 0 | $iP_0$ | 0 | $-iP_z$ | 0 | $iP_x$ | 0 | 0 | 0 | 0 |
| $P_z$ | 0 | 0 | $iP_0$ | $iP_y$ | $-iP_x$ | 0 | 0 | 0 | 0 | 0 |

The fact that the algebra given by (2.65)–(2.67) coincides with the algebra associated with the Poincaré group should not lead to the inference that $L_{\mu\nu}$ is necessarily identical to $J_{\mu\nu}$. All that is required is that both $L_{\mu\nu}$ and $J_{\mu\nu}$ satisfy the same algebra. Even the the $P_\mu$ appearing in (2.64) need not coincide with the generators of spacetime translations. Nevertheless, as is customary, we will concentrate on the algebraic aspect and will not explicitly differentiate between these distinctions notationally. The distinction will be obvious, and emphasised where necessary, from the physical context.

This establishes the connection of the Poincaré transformations $\{\Lambda, a\}$ in the ordinary spacetime and their effect in the Hilbert space of quantum mechanical systems determined by $U(\{\Lambda, a\})$. For a *finite* Poincaré transformation we have

$$U(\{\Lambda, a\}) = \exp\left[-\frac{i}{2}\lambda^{\mu\nu}J_{\mu\nu} + i\epsilon^{\mu}P_{\mu}\right]. \tag{2.68}$$

We asked the question: What constraints does the requirement of Poincaré covariance impose on quantum mechanical states and physical observables? The partial answer to the questions is that *physically acceptable quantum states* must transform under a Poincaré transformation $\{\Lambda, a\}$ as

$$|state\rangle' = \exp\left[-\frac{i}{2}\lambda^{\mu\nu}J_{\mu\nu} + i\epsilon^{\mu}P_{\mu}\right]|state\rangle. \tag{2.69}$$

While Eqs. (2.55) and (2.56) provide the Poincaré transformation properties of the generators, $J_{\mu\nu}$ and $P_{\mu}$; Eqs. (2.65)–(2.67) give the Lie algebra associated with these transformations. In the Appendix we will find what invariants specify physical states, $|state\rangle$'s.

# 3. $(j,0) \oplus (0,j)$ COVARIANT SPINORS

## 3.1 Boost for $(j,0) \oplus (0,j)$ Covariant Spinors

We begin with some useful definitions. Referring to Table II we note that the generators $\{\vec{J}, \vec{K}\}$ form a closed algebra. The Lie group generated by the generators $\{\vec{J}, \vec{K}\}$ is called[8] the "Lorentz group" and corresponds to the transformations (2.16) with $a^\mu$ set identically equal to zero. The term "representation" means a group of linear operators which is homomorphic to the group to be represented. The space of vectors on which these operators act is a complex vector space, and is called the "representation space" (cf. Ref. [42]). An operator, constructed out of the generators of a group, which commutes with *all* generators of the Lie group is called a "Casimir operator" associated with the group. Eigenvalues of the Casimir operators are called "Casimir invariants."

As argued by Weinberg [21], and discussed in the Appendix here, we will see that (with one exception, the scalar field) if one wishes to arrive at the particle interpretation within the framework of Poincaré covariant theory of quantum systems, one is forced to incorporate necessarily non–unitary finite dimensional representations of the Lorentz group. Since only unitary transformations of physical states allow for a probabilistic interpretation, the representation spaces of finite dimensional representations of the Lorentz group cannot be spanned by "physical states" defined via (2.69). The objects which span the finite dimensional representation spaces are called "matter fields," or just "fields." This last comment should not lead anyone to conclude that matter fields (Dirac spinors being one example) do not play a significant physical role in the description of quantum systems. In fact much of this work is devoted to the study of these fields, and extends Dirac's original work on spin one half particles to any spin. Because of historical reasons matter fields are also known as "covariant spinors."

The set of generators $\{\vec{J}, \vec{K}\}$ span a linear vector space with $\vec{J}$ and $\vec{K}$ as the basis vectors. The vector space *of the* generators should not be confused with

8 Strictly speaking this is the "proper Lorentz group" because the generators $\{\vec{J}, \vec{K}\}$ refer to the *continuous* spacetime transformations.

the vector space ($\equiv$ representation space) *on which* the generators act. Since the Lorentz group is "non-compact"[9]; all its finite dimensional representations are non–unitary according to a theorem in mathematics. To construct these finite dimensional representations, we explicitly note the Lie algebra associated with the Lorentz group

$$[K_i, J_i] = 0, \quad i = x, y, z \tag{3.1}$$

$$[J_x, J_y] = iJ_z, \quad [K_x, K_y] = -iJ_z, \quad [J_x, K_y] = iK_z, \quad [K_x, J_y] = iK_z, \tag{3.2}$$

and "cyclic permutations". Next we implement the standard rotation[10] by introducing a new basis:

$$\vec{S}_R = \frac{1}{2}(\vec{J} + i\vec{K}), \quad \vec{S}_L = \frac{1}{2}(\vec{J} - i\vec{K}). \tag{3.3}$$

The algebra associated with these generators reads:

$$[(S_R)_i, (S_L)_j] = 0, \quad i, j = x, y, z. \tag{3.4}$$

$$[(S_R)_x, (S_R)_y] = i(S_R)_z, \quad [(S_L)_x, (S_L)_y] = i(S_L)_z, \tag{3.5}$$

and "cyclic permutations."

As a result we see that each of $\vec{S}_R$ and $\vec{S}_L$ satisfies the algebra of a $SU(2)$ group. The finite dimensional irreducible representations of the Lorentz group are thus direct products of those for the sub-algebras $SU(2)_R$ and $SU(2)_L$. The $(2j_r+1)(2j_l+1)$ irreducible representations of $SU(2)_R \otimes SU(2)_L$ are labelled by two numbers $(j_r, j_l)$, $\quad j_r(j_r+1)$ and $j_l(j_l+1)$ being the eigenvalues of the two Casimir operators $(\vec{S}_R)^2$ and $(\vec{S}_L)^2$. The basis of $(2j_r+1)(2j_l+1)$ dimensional representation space containing the relativistic covariant spinors can be written as

$$\phi_{j_r,\sigma_r} \otimes \phi_{j_l,\sigma_l}, \tag{3.6}$$

where:

$$\begin{gathered}(\vec{S}_R)^2\ \phi_{j_r,\sigma_r} = j_r(j_r+1)\ \phi_{j_r,\sigma_r}, \quad (S_R)_z\ \phi_{j_r,\sigma_r} = \sigma_r\ \phi_{j_r,\sigma_r} \\ \sigma_r = j_r, j_r-1, j_r-2, \ldots, -j_r+1, -j_r \quad .\end{gathered} \tag{3.7}$$

$$\begin{gathered}(\vec{S}_L)^2\ \phi_{j_l,\sigma_l} = j_l(j_l+1)\ \phi_{j_l,\sigma_l}, \quad (S_L)_z\ \phi_{j_l,\sigma_l} = \sigma_l\ \phi_{j_l,\sigma_l} \\ \sigma_l = j_l, j_l-1, j_l-2, \ldots, -j_l+1, -j_l \quad .\end{gathered} \tag{3.8}$$

---

9 Ref. [53, p. 43]: "This corresponds roughly to the observation that velocities, which are parameters of Lorentz boosts, take on values along an open line, from $v/c = 0$ to $v/c = 1$, whereas angles of rotation extend from $\theta = 0$ to $\theta = 2\pi$, and these points are *identified*, so the line becomes a closed circle. The group space of the rotation group is finite, but that of the Lorentz group is infinite, so the Lorentz group is non–compact."

10 The introduction of $i \equiv \sqrt{-1}$, in the equations below, is a nontrivial construction, which allows us to construct the *finite* dimensional representations of the Lorentz group. Its introduction is ultimately justified by experimental observation that at least some of these finite dimensional representations are indeed physically realised in nature.

Since under *Parity:* $(j,0) \leftrightarrow (0,j)$, we introduce the $(j,0)\oplus(0,j)$ covariant spinors

$$\psi_{CH}(\vec{p}) = \begin{pmatrix} \phi^R(\vec{p}) \\ \\ \phi^L(\vec{p}) \end{pmatrix} \tag{3.9}$$

where $\phi^R(\vec{p})$ represents functions in the $(j,0)$ representation space, and $\phi^L(\vec{p})$ represents functions in the $(0,j)$ representation space.

Before we proceed further the distinction between the finite dimensional representation of $\{\vec{J},\vec{K}\}$ and infinite dimensional representations of $\{\vec{J},\vec{K}\}$ should be explicitly noted. For the $(j,0)$ representation $\vec{K} = -i\vec{J}$, since by definition for the $(j,0)$ representation $\vec{S}_R = \vec{J}$ and $\vec{S}_L = 0$. Similarly for the $(0,j)$ representation $\vec{K} = +i\vec{J}$. Covariant spinors for the Dirac field are the $(1/2,0)\oplus(0,1/2)$ matter fields. For the $(1/2,0)$ component of the field $\vec{K} = -i\vec{\sigma}/2$ and for the $(0,1/2)$ component $\vec{K} = +i\vec{\sigma}/2$. As such both $\vec{J} \pm i\vec{K}$, i.e. both $\vec{S}_R$ and $\vec{S}_L$, are hermitian. The same remains true for all other $(j,0)\oplus(0,j)$ representations. On the other hand for the infinite dimensional representations (in the $|x\rangle$ space), we have:

$$J_x \equiv -X_{\theta_x} = -i\left(y\frac{\partial}{\partial z} - z\frac{\partial}{\partial y}\right)$$

$$\vdots$$

$$K_z \equiv X_{\phi_z} = i\left(t\frac{\partial}{\partial z} + z\frac{\partial}{\partial t}\right). \tag{3.10}$$

Both $\vec{J}$ and $\vec{K}$, in the infinite dimensional representation, are hermitian. This makes $\vec{J} \pm i\vec{K}$, and hence $\vec{S}_R$ and $\vec{S}_L$, non-hermitian. Besides this observation there is a profound difference between the finite dimensional and the infinite dimensional representations, which is often not fully appreciated in the literature. This has to do with the following simple observation. The finite dimensional $\vec{J}$'s, such as $\vec{J} = \vec{\sigma}/2$ for the Dirac field, refer to the *internal* degrees of freedom, while the $\vec{J}$'s of infinite dimensional representation refer to the external spacetime degrees of freedom (that is they are interpreted as *orbital* angular momentum). Because of this simple fact even though both the finite and infinite dimensional representation $\vec{S}^{R,L}$ and $\vec{J}$ all satisfy the standard $SU(2)$ commutation relations, their commutation properties with the generators of spacetime translations are very different. For the infinite dimensional representations the full commutation relations are summarized in Table III. The same commutation relations for the finite dimensional representations are displayed in Table IV. For the purposes of comparison, the Poincaré algebra given by (2.65)–(2.67) is presented in a similar format in Table V. The difference between finite dimensional $\vec{S}$'s and the infinite dimensional $\vec{S}$'s is apparent. Also the commutation relations of infinite dimensional $\vec{S}$'s and infinite dimensional $\vec{J}$, both of which satisfy the same $SU(2)$ algebra, should be explicitly observed in reference to their commutation relations with $P_\mu$.

TABLE III

Lie algebra associated with the generators of the *infinite* dimensional representations of $SU(2)_R$, $SU(2)_L$ and the generators of the spacetime translations. [Generator in the first vertical column , Generator in the first horizontal row]=Entry at the intersection. For example: $[S_x^R, P_y] = \frac{i}{2}P_z$.

| | $S_x^R$ | $S_y^R$ | $S_z^R$ | $S_x^L$ | $S_y^L$ | $S_z^L$ | $P_0$ | $P_x$ | $P_y$ | $P_z$ |
|---|---|---|---|---|---|---|---|---|---|---|
| $S_x^R$ | 0 | $iS_z^R$ | $-iS_y^R$ | 0 | 0 | 0 | $\frac{1}{2}P_x$ | $\frac{1}{2}P_0$ | $\frac{i}{2}P_z$ | $\frac{-i}{2}P_y$ |
| $S_y^R$ | $-iS_z^R$ | 0 | $iS_x^R$ | 0 | 0 | 0 | $\frac{1}{2}P_y$ | $-\frac{i}{2}P_z$ | $\frac{1}{2}P_0$ | $\frac{i}{2}P_x$ |
| $S_z^R$ | $iS_y^R$ | $-iS_x^R$ | 0 | 0 | 0 | 0 | $\frac{1}{2}P_z$ | $\frac{i}{2}P_y$ | $-\frac{i}{2}P_x$ | $\frac{1}{2}P_0$ |
| $S_x^L$ | 0 | 0 | 0 | 0 | $iS_z^L$ | $-iS_y^L$ | $-\frac{1}{2}P_x$ | $-\frac{1}{2}P_0$ | $\frac{i}{2}P_z$ | $-\frac{i}{2}P_y$ |
| $S_y^L$ | 0 | 0 | 0 | $-iS_z^L$ | 0 | $iS_x^L$ | $-\frac{1}{2}P_y$ | $-\frac{i}{2}P_z$ | $-\frac{1}{2}P_0$ | $\frac{i}{2}P_x$ |
| $S_z^L$ | 0 | 0 | 0 | $iS_y^L$ | $-iS_x^L$ | 0 | $-\frac{1}{2}P_z$ | $\frac{i}{2}P_y$ | $-\frac{i}{2}P_x$ | $-\frac{1}{2}P_0$ |
| $P_0$ | $-\frac{1}{2}P_x$ | $-\frac{1}{2}P_y$ | $-\frac{1}{2}P_z$ | $\frac{1}{2}P_x$ | $\frac{1}{2}P_y$ | $\frac{1}{2}P_z$ | 0 | 0 | 0 | 0 |
| $P_x$ | $-\frac{1}{2}P_0$ | $\frac{i}{2}P_z$ | $-\frac{i}{2}P_y$ | $\frac{1}{2}P_0$ | $\frac{i}{2}P_z$ | $-\frac{i}{2}P_y$ | 0 | 0 | 0 | 0 |
| $P_y$ | $-\frac{i}{2}P_z$ | $-\frac{1}{2}P_0$ | $\frac{i}{2}P_x$ | $-\frac{i}{2}P_z$ | $\frac{1}{2}P_0$ | $\frac{i}{2}P_x$ | 0 | 0 | 0 | 0 |
| $P_z$ | $\frac{i}{2}P_y$ | $-\frac{i}{2}P_x$ | $-\frac{1}{2}P_0$ | $\frac{i}{2}P_y$ | $-\frac{i}{2}P_x$ | $\frac{1}{2}P_0$ | 0 | 0 | 0 | 0 |

TABLE IV

Lie algebra associated with the generators of the *finite* dimensional representations of $SU(2)_R$, $SU(2)_L$ and the generators of the space time translations. [Generator in the first vertical column , Generator in the first horizontal row]=Entry at the intersection. For example: $[S_x^R, P_y] = 0$.

| | $S_x^R$ | $S_y^R$ | $S_z^R$ | $S_x^L$ | $S_y^L$ | $S_z^L$ | $P_0$ | $P_x$ | $P_y$ | $P_z$ |
|---|---|---|---|---|---|---|---|---|---|---|
| $S_x^R$ | 0 | $iS_z^R$ | $-iS_y^R$ | 0 | 0 | 0 | 0 | 0 | 0 | 0 |
| $S_y^R$ | $-iS_z^R$ | 0 | $iS_x^R$ | 0 | 0 | 0 | 0 | 0 | 0 | 0 |
| $S_z^R$ | $iS_y^R$ | $-iS_x^R$ | 0 | 0 | 0 | 0 | 0 | 0 | 0 | 0 |
| $S_x^L$ | 0 | 0 | 0 | 0 | $iS_z^L$ | $-iS_y^L$ | 0 | 0 | 0 | 0 |
| $S_y^L$ | 0 | 0 | 0 | $-iS_z^L$ | 0 | $iS_x^L$ | 0 | 0 | 0 | 0 |
| $S_z^L$ | 0 | 0 | 0 | $iS_y^L$ | $-iS_x^L$ | 0 | 0 | 0 | 0 | 0 |
| $P_0$ | 0 | 0 | 0 | 0 | 0 | 0 | 0 | 0 | 0 | 0 |
| $P_x$ | 0 | 0 | 0 | 0 | 0 | 0 | 0 | 0 | 0 | 0 |
| $P_y$ | 0 | 0 | 0 | 0 | 0 | 0 | 0 | 0 | 0 | 0 |
| $P_z$ | 0 | 0 | 0 | 0 | 0 | 0 | 0 | 0 | 0 | 0 |

TABLE V

Lie algebra associated with the generators of the *infinite* dimensional representations of the Poincaré group [Generator in the first vertical column , Generator in the first horizontal row]=Entry at the intersection. For example: $[J_x, P_y] = iP_z$.

| | $K_x$ | $K_y$ | $K_z$ | $J_x$ | $J_y$ | $J_z$ | $P_0$ | $P_x$ | $P_y$ | $P_z$ |
|---|---|---|---|---|---|---|---|---|---|---|
| $K_x$ | 0 | $-iJ_z$ | $-iJ_y$ | 0 | $iK_z$ | $-iK_y$ | $-iP_x$ | $-iP_0$ | 0 | 0 |
| $K_y$ | $iJ_z$ | 0 | $-iJ_x$ | $-iK_z$ | 0 | $iK_x$ | $-iP_y$ | 0 | $-iP_0$ | 0 |
| $K_z$ | $iJ_y$ | $iJ_x$ | 0 | $iK_y$ | $-iK_x$ | 0 | $-iP_z$ | 0 | 0 | $-iP_0$ |
| $J_x$ | 0 | $iK_z$ | $-iK_y$ | 0 | $iJ_z$ | $-iJ_y$ | 0 | 0 | $iP_z$ | $-iP_y$ |
| $J_y$ | $-iK_z$ | 0 | $iK_x$ | $-iJ_z$ | 0 | $iJ_x$ | 0 | $-iP_z$ | 0 | $iP_x$ |
| $J_z$ | $iK_y$ | $-iK_x$ | 0 | $iJ_y$ | $-iJ_x$ | 0 | 0 | $iP_y$ | $-iP_x$ | 0 |
| $P_0$ | $iP_x$ | $iP_y$ | $iP_z$ | 0 | 0 | 0 | 0 | 0 | 0 | 0 |
| $P_x$ | $iP_0$ | 0 | 0 | 0 | $iP_z$ | $-iP_y$ | 0 | 0 | 0 | 0 |
| $P_y$ | 0 | $iP_0$ | 0 | $-iP_z$ | 0 | $iP_x$ | 0 | 0 | 0 | 0 |
| $P_z$ | 0 | 0 | $iP_0$ | $iP_y$ | $-iP_x$ | 0 | 0 | 0 | 0 | 0 |

Now if we wish to represent say $\Delta^{j=3/2}(1232)$ by a $(3/2,0)\oplus(0,3/2)$ relativistic covariant spinor we implicitly assume that the $j = 3/2$ belongs entirely to the *internal* degrees of freedom (i.e. quark spins) and does not contain in it any orbital angular momentum. Even though, in the present paper, we would treat the "spin" of the resonances as if it were an internal degree of freedom in the above sense, it must be remembered explicitly that this is in, general, only an approximation. It will be an interesting exercise to develop appropriate experimental means and the associated theoretical formalism to decompose the "$j$" of a given[11] resonance into orbital and internal part. For example a reference to Tables IV and V immediately tells us that if we were to first rotate a $|state\rangle$ with an orbital angular momentum

11 To fully explore this problem one needs a to extend the present work to relativistic composite particles.

about the x-axis and then translate it in the y-direction and compare the resulting $|state\rangle$ with the same operations interchanged then the two resulting $|state\rangle$'s need not be identical. On the other hand, if orbital angular momentum is replaced by the internal angular momentum (i.e. arising from the spin) in a similar system the above experiment would yield identical $|state\rangle$'s after performing the indicated operations in two different orders.

Having laid this background, we now proceed with the construction of the $(j,0)\oplus(0,j)$ boost operator. Since we are interested in constructing the boost operator we set $\vec{\theta}=\vec{0}$. Then if we consider the particle under consideration to be at rest in the unprimed frame, a Lorentz boost results in a particle with momentum $\vec{p}$. The boost connecting the $\vec{p}=\vec{0}$ wavefunctions with $\vec{p}\neq 0$ are readily obtained. From a formal point of view the matter fields also transform as the physical $|state\rangle$'s (see Eq. (2.69)), but with one difference. That the $J_{\mu\nu}$ is replaced by its finite dimensional counterpart and the unitary operator $U(\{\Lambda,a\})$ is replaced by the non-unitary $D(\{\Lambda,a\})$ satisfying the same condition of the Poincaré covariant description:

$$D(\{\overline{\Lambda},\overline{a}\})D(\{\Lambda,a\})=D(\{\overline{\Lambda}\Lambda,\overline{\Lambda}a+\overline{a}\}). \tag{3.11}$$

Then consistent with definitions given in Table I we obtain:

$$\phi^{R}(\vec{p})=\exp[\vec{J}\cdot\vec{\varphi}]\;\;\phi^{R}(\vec{0}) \tag{3.12}$$

$$\phi^{L}(\vec{p})=\exp[-\vec{J}\cdot\vec{\varphi}]\;\;\phi^{L}(\vec{0}). \tag{3.13}$$

As a consequence, the "chiral representation"[12] $(j,0)\oplus(0,j)$ relativistic covariant spinors defined by (3.9)transform as:

$$\psi_{CH}(\vec{p})=\begin{pmatrix}\exp(\vec{J}\cdot\vec{\varphi}) & 0\\ & \\ 0 & \exp(-\vec{J}\cdot\vec{\varphi})\end{pmatrix}\psi_{CH}(\vec{0}). \tag{3.14}$$

To make identifications with the historical work for $(1/2,0)\oplus(0,1/2)$ Dirac

12 We call this representation "chiral representation" because for $j=1/2$ the representation coincides with the "chiral" representation of the Dirac spin one half formalism.

covariant spinors it is convenient to introduce a "canonical representation". The connecting matrix A is given by

$$\psi_{CA}(\vec{p}) = A\ \psi_{CH}(\vec{p}), \quad A = \frac{1}{\sqrt{2}} \begin{pmatrix} I & I \\ I & -I \end{pmatrix}. \tag{3.15}$$

Each entry $I$ in the matrix $A$ represents a $(2j+1) \times (2j+1)$ identity matrix, and one is still free[13] to choose any representations for the $J_i$. In the canonical representation covariant spinors are

$$\psi_{CA}(\vec{p}) = \frac{1}{\sqrt{2}} \begin{pmatrix} \phi^{R}(\vec{p}) + \phi^{L}(\vec{p}) \\ \\ \phi^{R}(\vec{p}) - \phi^{L}(\vec{p}) \end{pmatrix}. \tag{3.16}$$

Referring to (3.14), we identify the chiral representation boost matrix as

$$M_{CH}(\vec{p}) = \begin{pmatrix} \exp(\vec{J} \cdot \vec{\varphi}) & 0 \\ \\ 0 & \exp(-\vec{J} \cdot \vec{\varphi}) \end{pmatrix}. \tag{3.17}$$

As a result the boost matrix in the canonical representation reads

$$M_{CA}(\vec{p}) = \begin{pmatrix} \cosh(\vec{J} \cdot \vec{\varphi}) & \sinh(\vec{J} \cdot \vec{\varphi}) \\ \\ \sinh(\vec{J} \cdot \vec{\varphi}) & \cosh(\vec{J} \cdot \vec{\varphi}) \end{pmatrix}. \tag{3.18}$$

If $\vec{J}$ is set equal to $\vec{\sigma}/2$ the boost matrix given by (3.18)coincides with the boost for Dirac spinors in the standard Bjorken and Drell [52] representation. $M_{CA}(\vec{p})$ contains all the essential information needed to construct any $(j,0) \oplus (0,j)$ relativistic covariant spinor. This we now show by an explicit example.

13 We will fix this freedom in the next section by choosing a representation in which $J_z$ is diagonal. This will define the "canonical representation" without ambiguity. If one wishes, one may call the canonical representation with $J_z$ diagonal to be the standard canonical representation, thus leaving the freedom for other "canonical" representations.

## 3.2 $(1,0)\oplus(0,1)$ Covariant Spinors

The representation space of the $(1,0)\oplus(0,1)$ matter fields is a six dimensional internal space whose basis vectors in the canonical representation can be chosen to be

$$u_{+1}(\vec{0}) = \begin{pmatrix} m \\ 0 \\ 0 \\ 0 \\ 0 \\ 0 \end{pmatrix}, \quad u_{0}(\vec{0}) = \begin{pmatrix} 0 \\ m \\ 0 \\ 0 \\ 0 \\ 0 \end{pmatrix}, \quad u_{-1}(\vec{0}) = \begin{pmatrix} 0 \\ 0 \\ m \\ 0 \\ 0 \\ 0 \end{pmatrix},$$

$$v_{+1}(\vec{0}) = \begin{pmatrix} 0 \\ 0 \\ 0 \\ m \\ 0 \\ 0 \end{pmatrix}, \quad v_{0}(\vec{0}) = \begin{pmatrix} 0 \\ 0 \\ 0 \\ 0 \\ m \\ 0 \end{pmatrix}, \quad v_{-1}(\vec{0}) = \begin{pmatrix} 0 \\ 0 \\ 0 \\ 0 \\ 0 \\ m \end{pmatrix}. \tag{3.19}$$

The indicated norm is dictated by the convenience introduced while considering the $m \to 0$ limit. This choice of the basis vectors, and the interpretation attached to them that $u_\sigma(\vec{0})$ represents a particle at rest with the z-component of its spin to be $\sigma$ ($\sigma = 0, \pm 1$) and $v_\sigma(\vec{0})$ represent the antiparticle at rest with the z-component of its spin to be $\sigma$, forces upon us the representation for the angular momentum operators $\vec{J}$. It requires that $J_z$ be diagonal. So in the canonical representation $J_i$ can be written as follows (see ref. [50] for the notational details)

$$J_x = \frac{1}{\sqrt{2}}\begin{pmatrix} 0 & 1 & 0 \\ 1 & 0 & 1 \\ 0 & 1 & 0 \end{pmatrix}, \quad J_y = \frac{1}{\sqrt{2}}\begin{pmatrix} 0 & -i & 0 \\ i & 0 & -i \\ 0 & i & 0 \end{pmatrix}, \quad J_z = \begin{pmatrix} 1 & 0 & 0 \\ 0 & 0 & 0 \\ 0 & 0 & -1 \end{pmatrix}. \tag{3.20}$$

The boost matrix $M_{CA}(\vec{p})$ takes the relativistic covariant spinor of a particle at rest, $\psi_{CA}(\vec{0})$, to $\psi_{CA}(\vec{p})$, the relativistic covariant spinor of the same particle

with momentum $\vec{p}$:

$$\psi_{CA}(\vec{p}) = M_{CA}(\vec{p})\ \psi_{CA}(\vec{0}). \tag{3.21}$$

The $\cosh(\vec{J}\cdot\vec{\varphi})$ which appears in the covariant spinor boost matrix[14] (3.18) can be expanded to yield

$$\cosh(\vec{J}\cdot\vec{\varphi}) = \cosh\left(2\vec{J}\cdot\frac{\vec{\varphi}}{2}\right) = 1 + 2(\vec{J}\cdot\hat{p}\,)(\vec{J}\cdot\hat{p}\,)\sinh^2\left(\frac{\varphi}{2}\right) \tag{3.22}$$

Now we note that

$$\sinh\left(\frac{\varphi}{2}\right) = \left(\frac{E-m}{2m}\right)^{\frac{1}{2}}, \tag{3.23}$$

and

$$\vec{J}\cdot\hat{p} = \frac{1}{|\vec{p}|}\vec{J}\cdot\vec{p} = \frac{1}{(E^2-m^2)^{\frac{1}{2}}}\left(J_x p_x + J_y p_y + J_z p_z\right). \tag{3.24}$$

Substituting for $J_i$ from (3.20) the matrix $\vec{J}\cdot\hat{p}$ reads

$$\vec{J}\cdot\hat{p} = \frac{1}{(E^2-m^2)^{\frac{1}{2}}}\begin{pmatrix} p_z & \frac{1}{\sqrt{2}}(p_x - ip_y) & 0 \\ \frac{1}{\sqrt{2}}(p_x + ip_y) & 0 & \frac{1}{\sqrt{2}}(p_x - ip_y) \\ 0 & \frac{1}{\sqrt{2}}(p_x + ip_y) & -p_z \end{pmatrix}. \tag{3.25}$$

Introducing

$$p_\pm \equiv p_x \pm ip_y, \tag{3.26}$$

and using the just obtained results and identities, we obtain

$$\cosh(\vec{J}\cdot\vec{\varphi}) = 1 + \frac{1}{m(E+m)}\begin{pmatrix} p_z^2 + \frac{1}{2}p_+p_- & \frac{1}{\sqrt{2}}p_zp_- & \frac{1}{2}p_-^2 \\ \frac{1}{\sqrt{2}}p_zp_+ & p_+p_- & -\frac{1}{\sqrt{2}}p_zp_- \\ \frac{1}{2}p_+^2 & -\frac{1}{\sqrt{2}}p_zp_+ & p_z^2 + \frac{1}{2}p_+p_- \end{pmatrix}. \tag{3.27}$$

Similarly $\sinh(\vec{J}\cdot\vec{\varphi})$ which appears in the boost matrix for the covariant spinors

14 See the last section of the Appendix for general expansions of $\cosh(\vec{J}\cdot\vec{\varphi})$ and $\sinh(\vec{J}\cdot\vec{\varphi})$

(3.18) can be expanded as

$$\sinh(\vec{J}\cdot\vec{\varphi}) = \sinh\left(2\vec{J}\cdot\frac{\vec{\varphi}}{2}\right) = 2(\vec{J}\cdot\hat{p})\cosh\left(\frac{\varphi}{2}\right)\sinh\left(\frac{\varphi}{2}\right). \tag{3.28}$$

We have already obtained the explicit expression for $\vec{J}\cdot\hat{p}$, so we only need to note that

$$\cosh\left(\frac{\varphi}{2}\right) = \left(\frac{E+m}{2m}\right)^{\frac{1}{2}}. \tag{3.29}$$

This yields:

$$\sinh(\vec{J}\cdot\vec{\varphi}) = \frac{1}{m}\begin{pmatrix} p_z & \frac{1}{\sqrt{2}}p_- & 0 \\ \frac{1}{\sqrt{2}}p_+ & 0 & \frac{1}{\sqrt{2}}p_- \\ 0 & \frac{1}{\sqrt{2}}p_+ & -p_z \end{pmatrix}. \tag{3.30}$$

Substituting $\sinh(\vec{J}\cdot\vec{\varphi})$ and $\cosh(\vec{J}\cdot\vec{\varphi})$ into (3.18) provides the specific boost for the $(1,0)\oplus(0,1)$ covariant spinors $M_{CA}^{(1,0)\oplus(0,1)}(\vec{p})$. The $(1,0)\oplus(0,1)$ covariant spinors, in the canonical representation, associated with momentum $\vec{p}$, are now immediately calculated by using $M_{CA}^{(1,0)\oplus(0,1)}(\vec{p})$ thus obtained and using (3.21) with $\psi_{CA}(\vec{0}) = u_\sigma(\vec{0}), v_\sigma(\vec{0})$ given by (3.19). The result is:

$$u_{+1}(\vec{p}) = \begin{pmatrix} m + \left[(2p_z^2 \ + \ p_+p_-)/2(E+m)\right] \\ \\ p_z p_+/\sqrt{2}(E+m) \\ \\ p_+^2/2(E+m) \\ \\ p_z \\ \\ p_+/\sqrt{2} \\ \\ 0 \end{pmatrix}, \tag{3.31}$$

$$u_0(\vec{p}) = \begin{pmatrix} p_z p_-/\sqrt{2}(E+m) \\ \\ m + [p_+p_-/(E+m)] \\ \\ -p_z p_+/\sqrt{2}(E+m) \\ \\ p_-/\sqrt{2} \\ \\ 0 \\ \\ p_+/\sqrt{2} \end{pmatrix}, \tag{3.32}$$

$$u_{-1}(\vec{p}) = \begin{pmatrix} p_-^2/2(E+m) \\ \\ -p_z p_-/\sqrt{2}(E+m) \\ \\ m + \left[(2p_z^2 \ + \ p_+ p_-)/2(E+m)\right] \\ \\ 0 \\ \\ p_-/\sqrt{2} \\ \\ -p_z \end{pmatrix}, \tag{3.33}$$

$$v_{+1}(\vec{p}) = \begin{pmatrix} p_z \\ \\ p_+/\sqrt{2} \\ \\ 0 \\ \\ m + \left[(2p_z^2 \ + \ p_+ p_-)/2(E+m)\right] \\ \\ p_z p_+/\sqrt{2}(E+m) \\ \\ p_+^2/2(E+m) \end{pmatrix}, \tag{3.34}$$

$$v_{0}(\vec{p}) = \begin{pmatrix} p_{-}/\sqrt{2} \\ 0 \\ p_{+}/\sqrt{2} \\ p_z p_{-}/\sqrt{2}(E+m) \\ m + [p_{+}p_{-}/(E+m)] \\ -p_z p_{+}/\sqrt{2}(E+m) \end{pmatrix}, \tag{3.35}$$

$$v_{-1}(\vec{p}) = \begin{pmatrix} 0 \\ p_{-}/\sqrt{2} \\ -p_z \\ p_{-}^2/2(E+m) \\ -p_z p_{-}/\sqrt{2}(E+m) \\ m + \left[(2p_z^2 \ + \ p_{+}p_{-})/2(E+m)\right] \end{pmatrix}. \tag{3.36}$$

The reader may wish to verify that this procedure when repeated with $\vec{J} = \vec{\sigma}/2$ reproduces the standard Dirac spinors. All other $(j, 0) \oplus (0, j)$ covariant spinors are obtained by following exactly the same procedure as above and using the appropriate identities given in the last section of the Appendix for the expansion of $\sinh(\vec{J} \cdot \vec{\varphi})$ and $\cosh(\vec{J} \cdot \vec{\varphi})$ which appear in (3.18).

Orthonormality of Relativistic $(1, 0) \oplus (0, 1)$ covariant spinors:

Introducing

$$\overline{u}_\sigma(\vec{p}) = u^\dagger_\sigma(\vec{p}) \gamma^{CA}_{00}, \tag{3.37}$$

where (see Eq. (7.1))

$$\gamma^{CA}_{00} = \begin{pmatrix} I & 0 \\ 0 & -I \end{pmatrix}, \tag{3.38}$$

it is readily verified that

$$\overline{u}_\sigma(\vec{p})\ u_{\sigma'}(\vec{p}) = m^2 \delta_{\sigma\sigma'} \tag{3.39}$$

$$\overline{v}_\sigma(\vec{p})\ v_{\sigma'}(\vec{p}) = -m^2 \delta_{\sigma\sigma'}. \tag{3.40}$$

Ultrarelativistic or Massless Limit:

Since we have chosen to work in a representation in which $J_z$ is diagonal we expect that, for a particle traveling along the $\hat{z}$ axis, only the $u_{\pm 1}(\vec{p})$ and $v_{\pm 1}(\vec{p})$ to survive. The $u_0(\vec{p})$ and $v_0(\vec{p})$ must vanish. In this ultrarelativistic or $m \to 0$ limit: $p^\mu = (E, 0, 0, p = E)$. Substituting $p^\mu = (E, 0, 0, p = E)$ in the $(1, 0) \oplus (0, 1)$ wave functions given by (3.31)–(3.36) we readily see that:

$$\lim_{m \to 0} u_{+1}(\vec{p}) = \begin{pmatrix} E \\ 0 \\ 0 \\ E \\ 0 \\ 0 \end{pmatrix}, \quad \lim_{m \to 0} u_0(\vec{p}) = \begin{pmatrix} 0 \\ 0 \\ 0 \\ 0 \\ 0 \\ 0 \end{pmatrix}, \quad \lim_{m \to 0} u_{-1}(\vec{p}) = \begin{pmatrix} 0 \\ 0 \\ E \\ 0 \\ 0 \\ -E \end{pmatrix}, \tag{3.41}$$

with

$$\lim_{m\to 0} \overline{u}_\sigma(E)\ u_\sigma(E) = 0. \tag{3.42}$$

Similarly,

$$\lim_{m\to 0} v_{+1}(\vec{p}) = \begin{pmatrix} E \\ 0 \\ 0 \\ E \\ 0 \\ 0 \end{pmatrix},\ \lim_{m\to 0} v_0(\vec{p}) = \begin{pmatrix} 0 \\ 0 \\ 0 \\ 0 \\ 0 \\ 0 \end{pmatrix},\ \lim_{m\to 0} u_{-1}(\vec{p}) = \begin{pmatrix} 0 \\ 0 \\ -E \\ 0 \\ 0 \\ E \end{pmatrix}, \tag{3.43}$$

with

$$\lim_{m\to 0} \overline{v}_\sigma(E)\ v_\sigma(E) = 0. \tag{3.44}$$

In the ultrarelativistic limit we thus see that we get only two non–null independent $u_{\pm 1}(E)$ and two non–null independent $v_{\pm 1}(E)$ . The $u_0(E)$ and $v_0(E)$ vanish identically. While $u_{+1}(E)$ and $v_{+1}(E)$ are identical, the $u_{-1}(E)$ $v_{-1}(E)$ differ by a relative phase of $\exp(i\pi)$.

### 3.3 $(3/2, 0) \oplus (0, 3/2)$ Covariant Spinors

For various spin 3/2 baryons [3], such as $\Delta^{\frac{3}{2}}(1232)$, we now introduce $(3/2, 0) \oplus (0, 3/2)$ covariant spinors. As for all $(j, 0) \oplus (0, j)$ covariant spinors, these covariant spinors for spin 3/2 have exactly the right degrees of spinorial and particle/antiparticle degrees of freedom, and have an elegance and general structure first seen in the Dirac spinors.

The canonical representation $(3/2) \oplus (0, 3/2)$ covariant spinors are obtained in a similar fashion as the $(1, 0) \oplus (0, 1)$ covariant spinors obtained in the last section. To obtain the boost

$$M_{CA}^{j=3/2}(\vec{p}) = \begin{pmatrix} \cosh(\vec{J}\cdot\vec{\varphi}) & \sinh(\vec{J}\cdot\vec{\varphi}) \\ & \\ \sinh(\vec{J}\cdot\vec{\varphi}) & \cosh(\vec{J}\cdot\vec{\varphi}) \end{pmatrix}. \tag{3.45}$$

we first note that Eqs. (B.3) and (B.4) for $j = 3/2$ yield:

$$\cosh(2\vec{J}\cdot\vec{\varphi}) = \cosh\varphi\Big[I + \frac{1}{2}\big\{(2\vec{J}\cdot\hat{p})^2 - I\big\}\sinh^2\varphi\Big], \tag{3.46}$$

$$\sinh(2\vec{J}\cdot\vec{\varphi}) = (2\vec{J}\cdot\hat{p})\sinh\varphi\Big[I + \frac{1}{6}\big\{(2\vec{J}\cdot\hat{p})^2 - I\big\}\sinh^2\varphi\Big], \tag{3.47}$$

Next letting

$$\varphi \to \frac{1}{2}\varphi \tag{3.48}$$

and using the identities

$$\sinh\frac{\varphi}{2} = \left(\frac{E-m}{2m}\right)^{1/2} \tag{3.49}$$

$$\cosh\frac{\varphi}{2} = \left(\frac{E+m}{2m}\right)^{1/2} \tag{3.50}$$

$$2\vec{J}\cdot\hat{p} = \frac{2\vec{J}\cdot\vec{p}}{|\vec{p}|} = \frac{2\vec{J}\cdot\vec{p}}{(E^2-m^2)^{1/2}}, \tag{3.51}$$

we obtain the desired expansions for the canonical representation boost. These expansions are:

$$\cosh(\vec{J}\cdot\vec{\varphi}) = \left(\frac{E+m}{2m}\right)^{1/2}\left[I + \frac{1}{2}\left\{\frac{(2\vec{J}\cdot\vec{p})^2}{(E^2-m^2)} - I\right\}\left(\frac{E-m}{2m}\right)\right] \tag{3.52}$$

$$\sinh(\vec{J}\cdot\vec{\varphi}) = \left(\frac{E+m}{2m}\right)^{1/2}\left[\frac{2\vec{J}\cdot\vec{p}}{(E+m)} + \frac{1}{6}\frac{2\vec{J}\cdot\vec{p}}{(E+m)}\left\{\frac{(2\vec{J}\cdot\vec{p})^2}{(E^2-m^2)} - I\right\}\left(\frac{E-m}{2m}\right)\right]. \tag{3.53}$$

These expansions when substituted in Eq. (3.45) provide the boost for the $(3/2,0)\oplus(0,3/2)$ covariant spinors. The boost matrix $M_{CA}^{j=3/2}(\vec{p})$ takes the relativistic covariant spinor of a particle at rest, $\psi_{CA}(\vec{0})$, to $\psi_{CA}(\vec{p})$, the covariant spinor of the same particle with momentum $\vec{p}$:

$$\psi_{CA}^{j=3/2}(\vec{p}) = M_{CA}^{j=3/2}(\vec{p})\ \psi_{CA}^{j=3/2}(\vec{0}). \tag{3.54}$$

In order that in the $m \rightarrow 0$ limit

(i) The rest covariant spinors vanish, and

(ii) The $m \rightarrow 0$ limit covariant spinors have a non–singular norm

we choose the following rest covariant spinors:

$$u_{+\frac{3}{2}}(\vec{0}) = \begin{pmatrix} m^{3/2} \\ 0 \\ 0 \\ 0 \\ 0 \\ 0 \\ 0 \\ 0 \end{pmatrix},\ u_{+\frac{1}{2}}(\vec{0}) = \begin{pmatrix} 0 \\ m^{3/2} \\ 0 \\ 0 \\ 0 \\ 0 \\ 0 \\ 0 \end{pmatrix},\ u_{-\frac{1}{2}}(\vec{0}) = \begin{pmatrix} 0 \\ 0 \\ m^{3/2} \\ 0 \\ 0 \\ 0 \\ 0 \\ 0 \end{pmatrix},$$

$$u_{-\frac{3}{2}}(\vec{0}) = \begin{pmatrix} 0 \\ 0 \\ 0 \\ m^{3/2} \\ 0 \\ 0 \\ 0 \\ 0 \end{pmatrix},$$

$$v_{+\frac{3}{2}}(\vec{0}) = \begin{pmatrix} 0 \\ 0 \\ 0 \\ 0 \\ m^{3/2} \\ 0 \\ 0 \\ 0 \end{pmatrix}, \; v_{+\frac{1}{2}}(\vec{0}) = \begin{pmatrix} 0 \\ 0 \\ 0 \\ 0 \\ 0 \\ m^{3/2} \\ 0 \\ 0 \end{pmatrix}, \; v_{-\frac{1}{2}}(\vec{0}) = \begin{pmatrix} 0 \\ 0 \\ 0 \\ 0 \\ 0 \\ 0 \\ m^{3/2} \\ 0 \end{pmatrix},$$

$$v_{-\frac{3}{2}}(\vec{0}) = \begin{pmatrix} 0 \\ 0 \\ 0 \\ 0 \\ 0 \\ 0 \\ 0 \\ m^{3/2} \end{pmatrix}. \tag{3.55}$$

If the usual interpretation is to be attached to these covariant spinors, then we must choose a representation for the spin 3/2 $J_i$ in which $J_z$ is diagonal. In this representation $\vec{J}$ is:

$$J_x = \frac{1}{2}\begin{pmatrix} 0 & \sqrt{3} & 0 & 0 \\ \sqrt{3} & 0 & 2 & 0 \\ 0 & 2 & 0 & \sqrt{3} \\ 0 & 0 & \sqrt{3} & 0 \end{pmatrix}, J_y = \frac{1}{2}\begin{pmatrix} 0 & -i\sqrt{3} & 0 & 0 \\ i\sqrt{3} & 0 & -2i & 0 \\ 0 & 2i & 0 & -i\sqrt{3} \\ 0 & 0 & i\sqrt{3} & 0 \end{pmatrix},$$

$$J_z = \frac{1}{2}\begin{pmatrix} 3 & 0 & 0 & 0 \\ 0 & 1 & 0 & 0 \\ 0 & 0 & -1 & 0 \\ 0 & 0 & 0 & -3 \end{pmatrix}. \tag{3.56}$$

The rest of the calculation[15] simply involves substituting these $J_i$ into Eqs. (3.52)

15 We have performed this part of the calculation using MACSYMA.

and (3.53), and the resulting block matrices into Eq. (3.45). The resulting boost $M_{CA}^{j=3/2}(\vec{p})$, in the form of a $8\times 8$ matrix, when applied to the basis rest covariant spinors (3.55) yields the $(3/2,0)\oplus(0,3/2)$ covariant spinors. The "particle" covariant spinors are:

$$u_{+\frac{3}{2}}(\vec{p}) = m^{\frac{1}{2}}\left(\frac{E+m}{2m}\right)^{\frac{1}{2}} \times \begin{pmatrix} (9p_z^2+3p_+p_-+5m^2+4Em-E^2)/4(m+E) \\ \sqrt{3}p_+p_z/(m+E) \\ \sqrt{3}p_+^2/2(m+E) \\ 0 \\ p_z(9p_z^2+7p_+p_-+13m^2+12Em-E^2)/4(m+E)^2 \\ \sqrt{3}p_+(13p_z^2+7p_+p_-+13m^2+12Em-E^2)/12(m+E)^2 \\ \sqrt{3}p_+^2p_z/2(m+E)^2 \\ p_+^3/2(m+E)^2 \end{pmatrix}, \tag{3.57}$$

$$u_{+\frac{1}{2}}(\vec{p}) = m^{\frac{1}{2}} \left(\frac{E+m}{2m}\right)^{\frac{1}{2}}$$

$$\times \begin{pmatrix} \sqrt{3}p_- p_z/(m+E) \\ (p_z^2 + 7p_+p_- + 5m^2 + 4Em - E^2)/4(m+E) \\ 0 \\ \sqrt{3}p_+^2/2(m+E) \\ \sqrt{3}p_-(13p_z^2 + 7p_+p_- + 13m^2 + 12Em - E^2)/12(m+E)^2 \\ p_z(p_z^2 + 19p_+p_- + 13m^2 + 12Em - E^2)/12(m+E)^2 \\ p_+(p_z^2 + 10p_+p_- + 13m^2 + 12Em - E^2)/6(m+E)^2 \\ -\sqrt{3}p_+^2 p_z/2(m+E)^2 \end{pmatrix}, \tag{3.58}$$

$$u_{-\frac{1}{2}}(\vec{p}) = m^{\frac{1}{2}} \left(\frac{E+m}{2m}\right)^{\frac{1}{2}} \times \begin{pmatrix} \sqrt{3}p_-^2/2(m+E) \\ 0 \\ (p_z^2 + 7p_+p_- + 5m^2 + 4Em - E^2)/4(m+E) \\ -\sqrt{3}p_+p_z/(m+E) \\ \sqrt{3}p_-^2 p_z/2(m+E)^2 \\ p_-(p_z^2 + 10p_+p_- + 13m^2 + 12Em - E^2)/6(m+E)^2 \\ -p_z(p_z^2 + 19p_+p_- + 13m^2 + 12Em - E^2)/12(m+E)^2 \\ \sqrt{3}p_+(13p_z^2 + 7p_+p_- + 13m^2 + 12Em - E^2)/12(m+E)^2 \end{pmatrix}, \tag{3.59}$$

$$u_{-\frac{3}{2}}(\vec{p}) = m^{\frac{1}{2}} \left(\frac{E+m}{2m}\right)^{\frac{1}{2}}$$

$$\times \begin{pmatrix} 0 \\ \sqrt{3}p_-^2/2(m+E) \\ -\sqrt{3}p_-p_z/(m+E) \\ (9p_z^2 + 3p_+p_- + 5m^2 + 4Em - E^2)/4(m+E) \\ p_-^3/2(m+E)^2 \\ -\sqrt{3}p_-^2 p_z/2(m+E)^2 \\ \sqrt{3}p_-(13p_z^2 + 7p_+p_- + 13m^2 + 12Em - E^2)/12(m+E)^2 \\ -p_z(9p_z^2 + 7p_+p_- + 13m^2 + 12Em - E^2)/4(m+E)^2 \end{pmatrix}. \tag{3.60}$$

Here

$$p_\pm = p_x \pm i p_y. \tag{3.61}$$

An inspection of the boost given by Eq. (3.45) immediately tells us that four $v_\sigma(\vec{p})'s$ are now readily obtained by flipping the four bottom elements with the four top elements of the respective $u_\sigma(\vec{p})'s$. That is:

$$v_\sigma(\vec{p}) = F\ u_\sigma(\vec{p}), \tag{3.62}$$

where the "flipping matrix" is

$$F = \begin{pmatrix} 0 & I \\ I & 0 \end{pmatrix} \tag{3.63}$$

In general the matrix $I$, appearing in $F$, is a $(2j+1)\times(2j+1)$ identity matrix. Hence for the spin 3/2 case under consideration $I = 4\times 4$ identity matrix.

<u>Orthonormality of Relativistic $(3/2,0)\oplus(0,3/2)$ covariant spinors:</u>

From Eq. (4.43) (See Sec. 4.4 below) we read off:

$$\gamma^{CH}_{000} = \begin{pmatrix} 0 & I \\ I & 0 \end{pmatrix},\ I = 4\times 4 \text{ identity matrix.} \tag{3.64}$$

The canonical representation $\gamma^{CA}_{000}$ is, by definition:

$$\gamma^{CA}_{000} = A\ \gamma^{CH}_{000}\ A^{-1} = \begin{pmatrix} I & 0 \\ 0 & -I \end{pmatrix},\quad A = \frac{1}{\sqrt{2}}\begin{pmatrix} I & I \\ I & -I \end{pmatrix}. \tag{3.65}$$

Introducing (in canonical representation)

$$\overline{u}_\sigma(\vec{p}) = u^\dagger_\sigma(\vec{p})\gamma^{CA}_{000}, \tag{3.66}$$

it is readily verified[16] that

$$\overline{u}_\sigma(\vec{p})\ u_{\sigma'}(\vec{p}) = m^3\delta_{\sigma\sigma'} \tag{3.67}$$

$$\overline{v}_\sigma(\vec{p})\ v_{\sigma'}(\vec{p}) = -m^3\delta_{\sigma\sigma'} \tag{3.68}$$

In the canonical representation the origin of the "minus" sign in the *rhs* of the orthonormality condition (3.68) can be readily traced back to the structure of $\gamma_{000}$,

16 Using MACSYMA.

and the fact that $v_\sigma(\vec{p})$ are obtained (due to the structure of $M(\vec{p})$, given by Eq. (3.45)) from the $v_\sigma(\vec{p})$ via the flipping matrix F. Symbolically:

$$u \sim \begin{pmatrix} a \\ b \end{pmatrix}, \quad \overline{u} \sim \begin{pmatrix} a^* & b^* \end{pmatrix} \begin{pmatrix} I & 0 \\ 0 & -I \end{pmatrix} = \begin{pmatrix} a^* & -b^* \end{pmatrix} \tag{3.69}$$

Hence

$$\overline{u}\ u \sim a^* a - b^* b. \tag{3.70}$$

Next

$$v \sim F\ u \Rightarrow v = \begin{pmatrix} b & a \end{pmatrix} \Rightarrow \overline{v}\ v \sim b^* b - a^* a = -\overline{u}\ u, \qquad QED. \tag{3.71}$$

We suspect that the (relative) minus sign in the $rhs$ of the orthonormality relations (3.67) and (3.68) is essential for the existence of the conserved charge constructed out of $\psi(x)$ for massive particles.

<u>Ultrarelativistic or Massless Limit:</u>

Since we have chosen to work in a representation in which $J_z$ is diagonal, we expect that for a particle traveling along the $\hat{z}$ axis, only the $u_{\pm\frac{3}{2}}(\vec{p})$ and $v_{\pm\frac{3}{2}}(\vec{p})$ to survive. The $u_{\pm\frac{1}{2}}(\vec{p})$ and $v_{\pm\frac{1}{2}}(\vec{p})$ must vanish. In this ultrarelativistic or $m \to 0$ limit: $p^\mu = (E, 0, 0, p = E)$. Substituting $p^\mu = (E, 0, 0, p = E)$ in the $(3/2, 0) \oplus (0, 3/2)$ wave functions we readily see that:

$$\lim_{m\to 0} u_{+\frac{3}{2}}(\vec{p}) = \sqrt{2} E^{\frac{3}{2}} \begin{pmatrix} 1 \\ 0 \\ 0 \\ 0 \\ 1 \\ 0 \\ 0 \\ 0 \end{pmatrix}, \qquad \lim_{m\to 0} u_{+\frac{1}{2}}(\vec{p}) = \begin{pmatrix} 0 \\ 0 \\ 0 \\ 0 \\ 0 \\ 0 \\ 0 \\ 0 \end{pmatrix} \tag{3.72}$$

$$\lim_{m\to 0} u_{-\frac{1}{2}}(\vec{p}) = \begin{pmatrix} 0 \\ 0 \\ 0 \\ 0 \\ 0 \\ 0 \\ 0 \\ 0 \end{pmatrix}, \quad \lim_{m\to 0} u_{-\frac{3}{2}}(\vec{p}) = \sqrt{2}E^{\frac{3}{2}} \begin{pmatrix} 0 \\ 0 \\ 0 \\ 1 \\ 0 \\ 0 \\ 0 \\ -1 \end{pmatrix}, \tag{3.73}$$

with

$$\lim_{m\to 0} \overline{u}_\sigma(E)\ u_\sigma(E) = 0. \tag{3.74}$$

Similarly

$$\lim_{m\to 0} v_{+\frac{3}{2}}(\vec{p}) = \sqrt{2}E^{\frac{3}{2}} \begin{pmatrix} 1 \\ 0 \\ 0 \\ 0 \\ 1 \\ 0 \\ 0 \\ 0 \end{pmatrix}, \quad \lim_{m\to 0} v_{+\frac{1}{2}}(\vec{p}) = \begin{pmatrix} 0 \\ 0 \\ 0 \\ 0 \\ 0 \\ 0 \\ 0 \\ 0 \end{pmatrix}, \tag{3.75}$$

$$\lim_{m\to 0} v_{-\frac{1}{2}}(\vec{p}) = \begin{pmatrix} 0 \\ 0 \\ 0 \\ 0 \\ 0 \\ 0 \\ 0 \\ 0 \end{pmatrix}, \quad \lim_{m\to 0} v_{-\frac{3}{2}}(\vec{p}) = \sqrt{2}E^{\frac{3}{2}} \begin{pmatrix} 0 \\ 0 \\ 0 \\ -1 \\ 0 \\ 0 \\ 0 \\ 1 \end{pmatrix}, \tag{3.76}$$

with

$$\lim_{m\to 0} \overline{v}_\sigma(E)\ v_\sigma(E) = 0. \tag{3.77}$$

In the ultrarelativistic limit we thus see that we get only two non–null independent $u_{\pm\frac{3}{2}}(E)$ and two non–null independent $v_{\pm\frac{3}{2}}(E)$ . The $u_{\pm\frac{1}{2}}(E)$ and $v_{\pm\frac{1}{2}}(E)$ vanish identically. While $u_{+\frac{3}{2}}(E)$ and $v_{+\frac{3}{2}}(E)$ are identical the $u_{-\frac{3}{2}}(E)$ $v_{-\frac{3}{2}}(E)$ differ by a relative phase of $\exp(i\pi)$.

## 3.4 $(2,0)\oplus(0,2)$ Covariant Spinors

For various spin 2 mesons, such as $f_2(1720)$, we now present the relativistic $(2,0)\oplus(0,2)$ covariant spinors. Again, the formal elegance is evident.

In order to calculate the $(2,0)\oplus(0,2)$ covariant spinors we follow the now familiar procedure. The boost is:

$$M_{CA}^{j=2}(\vec{p}) = \begin{pmatrix} \cosh(\vec{J}\cdot\vec{\varphi}) & \sinh(\vec{J}\cdot\vec{\varphi}) \\ & \\ \sinh(\vec{J}\cdot\vec{\varphi}) & \cosh(\vec{J}\cdot\vec{\varphi}) \end{pmatrix} . \tag{3.78}$$

with $\vec{J}$

$$J_x = \begin{pmatrix} 0 & 1 & 0 & 0 & 0 \\ 1 & 0 & \sqrt{3/2} & 0 & 0 \\ 0 & \sqrt{3/2} & 0 & \sqrt{3/2} & 0 \\ 0 & 0 & \sqrt{3/2} & 0 & 1 \\ 0 & 0 & 0 & 1 & 0 \end{pmatrix},$$

$$J_y = \begin{pmatrix} 0 & -i & 0 & 0 & 0 \\ i & 0 & -i\sqrt{3/2} & 0 & 0 \\ 0 & i\sqrt{3/2} & 0 & -i\sqrt{3/2} & 0 \\ 0 & 0 & i\sqrt{3/2} & 0 & -i \\ 0 & 0 & 0 & i & 0 \end{pmatrix},$$

$$J_z = \begin{pmatrix} 2 & 0 & 0 & 0 & 0 \\ 0 & 1 & 0 & 0 & 0 \\ 0 & 0 & 0 & 0 & 0 \\ 0 & 0 & 0 & -1 & 0 \\ 0 & 0 & 0 & 0 & -2 \end{pmatrix} \tag{3.79}$$

The expansions for $\cosh(\vec{J}\cdot\vec{\varphi})$ and $\sinh(\vec{J}\cdot\vec{\varphi})$ are now obtained from Eqs. (A406)

and (A407), and read:

$$\cosh(\vec{J}\cdot\vec{\varphi}) = I + \frac{(\vec{J}\cdot\vec{p})^2}{m(m+E)} + \frac{1}{6}\frac{(\vec{J}\cdot\vec{p})^2((\vec{J}\cdot\vec{p})^2 - \vec{p}^{\,2})}{m^2(m+E)^2} \tag{3.80}$$

$$\sinh(\vec{J}\cdot\vec{\varphi}) = \frac{\vec{J}\cdot\vec{p}}{m} + \frac{1}{3}\frac{\vec{J}\cdot\vec{p}\,((\vec{J}\cdot\vec{p})^2 - \vec{p}^{\,2})}{m^2(m+E)} \tag{3.81}$$

With $J_z$ diagonal and the requirements that the "rest" covariant spinors vanish in the $m \to 0$ limit and that the covariant spinors for massless particles (those corresponding to the ultrarelativistic limit) have a non–singular norm, we choose the following basis of rest covariant spinors:

$$u_{+2}(\vec{0}) = \begin{pmatrix} m^2 \\ 0 \\ 0 \\ 0 \\ 0 \\ 0 \\ 0 \\ 0 \\ 0 \\ 0 \end{pmatrix}, \quad u_{+1}(\vec{0}) = \begin{pmatrix} 0 \\ m^2 \\ 0 \\ 0 \\ 0 \\ 0 \\ 0 \\ 0 \\ 0 \\ 0 \end{pmatrix}, \quad u_{0}(\vec{0}) = \begin{pmatrix} 0 \\ 0 \\ m^2 \\ 0 \\ 0 \\ 0 \\ 0 \\ 0 \\ 0 \\ 0 \end{pmatrix}, \quad u_{-1}(\vec{0}) = \begin{pmatrix} 0 \\ 0 \\ 0 \\ m^2 \\ 0 \\ 0 \\ 0 \\ 0 \\ 0 \\ 0 \end{pmatrix},$$

$$u_{-2}(\vec{0}) = \begin{pmatrix} 0 \\ 0 \\ 0 \\ 0 \\ m^2 \\ 0 \\ 0 \\ 0 \\ 0 \\ 0 \end{pmatrix}, \tag{3.82}$$

$$v_{+2}(\vec{0}) = \begin{pmatrix} 0 \\ 0 \\ 0 \\ 0 \\ 0 \\ m^2 \\ 0 \\ 0 \\ 0 \\ 0 \end{pmatrix}, \quad v_{+1}(\vec{0}) = \begin{pmatrix} 0 \\ 0 \\ 0 \\ 0 \\ 0 \\ 0 \\ m^2 \\ 0 \\ 0 \\ 0 \end{pmatrix}, \quad v_{0}(\vec{0}) = \begin{pmatrix} 0 \\ 0 \\ 0 \\ 0 \\ 0 \\ 0 \\ 0 \\ m^2 \\ 0 \\ 0 \end{pmatrix}, \quad v_{-1}(\vec{0}) = \begin{pmatrix} 0 \\ 0 \\ 0 \\ 0 \\ 0 \\ 0 \\ 0 \\ 0 \\ m^2 \\ 0 \end{pmatrix},$$

$$v_{-2}(\vec{0}) = \begin{pmatrix} 0 \\ 0 \\ 0 \\ 0 \\ 0 \\ 0 \\ 0 \\ 0 \\ 0 \\ m^2 \end{pmatrix}. \tag{3.83}$$

With this skeleton of details, we now write down the $(2,0)\oplus(0,2)$. The reader who has gone through the construction of the $(1,0)\oplus(0,1)$ and the $(3/2,0)\oplus(0,3/2)$ covariant spinors would simply find any further details unnecessary.

$$u_{+2}(\vec{p}) = \begin{pmatrix} (8p_z^4 + 8(p_-p_+ + 2m^2 + 2Em)p_z^2 + p_-^2p_+^2 + 4m(m+E)p_-p_+ \\ +4m^2(m+E)^2)/4(m+E)^2 \\ \\ (4p_+p_z^3 + (3p_-p_+^2 + 6m(m+E)p_+)p_z)/2(m+E)^2 \\ \\ \sqrt{6}(2p_+^2p_z^2 + p_-p_+^3 + 2m(m+E)p_+^2)/4(m+E)^2 \\ \\ p_+^3p_z/2(m+E)^2 \\ \\ p_+^4/4(m+E)^2 \\ \\ (2p_z^3 + (p_-p_+ + 2m^2 + 2Em)p_z)/(m+E) \\ \\ (4p_+p_z^2 + p_-p_+^2 + 2m(m+E)p_+)/2(m+E) \\ \\ \sqrt{6}\ p_+^2p_z/2(m+E) \\ \\ p_+^3/2(m+E) \\ \\ 0 \end{pmatrix} \quad (3.84)$$

$$u_{+1}(\vec{p}) = \begin{pmatrix} (4p_-p_z^3 + 3(p_-^2p_+ + 2m(m+E)p_-)p_z)/2(m+E)^2 \\ \\ (2(2p_-p_+ + m^2 + Em)p_z^2 + 2p_-^2p_+^2 + 5m(m+E)p_-p_+ \\ +2m^2(m+E)^2)/2(m+E)^2 \\ \\ \sqrt{6}(p_-p_+^2 + m(m+E)p_+)p_z/2(m+E)^2 \\ \\ (2p_-p_+^3 + 3m(m+E)p_+^2)/2(m+E)^2 \\ \\ -p_+^3p_z/2(m+E)^2 \\ \\ (4p_-p_z^2 + p_-^2p_+ + 2m(m+E)p_-)/2(m+E) \\ \\ (2p_-p_+ + m(m+E))p_z/(m+E) \\ \\ \sqrt{6}(p_-p_+^2 + m(m+E)p_+)/2(m+E) \\ \\ 0 \\ \\ p_+^3/2(m+E) \end{pmatrix} \tag{3.85}$$

$$u_0(\vec{p}) = \begin{pmatrix} \sqrt{6}(2p_-^2 p_z^2 + p_-^3 p_+ + 2m(m+E)p_-^2)/4(m+E)^2 \\ \\ \sqrt{6}(p_-^2 p_+ + m(m+E)p_-)p_z/2(m+E)^2 \\ \\ (3p_-^2 p_+^2 + 6m(m+E)p_- p_+ + 2m^2(m+E)^2)/2(m+E)^2 \\ \\ -\sqrt{6}(p_- p_+^2 + m(m+E)p_+)p_z/2(m+E)^2 \\ \\ \sqrt{6}(2p_+^2 p_z^2 + p_- p_+^3 + 2m(m+E)p_+^2)/4(m+E)^2 \\ \\ \sqrt{6}p_-^2 p_z/2(m+E) \\ \\ \sqrt{6}(p_-^2 p_+ + m(m+E)p_-)/2(m+E) \\ \\ 0 \\ \\ \sqrt{6}(p_- p_+^2 + m(m+E)p_+)/2(m+E) \\ \\ -\sqrt{6}p_+^2 p_z/2(m+E) \end{pmatrix} \tag{3.86}$$

$$u_{-1}(\vec{p}) = \begin{pmatrix} p_-^3 p_z/2(m+E)^2 \\ \\ (2p_-^3 p_+ + 3m(m+E)p_-^2)/2(m+E)^2 \\ \\ -\sqrt{6}(p_-^2 p_+ + m(m+E)p_-)p_z/2(m+E)^2 \\ \\ (2(2p_-p_+ + m^2 + Em)p_z^2 + 2p_-^2 p_+^2 + 5m(m+E)p_-p_+ \\ +2m^2(m+E)^2)/2(m+E)^2 \\ \\ -(4p_+p_z^3 + 3(p_-p_+^2 + 2m(m+E)p_+)p_z/2(m+E)^2 \\ \\ p_-^3/2(m+E) \\ \\ 0 \\ \\ \sqrt{6}(p_-^2 p_+ + m(m+E)p_-)/2(m+E) \\ \\ -(2p_-p_+ + m(m+E))p_z/(m+E) \\ \\ (4p_+p_z^2 + p_-p_+^2 + 2m(m+E)p_+)/2(m+E) \end{pmatrix} \tag{3.87}$$

$$u_{-2}(\vec{p}) = \begin{pmatrix} p_-^4/4(m+E)^2 \\ \\ -p_-^3 p_z/2(m+E)^2 \\ \\ \sqrt{6}(2p_-^2 p_z^2 + p_-^3 p_+ + 2m(m+E)p_-^2)/4(m+E)^2 \\ \\ -(4p_- p_z^3 + 3(p_-^2 p_+ + 2m(m+E)p_-)p_z)/2(m+E)^2 \\ \\ (8p_z^4 + 8(p_- p_+ + 2m(m+E))p_z^2 + p_-^2 p_+^2 + 4m(m+E)p_- p_+ \\ +4m^2(m+E)^2)/4(m+E)^2 \\ \\ 0 \\ \\ p_-^3/2(m+E) \\ \\ -\sqrt{6}p_-^2 p_z/2(m+E) \\ \\ (4p_- p_z^2 + p_-^2 p_+ + 2m(m+E)p_-)/2(m+E) \\ \\ -(2p_z^3 + (p_- p_+ + 2m(m+E))p_z)/(m+E) \end{pmatrix} \qquad (3.88)$$

The "antiparticle" covariant spinors are

$$v_\sigma(\vec{p}) = F\ u_\sigma(\vec{p}), \tag{3.89}$$

where the "flipping matrix" is

$$F = \begin{pmatrix} 0 & I \\ I & 0 \end{pmatrix}, \quad I = 5\times 5 \text{ unit matrix.} \tag{3.90}$$

Using MACSYMA we obtain the expected orthonormality properties:

$$\overline{u}_\sigma(\vec{p})\ u_{\sigma'}(\vec{p}) = m^4\ \delta_{\sigma\sigma'} \tag{3.91}$$

$$\overline{v}_\sigma(\vec{p})\ v_{\sigma'}(\vec{p}) = -m^4\ \delta_{\sigma\sigma'}. \tag{3.92}$$

Where

$$\overline{u}_\sigma(\vec{p}) = u^\dagger_\sigma(\vec{p})\gamma^{CA}_{0000}, \tag{3.93}$$

$$\gamma^{CA}_{0000} = \begin{pmatrix} I & 0 \\ 0 & -I \end{pmatrix}, \quad I = 5\times 5 \text{ unit matrix.} \tag{3.94}$$

<u>Ultrarelativistic or Massless Limit:</u>

Since we have chosen to work in a representation in which $J_z$ is diagonal, we expect that for a particle traveling along the $\hat{z}$ axis, only the $u_{\pm 2}(\vec{p})$ and $v_{\pm 2}(\vec{p})$ to survive. In the $u_{\pm 1,0}(\vec{p})$ and $v_{\pm 1,0}(\vec{p})$ must vanish. The ultrarelativistic or $m \to 0$ limit: $p^\mu = (E, 0, 0, p = E)$. Substituting $p^\mu = (E, 0, 0, p = E)$ in the $(2,0)\oplus(0,2)$ wave functions given by (76a-e) we readily see that:

$$\lim_{m\to 0} u_{+2}(\vec{p}) = 2E^2 \begin{pmatrix} 1 \\ 0 \\ 0 \\ 0 \\ 0 \\ 1 \\ 0 \\ 0 \\ 0 \\ 0 \end{pmatrix},$$

$$\lim_{m\to 0} u_{+1}(\vec{p}) = \begin{pmatrix} 0 \\ 0 \\ 0 \\ 0 \\ 0 \\ 0 \\ 0 \\ 0 \\ 0 \\ 0 \end{pmatrix}, \lim_{m\to 0} u_{0}(\vec{p}) = \begin{pmatrix} 0 \\ 0 \\ 0 \\ 0 \\ 0 \\ 0 \\ 0 \\ 0 \\ 0 \\ 0 \end{pmatrix}, \lim_{m\to 0} u_{-1}(\vec{p}) = \begin{pmatrix} 0 \\ 0 \\ 0 \\ 0 \\ 0 \\ 0 \\ 0 \\ 0 \\ 0 \\ 0 \end{pmatrix},$$

$$\lim_{m\to 0} u_{-2}(\vec{p}) = 2E^2 \begin{pmatrix} 0 \\ 0 \\ 0 \\ 0 \\ 1 \\ 0 \\ 0 \\ 0 \\ 0 \\ -1 \end{pmatrix}, \tag{3.95}$$

with

$$\lim_{m\to 0} \overline{u}_\sigma(E)\ u_{\sigma'}(E) = 0. \tag{3.96}$$

Similarly:

$$\lim_{m\to 0} v_{+2}(\vec{p}) = 2E^2 \begin{pmatrix} 1 \\ 0 \\ 0 \\ 0 \\ 0 \\ 1 \\ 0 \\ 0 \\ 0 \\ 0 \end{pmatrix},$$

$$\lim_{m\to 0} v_{+1}(\vec{p}) = \begin{pmatrix} 0 \\ 0 \\ 0 \\ 0 \\ 0 \\ 0 \\ 0 \\ 0 \\ 0 \\ 0 \end{pmatrix},\ \lim_{m\to 0} v_{0}(\vec{p}) = \begin{pmatrix} 0 \\ 0 \\ 0 \\ 0 \\ 0 \\ 0 \\ 0 \\ 0 \\ 0 \\ 0 \end{pmatrix},\ \lim_{m\to 0} v_{-1}(\vec{p}) = \begin{pmatrix} 0 \\ 0 \\ 0 \\ 0 \\ 0 \\ 0 \\ 0 \\ 0 \\ 0 \\ 0 \end{pmatrix},$$

$$\lim_{m\to 0} v_{-2}(\vec{p}) = 2E^2 \begin{pmatrix} 0 \\ 0 \\ 0 \\ 0 \\ -1 \\ 0 \\ 0 \\ 0 \\ 0 \\ 1 \end{pmatrix}, \tag{3.97}$$

with

$$\lim_{m\to 0} \overline{v}_{\sigma}(E)\ v_{\sigma'}(E) = 0. \tag{3.98}$$

In the ultrarelativistic limit we thus see that we get only two non–null independent $u_{\pm 2}(E)$ and two non–null independent $v_{\pm 2}(E)$ . The $u_{\pm 1,0}(E)$ and $v_{\pm 1,0}(E)$ vanish identically. While $u_{+2}(E)$ and $v_{+2}(E)$ are identical the $u_{-2}(E)$ $v_{-2}(E)$ differ by a relative phase of $\exp(i\pi)$.

# 4. WAVE EQUATIONS SATISFIED BY $(j,0)\oplus(0,j)$ COVARIANT SPINORS

## 4.1 General Coupled Equations Satisfied by $(j,0)\oplus(0,j)$ Covariant Spinors

We begin with a simple observation that the $(j,0)\oplus(0,j)$ relativistic covariant spinors have been obtained purely from group theoretical considerations and not as solutions of specific equations. As a result, given a specific set of $(j,0)\oplus(0,j)$ covariant spinors, there may exist more than one equation which has these covariant spinors as their solutions.

In this chapter a class of these wave equations is obtained by a simple extension of a procedure described by Ryder [53] for the $(1/2,0)\oplus(0,1/2)$ case. A somewhat less transparent[17] procedure which yields the same wave equations is due to Weinberg [21].

The essential ingredient which enters in deriving the relativistic wave equations is the observation that for a particle at rest, owing to the isotropy of the null direction $\vec{p}=\vec{0}$, one cannot define its spin as either left or right handed. That is:

$$\phi^{R}(\vec{0}) = \phi^{L}(\vec{0}) \tag{4.1}$$

Hence equations Eqs. (4.1) and (3.12) yield

$$\phi^{R}(\vec{0}) = \exp[\vec{J}\cdot\vec{\varphi}]\;\;\phi^{L}(\vec{p}), \tag{4.2}$$

and similarly Eqs. (4.1) and (3.11) give

$$\phi^{L}(\vec{0}) = \exp[-\vec{J}\cdot\vec{\varphi}]\;\;\phi^{R}(\vec{p}). \tag{4.3}$$

Substitution of Eq. (4.2) in Eq. (3.11) and Eq. (4.3) in Eq. (3.12) results in the following coupled equations between the right and left handed matter fields $\phi^{R}(\vec{p})$

17 At least to the present author.

and $\phi^{L}(\vec{p})$:

$$\phi^{R}(\vec{p}) = \exp[2\vec{J}\cdot\vec{\varphi}\,] \;\; \phi^{L}(\vec{p}) \tag{4.4}$$

$$\phi^{L}(\vec{p}) = \exp[-2\vec{J}\cdot\vec{\varphi}\,] \;\; \phi^{R}(\vec{p}). \tag{4.5}$$

Expanding the exponentials, we obtain the coupled equations

$$\phi^{R}(\vec{p}) = \Big(\cosh(2\vec{J}\cdot\vec{\varphi}\,) + \sinh(2\vec{J}\cdot\vec{\varphi}\,)\Big) \;\; \phi^{L}(\vec{p}) \tag{4.6}$$

$$\phi^{L}(\vec{p}) = \Big(\cosh(2\vec{J}\cdot\vec{\varphi}\,) - \sinh(2\vec{J}\cdot\vec{\varphi}\,)\Big) \;\; \phi^{R}(\vec{p}), \tag{4.7}$$

from which we can obtain relativistic equations satisfied by the $(j,0)\oplus(0,j)$ covariant spinors as a simple algebraic exercises. This we now show by explicit examples.

## 4.2 Equation Satisfied by $(1/2,0)\oplus(0,1/2)$ Covariant Spinors: Dirac Equation

Before we take up the case of general fields for an arbritrary spin let us get acquainted with Eqs. (4.6) and (4.7) coupling $\phi^{R}(\vec{p})$ and $\phi^{L}(\vec{p})$ by considering the case

$$\vec{J} = \frac{\vec{\sigma}}{2} \qquad (4.8))$$

where $\vec{\sigma}$ = Pauli matrices. Then, for $\vec{J} = \vec{\sigma}/2$, we can write

$$\cosh[\,2\vec{J}\cdot\vec{\varphi}\,] = \left(I + (\vec{\sigma}\cdot\hat{p}\,)^2\frac{\varphi^2}{2!} + \cdots\cdots\right) = I \;\cosh\varphi \tag{4.9}$$

$$\sinh[\,2\vec{J}\cdot\vec{\varphi}\,] = \vec{\sigma}\cdot\hat{p}\left(1 + \frac{\varphi^3}{3!} + \cdots\cdots\right) = \vec{\sigma}\cdot\hat{p}\;\sinh\varphi, \tag{4.10}$$

where $\hat{p} = \vec{p}/|\vec{p}\,|$ and $I = 2\times 2$ identity matrix. Substituting expansions (4.9) and

(4.10) in the coupled Eqs. (4.6) and (4.7) we get

$$\phi^{R(j=1/2)}(\vec{p}) = (\cosh\varphi + \vec{\sigma}\cdot\hat{p}\ \sinh\varphi)\ \phi^{L(j=1/2)}(\vec{p}) \tag{4.11}$$

$$\phi^{L(j=1/2)}(\vec{p}) = (\cosh\varphi - \vec{\sigma}\cdot\hat{p}\ \sinh\varphi)\ \phi^{R(j=1/2)}(\vec{p}). \tag{4.12}$$

Since $\cosh\varphi = \gamma = (1-v^2)^{-\frac{1}{2}} = E/m = p_0/m$, and $\sinh\varphi = \gamma v = |\vec{p}|/m$ , Eqs. (4.11) and (4.12) can be written as

$$-m\ \phi^{R(j=1/2)}(\vec{p}) + (p_0 + \vec{\sigma}\cdot\vec{p})\ \phi^{L(j=1/2)}(\vec{p}) = 0 \tag{4.13}$$

$$(p_0 - \vec{\sigma}\cdot\vec{p})\ \phi^{R(j=1/2)}(\vec{p}) - m\ \phi^{L(j=1/2)}(\vec{p}) = 0. \tag{4.14}$$

These coupled equations can be combined to yield the equation for chiral representation $(1/2,0)\oplus(0,1/2)$ covariant spinors:

$$\psi^{(j=1/2)}(\vec{p}) = \begin{pmatrix} \phi^{R(j=1/2)}(\vec{p}) \\ \\ \phi^{L(j=1/2)}(\vec{p}) \end{pmatrix}. \tag{4.15}$$

When this is done, we obtain:

$$\begin{pmatrix} -m & (p_0 + \vec{\sigma}\cdot\vec{p}) \\ \\ (p_0 - \vec{\sigma}\cdot\vec{p}) & -m \end{pmatrix} \begin{pmatrix} \phi^{R(j=1/2)}(\vec{p}) \\ \\ \phi^{L(j=1/2)}(\vec{p}) \end{pmatrix} = 0 \quad . \tag{4.16}$$

That this has the formal form of the well known spin one half Dirac equation can be seen by introducing the $4\times4$ $\gamma$ matrices

$$\gamma^0 = \begin{pmatrix} 0 & 1 \\ 1 & 0 \end{pmatrix}, \quad \gamma^i = \begin{pmatrix} 0 & -\sigma^i \\ \sigma^i & 0 \end{pmatrix}$$

$$[Chiral\ Representation]. \tag{4.17}$$

Then Eq. (4.16) becomes[18]

$$(\gamma^\mu p_\mu - m\,I)\psi^{(j=1/2)}(\vec{p}) = 0. \tag{4.18}$$

We now notice that even though we have been motivated to consider the finite

18 Note: $p_\mu = (p_0 = E, -\vec{p})$

dimensional representations of the Lorentz group by quantum mechanical considerations , so far there is nothing very specifically quantum mechanical about our study of the matter fields. Our previous experience with quantum mechanical systems suggests that we postulate $\hbar\vec{J}$ as the operator corresponding to the *observable* associated with the quantum mechanical angular momentum, and $\hbar p_\mu = (\hbar i\partial/\partial t, -\{-i\hbar\vec{\nabla}\})$ be considered as the *observable* associated with the energy momentum vector. We implement these correspondences by interpreting the $p_\mu$ in Eq. (4.18) as the operator $P_\mu$. Thus in coordinate space, Eq. (4.18) translates to read

$$(i\gamma^\mu\partial_\mu \,-\, m\,I)\psi^{(j=1/2)}(x) = 0, \tag{4.19}$$

which is indeed the standard Dirac Equation. The linearity in $\partial_\mu$ of this field equation has its origin in the anti-commutator $\{\sigma^i, \sigma^j\} = 2\delta^{ij}$, which made the substitutions (9) and (10) possible. Put differently $(2\vec{J}\cdot\hat{p})(2\vec{J}\cdot\hat{p})$ appearing in the expansions of $\cosh(2\vec{J}\cdot\vec{\varphi})$ and $\sinh(2\vec{J}\cdot\vec{\varphi})$ equals

$$4\left(\frac{p_x^2}{|\vec{p}|^2}J_x^2 + \frac{p_y^2}{|\vec{p}|^2}J_y^2 + \frac{p_z^2}{|\vec{p}|^2}J_z^2 + \frac{p_x p_y}{|\vec{p}|^2}\{J_x, J_y\} + \frac{p_x p_z}{|\vec{p}|^2}\{J_x, J_z\} + \frac{p_y p_z}{|\vec{p}|^2}\{J_y, J_z\}\right) \tag{4.20}$$

which reduces to a simple *identity* matrix only for $\vec{J} = \vec{\sigma}/2$. In the absence of this property specific to the spin-$\frac{1}{2}$ fields the expansion of $\cosh(2\vec{J}\cdot\vec{\varphi})$ and $\sinh(2\vec{J}\cdot\vec{\varphi})$ will contain higher order terms in $\vec{p}$, and hence higher order $\partial_\mu$ in the coordinate space representation through the powers of $\vec{J}\cdot\hat{p} = \vec{J}\cdot\vec{p}/|\vec{p}|$.

## 4.3 A Equation Satisfied by $(1,0)\oplus(0,1)$ Covariant Spinors

Setting $j = 1$ in Eqs. (A406) and (A407) of the last section of the Appendix and taking $\vec{J}$ to be $3\times 3$ angular momentum 1 matrices we get

$$\cosh(2\vec{J}\cdot\vec{\varphi}) = 1 + 2(\vec{J}\cdot\hat{p})(\vec{J}\cdot\hat{p})\sinh^2\varphi, \tag{4.21}$$

$$\sinh(2\vec{J}\cdot\vec{\varphi}) = 2(\vec{J}\cdot\hat{p})\cosh\varphi\sinh\varphi. \tag{4.22}$$

Using these expansions in the coupled Eqs. (4.6) and (4.7) we obtain

$$\phi^{R(j=1)}(\vec{p}) = \left[1 + 2(\vec{J}\cdot\hat{p})(\vec{J}\cdot\hat{p})\sinh^2\varphi + 2(\vec{J}\cdot\hat{p})\cosh\varphi\sinh\varphi\right]\phi^{L(j=1)}(\vec{p}) \quad (4.23)$$

$$\phi^{L(j=1)}(\vec{p}) = \left[1 + 2(\vec{J}\cdot\hat{p})(\vec{J}\cdot\hat{p})\sinh^2\varphi - 2(\vec{J}\cdot\hat{p})\cosh\varphi\sinh\varphi\right]\phi^{R(j=1)}(\vec{p}). \quad (4.24)$$

By replacing $\sinh\varphi$ by $|\vec{p}|/m$ and $\cosh\varphi$ by $p^0/m$ the above expressions take the form

$$\phi^{R(j=1)}(\vec{p}) = \frac{1}{m^2}\left[m^2 + 2(\vec{J}\cdot\vec{p})(\vec{J}\cdot\vec{p}) + 2(\vec{J}\cdot\vec{p})p^0\right]\phi^{L(j=1)}(\vec{p}) \quad (4.25)$$

$$\phi^{L(j=1)}(\vec{p}) = \frac{1}{m^2}\left[m^2 + 2(\vec{J}\cdot\vec{p})(\vec{J}\cdot\vec{p}) - 2(\vec{J}\cdot\vec{p})p^0\right]\phi^{R(j=1)}(\vec{p}). \quad (4.26)$$

These coupled equations can be combined to yield the equation for the chiral representation six–component $(1,0)\oplus(0,1)$ covariant spinors:

$$\psi_{CH}^{(j=1)}(\vec{p}) = \begin{pmatrix} \phi^{R(j=1)}(\vec{p}) \\ \\ \phi^{L(j=1)}(\vec{p}) \end{pmatrix}. \quad (4.27)$$

This equation reads[19] :

$$\left[\begin{pmatrix} 0 & \left[\eta_{\mu\nu}p^\mu p^\nu + 2(\vec{J}\cdot\vec{p})(\vec{J}\cdot\vec{p}) + 2(\vec{J}\cdot\vec{p})p^0\right] \\ \left[\eta_{\mu\nu}p^\mu p^\nu + 2(\vec{J}\cdot\vec{p})(\vec{J}\cdot\vec{p}) - 2(\vec{J}\cdot\vec{p})p^0\right] & 0 \end{pmatrix} - m^2 I\right]\begin{pmatrix} \phi^{R(j=1)}(\vec{p}) \\ \\ \phi^{L(j=1)}(\vec{p}) \end{pmatrix} = 0. \quad (4.28)$$

19 Here we have used $\eta_{\mu\nu}p^\mu p^\nu = m^2$ in the first term on the *lhs* of the equation which follows. Similar substitutions will be made for the $(3/2,0)\oplus(0,3/2)$ and $(2,0)\oplus(0,2)$ cases without explicit mention. This is a non–trivial substitution and will receive detailed attention in Chapter 6.

Introducing[20] ten fully symmetric (in the Lorentz indices) $6\times 6$ spin–1 $\gamma$ matrices:

$$\gamma_{\mu\nu}p^\mu p^\nu = \begin{pmatrix} 0 & \Big[\eta_{\mu\nu}p^\mu p^\nu + 2(\vec{J}\cdot\vec{p})(\vec{J}\cdot\vec{p}) + 2(\vec{J}\cdot\vec{p})p^0\Big] \\ \Big[\eta_{\mu\nu}p^\mu p^\nu + 2(\vec{J}\cdot\vec{p})(\vec{J}\cdot\vec{p}) - 2(\vec{J}\cdot\vec{p})p^0\Big] & 0 \end{pmatrix} \tag{4.29}$$

we obtain[21]

$$\left(\gamma_{\mu\nu}p^\mu p^\nu \;-\; m^2\, I\right)\psi^{(j=1)}(\vec{p}) = 0. \tag{4.30}$$

The chiral representation expressions for the $j = 1$ $\gamma$ matrices are easily read from (4.29) to be:

$$\gamma_{00} = \begin{pmatrix} 0 & I \\ I & 0 \end{pmatrix} \tag{4.31}$$

$$\gamma_{i0} = \gamma_{0i} = \begin{pmatrix} 0 & J_i \\ -J_i & 0 \end{pmatrix} \tag{4.32}$$

$$\gamma_{ji} = \gamma_{ij} = \begin{pmatrix} 0 & I \\ I & 0 \end{pmatrix}\eta_{ij} + \begin{pmatrix} 0 & \{J_i, J_j\} \\ \{J_i, J_j\} & 0 \end{pmatrix}, \tag{4.33}$$

where $J_i$ are the $3\times 3$, $j = 1$ angular momentum matrices. In order that Eq. (4.30) be interpreted as a spin–1 quantum mechanical wave equation the reader is referred to the comments made at the end of the last section. In Eq. (4.33) $\eta_{ij}$ is the spacial part of $\eta_{\mu\nu}$. In Eqs. (4.30) to (4.33) $\hbar p^\mu$ should be interpreted as the *observable* energy momentum vector; and $\hbar\vec{J}$ as the standard *observable* operators associated with the angular momentum. It should be parenthetically noted that

$$\vec{J}_{Quantum\ Mechanical} = \hbar \times \vec{J}_{Here}. \tag{4.34}$$

So far the $\vec{J}$ in this work stands for the classical generators of rotation. The quantum mechanical angular momentum operator equals $\hbar$ times the classical

20 We drop the representation identifying subscripts $_{CH}$ or $_{CA}$ whenever no confusion is likely to occur.

21 Which first appeared in ref. [21].

generators of rotation. However it is most convenient to choose the units $\hbar = 1,\ c = 1$ then *numerically:* $\vec{J}_{Quantum\ Mechanical} = \vec{J}_{Here}$.

In coordinate space Eq. (4.30) becomes

$$\left(\gamma^{\mu\nu}\partial_\mu\partial_\nu + m^2\, I\right)\psi^{(j=1)}(x) = 0. \tag{4.35}$$

To obtain the above equation from Eq. (4.30) we first note $\gamma_{\mu\nu}p^\mu p^\nu = \gamma^{\mu\nu}p_\mu p_\nu$. Next, let $p_\mu \to i\partial_\mu$.

## 4.4 An Equation Satisfied by $(3/2,0)\oplus(0,3/2)$ Covariant Spinors

Setting $j = 3/2$ in Eqs. (A408) and (A409) of the last section of the Appendix and taking $\vec{J}$ to be $4\times4$ angular momentum 3/2 matrices we get

$$\cosh(2\vec{J}\cdot\vec{\varphi}) = \cosh\varphi\ \left[1 + \frac{1}{2!}(\eta^2-1)\sinh^2\varphi\right], \tag{4.36}$$

$$\sinh(2\vec{J}\cdot\vec{\varphi}) = \eta\sinh\varphi\ \left[1 + \frac{1}{3!}(\eta^2-1)\sinh^2\varphi\right]. \tag{4.37}$$

Substituting expansions (4.36) and (4.37) in the coupled Eqs. (4.6) and (4.7) we obtain

$$\phi^{R(j=3/2)}(\vec{p}) = \left[\cosh\varphi + \eta\sinh\varphi + \left(\frac{\eta^2-1}{6}\right)\left\{\eta\sinh^3\varphi + 3\cosh\varphi\sinh^2\varphi\right\}\right] \times\varphi^{L(j=3/2)}(\vec{p}), \tag{4.38}$$

$$\phi^{L(j=3/2)}(\vec{p}) = \left[\cosh\varphi - \eta\sinh\varphi - \left(\frac{\eta^2-1}{6}\right)\left\{\eta\sinh^3\varphi - 3\cosh\varphi\sinh^2\varphi\right\}\right] \times\phi^{R(j=3/2)}(\vec{p}). \tag{4.39}$$

Reintroducing $\eta = 2\vec{J}\cdot\hat{p}$ and $\sinh\varphi = |\vec{p}|/m$ and $\cosh\varphi = p^0/m$, the above

equations become

$$\phi^{R(j=3/2)}(\vec{p}) = \frac{1}{m^3}\left[m^2(p^0 + 2\vec{J}\cdot\vec{p}) + \frac{1}{6}\{(2\vec{J}\cdot\vec{p})^2 - p^2\}\{2\vec{J}\cdot\vec{p} + 3p^0\}\right] \times\phi^{L(j=3/2)}(\vec{p}), \tag{4.40}$$

$$\phi^{L(j=3/2)}(\vec{p}) = \frac{1}{m^3}\left[m^2(p^0 - 2\vec{J}\cdot\vec{p}) - \frac{1}{6}\{(2\vec{J}\cdot\vec{p})^2 - p^2\}\{2\vec{J}\cdot\vec{p} - 3p^0\}\right] \times\phi^{R(j=3/2)}(\vec{p}). \tag{4.41}$$

Finally, introducing the eight component $(3/2, 0)\oplus(0, 3/2)$ covariant spinor

$$\psi_{CH}^{(j=3/2)}(\vec{p}) = \begin{pmatrix} \phi^{R(j=3/2)}(\vec{p}) \\ \\ \phi^{L(j=3/2)}(\vec{p}) \end{pmatrix}. \tag{4.42}$$

and the $8\times 8$ spin–3/2 $\gamma$ matrices

$$\gamma_{\mu\nu\lambda}\, p^\mu\, p^\nu\, p^\lambda = \begin{pmatrix} 0 & \Big[\eta_{\mu\nu}\, p^\mu\, p^\nu(p^0 + 2\vec{J}\cdot\vec{p}) + \\ & \frac{1}{6}\{(2\vec{J}\cdot\vec{p})^2 - \vec{p}^{\,2}\}\{2\vec{J}\cdot\vec{p} + 3p^0\}\Big] \\ \\ \Big[\eta_{\mu\nu}\, p^\mu\, p^\nu(p^0 - 2\vec{J}\cdot\vec{p}) - & \\ \frac{1}{6}\{(2\vec{J}\cdot\vec{p})^2 - \vec{p}^{\,2}\}\{2\vec{J}\cdot\vec{p} - 3p^0\}\Big] & 0 \end{pmatrix}, \tag{4.43}$$

we obtain the $(3/2, 0)\oplus(0, 3/2)$ relativistic wave equation

$$(\gamma_{\mu\nu\lambda}\, p^\mu p^\nu p^\lambda \;-\; m^3\, I)\psi^{(j=3/2)}(\vec{p}) = 0. \tag{4.44}$$

In coordinate space it reads

$$(i\gamma^{\mu\nu\lambda}\,\partial_\mu\,\partial_\nu\,\partial_\lambda \;+\; m^3\, I)\psi^{(j=3/2)}(x) = 0. \tag{4.45}$$

## 4.5 An Equation Satisfied by $(2,0)\oplus(0,2)$ Covariant Spinors

To obtain a wave equation satisfied by the $(2,0)\oplus(0,2)$ covariant spinors using the coupled equations (4.6) and (4.7) we need the expansions of $\cosh(2\vec{J}\cdot\vec{\varphi}\,)$ and $\sinh(2\vec{J}\cdot\vec{\varphi}\,)$ with $\vec{J}$ as the $5\times 5$ angular momentum 2 matrices. These expansions are obtained by using Eqs. (A406) and (A407), and the standard identities

$$\sinh\varphi = \frac{|\vec{p}\,|}{m}, \quad \cosh\varphi = \frac{p^0}{m}. \tag{4.46}$$

The result is

$$\cosh(2\vec{J}\cdot\vec{\varphi}\,) = I + \frac{2(\vec{J}\cdot\vec{p}\,)^2}{m^2} + \frac{2}{3}\frac{(\vec{J}\cdot\vec{p}\,)^2\{(\vec{J}\cdot\vec{p}\,)^2-\vec{p}^{\,2}\}}{m^4} \tag{4.47}$$

$$\sinh(2\vec{J}\cdot\vec{\varphi}\,) = \frac{2(\vec{J}\cdot\vec{p}\,)p^0}{m^2} + \frac{4}{3}\frac{(\vec{J}\cdot\vec{p}\,)\{(\vec{J}\cdot\vec{p}\,)^2-\vec{p}^{\,2}\}p^0}{m^4} \tag{4.48}$$

A relativistic wave equation satisfied by the ten component $(2,0)\oplus(0,2)$ covariant spinors

$$\psi_{CH}^{(j=2)}(\vec{p}\,) = \begin{pmatrix} \phi^{R(j=2)}(\vec{p}\,) \\ \\ \phi^{L(j=2)}(\vec{p}\,) \end{pmatrix}, \tag{4.49}$$

is now readily obtained using the coupled equations (4.6) and (4.7). The result is

$$\left(\gamma_{\mu\nu\lambda\rho}\,p^\mu p^\nu p^\lambda p^\rho \;-\; m^4\,I\right)\psi(\vec{p}) = 0, \tag{4.50}$$

where in chiral representation the thirty five $10\times10$ fully symmetric (in the Lorentz indices) $\gamma$ matrices are:

$$\gamma_{\mu\nu\lambda\rho}\, p^\mu p^\nu p^\lambda p^\rho =$$

$$\begin{pmatrix} 0 & \begin{aligned}\Big[\eta_{\mu\nu}p^\mu p^\nu\, \eta_{\lambda\rho}p^\lambda p^\rho \\ +\, 2(\vec{J}\cdot\vec{p})\{(\vec{J}\cdot\vec{p}) + p^0\}\eta_{\mu\nu}p^\mu p^\nu \\ +\tfrac{2}{3}\,(\vec{J}\cdot\vec{p})\{(\vec{J}\cdot\vec{p})^2 - \vec{p}^{\,2}\}\{(\vec{J}\cdot\vec{p}) + 2p^0\}\Big]\end{aligned} \\ \begin{aligned}\Big[\eta_{\mu\nu}p^\mu p^\nu\, \eta_{\lambda\rho}p^\lambda p^\rho \\ +\, 2(\vec{J}\cdot\vec{p})\{(\vec{J}\cdot\vec{p}) - p^0\}\eta_{\mu\nu}p^\mu p^\nu \\ +\tfrac{2}{3}\,(\vec{J}\cdot\vec{p})\{(\vec{J}\cdot\vec{p})^2 - \vec{p}^{\,2}\}\{(\vec{J}\cdot\vec{p}) - 2p^0\}\Big]\end{aligned} & 0 \end{pmatrix}. \tag{4.51}$$

In the coordinate space we have the following expression for (4.50):

$$\left(\gamma^{\mu\nu\lambda\rho}\,\partial_\mu\partial_\nu\partial_\lambda\partial_\rho - m^4\, I\right)\psi(x) = 0. \tag{4.52}$$

# 5. CAUSAL PROPAGATORS FOR $(j,0)\oplus(0,j)$ MATTER FIELDS

## 5.1 Causality and Wave Equations Satisfied by $(j,0)\oplus(0,j)$ Covariant Spinors

A general wave equation, obtained from the coupled equations (4.6) and (4.7) , satisfied by the $(j,0)\oplus(0,j)$ relativistic covariant spinors is of the form

$$\left(\gamma_{\{\mu\}}\, p^{[\mu]} \,-\, m^{2j}\, I\right)\,\psi(\vec{p}) \,=\, 0, \tag{5.1}$$

where $\{\mu\}$ is a set of $2j$ Lorentz indices and $p^{[\mu]}$ is a set of $2j$ one Lorentz indexed contravariant energy momentum vectors. That is for $j=1/2$, $\gamma_{\{\mu\}}\,p^{[\mu]}=\gamma_\mu p^\mu$ and for $j=1$, $\gamma_{\{\mu\}}\,p^{[\mu]}=\gamma_{\mu\nu}p^\mu p^\nu$, and so on. $\gamma_{\{\mu\}}$ are a fully symmetric[22]

$$\frac{N\,(N\,+\,1)\,\cdots\,(N\,+\,S\,-\,1)}{S!} \tag{5.2}$$

set of $2(2j+1)\times 2(2j+1)$ spin–$j$ $\gamma$ matrices. Here $S=2j$, the number of indices on $\gamma_{\{\mu\}}$; and for the $(1,3)$ spacetime under consideration $N=4$. A basic requirement for *any* solutions to exist is

$$\text{Determinant}\;\left(\gamma_{\{\mu\}}\, p^{[\mu]} \,-\, m^{2j}\, I\right) \,=\, 0. \tag{5.3}$$

For a given $j$ this "existence requirement" can be interpreted as a $2j[2(2j+1)]th$ order equation in $E$. By working out specific examples, we will discover that there are

$$N_A(j) \,=\, 2j[2(2j+1)]-2(2j+1) \,=\, 2(2j-1)(2j+1) \tag{5.4}$$

"acausal" solutions – that is solutions for which:

$$E^2 \,\neq\, p^2 \,+\, m^2. \tag{5.5}$$

Even though we verify this relationship only for $j=1/2, 1, 3/2$ and 2, we expect it to be true in general. The remaining "causal" solutions, that is solutions for

22 In the Lorentz indices.

which $E^2 = p^2 + m^2$,

$$N_C(j) = 2(2j+1), \tag{5.6}$$

we identify with the already constructed $(j,0)\oplus(0,j)$ relativistic covariant spinors.

The observation

$$N_A\left(\frac{1}{2}\right) = 0; \text{ and } N_A\left(j > \frac{1}{2}\right) \neq 0 \tag{5.7}$$

leads us to the conclusion that the Green[23] functions $G^{(j>\frac{1}{2},0)\oplus(0,j>\frac{1}{2})}(x-x')$

$$\left( (i)^{2j}\,\gamma^{\{\mu\}}\,\partial_{[\mu]} - m^{2j} \right) G^{(j>\frac{1}{2},0)\oplus(0,j>\frac{1}{2})}(x-x') = \delta^4(x-x') \tag{5.8}$$

associated with wave equations satisfied by the $\left(j > \frac{1}{2}, 0\right) \oplus \left(0, j > \frac{1}{2}\right)$ covariant spinors, cease to be identical with the vacuum expectation value

$$\langle \quad | T[\Psi(x)\ \overline{\Psi}(x')]| \quad \rangle \tag{5.9}$$

where[24]

$$\Psi(x) = \sum_\sigma \int \frac{d^3p}{(2\pi)^{3/2}} \sqrt{\Omega} \\ \Big[ u(\vec{p},\sigma)\ a(\vec{p},\sigma)\ \exp(-ip\cdot x) + v(\vec{p},\sigma)\ b^\dagger(\vec{p},\sigma)\ \exp(+ip\cdot x) \Big], \tag{5.10}$$

However we note that the fundamental object which enters the canonical S–matrix calculations is not the Green function G(x-x') but the "propagator":

$$\langle \quad | T[\Psi(x)\ \overline{\Psi}(x')]| \quad \rangle. \tag{5.11}$$

The propagator can be constructed out of the $(j,0)\oplus(0,j)$ covariant spinors

$$u_\sigma(x) = u_\sigma(\vec{p}) \exp\left(-ip^\mu x_\mu\right), \tag{5.12}$$

$$v_\sigma(x) = v_\sigma(\vec{p}) \exp\left(+ip^\mu x_\mu\right), \tag{5.13}$$

and, unlike the Green function, contains only the physical "causal" solutions.

---

23 While working specific examples we may may not symbolically distinguish between the Green functions which arise from a source "$-\delta(x-x')$" rather than "$+\delta(x-x')$".

24 Here $\Omega = \Omega(m,\vec{p})$ is a $j$ dependent normalization factor.

For $j = 1/2$, $u_\sigma(\vec{p})$ and $v_\sigma(\vec{p})$ are the well known Dirac spinors. A general method of obtaining $u_\sigma(\vec{p})$ and $v_\sigma(\vec{p})$ for any spin has been discussed in Chapter 3. Explicit forms of $u_\sigma(\vec{p})$ and $v_\sigma(\vec{p})$ are worked out for $j = 1, 3/2$ and 2 in Secs. 3.2, 3.3 and 3.4 respectively.

We now explicitly examine the character of the solutions of the wave equations satisfied by the $(1,0)\oplus(0,1)$, $(3/2,0)\oplus(0,3/2)$, and $(2,0)\oplus(0,2)$ covariant spinors. Sec. 5.3 examines the $(1/2,0)\oplus(0,1/2)$ Dirac equation. We begin with $(3/2,0)\oplus(0,3/2)$ case.

## 5.2 Causal Propagator for $(3/2,0)\oplus(0,3/2)$ Matter Field

The momentum–space wave equation satisfied by the $(3/2,0)\oplus(0,3/2)$ covariant spinors, as derived in Sec. 4.4, reads:

$$\left(\gamma_{\mu\nu\lambda}\, p^\mu p^\nu p^\lambda \,-\, m^3\, I\right)\,\psi^{(j=3/2)}(\vec{p}) = 0. \tag{5.14}$$

We begin with the observation:

> We already know the eight physical solutions which satisfy this wave equation. These are:
>
> $$u_{\pm 3/2}(\vec{p}),\quad u_{\pm 1/2}(\vec{p}),\quad v_{\pm 3/2}(\vec{p}),\quad v_{\pm 1/2}(\vec{p}), \tag{5.15}$$
>
> Their explicit form is given in Sec. 3.3. These solutions were *not* obtained as solutions of any differential equations.

With this observation in mind, we note, that for the $(1/2,0)\oplus(0,1/2)$ case the Dirac propagator can be constructed either as a green function

$$(\,i\,\gamma^\mu\partial_\mu \,-\, m\, I\,)\;\, G^{\left(\frac{1}{2},0\right)\oplus\left(0,\frac{1}{2}\right)}(x-x') \;=\; \delta^4(x-x') \tag{5.16}$$

or, evaluated as the vacuum expectation value:

$$\langle\;\;|T[\Psi^{\left(\frac{1}{2},0\right)\oplus\left(0,\frac{1}{2}\right)}(x)\overline{\Psi}^{\left(\frac{1}{2},0\right)\oplus\left(0,\frac{1}{2}\right)}(x')]|\;\;\rangle. \tag{5.17}$$

For the $(1/2,0)\oplus(0,1/2)$ case the $G(x-x')$ and $\langle \quad |T[\Psi(x)\overline{\Psi}(x')]| \quad \rangle$ are identical to within a numerical factor of the order of unity.

In order to study the kinematical properties of (5.14) we note that a basic requirement for *any* solutions to exist is:

$$\text{Determinant}\left(\gamma_{\mu\nu\lambda}\,p^\mu p^\nu p^\lambda - m^3\,I\right) = 0. \tag{5.18}$$

For mathematical convenience, and without loss of generality, we now confine to $p^\mu = (E,0,0,p)$[25]. Equation (5.18) then becomes:

$$\left(p^2 + m^2 - E^2\right)^4 \left(p^4 - m^2\,p^2 - 2\,E^2\,p^2 + m^4 + E^2\,m^2 + E^4\right)^4 = 0. \tag{5.19}$$

Treating this equation as a twenty–fourth order equation in $E$, we obtain 24 solutions. These solutions are of the form $E = E(p,m)$ and are called "dispersion relations." The dispersion relations, the associated multiplicity (that is the number of times a particular solution occurs) and their interpretation, are tabulated in Table VI.

25 Throughout this chapter we will use the symbolic manipulation program "MACSYMA" to carry out various analytic calculations.

TABLE VI

Dispersion relations $E = E(p, m)$ associated with Eq. (5.14); obtained as solutions of Eq. (5.19): $N_C\left(\frac{3}{2}\right) = 8$, and $N_A\left(\frac{3}{2}\right) = 16$.

| (Multiplicity) Dispersion Relation | Interpretation |
|---|---|
| (4) $E = +\sqrt{p^2 + m^2}$ | Causal, "particle" $u_{\pm\frac{3}{2}}(\vec{p}),\ u_{\pm\frac{1}{2}}(\vec{p})$ |
| (4) $E = -\sqrt{p^2 + m^2}$ | Causal, "antiparticle" $v_{\pm\frac{3}{2}}(\vec{p}),\ v_{\pm\frac{1}{2}}(\vec{p})$ |
| (4) $E = +\left(\frac{2p^2 + i\sqrt{3}m^2 - m^2}{2}\right)^{1/2}$ | Acausal |
| (4) $E = -\left(\frac{2p^2 + i\sqrt{3}m^2 - m^2}{2}\right)^{1/2}$ | Acausal |
| (4) $E = +\left(\frac{2p^2 - i\sqrt{3}m^2 - m^2}{2}\right)^{1/2}$ | Acausal |
| (4) $E = -\left(\frac{2p^2 - i\sqrt{3}m^2 - m^2}{2}\right)^{1/2}$ | Acausal |

The term "causal" in Table VI refers to the fact that particle/antiparticle covariant spinors satisfy the correct $E = E(m, p)$ relationship

$$E^2 = p^2 + m^2. \tag{5.20}$$

On the other hand the solutions termed "acausal" emphasise the fact that they do *not* satisfy the correct $E = E(m, p)$ given by (5.20). It is because of the existence of "acausal" solutions admitted by (5.14) that the associated Green[26] function

26 In reference to footnote 23 note that the Green function defined here and that defined through Eq. (5.8) for $j = 3/2$ differ by a "$-$" sign in their source term. Similar care should be taken elsewhere.

$G^{\left(\frac{3}{2},0\right)\oplus\left(0,\frac{3}{2}\right)}(x-x')$

$$\left(i\,\gamma^{\mu\nu\lambda}\,\partial_\mu\partial_\nu\partial_\lambda\,+\,m^3\,I\right)\,G^{\left(\frac{3}{2},0\right)\oplus\left(0,\frac{3}{2}\right)}(x-x')\,=\,\delta^4(x-x'),\tag{5.21}$$

will propagate not only the physical "causal" solutions but also the unphysical "acausal" solutions.

The $(3/2,0)\oplus(0,3/2)$ causal propagator is now constructed via the known and causal $(3/2,0)\oplus(0,3/2)$ covariant spinors $u_\sigma(\vec{p})$ and $v_\sigma(\vec{p})$ by evaluating

$$\langle\ \ |T[\Psi^{\left(\frac{3}{2},0\right)\oplus\left(0,\frac{3}{2}\right)}(x)\ \overline{\Psi}^{\left(\frac{3}{2},0\right)\oplus\left(0,\frac{3}{2}\right)}(x')]|\ \ \rangle.\tag{5.22}$$

The eight causal solutions $u_\sigma(\vec{p})$ and $v_\sigma(\vec{p})$ needed to evaluate the above expression are given explicitly in Sec. 3.3. These solutions are independent of any specific wave equations which one may construct for phenomenological studies.

## 5.3 A Remark on $(1/2,0)\oplus(0,1/2)$ Matter Field

As a parenthetic remark, we note for the $(1/2,0)\oplus(0,1/2)$ Dirac case[27] that

$$\text{Determinant}\ \left(\gamma^\mu\,p_\mu\,-\,m\,I\,\right)=0\tag{5.23}$$

yields

$$E=+\sqrt{p^2+m^2},\qquad 2\ \text{times}\tag{5.24}$$

$$E=-\sqrt{p^2+m^2},\qquad 2\ \text{times}\tag{5.25}$$

without any acausal solutions.

27 Again taking $p^\mu=(E,0,0,p)$ to keep calculations simple.

## 5.4 Causal Propagator for $(1,0)\oplus(0,1)$ Matter Field

The $(1,0)\oplus(0,1)$ covariant spinors satisfy

$$\left(\gamma_{\mu\nu}\,p^\mu p^\nu\;-\;m^2\,I\right)\,\psi(\vec{p})\;=\;0, \tag{5.26}$$

as shown in Sec. 4.3. For this case, from a conceptual point of view, all of the above discussion still holds. Now

$$\text{Determinant}\;\left(\gamma_{\mu\nu}\,p^\mu p^\nu\;-\;m^2\,I\right)\;=\;0 \tag{5.27}$$

yields[28]

$$-\left(p^2\,-\,m^2\,-E^2\right)^3\,\left(p^2\,+\,m^2\,-\,E^2\right)^3\;=\;0. \tag{5.28}$$

Treating this equation as a twelfth order equation in $E$, we obtain 12 solutions. These solutions, the associated multiplicity and their interpretation, are tabulated in Table VII.

TABLE VII

Dispersion relations $E = E(p,m)$ associated with Eq. (5.26); obtained as solutions of Eq. (5.28): $N_C\,(1)\;=\;6$, and $N_A\,(1)\;=\;12$.

| (Multiplicity) Dispersion Relation | Interpretation |
|---|---|
| (3) $E=+\sqrt{p^2+m^2}$ | Causal, "particle" $u_{\pm 1}(\vec{p}),\;\;u_0(\vec{p})$ |
| (3) $E=-\sqrt{p^2+m^2}$ | Causal, "antiparticle" $v_{\pm 1}(\vec{p}),\;\;v_0(\vec{p})$ |
| (3) $E=+\sqrt{p^2-m^2}$ | Acausal, Tachyonic |
| (3) $E=-\sqrt{p^2-m^2}$ | Acausal, Tachyonic |

28 Calculations being performed with $p^\mu = (E,0,0,p)$.

In Table VII the term "techyonic" is used to indicate that these solutions propagate with velocities greater than light: $v > 1$. It is because of the existence of "acausal tachyonic" solutions admitted by (5.26) that the associated Green function $G^{(1,0)\oplus(0,1)}(x-x')$

$$\left(\gamma^{\mu\nu}\,\partial_\mu\partial_\nu \,+\, m^2\, I\right)\, G^{(1,0)\oplus(0,1)}(x-x') \,=\, \delta^4(x-x'), \tag{5.29}$$

will propagate not only the physical "causal" solutions but also the unphysical "acausal tachyonic" solutions.

The $(1,0)\oplus(0,1)$ causal propagator is now constructed via the known and causal $(1,0)\oplus(0,1)$ covariant spinors: $u_\sigma(\vec{p})$ and $v_\sigma(\vec{p})$ by evaluating

$$\langle\ \ |T[\Psi^{(1,0)\oplus(0,1)}(x)\ \overline{\Psi}^{(1,0)\oplus(0,1)}(x')]|\ \ \rangle. \tag{5.30}$$

The six causal solutions $u_\sigma(\vec{p})$ and $v_\sigma(\vec{p})$ needed to evaluate the above expression are given explicitly in Sec. 3.2. These solutions are independent of any specific wave equations which one may construct for phenomenological studies.

### 5.5 Causal Propagator for $(2,0)\oplus(0,2)$ Matter Field

The momentum–space wave equation satisfied by the $(2,0)\oplus(0,2)$ covariant spinors, as derived in Sec. 4.5 is:

$$\Big(\gamma_{\mu\nu\lambda\rho}\, p^\mu p^\nu p^\lambda p^\rho \ -\ m^4\, I\Big)\, \psi(\vec{p}) \,=\, 0. \tag{5.31}$$

As before, in order to study the kinematical properties of this equation we note that a basic requirement for *any* solutions to exist is

$$\text{Determinant}\ \Big(\gamma_{\mu\nu\lambda\rho}\, p^\mu p^\nu p^\lambda p^\rho \ -\ m^4\, I\Big) = 0\,. \tag{5.32}$$

For the mathematical convenience, and without loss of the generality, we now confine to $p^\mu = (E,0,0,p)$ and evaluate this determinant. Equation[29] (5.32) then

29 The long expression which follows is simply to emphasise the importance of every possible simplification which can be introduced, like taking $p^\mu = (E,0,0,p)$, to execute these high–spin calculations successfully.

becomes:

$$
\begin{aligned}
-E^{40} &+ 20\,p^2\,E^{38} - 190\,p^4\,E^{36} + 1140\,p^6\,E^{34} - \left(4845\,p^8 - 5\,m^8\right)\,E^{32} \\
&+ \left(15504\,p^{10} - 80\,m^8\,p^2\right)\,E^{30} - \left(38760\,p^{12} - 600\,m^8\,p^4\right)\,E^{28} \\
&+ \left(77520\,p^{14} - 2800\,m^8\,p^6\right)\,E^{26} - \left(125970\,p^{16} - 9100\,m^8\,p^8 + 10\,m^{16}\right)\,E^{24} \\
&- \left(21840\,m^8\,p^{10} - 120\,m^{16}\,p^2 - 167960\,p^{18}\right)\,E^{22} \\
&- \left(184756\,p^{20} - 40040\,m^8\,p^{12} + 660\,m^{16}\,p^4\right)\,E^{20} \\
&+ \left(167960\,p^{22} - 57200\,m^8\,p^{14} + 2200\,m^{16}\,p^6\right)\,E^{18} \\
&- \left(125970\,p^{24} - 64350\,m^8\,p^{16} + 4950\,m^{16}\,p^8 - 10\,m^{24}\right)\,E^{16} \\
&+ \left(77520\,p^{26} - 57200\,m^8\,p^{18} + 7920\,m^{16}\,p^{10} - 80\,m^{24}\,p^2\right)\,E^{14} \\
&- \left(38760\,p^{28} - 40040\,m^8\,p^{20} + 9240\,m^{16}\,p^{12} - 280\,m^{24}\,p^4\right)\,E^{12} \\
&+ \left(15504\,p^{30} - 21840\,m^8\,p^{22} + 7920\,m^{16}\,p^{14} - 560\,m^{24}\,p^6\right)\,E^{10} \\
&- \left(4845\,p^{32} - 9100\,m^8\,p^{24} + 4950\,m^{16}\,p^{16} - 700\,m^{24}\,p^8 + 5\,m^{32}\right)\,E^8 \\
&- \left(2800\,m^8\,p^{26} - 1140\,p^{34} - 2200\,m^{16}\,p^{18} + 560\,m^{24}\,p^{10} - 20\,m^{32}\,p^2\right)\,E^6 \\
&- \left(190\,p^{36} - 600\,m^8\,p^{28} + 660\,m^{16}\,p^{20} - 280\,m^{24}\,p^{12} + 30\,m^{32}\,p^4\right)\,E^4 \\
&+ \left(20\,p^{38} - 80\,m^8\,p^{30} + 120\,m^{16}\,p^{22} - 80\,m^{24}\,p^{14} + 20\,m^{32}\,p^6\right)\,E^2 \\
&\qquad - \left(p^{40} - 5\,m^8\,p^{32} + 10\,m^{16}\,p^{24} - 10\,m^{24}\,p^{16} + 5\,m^{32}\,p^8 - m^{40}\right) = 0.
\end{aligned}
\tag{5.33}
$$

It can be factored into the following simple expression:

$$
-\left(p^2 - m^2 - E^2\right)^5 \left(p^2 + m^2 - E^2\right)^5 \left(p^4 - 2\,E^2\,p^2 + m^4 + E^4\right)^5 = 0.
\tag{5.34}
$$

Treating this equation as a fortieth order equation in $E$, we obtain 40 solutions. These solutions, the associated multiplicity and their interpretation, are tabulated in Table VIII.

Once again, as already established for the $(1,0)\oplus(0,1)$ and the $(3/2,0)\oplus(0,3/2)$ matter fields, the wave equation (5.31) satisfied by the $(2,0)\oplus(0,2)$ covariant spinors also propagates acausal solutions via the Green function $G^{(2,0)\oplus(0,2)}(x-x')$:

$$\left( \gamma^{\mu\nu\lambda\rho} \, \partial_\mu \partial_\nu \partial_\lambda \, \partial_\rho \; - \; m^4 \, I \right) G^{(2,0)\oplus(0,2)}(x - x') = \; \delta^4(x - x'). \tag{5.35}$$

The physical "causal" propagator

$$\langle \;\; | T[\Psi(x)^{(2,0)\oplus(0,2)} \; \overline{\Psi}^{(2,0)\oplus(0,2)}(x')] | \;\; \rangle \tag{5.36}$$

is readily constructed using the known physical "causal" covariant spinors given in section 3.4.

TABLE VIII

Dispersion relations $E = E(p, m)$ associated with Eq. (5.31); obtained as solutions of Eq. (5.34): $N_C\,(2) \; = \; 10$, and $N_A\,(2) \; = \; 30$

| (Multiplicity) Dispersion Relation | Interpretation |
|---|---|
| (5) $E = +\sqrt{p^2 + m^2}$ | Causal, "particle" $u_{\pm 2}(\vec{p}),\;\; u_{\pm 1}(\vec{p}),\;\; u_0(\vec{p})$ |
| (5) $E = -\sqrt{p^2 + m^2}$ | Causal, "antiparticle" $v_{\pm 2}(\vec{p}),\;\; v_{\pm 1}(\vec{p}),\;\; v_0(\vec{p})$ |
| (5) $E = +\sqrt{p^2 - m^2}$ | Acausal, Tachyonic |
| (5) $E = -\sqrt{p^2 - m^2}$ | Acausal, Tachyonic |
| (5) $E = +\sqrt{p^2 + i\,m^2}$ | Acausal |
| (5) $E = -\sqrt{p^2 + i\,m^2}$ | Acausal |
| (5) $E = +\sqrt{p^2 - i\,m^2}$ | Acausal |
| (5) $E = -\sqrt{p^2 - i\,m^2}$ | Acausal |

## 5.6 A Remark on the Massless Matter Fields

Tables VI, VII and VIII summarize the dispersion relations associated with the wave equations associated with the $(1,0)\oplus(0,1)$, $(3/2,0)\oplus(0,3/2)$ and $(2,0)\oplus(0,2)$ matter fields. A quick reference to these tabulated results immediately leads to the following conclusion:

> For $m=0$ all "acausal" dispersion relations, $E=E(p,m)$, associated with the equations explicitly constructed in the last chapter cease to be acausal and reduce to $E=\pm p$.

In the next chapter we will find that a well known set of linear equations for massless particles of arbitrary spin have unexpected kinematical acausality for $j\geq 1$. On the other hand, as just noted, the $m\rightarrow 0$ limit of the wave equations satisfied by[30] $(j,0)\oplus(0,j)$ relativistic covariant spinors are free from *all* kinematical acausality. This paradoxical situation will be resolved, and corrected, by following some general considerations and working out a specific example associated with the $(3/2,0)\oplus(0,3/2)$ matter field. The chapter will begin with a review, and repeat some of the algebraic equations, in order to construct an appropriate logical environment for a rather subtle matter. This matter was ignored, so far, to keep the logic unconvoluted.

30 At least for $j=1,3/2$ and 2 where this claim is explicitly verified.

# 6. KINEMATICAL ACAUSALITY IN EQUATIONS FOR MASSLESS PARTICLES AND ITS RESOLUTION VIA INTRODUCTION OF A CONSTRAINING PRINCIPLE

## 6.1 Kinematical Acausality in Weinberg's Equations for Massless Particles

We begin with the observation that we have become accustomed [52] to thinking of the free particle covariant spinors, like the Dirac $(1/2,0)\oplus(0,1/2)$ spinors, as solutions of relativistic wave equations. However, inspired by works of Ryder[53] for spin one half particles and Weinberg[21] for any spin, we have explicitly constructed the $(1,0)\oplus(0,1)$, $(3/2,0)\oplus(0,3/2)$ and $(2,0)\oplus(0,2)$ covariant spinors without reference to any wave equations. These covariant spinors incorporate the correct particle/antiparticle and the spinorial degrees of freedom in the $2(2j+1)$ particle and antiparticle covariant spinors: $u_\sigma(\vec{p})$ and $v_\sigma(\vec{p})$, $\sigma = j, j-1, \cdots -j$. In the "canonical representation" $u_\sigma(\vec{p})$ and $v_\sigma(\vec{p})$ are constructed by the application (after appropriate algebraic expansions) of the boost:

$$M_{_{CA}}(\vec{p}) = \begin{pmatrix} \cosh(\vec{J}\cdot\vec{\varphi}\,) & \sinh(\vec{J}\cdot\vec{\varphi}\,) \\ & \\ \sinh(\vec{J}\cdot\vec{\varphi}\,) & \cosh(\vec{J}\cdot\vec{\varphi}\,) \end{pmatrix}. \tag{6.1}$$

on the $2(2j+1)$ "rest covariant spinors" in the form of the $2(2j+1)$–dimensional column vectors

$$u_{+j}(\vec{0}) = \begin{pmatrix} N(j) \\ 0 \\ 0 \\ \vdots \\ 0 \end{pmatrix}, \quad \cdots, \quad v_{-j}(\vec{0}) = \begin{pmatrix} 0 \\ 0 \\ \vdots \\ 0 \\ N(j) \end{pmatrix}. \tag{6.2}$$

Here $N(j)$ is a convenient spin–dependent factor required to satisfy the requirements that in the $m \to 0$ limit: (i) The "rest covariant spinors" vanish, and (ii) The $m \to 0$ covariant spinors have a non-singular norm. As an example, we choose $N(j) = m^{3/2}$ for the $(3/2,0)\oplus(0,3/2)$ covariant spinors. The parameter $\varphi$ is defined as

$$\cosh(\varphi\,) \,=\, \gamma \,=\, \frac{1}{\sqrt{1-v^2}} \,=\, \frac{E}{m},\;\; \sinh(\varphi\,) \,=\, v\gamma \,=\, \frac{|\vec{p}\,|}{m}, \tag{6.3}$$

where $\vec{v}$ is the velocity which a particle at rest in the unprimed frame acquires when seen from the primed frame. In the canonical representation the angular momentum matrices $J_i$ have $J_z$ diagonal. The physical "propagators" needed in standard S–matrix calculations are constructed[23] using these $u_\sigma(\vec{p})$ and $v_\sigma(\vec{p})$ in evaluating appropriate vacuum expectation values of the time ordered product of relevant field operators.

While the $(j,0)\oplus(0,j)$ covariant spinors are obtained purely from group theoretical arguments , phenomenological studies cannot be carried out without an interaction Lagrangian density. A large class of interaction Lagrangian densities are suggested if we construct wave equations which these covariant spinors satisfy. A class of relativistic wave equations satisfied by the $(j,0)\oplus(0,j)$ covariant spinors are obtained from the coupled equations (in the "chiral representation" ):

$$\begin{aligned} \phi^{R}(\vec{p}) &= \Big(\cosh(2\vec{J}\cdot\vec{\varphi}\,) + \sinh(2\vec{J}\cdot\vec{\varphi}\,)\Big) \;\; \phi^{L}(\vec{p}) \\ \phi^{L}(\vec{p}) &= \Big(\cosh(2\vec{J}\cdot\vec{\varphi}\,) - \sinh(2\vec{J}\cdot\vec{\varphi}\,)\Big) \;\; \phi^{R}(\vec{p}), \end{aligned} \tag{6.4}$$

where $\phi^R(\vec{p})$ are the matter fields associated with the $SU(2)_R$ generated by $\vec{S}_R = \frac{1}{2}(\vec{J}+i\vec{K})$, and $\phi^L(\vec{p})$ are the matter fields associated with the $SU(2)_L$ generated by $\vec{S}_L = \frac{1}{2}(\vec{J}-i\vec{K})$. The $\vec{J}$ are the three spin dependent $(2j+1)\times(2j+1)$ matrices of angular momentum and $\vec{K}$ are the generators of the Lorentz boosts. The "chiral representation" covariant spinors

$$\psi_{CH}(\vec{p}) = \begin{pmatrix} \phi^{R}(\vec{p}) \\ \\ \phi^{L}(\vec{p}) \end{pmatrix} \tag{6.5}$$

are related to the canonical representation covariant spinors by the expression

$$\psi_{CA}(\vec{p}) = A\;\psi_{CH}(\vec{p}),\quad A = \frac{1}{\sqrt{2}}\begin{pmatrix} I & I \\ I & -I \end{pmatrix}\,. \tag{6.6}$$

By imposing a physical criterion, which has not been explicitly stated so far, a general wave equation satisfied by the $(j,0)\oplus(0,j)$ covariant spinors is obtained from the coupled Eqs. (6.4). It has the general form:

$$\left(\gamma_{\{\mu\}}\,p^{[\mu]}\;-\;m^{2j}\,I\right)\,\psi(\vec{p})\;=\;0, \tag{6.7}$$

where $\{\mu\}$ is a set of $2j$ Lorentz indices. $\gamma_{\{\mu\}}$ are fully symmetric (in the Lorentz indices)

$$\frac{N\,(N\,+\,1)\,\cdots\,(N\,+\,S\,-\,1)}{S!} \tag{6.8}$$

set of $2(2j+1)\times 2(2j+1)$ spin–$j$ $\gamma$ matrices. Here $S=2j$, the number of indices on $\gamma_{\{\mu\}}$; and for the $(1,3)$ spacetime under consideration $N=4$. A basic requirement for *any* solutions to exist is:

$$\text{Determinant}\;\left(\gamma_{\{\mu\}}\,p^{[\mu]}\;-\;m^{2j}\,I\right)\;=\;0, \tag{6.9}$$

For a given $j$ this "existence requirement" can be interpreted as a $2j[2(2j+1)]th$ order equation in $E$. By working out specific examples, we discover that there are

$$N_A(j)\;=\;2j[2(2j+1)]-2(2j+1)\;=\;2(2j-1)(2j+1) \tag{6.10}$$

"acausal" solutions – that is solutions for which:

$$E^2\;\neq\;p^2\;+\;m^2. \tag{6.11}$$

The remaining "causal" solutions, that is solutions for which $E^2\;=\;p^2\;+\;m^2$,

$$N_C(j)\;=\;2(2j+1), \tag{6.12}$$

we identify with the $(j,0)\oplus(0,j)$ relativistic covariant spinors.

For the $(1,0)\oplus(0,1)$, $(3/2,0)\oplus(0,3/2)$ and $(2,0)\oplus(0,2)$ matter fields we have verified through explicit calculations that in the $m\rightarrow 0$ limit only $u_{\pm j}(\vec{p})$ and $v_{\pm j}(\vec{p})$ are *non-null.* In addition *all* "acausal" solutions turn into "causal" solutions, as we noted in Section 5.6.

On the other hand it is generally recognised[33] that in the $m \to 0$ limit the right and left handed matter fields decouple and satisfy the the following set of linear equations

$$\begin{aligned} \left(\vec{J}\cdot\vec{p} + j\,p^0\right)\,\phi^L(\vec{p}) \,&=\, 0 \\ \left(\vec{J}\cdot\vec{p} - j\,p^0\right)\,\phi^R(\vec{p}) \,&=\, 0. \end{aligned} \tag{6.13}$$

That such a limit is allowed for the $(j,0)\oplus(0,j)$ matter field was shown by Weinberg in Sec. III of Ref. 33, and is shown here in Appendix. As before, the kinematical causality of these equations is readily studied by examining the solutions of

$$\text{Determinant}\,\left(\vec{J}\cdot\vec{p} \pm j\,p^0\right)\,=\,0. \tag{6.14}$$

When this is done we find that for $j\geq 1$ Eq. (6.14) not only admits the causal solutions $E\,=\,\pm p$, but also one or more acausal solutions $E\,\neq\,\pm p$. This *contradicts* our earlier result obtained by taking the $m\to 0$ limit of the massive case. For the sake of concreteness, the dispersion relations implied by (6.14) are tabulated in Table IX for $j\,=1,\,3/2,\,2$.

TABLE IX

Dispersion relations $E=E(p)$ associated with Eqs. (6.13)

| Spin $j$ | Dispersion Relation |
|---|---|
| 1 | $E\,=\pm p,\;E\,=0$ |
| 3/2 | $E\,=\pm p,\;E\,=\pm(p/3)$ |
| 2 | $E\,=\pm p,\;E\,=\pm(p/2),\;E\,=0$ |

## 6.2 A Constraining Principle

The resolution of the paradoxical situation created above is now studied by carefully examining the transition from the essentially classical group theoretical arguments which yield the coupled Eqs. (6.4) and the quantum mechanically interpreted Eqs. (6.7). The physical criterion which we now impose is in the form of a constraining principle. It reads:

> The freedom provided by the classical c–number equivalence of the substitution $m^2 \leftrightarrow \eta_{\mu\nu}\, p^\mu p^\nu$ is constrained in the construction of the quantum mechanical equations of motion so that the resulting equations are free from *all* kinematical "acausality" in the $m \to 0$ limit.

The equations which violate this constraining principle will be termed "pathological."

For the sake of simplicity we shall do this by studying the $(3/2,0)\oplus(0,3/2)$ matter field. We begin with $m \neq 0$ case. Using the standard expansions for $\cosh(\vec{J}\cdot\vec{\varphi})|_{(j=3/2)}$ and $\sinh(\vec{J}\cdot\vec{\varphi})|_{(j=3/2)}$ and freely exploiting the identities (6.3), the coupled Eqs. (6.4) can be written in the form:

$$\phi^{R(j=3/2)}(\vec{p}) = \frac{1}{m^3}\Big[m^2(p^0 + 2\vec{J}\cdot\vec{p}) + \frac{1}{6}\{(2\vec{J}\cdot\vec{p})^2 - p^2\}\{2\vec{J}\cdot\vec{p} + 3p^0\}\Big] \times \phi^{L(j=3/2)}(\vec{p}), \tag{6.15}$$

$$\phi^{L(j=3/2)}(\vec{p}) = \frac{1}{m^3}\Big[m^2(p^0 - 2\vec{J}\cdot\vec{p}) - \frac{1}{6}\{(2\vec{J}\cdot\vec{p})^2 - p^2\}\{2\vec{J}\cdot\vec{p} - 3p^0\}\Big] \times \phi^{R(j=3/2)}(\vec{p}). \tag{6.16}$$

Introducing the eight component $(3/2,0)\oplus(0,3/2)$ relativistic covariant spinor

$$\psi_{CH}^{(j=3/2)}(\vec{p}) = \begin{pmatrix} \phi^{R(j=3/2)}(\vec{p}) \\ \\ \phi^{L(j=3/2)}(\vec{p}) \end{pmatrix}. \tag{6.17}$$

and the $8 \times 8$ spin–3/2 $\gamma$ matrices

$$\gamma_{\mu\nu\lambda}\, p^\mu\, p^\nu\, p^\lambda = \begin{pmatrix} 0 & \Big[\eta_{\mu\nu}\, p^\mu\, p^\nu (p^0 + 2\vec{J}\cdot\vec{p}) + \\ & \frac{1}{6}\{(2\vec{J}\cdot\vec{p})^2 - \vec{p}^{\,2}\}\{2\vec{J}\cdot\vec{p} + 3p^0\}\Big] \\ \Big[\eta_{\mu\nu}\, p^\mu\, p^\nu (p^0 - 2\vec{J}\cdot\vec{p}) - & \\ \frac{1}{6}\{(2\vec{J}\cdot\vec{p})^2 - \vec{p}^{\,2}\}\{2\vec{J}\cdot\vec{p} - 3p^0\}\Big] & 0 \end{pmatrix}, \tag{6.18}$$

we obtain the $(3/2, 0) \oplus (0, 3/2)$ relativistic wave equation

$$(\gamma_{\mu\nu\lambda}\, p^\mu p^\nu p^\lambda \,-\, m^3\, I)\psi^{(j=3/2)}(\vec{p}) = 0. \tag{6.19}$$

The $\eta_{\mu\nu}\, p^\mu p^\nu$ which appears in the $\gamma_{\mu\nu\lambda}$ is the crucial factor. It has been obtained by the substitution $m^2 \,\to\, \eta_{\mu\nu}\, p^\mu p^\nu$, based on the criterion above. It is the *only* equation which can be constructed out of the coupled Eqs. (6.4) which satisfies the criterion: The $m \to 0$ limit yield the causal solutions, $E \,=\, \pm p$, for *all* the 24 solutions[30]. The freedom in the form of four choices which apparently seem to exist via the substitution: $m^2 \,\leftrightarrow\, \eta_{\mu\nu}\, p^\mu p^\nu$, disappear with the physical constraint we impose. As a result of invoking this requirement we are left with one unique equation, that is Eq. (6.19) with $\gamma_{\mu\nu\lambda}$ given by (6.18).

Besides Eq. (6.19) there are three alternate equations, which can be constructed exploiting the ambiguity mentioned above. These are:

$$\left(\Gamma^{(1)}_{\mu\nu\lambda}\, p^\mu p^\nu p^\lambda \,-\, \eta_{\mu\nu}\, p^\mu p^\nu\, m\, I\right)\psi^{(j=3/2)}(\vec{p}) = 0, \tag{6.20}$$

$$\left(\Gamma^{(2)}_{\mu\nu\lambda}\, p^\mu p^\nu p^\lambda \,-\, m^3\, I\right)\psi^{(j=3/2)}(\vec{p}) = 0, \tag{6.21}$$

$$\left(\Gamma^{(3)}_{\mu\nu\lambda}\, p^\mu p^\nu p^\lambda \,-\, \eta_{\mu\nu}\, p^\mu p^\nu\, m\, I\right)\psi^{(j=3/2)}(\vec{p}) = 0, \tag{6.22}$$

30 For the massive case the spin 3/2 equation satisfying the physical criterion also allows for exactly $2(2 \times \frac{3}{2} \,+\, 1) = 8$ physical “causal” solutions.

with

$$\Gamma^{(1)}_{\mu\nu\lambda}\, p^\mu\, p^\nu\, p^\lambda = \Gamma^{(2)}_{\mu\nu\lambda}\, p^\mu\, p^\nu\, p^\lambda$$

$$= \begin{pmatrix} 0 & \Big[m^2\,(p^0 + 2\vec{J}\cdot\vec{p}) + \frac{1}{6}\{(2\vec{J}\cdot\vec{p})^2 - \vec{p}^{\;2}\}\{2\vec{J}\cdot\vec{p} + 3p^0\}\Big] \\ \Big[m^2(p^0 - 2\vec{J}\cdot\vec{p}) - \frac{1}{6}\{(2\vec{J}\cdot\vec{p})^2 - \vec{p}^{\;2}\}\{2\vec{J}\cdot\vec{p} - 3p^0\}\Big] & 0 \end{pmatrix}, \tag{6.23}$$

and

$$\Gamma^{(3)}_{\mu\nu\lambda}\, p^\mu\, p^\nu\, p^\lambda = \gamma_{\mu\nu\lambda}\, p^\mu\, p^\nu\, p^\lambda$$

$$= \begin{pmatrix} 0 & \Big[\eta_{\mu\nu}\, p^\mu p^\nu\,(p^0 + 2\vec{J}\cdot\vec{p}) + \frac{1}{6}\{(2\vec{J}\cdot\vec{p})^2 - \vec{p}^{\;2}\}\{2\vec{J}\cdot\vec{p} + 3p^0\}\Big] \\ \Big[\eta_{\mu\nu}\, p^\mu\, p^\nu(p^0 - 2\vec{J}\cdot\vec{p}) - \frac{1}{6}\{(2\vec{J}\cdot\vec{p})^2 - \vec{p}^{\;2}\}\{2\vec{J}\cdot\vec{p} - 3p^0\}\Big] & 0 \end{pmatrix}. \tag{6.24}$$

Tables X, XI and XII provide us dispersion relations associated with Eqs. (6.20), (6.21), and (6.22). A careful and detailed study of these tabulated dispersion relations reveals that none

of the alternate equations satisfy the the physical criteria defined above. In fact for two of the three alternate equations we find that since

$$\begin{aligned}
&\text{Determinant}\ \left(\Gamma^{(1)}_{\mu\nu\lambda}\, p^\mu p^\nu p^\lambda \;-\; \eta_{\mu\nu}\, p^\mu p^\nu\, m\, I\right) \\
&= m^4\,(p-E)^2(p+E)^2\left(p^2+m^2-E^2\right)^4\left(16\,p^4+9\,m^2\,p^2-E^2m^2\right)^2 \;=\; 0,
\end{aligned} \tag{6.25}$$

$$\begin{aligned}
&\text{Determinant}\ \left(\Gamma^{(2)}_{\mu\nu\lambda}\, p^\mu p^\nu p^\lambda \;-\; m^3\, I\right) \\
&= m^8\,\left(p^2+m^2-E^2\right)^2\left(16\,p^6-16\,E^2p^4-16\,Em^2p^3-9\,m^4p^2-m^6+E^2m^4\right) \\
&\times\left(16\,p^6-16\,E^2p^4+16\,Em^2p^3-9\,m^4p^2-m^6+E^2m^4\right)\;=0,
\end{aligned} \tag{6.26}$$

vanish *identically* for $m \to 0$ , *no* solutions exist for the massless matter fields for these two equation. Parenthetically, it should be noted that even though Eq. (6.20) seems to admit $E = \pm p$, with the correct multiplicity of 2, the associated solution is null because of the observation just made:

$$\text{Determinant}\ \left(\Gamma^{(1)}_{\mu\nu\lambda}\, p^\mu p^\nu p^\lambda \;-\; \eta_{\mu\nu}\, p^\mu p^\nu\, m\, I\right)\Big|_{m\to 0} = 0. \tag{6.27}$$

TABLE X

Dispersion relations $E = E(p,m)$ associated with Eq. (6.20)

| (Multiplicity) Dispersion Relation | Interpretation |
|---|---|
| (4) $E = +\sqrt{p^2+m^2}$ | Causal, "particle." $u_{\pm\frac{3}{2}}(\vec{p})$, $u_{\pm\frac{1}{2}}(\vec{p})$. |
| (4) $E = -\sqrt{p^2+m^2}$ | Causal, "antiparticle." $v_{\pm\frac{3}{2}}(\vec{p})$, $v_{\pm\frac{1}{2}}(\vec{p})$. |
| (2) $E = +p$ | $m \Rightarrow 0$ (But, see text.) |
| (2) $E = -p$ | $m \Rightarrow 0$ (But, see text.) |

4 Additional acausal dispersion relations, with two fold degeneracy in $p$

$$
\begin{aligned}
E &= \pm\frac{p\sqrt{16p^2+9m^2}}{m}, \\
p &= -\frac{1}{4\sqrt{2}}\left(m\sqrt{81\,m^2+64\,E^2}-9\,m^2\right)^{1/2}, \\
p &= +\frac{1}{4\sqrt{2}}\left(m\sqrt{81\,m^2+64\,E^2}-9\,m^2\right)^{1/2}, \\
p &= -\frac{1}{4\sqrt{2}}\left(-m\sqrt{81\,m^2+64\,E^2}-9\,m^2\right)^{1/2}, \\
p &= +\frac{1}{4\sqrt{2}}\left(-m\sqrt{81\,m^2+64\,E^2}-9\,m^2\right)^{1/2}.
\end{aligned}
$$

TABLE XI

Dispersion relations $E = E(p,m)$ associated with Eq. (6.21)

| (Multiplicity) Dispersion Relation | Interpretation |
|---|---|
| (2) $E = +\sqrt{p^2+m^2}$ | Causal, "particle." Not all: $u_{\pm\frac{3}{2}}(\vec{p})$, $u_{\pm\frac{1}{2}}(\vec{p})$ , allowed. |
| (2) $E = -\sqrt{p^2+m^2}$ | Causal, "antiparticle." Not all: $v_{\pm\frac{3}{2}}(\vec{p})$, $v_{\pm\frac{1}{2}}(\vec{p})$, allowed. |

4 Acausal dispersion relations, with three fold degeneracy in $p$

$$
\begin{aligned}
E &= +\frac{\left(256\,p^{10} - 96\,m^4\,p^6 - 16\,m^6p^4 + 9\,m^8p^2 + m^{10}\right)^{1/2} + 8\,m^2p^3}{(16\,p^4 - m^4)} \\
E &= -\frac{\left(256\,p^{10} - 96\,m^4\,p^6 - 16\,m^6p^4 + 9\,m^8p^2 + m^{10}\right)^{1/2} - 8\,m^2p^3}{(16\,p^4 - m^4)} \\
E &= +\frac{\left(256\,p^{10} - 96\,m^4\,p^6 - 16\,m^6p^4 + 9\,m^8p^2 + m^{10}\right)^{1/2} - 8\,m^2p^3}{(16\,p^4 - m^4)} \\
E &= -\frac{\left(256\,p^{10} - 96\,m^4\,p^6 - 16\,m^6p^4 + 9\,m^8p^2 + m^{10}\right)^{1/2} + 8\,m^2p^3}{(16\,p^4 - m^4)}.
\end{aligned}
$$

The three fold degeneracy in $p$ arises because the above four $E's$ are obtained as solutions of

$$16\,p^6 - 16\,E^2p^4 \pm 16\,Em^2p^3 - 9\,m^4p^2 - m^6 + E^2m^4 \;= 0$$

TABLE XII

Dispersion relations $E = E(p,m)$ associated with Eq. (6.22)

| (Multiplicity) Dispersion Relation | Interpretation |
|---|---|
| (2) $E = +\sqrt{p^2 + m^2}$ | Causal, "particle." Not all: $u_{\pm\frac{3}{2}}(\vec{p})$, $u_{\pm\frac{1}{2}}(\vec{p})$, allowed. |
| (2) $E = -\sqrt{p^2 + m^2}$ | Causal, "antiparticle." Not all: $v_{\pm\frac{3}{2}}(\vec{p})$, $v_{\pm\frac{1}{2}}(\vec{p})$, allowed. |
| (7) $E = +p$ | $\Rightarrow\ m = 0$ |
| (7) $E = -p$ | $\Rightarrow\ m = 0$ |

For 6 acausal dispersion relations, see continuation of this table on the next page.

Table XII continued.

| 6 Acausal dispersion relations associated with Eq. (6.22) |
|---|

$$E = \pm\frac{p}{3} - \left(\frac{\sqrt{3}\,i+1}{2}\right)\left(\frac{i\,\sqrt{128\,p^6+352\,m^2\,p^4+25\,m^4\,p^2+m^6}}{3\,\sqrt{3}} \mp \frac{9\,m^2\,p-136\,p^3}{27}\right)^{1/3}$$
$$+ \left(\frac{\sqrt{3}\,i-1}{2}\right)\left(\frac{28\,p^2+3\,m^2}{9}\right)\left(\frac{i\,\sqrt{128\,p^6+352\,m^2\,p^4+25\,m^4\,p^2+m^6}}{3\,\sqrt{3}} \mp \frac{9\,m^2\,p-136\,p^3}{27}\right)^{-1/3}$$

$$E = \pm\frac{p}{3} + \left(\frac{\sqrt{3}\,i-1}{2}\right)\left(\frac{i\,\sqrt{128\,p^6+352\,m^2\,p^4+25\,m^4\,p^2+m^6}}{3\,\sqrt{3}} \mp \frac{9\,m^2\,p-136\,p^3}{27}\right)^{1/3}$$
$$- \left(\frac{\sqrt{3}\,i+1}{2}\right)\left(\frac{28\,p^2+3\,m^2}{9}\right)\left(\frac{i\,\sqrt{128\,p^6+352\,m^2\,p^4+25\,m^4\,p^2+m^6}}{3\,\sqrt{3}} \mp \frac{9\,m^2\,p-136\,p^3}{27}\right)^{-1/3}$$

Continued to the next page.

Table XII continued.

| Continued tabulation of 6 Acausal dispersion relations associated with Eq. (6.22) |
| --- |
| $$E = \pm\frac{p}{3} + \left(\frac{i\,\sqrt{128\,p^6 + 352\,m^2\,p^4 + 25\,m^4\,p^2 + m^6}}{3\,\sqrt{3}} \mp \frac{9\,m^2\,p - 136\,p^3}{27}\right)^{1/3} + \left(\frac{28\,p^2 + 3\,m^2}{9}\right)\left(\frac{i\,\sqrt{128\,p^6 + 352\,m^2\,p^4 + 25\,m^4\,p^2 + m^6}}{3\,\sqrt{3}} \mp \frac{9\,m^2\,p - 136\,p^3}{27}\right)^{-1/3}$$ |

The $m = 0$ equations, which have *no* "acausal" solutions, for the uncoupled right and left handed matter fields follow directly from Eqs. (6.18) and (6.19). These equations read:

$$\begin{aligned}
\Big[\eta_{\mu\nu}\,p^\mu\,p^\nu(p^0 + 2\vec{J}\cdot\vec{p}) \;+\; \frac{1}{6}\{(2\vec{J}\cdot\vec{p})^2 - \vec{p}^{\;2}\}\{2\vec{J}\cdot\vec{p} + 3p^0\}\Big]\,\phi^{L(j=3/2)}(\vec{p}) \;&= 0 \\
\Big[\eta_{\mu\nu}\,p^\mu\,p^\nu(p^0 - 2\vec{J}\cdot\vec{p}) \;-\; \frac{1}{6}\{(2\vec{J}\cdot\vec{p})^2 - \vec{p}^{\;2}\}\{2\vec{J}\cdot\vec{p} - 3p^0\}\Big]\,\phi^{R(j=3/2)}(\vec{p}) \;&= 0.
\end{aligned} \tag{6.28}$$

Finally we note that one may be tempted to argue that Eq. (6.13) for $j = 3/2$ can be arrived at, in the pathological case violating the physical criterion defined above, by merely replacing $\eta_{\mu\nu}\,p^\mu p^\nu$ by $m^2$ in (6.18), in the $m \to 0$ limit. However, this is not so. The reason is quite simple. The operator

$$\Big((2\vec{J}\cdot\vec{p})^2 \;-\; \vec{p}^{2}\Big)\Big|_{j=3/2} \tag{6.29}$$

is singular.

We conclude that on careful examination the linear Eqs. (6.13) are inconsistent with the relativistic wave equation satisfied by the massive $(j,0)\oplus(0,j)$ matter field in the $m\to 0$ limit. We have discovered the origin of this contradiction, and resolved the apparently paradoxical situation by invoking a set of physical criterion.

## 6.3 A Stronger Form of the Constraining Principle

For spin 1/2 we see that the constraining principle automatically yields us an equation which has $2(2j+1)$ causal particle/antiparticle solutions for the massive matter fields. In the next section we will verify that same is true for the $(1,0)\oplus(0,1)$ matter field. However, for some $j\geq 2$ it *may* happen that principle we have proposed need to be strengthened to accomplish this. For this reason we now introduce its stronger form in anticipation. It reads:

> The freedom provided by the classical c–number equivalence of the substitution $m^2 \leftrightarrow \eta_{\mu\nu}\,p^\mu p^\nu$ is constrained in the construction of the quantum mechanical equations of motion so that the resulting equations are free from *all* kinematical "acausality" in the $m\to 0$ limit *and* allow exactly $2(2j+1)$ "causal" particle/antiparticle solutions for the massive $(j,0)\oplus(0,j)$ matter fields.

## 6.4 Some Remarks on the $m\to 0$ Limit for Spin One, Alternate Equations, and an Observation

*Remarks:* We note that for spin 1, the Eqs. (6.13) contain acausality only of the form $E=0$. One *may* be able to live with this "acausality" because all it says is that there exists a solution which has no energy content. However, (6.13) for $j=1$ does not follow from the $m\to 0$ limit of the equation satisfying the physical principle introduced in the last section in any straight forward fashion. The $m\to 0$ limit of (4.28) reads:

$$\begin{aligned}\Big[\eta_{\mu\nu}p^\mu p^\nu\ +2(\vec{J}\cdot\vec{p})\big\{(\vec{J}\cdot\vec{p})+p^0\big\}\Big]\,\phi^{L(j=1)}=0\\ \Big[\eta_{\mu\nu}p^\mu p^\nu\ +2(\vec{J}\cdot\vec{p})\big\{(\vec{J}\cdot\vec{p})-p^0\big\}\Big]\,\phi^{R(j=1)}=0.\end{aligned}\tag{6.30}$$

In order to see how close we can get to Weinberg's equations, we may argue that at least for the massless causal matter fields we may postulate:

$$(\eta_{\mu\nu}\, p^\mu p^\nu)\, \phi^{rel.} = 0, \tag{6.31}$$

and hence Eqs. (6.30) are equivalent to Weinberg's Eq. (6.13). But for that to be true we must have:

$$\text{Determinant } \left(2(\vec{J}\cdot\vec{p})\right)\Big|_{j=1},\neq 0 \tag{6.32}$$

However, this is obviously *not* true.

*Alternate Equations:* We already have a equation satisfied by the $(1,0)\oplus(0,1)$ covariant spinors which satisfies the constraining principle introduced above. It reads:

$$\left(\gamma_{\mu\nu}p^\mu p^\nu \;-\; m^2\, I\right)\psi^{(j=1)}(\vec{p}) = 0. \tag{6.33}$$

where (in chiral representation) the $6\times 6$ spin–1 $\gamma$ matrices are

$$\gamma_{\mu\nu}p^\mu p^\nu = \begin{pmatrix} 0 & \Big[\eta_{\mu\nu}p^\mu p^\nu + 2(\vec{J}\cdot\vec{p})(\vec{J}\cdot\vec{p}) + 2(\vec{J}\cdot\vec{p})p^0\Big] \\ \Big[\eta_{\mu\nu}p^\mu p^\nu + 2(\vec{J}\cdot\vec{p})(\vec{J}\cdot\vec{p}) - 2(\vec{J}\cdot\vec{p})p^0\Big] & 0 \end{pmatrix}. \tag{6.34}$$

We now enumerate the three alternate equations, and verify that none of them satisfies the required physical criterion thus establishing the uniqueness[31] of the

31 "Uniquness", as far as the set of equations which can be directly derived from the coupled Eqs. (6.4) is concerned.

wave equation (6.33) satisfied by the $(1,0)\oplus(0,1)$ matter field.

$$\left(\Gamma^{(1)}_{\mu\nu}p^\mu p^\nu \;-\; \eta_{\mu\nu}\,p^\mu p^\nu\, I\right)\psi^{^{(j=1)}}(\vec{p}) = 0. \tag{6.35}$$

$$\left(\Gamma^{(2)}_{\mu\nu}p^\mu p^\nu \;-\; m^2\, I\right)\psi^{^{(j=1)}}(\vec{p}) = 0. \tag{6.36}$$

$$\left(\Gamma^{(3)}_{\mu\nu}p^\mu p^\nu \;-\; \eta_{\mu\nu}\,p^\mu p^\nu\, I\right)\psi^{^{(j=1)}}(\vec{p}) = 0. \tag{6.37}$$

with

$$\Gamma^{(1)}_{\mu\nu}p^\mu p^\nu \;=\; \Gamma^{(2)}_{\mu\nu}p^\mu p^\nu$$

$$\begin{pmatrix} 0 & \Big[m^2 + 2(\vec{J}\cdot\vec{p})(\vec{J}\cdot\vec{p}) + 2(\vec{J}\cdot\vec{p})p^0\Big] \\ \Big[m^2 + 2(\vec{J}\cdot\vec{p})(\vec{J}\cdot\vec{p}) - 2(\vec{J}\cdot\vec{p})p^0\Big] & 0 \end{pmatrix} \tag{6.38}$$

and

$$\Gamma^{(3)}_{\mu\nu}p^\mu p^\nu \;=\; \gamma_{\mu\nu}p^\mu p^\nu$$

$$\begin{pmatrix} 0 & \Big[\eta_{\mu\nu}p^\mu p^\nu + 2(\vec{J}\cdot\vec{p})(\vec{J}\cdot\vec{p}) + 2(\vec{J}\cdot\vec{p})p^0\Big] \\ \Big[\eta_{\mu\nu}p^\mu p^\nu + 2(\vec{J}\cdot\vec{p})(\vec{J}\cdot\vec{p}) - 2(\vec{J}\cdot\vec{p})p^0\Big] & 0 \end{pmatrix}. \tag{6.39}$$

A simple algebraic exercise yields:

$$\begin{aligned}\text{Determinant }&\Big(\Gamma^{(2)}_{\mu\nu}p^\mu p^\nu \;-\; \eta_{\mu\nu}\,p^\mu p^\nu\, I \;=\; \{0\}^I \;= \\ &\qquad\text{Determinant }\left(\Gamma^{(3)}_{\mu\nu}p^\mu p^\nu \;-\; \eta_{\mu\nu}\,p^\mu p^\nu\, I\right).\end{aligned} \tag{6.40}$$

Here symbol $\{0\}^I$ means "identically equal to zero." Consequently alternate Eqs. (6.36) and (6.37) have *no* solutions. The dispersion relations associated with the only remaining equation, that is Eq. (6.35), are tabulated in table XIII. An inspection of this table immediately confirms the result claimed earlier, that this equation too does not meet the physical criterion of Sec. 6.1.

TABLE XIII

Dispersion relations $E = E(p, m)$ associated with Eq. (6.35)

| (Multiplicity) Dispersion Relation | Interpretation |
|---|---|
| (3) $E = +\sqrt{p^2 + m^2}$ | Causal, "particle." $u_{\pm 1}(\vec{p}),\ u_0(\vec{p})$. |
| (3) $E = -\sqrt{p^2 + m^2}$ | Causal, "antiparticle." $v_{\pm 1}(\vec{p}),\ v_0(\vec{p})$. |
| (1) $E = +\sqrt{p^2 - m^2}$ | Acausal, Tachyonic |
| (1) $E = -\sqrt{p^2 - m^2}$ | Acausal, Tachyonic |
| (2) $E = +\sqrt{-3p^2 - m^2}$ | Acausal |
| (2) $E = -\sqrt{-3p^2 - m^2}$ | Acausal |

*An Observation:* Note that alternate Eq. (6.37) has *no* reference to "mass." Hence what does $m \to 0$ limit mean for this case. This circumstance makes it a rather interesting equation. But, as we have seen this equation has no solution. An inspection of (6.7) should convince the reader that such equations exist for *all* bosonic $j \geq 1$. It may turn out that for some bosonic $j > 1$

$$\text{Determinant} \left( \gamma_{\{\mu\}}\, p^{[\mu]} \,-\, m^{2j}\, I \right) \{\neq 0\}^I . \tag{6.41}$$

The symbol $\{\neq 0\}^I$ means "not identically equal to zero." Such an equation would invite a physical interpretation if it also happens to satisfy the physical criterion of Sec. 6.1.

# 7. CONSERVED CURRENT, TWO PHOTON MEDIATED PARTICLE PRODUCTION AND CONCLUSIONS

## 7.1 Algebra of Spin–1 Gamma Matrices and Conserved Current Density

Since we have obtained the $(1,0)\oplus(0,1)$ covariant spinors in the canonical representation we begin by transforming the chiral representation $\gamma^{\mu\nu}$, given in Chap. 4, into the canonical representation $\gamma^{\mu\nu}$ using the matrix $A$ defined in Chap. 6. The result is:

[canonical representation]

$$\gamma_{00} = \begin{pmatrix} I & 0 \\ 0 & -I \end{pmatrix}, \tag{7.1}$$

$$\gamma_{i0} = \gamma_{0i} = \begin{pmatrix} 0 & -J_i \\ J_i & 0 \end{pmatrix}, \tag{7.2}$$

$$\gamma_{ij} = \gamma_{ji} = \begin{pmatrix} I & 0 \\ 0 & -I \end{pmatrix} \eta_{ij} + \begin{pmatrix} \{J_i, J_j\} & 0 \\ 0 & -\{J_i, J_j\} \end{pmatrix}. \tag{7.3}$$

The "commutator algebra" satisfied by these matrices, is as follows:

$$[\gamma_{00}, \gamma_{ij}] = [\gamma_{00}, \gamma_{ji}] = 0, \tag{7.4}$$

$$[\gamma_{00}, \gamma_{0i}] = [\gamma_{00}, \gamma_{i0}] = -2 \begin{pmatrix} 0 & J_i \\ J_i & 0 \end{pmatrix}, \tag{7.5}$$

$$[\gamma_{i0}, \gamma_{j0}] = [\gamma_{i0}, \gamma_{0j}] = [\gamma_{0i}, \gamma_{j0}] = [\gamma_{0i}, \gamma_{0j}] = - \begin{pmatrix} [J_i, J_j] & 0 \\ 0 & [J_i, J_j] \end{pmatrix}, \tag{7.6}$$

$$\begin{aligned}[\gamma_{i0}, \gamma_{kl}] = [\gamma_{i0}, \gamma_{lk}] =& [\gamma_{0i}, \gamma_{kl}] = [\gamma_{i0}, \gamma_{lk}] = \begin{pmatrix} 0 & J_i \\ J_i & 0 \end{pmatrix} 2\eta_{kl} + \\ & \begin{pmatrix} 0 & \{J_i, \{J_k, J_l\}\} \\ \{J_i, \{J_k, J_l\}\} & 0 \end{pmatrix},\end{aligned} \tag{7.7}$$

$$[\gamma_{ij},\gamma_{kl}] = [\gamma_{ij},\gamma_{lk}] = [\gamma_{ji},\gamma_{kl}] = [\gamma_{ji},\gamma_{lk}] = \begin{pmatrix} [\{J_i,J_j\},\{J_k,J_l\}] & 0 \\ 0 & [\{J_i,J_j\},\{J_k,J_l\}] \end{pmatrix}. \tag{7.8}$$

The "anticommutator algebra" is similarly seen to be:

$$\{\gamma_{00},\gamma_{0i}\} = \{\gamma_{00},\gamma_{i0}\} = 0, \tag{7.9}$$

$$\{\gamma_{00},\gamma_{ij}\} = \{\gamma_{00},\gamma_{ji}\} = 2\left[\begin{pmatrix} 1 & 0 \\ 0 & 1 \end{pmatrix}\eta_{ij} + \begin{pmatrix} \{J_i,J_j\} & 0 \\ 0 & \{J_i,J_j\} \end{pmatrix}\right], \tag{7.10}$$

$$\{\gamma_{i0},\gamma_{j0}\} = \{\gamma_{i0},\gamma_{0j}\} = \{\gamma_{0i},\gamma_{j0}\} = \{\gamma_{0i},\gamma_{0j}\} = -\begin{pmatrix} \{J_i,J_j\} & 0 \\ 0 & \{J_i,J_j\} \end{pmatrix}, \tag{7.11}$$

$$\{\gamma_{i0},\gamma_{kl}\} = \{\gamma_{i0},\gamma_{lk}\} = \{\gamma_{0i},\gamma_{kl}\} = \{\gamma_{i0},\gamma_{lk}\} = \begin{pmatrix} 0 & [J_i,\{J_k,J_l\}] \\ [J_i,\{J_k,J_l\}] & 0 \end{pmatrix}, \tag{7.12}$$

$$\{\gamma_{ij},\gamma_{kl}\} = \{\gamma_{ij},\gamma_{lk}\} = \{\gamma_{ji},\gamma_{kl}\} = \{\gamma_{ji},\gamma_{lk}\} = \begin{pmatrix} \{\{J_i,J_j\},\{J_k,J_l\}\} & 0 \\ 0 & \{\{J_i,J_j\},\{J_k,J_l\}\} \end{pmatrix} + 2\left[\begin{pmatrix} 1 & 0 \\ 0 & 1 \end{pmatrix}\eta_{ij}\eta_{kl} + \begin{pmatrix} \{J_i,J_j\} & 0 \\ 0 & \{J_i,J_j\} \end{pmatrix}\eta_{kl} + \begin{pmatrix} \{J_k,J_l\} & 0 \\ 0 & \{J_k,J_l\} \end{pmatrix}\eta_{ij}\right], \tag{7.13}$$

In preparation for the construction of the $(1,0)\oplus(0,1)$ conserved current density we point to a noteworthy feature of the spin 1 gamma matrices. While $\gamma_{00}$ *commutes* with $\gamma_{ij}$, it *anticommutes* with $\gamma_{0i}$. It is precisely this property which allows us to introduce the conserved current density for the $(1,0)\oplus(0,1)$ matter field. In addition we note the following hermiticity properties

$${\gamma_{00}}^\dagger = \gamma_{00}, \quad {\gamma_{0i}}^\dagger = -\gamma_{0i}, \quad {\gamma_{i0}}^\dagger = -\gamma_{i0}, \quad {\gamma_{ij}}^\dagger = \gamma_{ij}. \tag{7.14}$$

Conserved Current Density:

Taking the hermitian conjugate of the wave equation (which satisfies the constraining principle introduced in Chap. 6) satisfied by the $(1,0)\oplus(0,1)$ covariant spinors, and using the hermiticity of $\gamma^{00}$ and $\gamma^{ij}$, and the anti–hermiticity of the $\gamma^{0i}$ and $\gamma^{i0}$, we obtain

$$\psi^\dagger(x)\left[\gamma^{00}\overleftarrow{\partial}_0\overleftarrow{\partial}_0-\gamma^{0i}\overleftarrow{\partial}_0\overleftarrow{\partial}_i-\gamma^{i0}\overleftarrow{\partial}_i\overleftarrow{\partial}_0+\gamma^{ij}\overleftarrow{\partial}_i\overleftarrow{\partial}_j+m^2\right]=0. \tag{7.15}$$

Since $\gamma^{00}$ commutes with $\gamma^{ij}$ and anticommutes with $\gamma^{0i}$ and $\gamma^{i0}$, multiplication from the right by $\gamma^{00}$ yields

$$\psi^\dagger(x)\gamma^{00}\left[\gamma^{00}\overleftarrow{\partial}_0\overleftarrow{\partial}_0+\gamma^{0i}\overleftarrow{\partial}_0\overleftarrow{\partial}_i+\gamma^{i0}\overleftarrow{\partial}_i\overleftarrow{\partial}_0+\gamma^{ij}\overleftarrow{\partial}_i\overleftarrow{\partial}_j+m^2\right]=0. \tag{7.16}$$

On introducing

$$\overline{\psi}(x)\equiv\psi^\dagger(x)\gamma^{00}, \tag{7.17}$$

the above equation can be written as

$$\overline{\psi}(x)\left(\gamma^{\mu\nu}\overleftarrow{\partial}_\mu\overleftarrow{\partial}_\nu+m^2\right)=0. \tag{7.18}$$

Multiplying the above equation by $\psi(x)$ from the right, and subtracting the left $\overline{\psi}(x)$–multiplied (7.15), we obtain

$$(\partial_\mu\partial_\nu\overline{\psi})\gamma^{\mu\nu}\psi-\overline{\psi}\gamma^{\mu\nu}(\partial_\mu\partial_\nu\psi)=0. \tag{7.19}$$

Now because $\gamma^{\mu\nu}$ is symmetric in its Lorentz indices, we have the following identity

$$(\partial_\nu\overline{\psi})\gamma^{\mu\nu}(\partial_\mu\psi)-(\partial_\mu\overline{\psi})\gamma^{\mu\nu}(\partial_\nu\psi)=0. \tag{7.20}$$

Combining (7.19) and (7.20) additively, yields

$$(\partial_\mu\partial_\nu\overline{\psi})\gamma^{\mu\nu}\psi+(\partial_\nu\overline{\psi})\gamma^{\mu\nu}(\partial_\mu\psi)-(\partial_\mu\overline{\psi})\gamma^{\mu\nu}(\partial_\nu\psi)-\overline{\psi}\gamma^{\mu\nu}(\partial_\mu\partial_\nu\psi)=0. \tag{7.21}$$

This implies that the we have the *conserved current density*

$$j^\mu(x) \equiv (\partial_\nu \overline{\psi}(x))\gamma^{\mu\nu}\psi(x) - \overline{\psi}(x)\gamma^{\mu\nu}(\partial_\nu \psi(x)), \tag{7.22}$$

which satisfies the *continuity equation*

$$\partial_\mu j^\mu(x) = 0. \tag{7.23}$$

Introducing the notation

$$A(x)\overleftrightarrow{\partial}_\mu B(x) \equiv A(x)(\partial_\mu B(x)) - (\partial_\mu A(x))B(x), \tag{7.24}$$

we have the conserved current density in a more compact form

$$j^\mu(x) = -\overline{\psi}(x)\gamma^{\mu\nu}\overleftrightarrow{\partial}_\nu \psi(x). \tag{7.25}$$

Note: Some authors, for example Ryder [53], prefer to have a factor of 1/2 on the *r.h.s.* of the definition (7.24) for their convenience.

## 7.2 A Remark on Spin–1 Algebra

The the algebra associated with $\gamma_{\mu\nu}$ can also be written as

$$\{\gamma^{\mu\nu}, \gamma^{\rho\sigma}\} + \{\gamma^{\mu\sigma}, \gamma^{\nu\rho}\} + \{\gamma^{\mu\rho}, \gamma^{\sigma\nu}\} = 2(\eta^{\mu\nu}\eta^{\rho\sigma} + \eta^{\mu\sigma}\eta^{\nu\rho} + \eta^{\mu\rho}\eta^{\sigma\nu})I, \tag{7.26}$$

with $I = 6 \times 6$ unit matrix. For the sake of convenience we take the liberty of naming (7.26) as the *Weinberg Algebra* because, to the best of our knowledge, it first appeared as equation (B12) in the work of Weinberg (1964), for the $(1,0) \oplus (0,1)$ matter field.

A trivial solution of the Weinberg algebra is

$$\gamma^{\mu\nu} = \eta^{\mu\nu} I \tag{7.27}$$

with $I = 1$, rather than a $6 \times 6$ unit matrix. Substituting this solution in the formal equation

$$\left(\gamma^{\mu\nu}\partial_\mu\partial_\nu + m^2 I\right)\psi(x) = 0, \tag{7.28}$$

which with the spin–1 $\gamma^{\mu\nu}$ is the relativistic wave equation for $(1,0)\oplus(0,1)$ matter field, yields the Klein Gordon equation for the *scalar field:*

$$\left(\eta^{\mu\nu}\partial_\mu\partial_\nu + m^2\right)\psi(x) = \left(\Box + m^2\right)\psi(x) = 0. \tag{7.29}$$

## 7.3 $(\vec{J}\cdot\vec{p} \pm jp^0)\phi^{L,R}(\vec{p}) = 0$ for $j=1$ and Maxwell Equations

The acausality associated with $(\vec{J}\cdot\vec{p} \pm jp^0)\phi^{L,R}(\vec{p}) = 0$ equations for $j = 1$ was seen to be of the form $E = 0$. Here we show that source free Maxwell Equations and $(\vec{J}\cdot\vec{p} \pm jp^0)\phi^{L,R}(\vec{p}) = 0$ for $j = 1$ are identical, or so it seems in our derivation. In our opinion the connection between $(\vec{J}\cdot\vec{p} \pm p^0)\phi^{L,R}(\vec{p}) = 0$ and Maxwell Equations needs a study beyond what we present below. Specifically, we are unable to answer the following questions at present:

1.) How does the dispersion relation $E = 0$, allowed by $(\vec{J}\cdot\vec{p} \pm p^0)\phi^{L,R}(\vec{p}) = 0$, manifest itself in the Maxwell equations?

2.) What is the connection between the Maxwell equations and the acausality free $m \to 0$ limit of the wave equation satisfied by the $(1,0)\oplus(0,1)$ covariant spinors?

In the $\vec{x}$-representation we have

$$\vec{p} = -i\vec{\nabla}, \qquad p^0 = i\frac{\partial}{\partial t}. \tag{7.30}$$

As a result in the $\vec{x}$-representation

$$(\vec{J}\cdot\vec{p} \pm p^0)\phi^{L,R}(\vec{p}) = 0 \tag{7.31}$$

become

$$\left(\vec{J}\cdot\vec{\nabla}-\frac{\partial}{\partial t}\right)\phi^{L}(x)=0 \tag{7.32}$$

$$\left(\vec{J}\cdot\vec{\nabla}+\frac{\partial}{\partial t}\right)\phi^{R}(x)=0 \tag{7.33}$$

The specific components $\phi_i^{L}(\vec{p})$ and $\phi_i^{R}(\vec{p})$ depend on the choice of representation for the $\vec{J}$ matrices. We define *chiral representation* by choosing the following representation for the $\vec{J}$ operators

[Chiral Representation]

$$J_x=\begin{pmatrix}0&0&0\\0&0&-i\\0&i&0\end{pmatrix},\quad J_y=\begin{pmatrix}0&0&i\\0&0&0\\-i&0&0\end{pmatrix},\quad J_z=\begin{pmatrix}0&-i&0\\i&0&0\\0&0&0\end{pmatrix}. \tag{7.34}$$

In the next section section we will explicitly construct a unitary matrix which connects these matrices, $\vec{J}_{CH}$, with the canonical, or standard, $j=1$ matrices $\vec{J}_{CA}$ in which $J_z$ is diagonal.

Next we introduce the even and odd parity linear combinations:

$$X_i(x)=\frac{1}{2}\left(\phi_i^{R}(x)+\phi_i^{L}(x)\right) \tag{7.35}$$

$$Y_i(x)=\frac{1}{2i}\left(\phi_i^{R}(x)-\phi_i^{L}(x)\right),\quad i=1,2,3. \tag{7.36}$$

so that

$$\phi_i^{R}(x)=X_i(x)+iY_i(x) \tag{7.37}$$

$$\phi_i^{L}(x)=X_i(x)-iY_i(x). \tag{7.38}$$

Using (7.34) we can write $\vec{J} \cdot \vec{\nabla}$ as

$$\vec{J} \cdot \vec{\nabla} = i \begin{pmatrix} 0 & -\partial_z & \partial_y \\ \partial_z & 0 & -\partial_x \\ -\partial_y & \partial_x & 0 \end{pmatrix}. \tag{7.39}$$

Or in the component form

$$(\vec{J} \cdot \vec{\nabla})_{ij} = -i\epsilon_{ijk}\partial_k. \tag{7.40}$$

Here $\epsilon_{ijk}$ is the completely antisymmetric tensor with $\epsilon_{123} = 1$. With these observations and definitions the uncoupled equations (7.32) and (7.33) take the form

$$\left(-\epsilon_{ijk}\partial_k - \frac{\partial}{\partial t}\right)(X_j - iY_j) = 0 \tag{7.41}$$

$$\left(-\epsilon_{ijk}\partial_k + \frac{\partial}{\partial t}\right)(X_j + iY_j) = 0 \tag{7.42}$$

Since $-\epsilon_{ijk}A_kB_j = (\vec{A} \times \vec{B})_i$, the above set of equations can be rewritten as

$$\vec{\nabla} \times (\vec{X} - i\vec{Y}) - i\frac{\partial}{\partial t}(\vec{X} - i\vec{Y}) = 0 \tag{7.43}$$

$$\vec{\nabla} \times (\vec{X} + i\vec{Y}) + i\frac{\partial}{\partial t}(\vec{X} + i\vec{Y}) = 0 \tag{7.44}$$

Adding (7.43) and (7.44) gives

$$\vec{\nabla} \times \vec{X} - \frac{\partial \vec{Y}}{\partial t} = 0 \tag{7.45}$$

Similarly subtracting (7.43) from (7.44) yields

$$\vec{\nabla} \times \vec{Y} + \frac{\partial \vec{X}}{\partial t} = 0 \tag{7.46}$$

Finally taking the divergence of these two equations provides two additional equa-

tions

$$\vec{\nabla} \cdot \vec{Y} = 0, \quad \vec{\nabla} \cdot \vec{X} = 0. \tag{7.47}$$

These four equations can be put in a more compact form by introducing

$$Z^{\mu\nu} = \begin{pmatrix} 0 & -Y_x & -Y_y & -Y_z \\ Y_x & 0 & -X_z & X_y \\ Y_y & X_z & 0 & -X_x \\ Y_z & -X_y & X_x & 0 \end{pmatrix} \tag{7.48}$$

We now assume: "what looks like electromagnetic field must be the electromagnetic field". Or put differently we assume that the $Z^{\mu\nu}$ couples with matter with the same strength as $F^{\mu\nu}$. Identifying $\vec{Y}$ and $\vec{X}$ with electric and magnetic fields of the electromagnetic field:

$$\vec{Y} = \vec{E}, \quad \vec{X} = \vec{B}, \tag{7.49}$$

we have

$$Z^{\mu\nu} = F^{\mu\nu}. \tag{7.50}$$

Equations (7.45), (7.46) and (7.47) are now readily seen to take the simple Poincaré covariant form

$$\partial_\mu F_{\mu\nu} = 0, \tag{7.51}$$

$$\partial^\mu F^{\nu\lambda} + \partial^\lambda F^{\mu\nu} + \partial^\nu F^{\lambda\mu} = 0. \tag{7.52}$$

Referring to the conventional definitions of $\vec{E}(x)$ and $\vec{B}(x)$, in terms of $A^\mu(x) = (\ \phi(x),\ \vec{A}(x)\ )$ such as found in Jackson [44], the connection between $F^{\mu\nu}(x)$

$$B_i(x) = \frac{1}{2}\left(\phi_i^{R}(x) + \phi_i^{L}(x)\right) \tag{7.53}$$

$$E_i(x) = \frac{1}{2i}\left(\phi_i^{R}(x) - \phi_i^{L}(x)\right), \quad i = 1, 2, 3. \tag{7.54}$$

and the gauge vector potential $A^\mu(x)$ for the electromagnetic field is

$$F^{\mu\nu}(x) = \partial^\mu A^\nu(x) - \partial^\nu A^\mu(x). \tag{7.55}$$

The freedom to choose any representation for the left and right handed matter fields exhibits itself as the freedom to choose $A^\mu(x)$ in any "gauge."

To be able to see this connection in the canonical representation it will be useful to explicitly know the connection between the $j = 1$ angular momentum operators in the canonical representation (representation in which $j_z$ is diagonal), and the same in the chiral representation provided by (7.34) here.

## 7.4 Construction of the Unitary Transformation which Connects $\vec{J}_{CH}$ and $\vec{J}_{CA}$

We start with

$$(J_z)_{CH} = \begin{pmatrix} 0 & -i & 0 \\ i & 0 & 0 \\ 0 & 0 & 0 \end{pmatrix}, \tag{7.56}$$

and note that the normalised eigenvectors corresponding to the three eigenvalues $\sigma = +1, 0, -1$ are

$$|+\rangle = \frac{1}{\sqrt{2}}\begin{pmatrix} 1 \\ i \\ 0 \end{pmatrix}, \quad |0\rangle = e^{i\theta}\begin{pmatrix} 0 \\ 0 \\ 1 \end{pmatrix}, \quad |-\rangle = \frac{e^{i\vartheta}}{\sqrt{2}}\begin{pmatrix} i \\ 1 \\ 0 \end{pmatrix} \tag{7.57}$$

In writing down these eigenvectors we have ignored a global phase factor. Next

we construct

$$U = \begin{pmatrix} \frac{1}{\sqrt{2}} & 0 & \frac{i}{\sqrt{2}}e^{i\vartheta} \\ \\ \frac{i}{\sqrt{2}} & 0 & \frac{1}{\sqrt{2}}e^{i\vartheta} \\ \\ 0 & e^{i\theta} & 0 \end{pmatrix}, \tag{7.58}$$

and verify that $U^\dagger U$ is indeed unity. The phase factors $\theta$ and $\vartheta$ will now be so fixed as to yield the identity

$$(J_i)_{CA} = U^\dagger \ (J_i)_{CH} \ U, \quad i = x, y, z \tag{7.59}$$

For $i = z$, we obtain

$$U^\dagger \ (J_z)_{CH} \ U = \begin{pmatrix} 1 & 0 & 0 \\ 0 & 0 & 0 \\ 0 & 0 & -1 \end{pmatrix} = (J_z)_{CA} \tag{7.60}$$

which puts no constraints on $\theta$ and $\vartheta$ as expected. For $i = x$, we get

$$U^\dagger \ (J_x)_{CH} \ U = \frac{1}{\sqrt{2}} \begin{pmatrix} 0 & -e^{i\theta} & 0 \\ \\ -e^{-i\theta} & 0 & ie^{i\vartheta}e^{-i\theta} \\ \\ 0 & -ie^{-i\vartheta}e^{i\theta} & 0 \end{pmatrix}. \tag{7.61}$$

Equating the *rhs* of the above expression to $(J_x)_{CA}$ we obtain

$$\theta = \pi, \quad \vartheta = \pi/2. \tag{7.62}$$

The unitary transformation is now completely fixed and reads

$$U = \begin{pmatrix} \frac{1}{\sqrt{2}} & 0 & -\frac{1}{\sqrt{2}} \\ \\ \frac{i}{\sqrt{2}} & 0 & \frac{i}{\sqrt{2}} \\ \\ 0 & -1 & 0 \end{pmatrix}. \tag{7.63}$$

That

$$U^\dagger\ (J_y)_{CH}\ U = (J_y)_{CA} \tag{7.64}$$

is easily verified.

## 7.5 Construction of the Proca Vector Potential From (1,0) and (0,1) Matter Fields Via "Spinorial Summation"

The basis of the $(1,0)$ representation space can be chosen to be

$$\overset{R}{\phi}_{1,+1}(\vec{0}),\quad \overset{R}{\phi}_{1,0}(\vec{0}),\quad \overset{R}{\phi}_{1,-1}(\vec{0}). \tag{7.65}$$

Similarly the basis for the $(0,1)$ representation space reads:

$$\overset{L}{\phi}_{1,+1}(\vec{0}),\quad \overset{L}{\phi}_{1,0}(\vec{0}),\quad \overset{L}{\phi}_{1,-1}(\vec{0}). \tag{7.66}$$

The effect of Lorentz boosts on these basis vectors is given by Eqs. (3.12) and (3.13). Using these transformation properties, it is readily seen that the effect of boosting along the $\hat{z}$ direction is given by

$$\overset{R}{\phi}_{1,+1}(\vec{p}=p\hat{z}) = \exp(\varphi)\ \overset{R}{\phi}_{1,+1}(\vec{0}) \tag{7.67}$$

$$\overset{R}{\phi}_{1,0}(\vec{p}=p\hat{z}) = \overset{R}{\phi}_{1,0}(\vec{0}) \tag{7.68}$$

$$\overset{R}{\phi}_{1,-1}(\vec{p}=p\hat{z}) = \exp(-\varphi)\ \overset{R}{\phi}_{1,-1}(\vec{0}) \tag{7.69}$$

and:

$$\overset{L}{\phi}_{1,+1}(\vec{p}=p\hat{z}) = \exp(-\varphi)\ \overset{L}{\phi}_{1,+1}(\vec{0}) \tag{7.70}$$

$$\overset{L}{\phi}_{1,0}(\vec{p}=p\hat{z}) = \overset{L}{\phi}_{1,0}(\vec{0}) \tag{7.71}$$

$$\overset{L}{\phi}_{1,-1}(\vec{p}=p\hat{z}) = \exp(\varphi)\ \overset{L}{\phi}_{1,-1}(\vec{0}) \tag{7.72}$$

Since the completely antisymmetric tensor $F^{\mu\nu} \equiv \partial^\mu A^\nu - \partial^\nu A^\mu$ has six independent elements and the representation space of the $(1,0)\oplus(0,1)$ matter field also has six independent basis vectors, we suspect some linear combination of $\phi^{R}_{1,\sigma}$ and $\phi^{L}_{1,\sigma}$ to transform as an antisymmetric tensor. However, at the outset, it should be noticed that each of the $\phi(\vec{p})$ is a *three column.* This "spinorial" degree of freedom implies that we should suspect some linear combination of $\phi^{R}_{1,\sigma}$ and $\phi^{L}_{1,\sigma}$ to map onto some yet undetermined: $\mathcal{F}^{\mu\nu}_{(\alpha)} = \partial^\mu \mathcal{A}^\nu_{(\alpha)} - \partial^\nu \mathcal{A}^\mu_{(\alpha)}$. The index $(\alpha)$ runs over the elements of a three column.

To begin note that if the frame in which the particle has momentum $\vec{p} = p\hat{z}$ be called the "primed" frame, then we can write

$$F'^{\mu\nu} = \Lambda^\mu{}_\lambda \Lambda^\nu{}_\sigma F^{\lambda\sigma}, \quad F^{\mu\nu} = \partial^\mu A^\nu - \partial^\nu A^\mu. \tag{7.73}$$

Using the boosts defined in Chap. 2, the effect of a boost in the $\hat{z}$ direction, then yields

$$F'^{01} = \cosh\varphi \; F^{01} - \sinh\varphi \; F^{13} \tag{7.74}$$

$$F'^{02} = \cosh\varphi \; F^{02} + \sinh\varphi \; F^{32} \tag{7.75}$$

$$F'^{03} = F^{03} \tag{7.76}$$

$$F'^{32} = \cosh\varphi \; F^{32} + \sinh\varphi \; F^{02} \tag{7.77}$$

$$F'^{13} = \cosh\varphi \; F^{13} - \sinh\varphi \; F^{01} \tag{7.78}$$

$$F'^{21} = F^{21}. \tag{7.79}$$

These transformation properties of the antisymmetric tensor $F^{\mu\nu}$ encourage us to study the Lorentz transformation characteristics of the odd and even (under parity) linear combinations: $(\phi^{R}_{1,\sigma} \pm \phi^{R}_{1,\sigma})$. From the transformation properties given by Eqs. ((7.67)-(7.72)) we find:

$$\begin{aligned}&\left[\overset{R}{\phi}_{1,+1}(\vec{p}=p\hat{z})+\overset{L}{\phi}_{1,+1}(\vec{p}=p\hat{z})\right]\\&\qquad=\cosh\varphi\left[\overset{R}{\phi}_{1,+1}(\vec{0})+\overset{L}{\phi}_{1,+1}(\vec{0})\right]+\sinh\varphi\left[\overset{R}{\phi}_{1,+1}(\vec{0})-\overset{L}{\phi}_{1,+1}(\vec{0})\right]\end{aligned}\tag{7.80}$$

$$\left[\overset{R}{\phi}_{1,0}(\vec{p}=p\hat{z})+\overset{L}{\phi}_{1,0}(\vec{p}=p\hat{z})\right]=\left[\overset{R}{\phi}_{1,0}(\vec{0})+\overset{L}{\phi}_{1,0}(\vec{0})\right]\tag{7.81}$$

$$\begin{aligned}&\left[\overset{R}{\phi}_{1,-1}(\vec{p}=p\hat{z})+\overset{L}{\phi}_{1,-1}(\vec{p}=p\hat{z})\right]\\&\qquad=\cosh\varphi\left[\overset{R}{\phi}_{1,-1}(\vec{0})+\overset{L}{\phi}_{1,-1}(\vec{0})\right]-\sinh\varphi\left[\overset{R}{\phi}_{1,-1}(\vec{0})-\overset{L}{\phi}_{1,-1}(\vec{0})\right],\end{aligned}\tag{7.82}$$

and:

$$\begin{aligned}&\left[\overset{R}{\phi}_{1,+1}(\vec{p}=p\hat{z})-\overset{L}{\phi}_{1,+1}(\vec{p}=p\hat{z})\right]\\&\qquad=\cosh\varphi\left[\overset{R}{\phi}_{1,+1}(\vec{0})-\overset{L}{\phi}_{1,+1}(\vec{0})\right]+\sinh\varphi\left[\overset{R}{\phi}_{1,+1}(\vec{0})+\overset{L}{\phi}_{1,+1}(\vec{0})\right]\end{aligned}\tag{7.83}$$

$$\left[\overset{R}{\phi}_{1,0}(\vec{p}=p\hat{z})-\overset{L}{\phi}_{1,0}(\vec{p}=p\hat{z})\right]=\left[\overset{R}{\phi}_{1,0}(\vec{0})-\overset{L}{\phi}_{1,0}(\vec{0})\right]\tag{7.84}$$

$$\begin{aligned}&\left[\overset{R}{\phi}_{1,-1}(\vec{p}=p\hat{z})-\overset{L}{\phi}_{1,-1}(\vec{p}=p\hat{z})\right]\\&\qquad=\cosh\varphi\left[\overset{R}{\phi}_{1,-1}(\vec{0})-\overset{L}{\phi}_{1,-1}(\vec{0})\right]-\sinh\varphi\left[\overset{R}{\phi}_{1,-1}(\vec{0})+\overset{L}{\phi}_{1,-1}(\vec{0})\right].\end{aligned}\tag{7.85}$$

Comparison of Eqs. ((7.74)-(7.79)) and ((7.80)-(7.85)) implies existence of an object

$$\mathcal{F}^{\mu\nu}_{(\alpha)}(\vec{p})=\partial^\mu\mathcal{A}^\nu_{(\alpha)}(\vec{p})-\partial^\nu\mathcal{A}^\mu_{(\alpha)}(\vec{p}),\tag{7.86}$$

with

$$\mathcal{F}^{01}_{(\alpha)}(\vec{p})=\left[\overset{R}{\phi}_{1,-1}(\vec{p})-\overset{L}{\phi}_{1,-1}(\vec{p})\right]_{(\alpha)}\tag{7.87}$$

$$\mathcal{F}^{02}_{(\alpha)}(\vec{p})=\left[\overset{R}{\phi}_{1,+1}(\vec{p})-\overset{L}{\phi}_{1,+1}(\vec{p})\right]_{(\alpha)}\tag{7.88}$$

$$\mathcal{F}^{03}_{(\alpha)}(\vec{p})=\left[\overset{R}{\phi}_{1,0}(\vec{p})-\overset{L}{\phi}_{1,0}(\vec{p})\right]_{(\alpha)}\tag{7.89}$$

$$\mathcal{F}^{32}_{(\alpha)}(\vec{p}) = \left[\overset{R}{\phi}_{1,+1}(\vec{p}) + \overset{L}{\phi}_{1,+1}(\vec{p})\right]_{(\alpha)} \tag{7.90}$$

$$\mathcal{F}^{13}_{(\alpha)}(\vec{p}) = \left[\overset{R}{\phi}_{1,-1}(\vec{p}) + \overset{L}{\phi}_{1,-1}(\vec{p})\right]_{(\alpha)} \tag{7.91}$$

$$\mathcal{F}^{21}_{(\alpha)}(\vec{p}) = \left[\overset{R}{\phi}_{1,0}(\vec{p}) + \overset{L}{\phi}_{1,0}(\vec{p})\right]_{(\alpha)}. \tag{7.92}$$

The "spinorial" index $(\alpha)$ runs over $I, II, III$ the elements of columns $\overset{R}{\phi}_{1,\sigma}(\vec{p})$ and $\overset{L}{\phi}_{1,\sigma}(\vec{p})$.

Next we introduce the operation of "raising" the spinorial index as follows:

$$\mathcal{F}^{01\ (\alpha)}(\vec{p}) = \left\{\left[\overset{R}{\phi}_{1,-1}(\vec{p}) - \overset{L}{\phi}_{1,-1}(\vec{p})\right]^{\dagger}\right\}_{(\alpha)}, \ \cdots. \tag{7.93}$$

In configuration space $\mathcal{F}^{\mu\nu}_{(\alpha)}$ is

$$\mathcal{F}^{\mu\nu}_{(\alpha)}(x) = \partial^{\mu}\mathcal{A}^{\nu}_{(\alpha)}(x) - \partial^{\nu}\mathcal{A}^{\mu}_{(\alpha)}(x), \tag{7.94}$$

with

$$\mathcal{F}^{01}_{(\alpha)}(x) = \left[\overset{R}{\phi}_{1,-1}(x) - \overset{L}{\phi}_{1,-1}(x)\right]_{(\alpha)}, \ \cdots \tag{7.95}$$

$$\mathcal{F}^{01\ (\alpha)}(x) = \left\{\left[\overset{R}{\phi}_{1,-1}(x) - \overset{L}{\phi}_{1,-1}(x)\right]^{\dagger}\right\}_{(\alpha)}, \ \cdots. \tag{7.96}$$

The Proca vector potential is now obtained by summing over the "spinorial" index as follows

$$\mathcal{F}^{(\alpha)}_{\hat{\mu}\hat{\nu}}(x)\mathcal{F}^{\mu\nu}_{(\alpha)}(x) \equiv F_{\hat{\mu}\hat{\nu}}(x)F^{\mu\nu}(x) \tag{7.97}$$

$$\mathcal{A}^{(\alpha)}_{\hat{\mu}}(x)\mathcal{A}^{\mu}_{(\alpha)}(x) \equiv A_{\hat{\mu}}(x)A^{\mu}(x). \tag{7.98}$$

The $F^{\mu\nu}(x)$ and $A^{\mu}(x)$ so obtained contain no spinorial degrees of freedom. The

Lagrangian density

$$\mathcal{L}(x) = -\frac{1}{4}F_{\mu\nu}(x)F^{\mu\nu}(x) + \frac{1}{2}m^2 A_\mu(x)A^\mu(x) \tag{7.99}$$

yields the Proca equation:

$$\partial_\mu F^{\mu\nu}(x) + m^2 A^\nu(x) = 0. \tag{7.100}$$

Taking the four divergence of the above equation and utilizing the antisymmetry of $F^{\mu\nu}(x)$ we get

$$\partial_\nu A^\nu(x) = 0. \tag{7.101}$$

As a result of the vanishing $\partial_\mu A^\mu$, $A^\mu(x)$ has only *three* degrees of freedom. Further since $\partial_\nu A^\nu = 0$, $\partial_\mu F^{\mu\nu} = \partial_\mu(\partial^\mu A^\nu - \partial^\nu A^\mu) = \partial_\mu\partial^\mu A^\nu - \partial^\nu\partial_\mu A^\mu = \partial_\mu\partial^\mu A^\nu = \Box A^\nu$; where we have interchanged the order of differentiation: $\partial_\mu\partial^\nu \to \partial^\nu\partial_\mu$. As such the Proca vector potential $A^\mu(x)$ satisfies:

$$(\Box + m^2)A^\mu(x) = 0. \tag{7.102}$$

It should be noted that we have called $A^\mu(x)$ constructed via "spinorial summation" (see Eq. (7.98)) on the $(1,0)$ and $(0,1)$ spinorial indices as the *Proca* vector potential. This has been done to emphasize how we reached at $A^\mu(x)$ from the $(1,0)$ and $(0,1)$ matter fields. It is not clear to the present authors what physical consequences lie behind the existence of $\mathcal{A}^\mu_{(\alpha)}(x)$ and similar other objects. Moreover, the relation between the $m \to 0$ of this construction and the similar construction done in the section on Maxwell Equations remains open to further study.

## 7.6 $\{(1/2,0)\oplus(0,1/2)\}\otimes\{(1/2,0)\oplus(0,1/2)\}$: Construction of the Spin–1 Proca Equation from the Bargmann Wigner Equations

The purpose of this section is to exhibit why the study of matter fields for arbitrary spin becomes increasingly complicated if one begins with the Bargmann Wigner formalism. This is done by studying the simple example associated with spin one. The arguments presented here closely follow Lurié [4]. For spin three half details can be found in Ref. [4]. We do not know of any similar constructions beyond spin 3/2 where the Bargmann Wigner formalism provides the Rarita–Schwinger type equation, and the associated constraints. On the other hand the $(j,0)\oplus(0,j)$ formalism presented in this work has no similar limitation.

We first construct the basis for the $\{(1/2,0)\oplus(0,1/2)\}\otimes\{(1/2,0)\oplus(0,1/2)\}$ representation. Since the basis vectors for the $\{(1/2,0)\oplus(0,1/2)\}$ representation are the four Dirac spinors

$$\psi_\alpha = \left\{ \begin{pmatrix} \phi^R_{\frac{1}{2},\frac{1}{2}} \\ \\ \phi^L_{\frac{1}{2},\frac{1}{2}} \end{pmatrix}, \begin{pmatrix} \phi^R_{\frac{1}{2},\frac{1}{2}} \\ \\ \phi^L_{\frac{1}{2},-\frac{1}{2}} \end{pmatrix}, \begin{pmatrix} \phi^R_{\frac{1}{2},-\frac{1}{2}} \\ \\ \phi^L_{\frac{1}{2},\frac{1}{2}} \end{pmatrix}, \begin{pmatrix} \phi^R_{\frac{1}{2},-\frac{1}{2}} \\ \\ \phi^L_{\frac{1}{2},-\frac{1}{2}} \end{pmatrix} \right\}, \tag{7.103}$$

the basis vectors for $\{(1/2,0)\oplus(0,1/2)\}\otimes\{(1/2,0)\oplus(0,1/2)\}$ are

$$\begin{aligned} \psi_{\alpha\beta} = \Big\{ & \psi_1\otimes\psi_1, \psi_1\otimes\psi_2, \psi_1\otimes\psi_3, \psi_1\otimes\psi_4, \\ & \psi_2\otimes\psi_1, \psi_2\otimes\psi_2, \psi_2\otimes\psi_3, \psi_2\otimes\psi_4, \\ & \psi_3\otimes\psi_1, \psi_3\otimes\psi_2, \psi_3\otimes\psi_3, \psi_3\otimes\psi_4, \\ & \psi_4\otimes\psi_1, \psi_4\otimes\psi_2, \psi_4\otimes\psi_3, \psi_4\otimes\psi_4 \Big\}. \end{aligned} \tag{7.104}$$

To describe this field we have a choice of either considering 16–component spinors, or a $4\times4$ symmetric bi–spinors $\psi_{\alpha\beta}$ which by construction satisfy the Bargmann Wigner equations (the origin of these equations goes back to Dirac [2], Bargmann and Wigner [9], see also Lurié [4] for a more detailed treatment). We choose to interpret this field as a bi-spinor satisfying [$\mu, \nu$ are the Lorentz indices $0, 1, 2, 3$ and $\alpha, \beta$ are the spinorial indices $1, 2, 3, 4$]

$$(i\gamma^\mu \partial_\mu - m)_{\alpha\alpha'} \;\; \psi_{\alpha'\beta}(x) = 0, \tag{7.105}$$

$$(i\gamma^\mu \partial_\mu - m)_{\beta\beta'} \;\; \psi_{\alpha\beta'}(x) = 0. \tag{7.106}$$

We now seek a *single* wave equation satisfied by the field described by these Bargmann Wigner equations. It will be a rather lengthy exercise. To begin we write Eqs. (7.105) and (7.106) in the matrix form [where $\psi$ is a $4 \times 4$ symmetric matrix]

$$(i\gamma^\mu \partial_\mu - m)\, \psi(x) = 0, \tag{7.107}$$

$$\psi \left( i\gamma^{\mu T} \overleftarrow{\partial}_\mu - m \right) = 0. \tag{7.108}$$

In the above expression $T$ stands for 'transpose' and $\leftarrow$ on $\partial_\mu$ in Eq. (7.108) indicates that $\partial_\mu$ acts on the $\psi(x)$ appearing on the left. We now wish to express the bi-spinor $\psi_{\alpha\beta}$ in terms of appropriate functions of $x$, and a complete set of symmetric $4 \times 4$ matrices. Towards this end we note that since on taking the transpose of

$$\gamma^\mu \gamma^\nu + \gamma^\nu \gamma^\mu = 2\delta^{\mu\nu}, \tag{7.109}$$

we get

$$\gamma^{\nu T} \gamma^{\mu T} + \gamma^{\mu T} \gamma^{\nu T} = 2\delta^{\mu\nu}, \tag{7.110}$$

the fundamental theorem of Pauli requires that there exist a non-singular matrix $B$ such that

$$\gamma^{\mu T} = B \gamma^\mu B^{-1}. \tag{7.111}$$

We now demonstrate that $B$ is anti-symmetric. First take the transpose of (7.111),

and then substitute (7.111) into the result

$$\begin{aligned}\gamma^\mu &= (B^{-1})^T\ \gamma^{\mu T}\ B^T \\ &= (B^{-1})^T\ B\gamma^\mu B^{-1}\ B^T \\ &= (B^{-1}B^T)^{-1}\ \gamma^\mu\ (B^{-1}B^T).\end{aligned} \tag{7.112}$$

Multiplying this equation by $B^{-1}B^T$ from the left yields

$$\Big[(B^{-1}B^T), \gamma^\mu\Big] = 0. \tag{7.113}$$

However (see Good [45] for the detailed arguments) any $4 \times 4$ matrix which commutes with each of the $\gamma^\mu$ must be a multiple of the unit matrix

$$B^{-1}B^T = k, \tag{7.114}$$

which implies $B^T = kB$. Taking transpose of it gives $kB^T = B$; putting $B^T = kB$ back in the preceding equation gives $k^2 = 1$. Good [45] chooses the possibility $k = \pm 1$ (why *not* $\exp(i\theta)$ ? For an answer see Ref. [38]), and following Pauli and Haantjes, argues that the possibility $k = +1$ leads to *ten* independent anti-symmetric $4 \times 4$ matrices, which is an impossibility. The other possibility $k = -1$ then implies that $B$ is anti-symmetric. We thus have

$$\gamma^{\mu T} = B\gamma^\mu B^{-1}, \qquad\qquad B^T = -B. \tag{7.115}$$

Now introducing

$$C \equiv iB^{-1}, \tag{7.116}$$

we find that

$$\Gamma^\mu \equiv \gamma^\mu C, \tag{7.117}$$

and

$$\Gamma^{\mu\nu} \equiv \frac{1}{2i}[\gamma^\mu, \gamma^\nu]C, \tag{7.118}$$

form the needed set of the ten $4 \times 4$ symmetric matrices. To show this first note that Eq. (7.116) when substituted in Eq. (7.115) yields

$$C^{-1}\gamma^\mu C = -\gamma^{\mu T} \qquad \Longleftrightarrow \qquad \gamma^\mu C = -C\gamma^{\mu T}, \tag{7.119}$$

and

$$C^T = C^{-1}. \tag{7.120}$$

Next consider

$$\Gamma^{\mu T} = (\gamma^\mu C)^T = C^T\gamma^{\mu T} = -C\gamma^{\mu T} = \gamma^\mu C = \Gamma^\mu. \tag{7.121}$$

(Here we have freely used Eqs. (7.119) and (7.120)) This establishes that $\Gamma^\mu$ are symmetric matrices. Similarly the symmetry of $\Gamma^{\mu\nu}$ is proved by appropriate use of Eqs. (7.119) and (7.120) as follows:

$$\begin{aligned}(\Gamma^{\mu\nu})^T &= \left\{\frac{1}{2i}(\gamma^\mu\gamma^\nu - \gamma^\nu\gamma^\mu)C\right\}^T = \frac{1}{2i}C^T(\gamma^{\nu T}\gamma^{\mu T} - \gamma^{\mu T}\gamma^{\nu T}) \\ &= -\frac{1}{2i}(C\gamma^{\nu T}\gamma^{\mu T} - C\gamma^{\mu T}\gamma^{\nu T}) = -\frac{1}{2i}(-\gamma^\nu C\gamma^{\mu T} + \gamma^\mu C\gamma^{\nu T}) \\ &= -\frac{1}{2i}(\gamma^\nu\gamma^\mu C - \gamma^\mu\gamma^\nu C) = \frac{1}{2i}(\gamma^\mu\gamma^\nu - \gamma^\nu\gamma^\mu)C = \Gamma^{\mu\nu}.\end{aligned} \tag{7.122}$$

We thus have

$$\textit{set}\ \{\Gamma^\mu, \Gamma^{\mu\nu}\} = \textit{set}\ \{\Gamma^0, \Gamma^1, \Gamma^2, \Gamma^3;\ \Gamma^{01}, \Gamma^{02}, \Gamma^{03}, \Gamma^{12}, \Gamma^{13}, \Gamma^{23}\} \tag{7.123}$$

as the complete set of ten $4 \times 4$ symmetric matrices.

*Reminder:* Each of these ten matrices is labelled by Lorentz indices. By definition *matrix* $\Gamma^{\mu\nu} = -$ *matrix* $\Gamma^{\nu\mu}$, as such we note the *anti-symmetry* in the Lorentz indices which by convention run as $0, 1, 2, 3$. The *symmetry* refers to the fact that $\{\Gamma^{\mu\nu}\}^T = \Gamma^{\mu\nu}$, that is: *matrix element* $\{\Gamma^{\mu\nu}\}_{\alpha\beta} =$ *matrix element* $\{\Gamma^{\mu\nu}\}_{\beta\alpha}$. The spinorial indices $\alpha, \beta$ run as $1, 2, 3, 4$.

Having thus discovered the complete set (7.123), we now express the Bargmann Wigner bispinor in terms of a yet unspecified [Lorentz–]four vector field $A_\mu(x)$ and [Lorentz–] anti-symmetric second rank tensor field $F_{\mu\nu}(x)$

$$\psi(x) = c_1\ \Gamma^\mu A_\mu(x)\ +\ c_2\ \Gamma^{\mu\nu} F_{\mu\nu}(x), \tag{7.124}$$

where $c_1$ and $c_2$ are still unknown numerical factors. Substituting $\psi(x)$ from above in (7.107), similarly substituting $\psi(x)$ in (7.108) and then replacing $C\gamma^{\mu T}$ by $-\gamma^\mu C$ (see (7.119)), and adding the resulting equations we obtain, where for simplicity we introduced $\Sigma^{\mu\nu} = (1/2i)[\gamma^\mu, \gamma^\nu]$ thus $\Gamma^{\mu\nu} = \Sigma^{\mu\nu} C$,

$$ic_1[\gamma^\mu, \gamma^\nu]C\partial_\mu A_\nu + ic_2[\gamma^\mu, \Sigma^{\nu\lambda}]C\partial_\mu F_{\nu\lambda} - 2mc_1\gamma^\nu C A_\nu - 2mc_2\Sigma^{\nu\lambda} C F_{\nu\lambda} = 0. \tag{7.125}$$

Next using Eq. (7.109) and the standard identity (see, for example, Dirac [54]) $[u, v_1 v_2] = [u, v_1]v_2 = v_1[u, v_2]$ we replace the commutator

$$[\gamma^\mu, \Sigma^{\nu\lambda}] \text{ by } 2i(\eta^{\lambda\mu}\gamma^\nu - \eta^{\nu\mu}\gamma^\lambda) \tag{7.126}$$

and substitute $[\gamma^\mu, \gamma^\nu]$ by $2i\Sigma^{\mu\nu}$ in Eq. (7.125) to get, on dividing the resulting equation by $-2c_2$

$$\Sigma^{\mu\nu}\left\{\frac{c_1}{c_2} C\partial_\mu A_\nu + mCF_{\mu\nu}\right\} + \gamma^\nu\left\{-2C\partial^\lambda F_{\lambda\nu} + m\frac{c_1}{c_2} C A^\nu\right\} = 0, \tag{7.127}$$

where in writing the second term we exploited the anti-symmetry of $F_{\nu\lambda}$ and replaced it by $-F_{\lambda\nu}$ apart from renaming the dummy indices to collect terms together. Now we break $\partial_\mu A_\nu$ in its symmetric and anti-symmetric parts by

$$\partial_\mu A_\nu \quad\rightarrow\quad \frac{1}{2}(\partial_\mu A_\nu - \partial_\nu A_\mu) + \frac{1}{2}(\partial_\mu A_\nu + \partial_\nu A_\mu), \tag{7.128}$$

and note

$$\Sigma^{\mu\nu}(\partial_\mu A_\nu + \partial_\nu A_\mu) = 0, \tag{7.129}$$

because of the anti-symmetry of $\Sigma^{\mu\nu}$ in the Lorentz indices. Having done this, we set the coefficients of $\Sigma^{\mu\nu}$and $\gamma^\nu$ equal to zero and get

$$\left(\frac{c_1}{2c_2}\right)(\partial_\mu A_\nu - \partial_\nu A_\mu) + mF_{\mu\nu} = 0, \tag{7.130}$$

$$-\partial^\lambda F_{\lambda\nu} + m\left(\frac{c_1}{2c_2}\right) A_\nu = 0. \tag{7.131}$$

Written in this form it is apparent that $c_1/2c_2$ can be absorbed in the definition (by re–defining the units) of $A_\mu$, or alternately we may so choose the units of $A_\mu$ such that

$$\left(\frac{c_1}{2c_2}\right) = -m. \tag{7.132}$$

With the choice (7.132), Eqs.(7.130) and (7.131) read

$$F_{\mu\nu} = \partial_\mu A_\nu - \partial_\nu A_\mu, \tag{7.133}$$

$$\partial_\mu F^{\mu\nu} + m^2 A^\nu = 0. \tag{7.134}$$

Equation (7.134) is called the Proca Equation and describes the $\{(1/2,0)\oplus(0,1/2)\}\otimes\{(1/2,0)\oplus(0,1/2)\}$ spin–1 field. Applying $\partial_\nu$ from the left on (81) yields the *constraint equation*

$$\partial_\nu A^\nu = 0, \tag{7.135}$$

due to the anti–symmetry of $F^{\mu\nu}$. Thus, only *three* out of the *four* components of $A^\mu$ are independent, as required for a spin–1 field.

## 7.7 Two Photon Mediated Production of Electrically Neutral Pseudoscalar Particles in High Energy Scattering of Two Charged Particles

FORMAL EXPRESSION FOR THE S–MATRIX

We will designate the $|\text{in}\rangle$ particles as $\chi_1$ and $\chi_2$, the $|$ states will have, in addition to $\chi_1$ and $\chi_2$, an additional particle: $\chi_3$. We will use the following notation

$$|m;\vec{p},\sigma_z\rangle \equiv |\text{mass; three momentum, z} - \text{component of } \vec{\text{J}}\rangle, \tag{7.136}$$

associated with the particle $\chi$. To systematically study the two photon mediated production of particles perturbatively within the framework of the S–matrix theory, outlined in the Appendix, we begin with the single particle production of neutral pseudoscalar particles.

One of the simplest of such processes can be defined by making the following specific choice for various particles $\chi_i$ involved in the scattering process

$\chi_1 =$ A scalar particle of mass $m_1$ and charge $Q_1$,

$\chi_2 =$ A scalar particle of mass $m_2$ and charge $Q_2$,

$\chi_3 =$ A *neutral pseudoscalar* particle of mass $m_3$.

Such a process approximates, according to our discussion in the Appendix, the two photon mediated production of a neutral pseudoscalars, such as the $\pi^0$, $\eta$ or $\eta'(958)$, in the scattering of two spin–zero nuclei. The matter field operators associated with the three particles involved are

$$\chi_1: \quad \Sigma(x) = \int \frac{d^3p_1''}{(2\pi)^3 2\omega_{p_1''}}\left[a(\vec{p}_1{}'')e^{-ip_1''\cdot x} + a^\dagger(\vec{p}_1{}'')e^{ip_1''\cdot x}\right] \tag{7.137}$$

$$\chi_2: \quad \Theta(x) = \int \frac{d^3p_2''}{(2\pi)^3 2\omega_{p_2''}}\left[a(\vec{p}_2{}'')e^{-ip_2''\cdot x} + a^\dagger(\vec{p}_2{}'')e^{ip_2''\cdot x}\right] \tag{7.138}$$

$$\chi_3: \quad \Phi(x) = \int \frac{d^3p_3''}{(2\pi)^3 2\omega_{p_3''}} \Big[a(\vec{p_3}'')e^{-ip_3''\cdot x} + a^\dagger(\vec{p_3}'')e^{ip_3''\cdot x}\Big] \tag{7.139}$$

with

$$\omega_p = (m^2 + \vec{p}^{\;2})^{1/2}. \tag{7.140}$$

The interaction Lagrangian density operator for the two photon mediated production under consideration is

$$\begin{aligned}\mathcal{L}_{int.}(x) = g_{s1} \;\; \Sigma(x)^\dagger \; \partial_\mu \Sigma(x) \; A^\mu(x) + g_{s2} \;\; \Theta(x)^\dagger \; \partial_\mu \Theta(x) \; A^\mu(x) \\ + g_{ps} \;\; \epsilon_{\mu\nu\rho\sigma} F^{\mu\nu}(x) F^{\rho\sigma}(x) \Phi(x),\end{aligned} \tag{7.141}$$

where we have suppressed the *Normal ordering.* Under the assumption

$$\mathcal{H}_{int.}(x) = -\mathcal{L}_{int.}(x), \tag{7.142}$$

the Dyson formula derived in the Appendix yields the following expression for the S–matrix

$$S = 1 + \sum_{n=1}^{\infty} \frac{(i)^n}{n!} \int_{-\infty}^{\infty} d^4x_1 \ldots \int_{-\infty}^{\infty} d^4x_n \; T\Big[\mathcal{L}_{int.}(x_1) \ldots \mathcal{L}_{int.}(x_n)\Big]. \tag{7.143}$$

We wish to calculate the following *amplitude of transition*

$$|i\rangle = |m_1, \vec{p}_1; \;\, m_2, \vec{p}_2\rangle \rightarrow |f\rangle = |m_1, \vec{p_1}'; \;\, m_2, \vec{p_2}'; m_3, \vec{p_3}'\rangle. \tag{7.144}$$

Formally, this transition amplitude is

$$\begin{aligned}\langle f|S|i\rangle = \langle m_1, \vec{p}_1{}'; \; m_2, \vec{p}_2{}'; \; m_3, \vec{p}_3{}'| \Bigg\{ 1 + i \int_{-\infty}^{\infty} d^4x_1 \mathcal{L}_{int.}(x_1) \\ - \frac{1}{2} \int_{-\infty}^{\infty} d^4x_1 \int_{-\infty}^{\infty} d^4x_2 \; T\Big[\mathcal{L}_{int.}(x_1)\mathcal{L}_{int.}(x_2)\Big] \\ - \frac{i}{6} \int_{-\infty}^{\infty} d^4x_1 \int_{-\infty}^{\infty} d^4x_2 \int_{-\infty}^{\infty} d^4x_3 \; T\Big[\mathcal{L}_{int.}(x_1)\mathcal{L}_{int.}(x_2)\mathcal{L}_{int.}(x_3)\Big] \\ + \cdots + \frac{(i)^n}{n!} \int_{-\infty}^{\infty} d^4x_1 \cdots \int_{-\infty}^{\infty} d^4x_n \; T\Big[\mathcal{L}_{int.}(x_1) \cdots\cdots \mathcal{L}_{int.}(x_n)\Big] \\ + \cdots\cdots \Bigg\} |m_1, \vec{p}_1; \; m_2, \vec{p}_2\rangle.\end{aligned} \tag{7.145}$$

Referring to the right hand side of the above expression note that the orthonormality of the initial and final states under consideration yields

$$\langle m_1, \vec{p}_1{}'; \; m_2, \vec{p}_2{}'; \; m_3, \vec{p}_3{}'|m_1, \vec{p}_1; \; m_2, \vec{p}_2\rangle = 0. \tag{7.146}$$

In the same context the matrix element

$$\langle m_1, \vec{p}_1{}'; \; m_2, \vec{p}_2{}'; \; m_3, \vec{p}_3{}'|\mathcal{L}_{int.}(x)|m_1, \vec{p}_1; \; m_2, \vec{p}_2\rangle. \tag{7.147}$$

consists of three terms:

($i$) *Two terms* arising from the coupling of the scalar matter fields, $\Sigma(x)$ and $\Theta(x)$, with the gauge vector potential $A^\mu(x)$, and

($ii$) *One* term corresponding to the coupling between the pseudoscalar matter field $\Phi(x)$ with $A^\mu(x)$.

The former *vanish* because neither the initial nor the final state involves a photon, whereas the part of the Lagrangian density operator associated with these two terms involves $A^\mu(x)$ *linearly.* The latter and the remaining term is

$$ig_{ps}\int_{-\infty}^{\infty} d^4x\langle m_1,\vec{p}_1{}';\ m_2,\vec{p}_2{}';\ m_3,\vec{p}_3{}'|\epsilon_{\mu\nu\rho\sigma}F^{\mu\nu}(x) \\ F^{\rho\sigma}(x)\Phi(x)|m_1,\vec{p}_1;\ m_2,\vec{p}_2\rangle, \tag{7.148}$$

and equals

$$ig_{ps}\int_{-\infty}^{\infty} d^4x\langle m_1,\vec{p}_1{}';\ m_2,\vec{p}_2{}';\ m_3,\vec{p}_3{}'|\epsilon^{\mu\nu\rho\sigma}(\partial_\mu A_\nu - \partial_\nu A_\mu) \\ (\partial_\rho A_\sigma - \partial_\sigma A_\rho)\Phi(x)|m_1,\vec{p}_1;\ m_2,\vec{p}_2\rangle. \tag{7.149}$$

The gauge vector potential $A_\mu(x)$ in the Lorentz gauge is (see [53, Sec. 4.4])

$$A_\mu(x) = \int \frac{d^3k}{(2\pi)^3 2k_0}\sum_{\lambda=0}^{3}\epsilon_\mu^{(\lambda)}(k)\left[a^{(\lambda)}(k)e^{-ik\cdot x} + a^{(\lambda)\dagger}(k)e^{ik\cdot x}\right]. \tag{7.150}$$

Substituting this and $\Phi(x)$ from (7.139) we obtain

$$
\begin{aligned}
(7.149) = ig_{ps} \int_{-\infty}^{\infty} d^4x \langle m_1, \vec{p}_1{}'; \; m_2, \vec{p}_2{}'; \; m_3, \vec{p}_3{}' | \epsilon^{\mu\nu\rho\sigma} \\
&\Bigg[ \partial_\mu \Big\{ \int \frac{d^3k}{(2\pi)^3 2k_0} \sum_{\lambda=0}^{3} \epsilon_\nu^{(\lambda)}(k) \Big[ a^{(\lambda)}(k) e^{-ik\cdot x} + a^{(\lambda)\dagger}(k) e^{ik\cdot x} \Big] \Big\} \\
&- \partial_\nu \Big\{ \int \frac{d^3k'}{(2\pi)^3 2k'_0} \sum_{\lambda'=0}^{3} \epsilon_\mu^{(\lambda')}(k') \Big[ a^{(\lambda')}(k') e^{-ik'\cdot x} + a^{(\lambda')\dagger}(k') e^{ik'\cdot x} \Big] \Big\} \Bigg] \\
&\Bigg[ \partial_\rho \Big\{ \int \frac{d^3k''}{(2\pi)^3 2k''_0} \sum_{\lambda''=0}^{3} \epsilon_\sigma^{(\lambda'')}(k'') \Big[ a^{(\lambda'')}(k'') e^{-ik''\cdot x} + a^{(\lambda'')\dagger}(k'') e^{ik''\cdot x} \Big] \Big\} \\
&- \partial_\sigma \Big\{ \int \frac{d^3k'''}{(2\pi)^3 2k'''_0} \sum_{\lambda'''=0}^{3} \epsilon_\rho^{(\lambda''')}(k''') \Big[ a^{(\lambda''')}(k''') e^{-ik'''\cdot x} \\
&\qquad + a^{(\lambda''')\dagger}(k''') e^{ik'''\cdot x} \Big] \Big\} \Bigg] \\
&\int \frac{d^3p''_3}{(2\pi)^3 2\omega_{p_3{}''}} \Big[ a(\vec{p}_3{}'') e^{-ip_3{}''\cdot x} + a^\dagger(\vec{p}_3{}'') e^{ip_3{}''\cdot x} \Big] | m_1, \vec{p}_1; \; m_2, \vec{p}_2 \rangle .
\end{aligned}
\tag{7.151}
$$

Since

$$
\langle m_1, \vec{p}_1{}'; \; m_2, \vec{p}_2{}'; \; m_3, \vec{p}_3{}' | a^{(\lambda)\dagger}(k) = 0 \tag{7.152}
$$

$$
a^{(\lambda)}(k) | m_1, \vec{p}_1; \; m_2, \vec{p}_2 \rangle = 0, \tag{7.153}
$$

we have

$$
\begin{aligned}
(7.149) = ig_{ps} \int_{-\infty}^{\infty} & d^4x \langle m_1, \vec{p_1}'; \; m_2, \vec{p_2}'; \; m_3, \vec{p_3}' | \epsilon^{\mu\nu\rho\sigma} \\
& \left[ \int \frac{d^3k}{(2\pi)^3 2k_0} \sum_{\lambda=0}^{3} \epsilon_\nu^{(\lambda)}(k)(-ik_\mu) a^{(\lambda)}(k) e^{-ik\cdot x} \right. \\
& \qquad \left. - \int \frac{d^3k'}{(2\pi)^3 2k'_0} \sum_{\lambda'=0}^{3} \epsilon_\mu^{(\lambda')}(k')(-ik'_\nu) a^{(\lambda')}(k') e^{-ik'\cdot x} \right] \\
& \left[ \int \frac{d^3k''}{(2\pi)^3 2k''_0} \sum_{\lambda''=0}^{3} \epsilon_\sigma^{(\lambda'')}(k'')(ik''_\rho) a^{(\lambda'')\dagger}(k'') e^{ik''\cdot x} \right. \\
& \qquad \left. - \int \frac{d^3k'''}{(2\pi)^3 2k'''_0} \sum_{\lambda'''=0}^{3} \epsilon_\rho^{(\lambda''')}(k''')(ik'''_\sigma) a^{(\lambda''')\dagger}(k''') e^{ik'''\cdot x} \right] \\
& \qquad\qquad \left[ \int \frac{d^3p''_3}{(2\pi)^3 2\omega_{p_3''}} a^\dagger(\vec{p_3}'') e^{ip_3''\cdot x} \right] | m_1, \vec{p}_1; \; m_2, \vec{p}_2 \rangle .
\end{aligned}
\tag{7.154}
$$

To calculate this we can rewrite the above expression as

$$
\begin{aligned}
(7.149) = ig_{ps} \int_{-\infty}^{\infty} & d^4x \langle m_1, \vec{p_1}'; \; m_2, \vec{p_2}'; \; m_3, \vec{p_3}' | \epsilon^{\mu\nu\rho\sigma} \\
& \int \frac{d^3k}{(2\pi)^3 2k_0} \int \frac{d^3k''}{(2\pi)^3 2k''_0} \int \frac{d^3p''_3}{(2\pi)^3 2\omega_{p_3''}} \sum_{\lambda'=0}^{3} \sum_{\lambda''=0}^{3} \\
& \epsilon_\nu^{(\lambda)}(k) \epsilon_\sigma^{(\lambda'')}(k'')(-ik_\mu)(-ik_\rho) \\
& \exp[i(-k + k'' + p''_3)\cdot x] a^{(\lambda)}(k) a^{(\lambda'')\dagger}(k'') a^\dagger(p''_3) \\
& \qquad | m_1, \vec{p}_1; \; m_2, \vec{p}_2 \rangle + three\ similar\ terms\ (t.s.t).
\end{aligned}
\tag{7.155}
$$

The $x$–integration is immediately performed by taking note of the fact that

$$
\int \frac{d^4x}{(2\pi)^4} e^{ik\cdot x} = \delta^4(k). \tag{7.156}
$$

This yields

$$
\begin{aligned}
(7.149) = {} & ig_{ps}\langle m_1, \vec{p_1}';\ m_2, \vec{p_2}';\ m_3, \vec{p_3}'|\epsilon^{\mu\nu\rho\sigma} \\
& \int \frac{d^3k}{(2\pi)^3 2k_0} \int \frac{d^3k''}{(2\pi)^3 2k_0''} \int \frac{d^3p_3''}{(2\pi)^3 2\omega_{p_3''}} \sum_{\lambda'=0}^{3} \sum_{\lambda''=0}^{3} \\
& \epsilon_\nu^{(\lambda)}(k)\epsilon_\sigma^{(\lambda'')}(k'')(-ik_\mu)(-ik_\rho) \\
& (2\pi)^4\delta^4(-k + k'' + p_3'')a^{(\lambda)}(k)a^{(\lambda'')\dagger}(k'')a^\dagger(p_3'') \\
& \qquad |m_1, \vec{p}_1;\ m_2, \vec{p}_2\rangle + t.s.t.
\end{aligned}
\tag{7.157}
$$

Now note that we have suppressed Normal ordering of the $\mathcal{L}_{int.}$ Normal ordering has the effect of moving $a^{(\lambda)}(k)$ to the right of the $a^\dagger$s in the first term of the above expression. But since

$$
a^{(\lambda)}(k)|m_1, \vec{p}_1;\ m_2, \vec{p}_2\rangle = 0 \tag{7.158}
$$

the first term in (7.157). Similarly the $t.s.t.$ in (7.157) are zero, with the result

$$
(7.149) = 0. \tag{7.159}
$$

Because of this result and the observations noted down after equation (7.147) we arrive at the result *that the first order contribution to the transition amplitude* $\langle f|S|i\rangle$ *for the process under consideration vanishes.* From a physical point of view this should not be a surprising result because the transition amplitude for the process under consideration behaves as $\sim g_{s1}\ g_{s2}\ g_{ps}$ to the lowest order. We assume, without explicit calculations, that the second order contribution is zero and proceed to calculate the first non–zero contribution to the transition amplitude given in (7.145).

In order that the subscripts on the spacetime volume elements $d^4x_i$ in equation (7.145) are not confused with the subscripts on the *in* and *out* state momenta we change the notation as follows:

$$
\begin{aligned}
x_1 &\to x' \\
x_2 &\to x'' \\
x_3 &\to x''',
\end{aligned}
\tag{7.160}
$$

and extract third order contribution to $\langle f|S|i\rangle$. This contribution reads

$$
\begin{aligned}
\langle 1', 2', 3'|S|1, 2\rangle^{III} = -\frac{i}{6} \int_{-\infty}^{\infty} d^4x' \int_{-\infty}^{\infty} d^4x'' \int_{-\infty}^{\infty} d^4x''' \langle 1', 2', 3'| \\
T\Big[\mathcal{L}_{int.}(x')\mathcal{L}_{int.}(x'')\mathcal{L}_{int.}(x''')\Big]|1, 2\rangle,
\end{aligned}
\tag{7.161}
$$

where we have introduced the following abbreviations:

$$
\begin{aligned}
\langle m_1, \vec{p}_1{}'; \; m_2, \vec{p}_2{}'; \; m_3, \vec{p}_3{}'| &= \langle 1', 2', 3'| \\
|m_1, \vec{p}_1; \; m_2, \vec{p}_2\rangle &= |1, 2\rangle.
\end{aligned}
\tag{7.162}
$$

The only non–vanishing contributions to $\langle 1', 2', 3'|S|1, 2\rangle^{III}$ come from terms $\sim$ $g_{s1}\ g_{s2}\ g_{ps}$. As such

we have (using (7.141))

$$\langle 1', 2', 3'|S|1, 2\rangle^{III} = -\frac{i}{6} g_{s1}\ g_{s2}\ g_{ps} \int\limits_{-\infty}^{\infty} d^4x' \int\limits_{-\infty}^{\infty} d^4x'' \int\limits_{-\infty}^{\infty} d^4x''' \langle 1', 2', 3'|$$

$$T\Big[\Sigma(x')^\dagger \partial_\epsilon \Sigma(x') A^\epsilon(x') \Theta(x'')^\dagger \partial_\eta \Theta(x'') A^\eta(x'') \epsilon_{\mu\nu\rho\sigma} F^{\mu\nu}(x''') F^{\rho\sigma}(x''') \Phi(x''')$$
$$+ \Sigma(x')^\dagger \partial_\epsilon \Sigma(x') A^\epsilon(x') \Theta(x''')^\dagger \partial_\eta \Theta(x''') A^\eta(x''') \epsilon_{\mu\nu\rho\sigma} F^{\mu\nu}(x'') F^{\rho\sigma}(x'') \Phi(x'')$$
$$+ \Sigma(x'')^\dagger \partial_\epsilon \Sigma(x'') A^\epsilon(x'') \Theta(x')^\dagger \partial_\eta \Theta(x') A^\eta(x') \epsilon_{\mu\nu\rho\sigma} F^{\mu\nu}(x''') F^{\rho\sigma}(x''') \Phi(x''')$$
$$+ \Sigma(x'')^\dagger \partial_\epsilon \Sigma(x) A^\epsilon(x'') \Theta(x''')^\dagger \partial_\eta \Theta(x''') A^\eta(x''') \epsilon_{\mu\nu\rho\sigma} F^{\mu\nu}(x') F^{\rho\sigma}(x') \Phi(x')$$
$$+ \Sigma(x''')^\dagger \partial_\epsilon \Sigma(x''') A^\epsilon(x''') \Theta(x')^\dagger \partial_\eta \Theta(x') A^\eta(x') \epsilon_{\mu\nu\rho\sigma} F^{\mu\nu}(x'') F^{\rho\sigma}(x'') \Phi(x'')$$
$$+ \Sigma(x''')^\dagger \partial_\epsilon \Sigma(x''') A^\epsilon(x''') \Theta(x'')^\dagger \partial_\eta \Theta(x'') A^\eta(x'') \epsilon_{\mu\nu\rho\sigma} F^{\mu\nu}(x') F^{\rho\sigma}(x') \Phi(x')\Big]$$
$$|1, 2\rangle. \tag{7.163}$$

Reintroducing the suppressed normal ordering of the Lagrangian density operator, we have [also see the observations below]

$$\langle 1', 2', 3'|S|1, 2\rangle^{III} = -\frac{i}{6}\ g_{s1}\ g_{s2}\ g_{ps} \int\limits_{-\infty}^{\infty} d^4x' \int\limits_{-\infty}^{\infty} d^4x'' \int\limits_{-\infty}^{\infty} d^4x'''$$
$$\langle\ \ |a(\vec{p}_1{}')a(\vec{p}_2{}')a(\vec{p}_3{}')$$
$$T\Big[N\Big(\Sigma^\dagger\ \partial_\epsilon \Sigma\ A^\epsilon\Big)_{x'} N\Big(\Theta^\dagger\ \partial_\eta \Theta\ A^\eta\Big)_{x''} N\Big(\epsilon_{\mu\nu\rho\sigma} F^{\mu\nu} F^{\rho\sigma} \Phi\Big)_{x'''} \tag{7.164}$$
$$+ (x'' \to x''',\ x''' \to x'') + (x' \to x'',\ x'' \to x')+$$
$$(x' \to x'',\ x'' \to x''',\ x''' \to x') + (x' \to x''',\ x'' \to x', x''' \to x'')$$
$$+ (x' \to x''', x''' \to x')\Big] a^\dagger(\vec{p}_1) a^\dagger(\vec{p}_2)|\ \ \rangle.$$

The non–explicitly written terms in the above expression are obtained from the *explicitly* written first term by the indicated substitutions. In addition recall that:

$$a^\dagger(\vec{p}_1)|\ \ \rangle = |m_1, \vec{p}_1\rangle \tag{7.165}$$

$$a^\dagger(\vec{p}_2)|\ \ \rangle = |m_2, \vec{p}_2\rangle \tag{7.166}$$

$$a^\dagger(\vec{p}_3)|\ \ \rangle = |m_3, \vec{p}_3\rangle. \tag{7.167}$$

In addition note the standard abbreviation

$$A(x)\ B(x)\ C(x) = \Big(A\ B\ C\ \Big)_x. \tag{7.168}$$

To proceed further with our calculations we consider the

<u>first time ordered term in (7.164)</u>

$$T\Big[N\Big(\Sigma^\dagger\ \partial_\epsilon \Sigma\ A^\epsilon\Big)_{x'} N\Big(\Theta^\dagger\ \partial_\eta \Theta\ A^\eta\Big)_{x''} N\Big(\epsilon_{\mu\nu\rho\sigma} F^{\mu\nu} F^{\rho\sigma} \Phi\Big)_{x'''}\Big], \tag{7.169}$$

and use the definition

$$F^{\mu\nu} = \partial^\mu A^\nu - \partial^\nu A^\mu \tag{7.170}$$

to get

$$\begin{aligned}
&T\Big[N\Big(\Sigma^\dagger\ \partial_\epsilon \Sigma\ A^\epsilon\Big)_{x'} N\Big(\Theta^\dagger\ \partial_\eta \Theta\ A^\eta\Big)_{x''} N\Big(\epsilon_{\mu\nu\rho\sigma} F^{\mu\nu} F^{\rho\sigma} \Phi\Big)_{x'''}\Big] \\
&\quad = T\Big[N\Big(\Sigma^\dagger\ \partial_\epsilon \Sigma\ A^\epsilon\Big)_{x'} N\Big(\Theta^\dagger\ \partial_\eta \Theta\ A^\eta\Big)_{x''} \\
&\qquad\qquad\qquad N\Big(\epsilon_{\mu\nu\rho\sigma}\{\partial^\mu A^\nu - \partial^\nu A^\mu\}\{\partial^\rho A^\sigma - \partial^\sigma A^\rho\}\Phi\Big)_{x'''}\Big] \\
&\quad = T\Big[N\Big(\Sigma^\dagger\ \partial_\epsilon \Sigma\ A^\epsilon\Big)_{x'} N\Big(\Theta^\dagger\ \partial_\eta \Theta\ A^\eta\Big)_{x''} N\Big(\epsilon_{\mu\nu\rho\sigma} \partial^\mu A^\nu \partial^\rho A^\sigma \Phi\Big)_{x'''}\Big] \\
&\quad - T\Big[N\Big(\Sigma^\dagger\ \partial_\epsilon \Sigma\ A^\epsilon\Big)_{x'} N\Big(\Theta^\dagger\ \partial_\eta \Theta\ A^\eta\Big)_{x''} N\Big(\epsilon_{\mu\nu\rho\sigma} \partial^\mu A^\nu \partial^\sigma A^\rho \Phi\Big)_{x'''}\Big] \\
&\quad - T\Big[N\Big(\Sigma^\dagger\ \partial_\epsilon \Sigma\ A^\epsilon\Big)_{x'} N\Big(\Theta^\dagger\ \partial_\eta \Theta\ A^\eta\Big)_{x''} N\Big(\epsilon_{\mu\nu\rho\sigma} \partial^\nu A^\mu \partial^\rho A^\sigma \Phi\Big)_{x'''}\Big] \\
&\quad + T\Big[N\Big(\Sigma^\dagger\ \partial_\epsilon \Sigma\ A^\epsilon\Big)_{x'} N\Big(\Theta^\dagger\ \partial_\eta \Theta\ A^\eta\Big)_{x''} N\Big(\epsilon_{\mu\nu\rho\sigma} \partial^\nu A^\mu \partial^\sigma A^\rho \Phi\Big)_{x'''}\Big].
\end{aligned} \tag{7.171}$$

We now use the <u>second Wick's theorem</u> (see: Ref. [51, p.167]) to evaluate the time ordered product $T[\cdots]$ which appears in (7.164). Towards this end we start with the <u>first time ordered term in the *r.h.s* of (7.171)</u> (Note: *l.h.s.* of (7.171) equals the first time ordered term in the *r.h.s.* of (7.164))

$$
\begin{aligned}
T\Big[&N\Big(\Sigma^\dagger\,\partial_\epsilon\Sigma\,A^\epsilon\Big)_{x'}N\Big(\Theta^\dagger\,\partial_\eta\Theta\,A^\eta\Big)_{x''}N\Big(\epsilon_{\mu\nu\rho\sigma}\partial^\mu A^\nu\partial^\rho A^\sigma\Phi\Big)_{x'''}\Big]\\
&=N\Big[\Big(\Sigma^\dagger\,\partial_\epsilon\Sigma\,A^\epsilon\Big)_{x'}\Big(\Theta^\dagger\,\partial_\eta\Theta\,A^\eta\Big)_{x''}\Big(\epsilon_{\mu\nu\rho\sigma}\partial^\mu A^\nu\partial^\rho A^\sigma\Phi\Big)_{x'''}\Big]\\
&+N\Big[\Big(\Sigma^\dagger\,\partial_\epsilon\Sigma\,\underbrace{A^\epsilon\Big)_{x'}\Big(\Theta^\dagger\,\partial_\eta\Theta\,A^\eta}\Big)_{x''}\Big(\epsilon_{\mu\nu\rho\sigma}\partial^\mu A^\nu\partial^\rho A^\sigma\Phi\Big)_{x'''}\Big]\\
&+N\Big[\Big(\Sigma^\dagger\,\partial_\epsilon\Sigma\,\underbrace{A^\epsilon\Big)_{x'}\Big(\Theta^\dagger\,\partial_\eta\Theta\,A^\eta\Big)_{x''}\Big(\epsilon_{\mu\nu\rho\sigma}\partial^\mu A^\nu}\,\partial^\rho A^\sigma\Phi\Big)_{x'''}\Big]\\
&+N\Big[\Big(\Sigma^\dagger\,\partial_\epsilon\Sigma\,\underbrace{A^\epsilon\Big)_{x'}\Big(\Theta^\dagger\,\partial_\eta\Theta\,A^\eta\Big)_{x''}\Big(\epsilon_{\mu\nu\rho\sigma}\partial^\mu A^\nu\partial^\rho A^\sigma}\,\Phi\Big)_{x'''}\Big]\\
&+N\Big[\Big(\Sigma^\dagger\,\partial_\epsilon\Sigma\,A^\epsilon\Big)_{x'}\Big(\Theta^\dagger\,\partial_\eta\Theta\,\underbrace{A^\eta\Big)_{x''}\Big(\epsilon_{\mu\nu\rho\sigma}\partial^\mu A^\nu}\,\partial^\rho A^\sigma\Phi\Big)_{x'''}\Big]\\
&+N\Big[\Big(\Sigma^\dagger\,\partial_\epsilon\Sigma\,A^\epsilon\Big)_{x'}\Big(\Theta^\dagger\,\partial_\eta\Theta\,\underbrace{A^\eta\Big)_{x''}\Big(\epsilon_{\mu\nu\rho\sigma}\partial^\mu A^\nu\partial^\rho A^\sigma}\,\Phi\Big)_{x'''}\Big]\\
&+N\Big[\Big(\Sigma^\dagger\,\partial_\epsilon\Sigma\,\underbrace{A^\epsilon\Big)_{x'}\Big(\Theta^\dagger\,\partial_\eta\Theta\,A^\eta\Big)_{x''}\Big(\epsilon_{\mu\nu\rho\sigma}\partial^\mu A^\nu}\,\partial^\rho A^\sigma\Phi\Big)_{x'''}\Big]\\
&+N\Big[\Big(\Sigma^\dagger\,\partial_\epsilon\Sigma\,\underbrace{A^\epsilon\Big)_{x'}\Big(\Theta^\dagger\,\partial_\eta\Theta\,\overbrace{A^\eta\Big)_{x''}\Big(\epsilon_{\mu\nu\rho\sigma}\partial^\mu A^\nu}\,\partial^\rho A^\sigma}\,\Phi\Big)_{x'''}\Big].
\end{aligned}
\tag{7.172}
$$

Where we omitted all contractions between the *different* field operators, such as $\underbrace{\Sigma(x')\Theta(x'')}$, which give a zero contribution.

Next we calculate the needed matrix element of the above time ordered product

$$
\begin{aligned}
&\langle\ \ |a(\vec{p}_1{}')a(\vec{p}_2{}')a(\vec{p}_3{}')T\Big[N\Big(\Sigma^\dagger\,\partial_\epsilon\Sigma\,A^\epsilon\Big)_{x'}N\Big(\Theta^\dagger\,\partial_\eta\Theta\,A^\eta\Big)_{x''}\\
&N\Big(\epsilon_{\mu\nu\rho\sigma}\partial^\mu A^\nu\partial^\rho A^\sigma\Phi\Big)_{x'''}\Big]a^\dagger(\vec{p}_1)a^\dagger(\vec{p}_2)|\ \ \rangle.
\end{aligned}
\tag{7.173}
$$

*Since neither the initial state* $|i\rangle = a^\dagger(\vec{p}_1)a^\dagger(\vec{p}_2)|\ \ \rangle$ *nor the final state* $|f\rangle = a^\dagger(\vec{p}_1{}')a^\dagger(\vec{p}_2{}')a^\dagger(\vec{p}_3{}')|\ \ \rangle$ *contains a photon, the first six terms on the r.h.s. contribute*

*zero to the matrix element* (7.173). As such we have

$$\begin{aligned}
&\langle\ \ |a(\vec{p}_1{}')a(\vec{p}_2{}')a(\vec{p}_3{}')T\Big[N\Big(\Sigma^\dagger\ \partial_\epsilon\Sigma\ A^\epsilon\Big)_{x'}N\Big(\Theta^\dagger\ \partial_\eta\Theta\ A^\eta\Big)_{x''}\\
&\qquad N\Big(\epsilon_{\mu\nu\rho\sigma}\partial^\mu A^\nu\partial^\rho A^\sigma\Phi\Big)_{x'''}\Big]a^\dagger(\vec{p}_1)a^\dagger(\vec{p}_2)|\ \ \rangle\\
&=\langle\ \ |a(\vec{p}_1{}')a(\vec{p}_2{}')a(\vec{p}_3{}')\\
&N\Big[\Big(\Sigma^\dagger\ \partial_\epsilon\Sigma\ \underbrace{A^\epsilon\Big)_{x'}\Big(\Theta^\dagger\ \partial_\eta\Theta\ A^\eta\Big)_{x''}\Big(\epsilon_{\mu\nu\rho\sigma}\partial^\mu A^\nu}\ \partial^\rho A^\sigma\Phi\Big)_{x'''}\Big]\\
&\qquad\qquad a^\dagger(\vec{p}_1)a^\dagger(\vec{p}_2)|\ \ \rangle\\
&+\langle\ \ |a(\vec{p}_1{}')a(\vec{p}_2{}')a(\vec{p}_3{}')\\
&N\Big[\Big(\Sigma^\dagger\ \partial_\epsilon\Sigma\ \underbrace{A^\epsilon\Big)_{x'}\Big(\Theta^\dagger\ \partial_\eta\Theta\ \overbrace{A^\eta\Big)_{x''}\Big(\epsilon_{\mu\nu\rho\sigma}\partial^\mu A^\nu}\ \partial^\rho A^\sigma}\ \Phi\Big)_{x'''}\Big].\\
&\qquad\qquad a^\dagger(\vec{p}_1)a^\dagger(\vec{p}_2)|\ \ \rangle.
\end{aligned} \tag{7.174}$$

Since contractions are c–numbers we may take the contractions on the *r.h.s.* of (7.174) out of the Normal product and write the above result as follows:

$$\begin{aligned}
&\langle\ \ |a(\vec{p}_1{}')a(\vec{p}_2{}')a(\vec{p}_3{}')T\Big[N\Big(\Sigma^\dagger\ \partial_\epsilon\Sigma\ A^\epsilon\Big)_{x'}N\Big(\Theta^\dagger\ \partial_\eta\Theta\ A^\eta\Big)_{x''}\\
&\qquad N\Big(\epsilon_{\mu\nu\rho\sigma}\partial^\mu A^\nu\partial^\rho A^\sigma\Phi\Big)_{x'''}\Big]a^\dagger(\vec{p}_1)a^\dagger(\vec{p}_2)|\ \ \rangle\\
&=\underbrace{A^\epsilon(x')\partial^\mu A^\nu(x''')}\ \ \underbrace{A^\eta(x'')\partial^\rho A^\sigma(x''')}\epsilon_{\mu\nu\rho\sigma}\\
&\langle\ \ |a(\vec{p}_1{}')a(\vec{p}_2{}')a(\vec{p}_3{}')N\Big[\Big(\Sigma^\dagger\partial_\epsilon\Sigma\Big)_{x'}\Big(\Theta^\dagger\partial_\eta\Theta\Big)_{x''}\Phi(x''')\Big]a^\dagger(\vec{p}_1)a^\dagger(\vec{p}_2)|\ \ \rangle\\
&+\underbrace{A^\epsilon(x')\partial^\rho A^\sigma(x''')}\ \ \underbrace{A^\eta(x'')\partial^\mu A^\nu(x''')}\epsilon_{\mu\nu\rho\sigma}\\
&\langle\ \ |a(\vec{p}_1{}')a(\vec{p}_2{}')a(\vec{p}_3{}')N\Big[\Big(\Sigma^\dagger\partial_\epsilon\Sigma\Big)_{x'}\Big(\Theta^\dagger\partial_\eta\Theta\Big)_{x''}\Phi(x''')\Big]a^\dagger(\vec{p}_1)a^\dagger(\vec{p}_2)|\ \ \rangle\\
&=\epsilon_{\mu\nu\rho\sigma}\\
&\Big\{\underbrace{A^\epsilon(x')\partial^\mu A^\nu(x''')}\ \ \underbrace{A^\eta(x'')\partial^\rho A^\sigma(x''')}+\underbrace{A^\epsilon(x')\partial^\rho A^\sigma(x''')}\ \ \underbrace{A^\eta(x'')\partial^\mu A^\nu(x''')}\Big\}\\
&\langle\ \ |a(\vec{p}_1{}')a(\vec{p}_2{}')a(\vec{p}_3{}')N\Big[\Big(\Sigma^\dagger\partial_\epsilon\Sigma\Big)_{x'}\Big(\Theta^\dagger\partial_\eta\Theta\Big)_{x''}\Phi(x''')\Big]a^\dagger(\vec{p}_1)a^\dagger(\vec{p}_2)|\ \ \rangle.
\end{aligned} \tag{7.175}$$

Referring to this result, and equation (7.171), we can finally write down the following fundamental matrix element (see (7.169)) needed to evaluate

$\langle 1', 2', 3'|S|1, 2\rangle^{III}$ given by (7.163) and (7.164):

$$
\begin{aligned}
&\langle\ |a(\vec{p}_1{}')a(\vec{p}_2{}')a(\vec{p}_3{}') \\
&\quad T\Big[N\Big(\Sigma^\dagger\ \partial_\epsilon\Sigma\ A^\epsilon\Big)_{x'}N\Big(\Theta^\dagger\ \partial_\eta\Theta\ A^\eta\Big)_{x''}N\Big(\epsilon_{\mu\nu\rho\sigma}F^{\mu\nu}F^{\rho\sigma}\Phi\Big)_{x'''}\Big] \\
&\qquad\qquad\qquad\qquad a^\dagger(\vec{p}_1)a^\dagger(\vec{p}_2)|\ \rangle \\
&\quad = \epsilon_{\mu\nu\rho\sigma} \\
&\quad \Big\{\underbrace{A^\epsilon(x')\partial^\mu A^\nu(x''')}\ \ \underbrace{A^\eta(x'')\partial^\rho A^\sigma(x''')} \\
&\qquad\qquad + \underbrace{A^\epsilon(x')\partial^\rho A^\sigma(x''')}\ \ \underbrace{A^\eta(x'')\partial^\mu A^\nu(x''')} \\
&\quad - \underbrace{A^\epsilon(x')\partial^\mu A^\nu(x''')}\ \ \underbrace{A^\eta(x'')\partial^\sigma A^\rho(x''')} \\
&\qquad\qquad - \underbrace{A^\epsilon(x')\partial^\sigma A^\rho(x''')}\ \ \underbrace{A^\eta(x'')\partial^\mu A^\nu(x''')} \\
&\quad - \underbrace{A^\epsilon(x')\partial^\nu A^\mu(x''')}\ \ \underbrace{A^\eta(x'')\partial^\rho A^\sigma(x''')} \\
&\qquad\qquad - \underbrace{A^\epsilon(x')\partial^\rho A^\sigma(x''')}\ \ \underbrace{A^\eta(x'')\partial^\nu A^\mu(x''')} \\
&\quad + \underbrace{A^\epsilon(x')\partial^\nu A^\mu(x''')}\ \ \underbrace{A^\eta(x'')\partial^\sigma A^\rho(x''')} \\
&\qquad\qquad + \underbrace{A^\epsilon(x')\partial^\sigma A^\rho(x''')}\ \ \underbrace{A^\eta(x'')\partial^\nu A^\mu(x''')}\Big\} \\
&\quad \langle\ |a(\vec{p}_1{}')a(\vec{p}_2{}')a(\vec{p}_3{}')N\Big[\Big(\Sigma^\dagger\partial_\epsilon\Sigma\Big)_{x'}\Big(\Theta^\dagger\partial_\eta\Theta\Big)_{x''}\Phi(x''')\Big]a^\dagger(\vec{p}_1)a^\dagger(\vec{p}_2)|\ \rangle.
\end{aligned}
\tag{7.176}
$$

After the $T$ ordering is performed, as just done, the $x', x'', x'''$ are dummy variables of integration. Consequently we need not calculate the other *five* matrix elements of the $T$ ordered product which appear in (7.164). With this observation we have our final gauge independent expression for the scattering amplitude under study. This expression reads:

$$\langle 1', 2', 3'|S|1, 2\rangle^{III} = -ig_{s1}\ g_{s2}\ g_{ps} \int_{-\infty}^{\infty} d^4x' \int_{-\infty}^{\infty} d^4x'' \int_{-\infty}^{\infty} d^4x'''\ \ \epsilon_{\mu\nu\rho\sigma}$$

$$\Big\{ \underbrace{A^{\epsilon}(x')\partial^{\mu}A^{\nu}(x''')}\ \ \underbrace{A^{\eta}(x'')\partial^{\rho}A^{\sigma}(x''')}$$

$$+ \underbrace{A^{\epsilon}(x')\partial^{\rho}A^{\sigma}(x''')}\ \ \underbrace{A^{\eta}(x'')\partial^{\mu}A^{\nu}(x''')}$$

$$- \underbrace{A^{\epsilon}(x')\partial^{\mu}A^{\nu}(x''')}\ \ \underbrace{A^{\eta}(x'')\partial^{\sigma}A^{\rho}(x''')}$$

$$- \underbrace{A^{\epsilon}(x')\partial^{\sigma}A^{\rho}(x''')}\ \ \underbrace{A^{\eta}(x'')\partial^{\mu}A^{\nu}(x''')}$$

$$- \underbrace{A^{\epsilon}(x')\partial^{\nu}A^{\mu}(x''')}\ \ \underbrace{A^{\eta}(x'')\partial^{\rho}A^{\sigma}(x''')}$$

$$- \underbrace{A^{\epsilon}(x')\partial^{\rho}A^{\sigma}(x''')}\ \ \underbrace{A^{\eta}(x'')\partial^{\nu}A^{\mu}(x''')}$$

$$+ \underbrace{A^{\epsilon}(x')\partial^{\nu}A^{\mu}(x''')}\ \ \underbrace{A^{\eta}(x'')\partial^{\sigma}A^{\rho}(x''')}$$

$$+ \underbrace{A^{\epsilon}(x')\partial^{\sigma}A^{\rho}(x''')}\ \ \underbrace{A^{\eta}(x'')\partial^{\nu}A^{\mu}(x''')} \Big\}$$

$$\langle\ \ |a(\vec{p}_1^{\,\prime})a(\vec{p}_2^{\,\prime})a(\vec{p}_3^{\,\prime})N\Big[\Big(\Sigma^{\dagger}\partial_{\epsilon}\Sigma\Big)_{x'}\Big(\Theta^{\dagger}\partial_{\eta}\Theta\Big)_{x''}\Phi(x''')\Big]a^{\dagger}(\vec{p}_1)a^{\dagger}(\vec{p}_2)|\ \ \rangle. \tag{7.177}$$

Note that the factor of 6, arising from the 3!, has now disappeared because of six equal contributions from the six time ordered terms on the *r.h.s.* of (7.164). This expression can be further simplified by exploiting the antisymmetry of $\epsilon_{\mu\nu\rho\sigma}$ as follows

$$\begin{aligned} &\epsilon_{\mu\nu\rho\sigma}\Big\{ \underbrace{A^{\epsilon}(x')\partial^{\mu}A^{\nu}(x''')}\ \ \underbrace{A^{\eta}(x'')\partial^{\rho}A^{\sigma}(x''')} \\ &\qquad - \underbrace{A^{\epsilon}(x')\partial^{\mu}A^{\nu}(x''')}\ \ \underbrace{A^{\eta}(x'')\partial^{\sigma}A^{\rho}(x''')} \Big\} \\ &= 2\epsilon_{\mu\nu\rho\sigma}\underbrace{A^{\epsilon}(x')\partial^{\mu}A^{\nu}(x''')}\ \ \underbrace{A^{\eta}(x'')\partial^{\rho}A^{\sigma}(x''')}. \end{aligned} \tag{7.178}$$

Using this identity in (7.177) we obtain

$$\langle 1',2',3'|S|1,2\rangle^{III} = -ig_{s1}\ g_{s2}\ g_{ps} \int\limits_{-\infty}^{\infty} d^4x' \int\limits_{-\infty}^{\infty} d^4x'' \int\limits_{-\infty}^{\infty} d^4x'''\ \ \epsilon_{\mu\nu\rho\sigma}$$

$$2\Big\{ \underbrace{A^{\epsilon}(x')\partial^{\mu}A^{\nu}(x''')}\ \ \underbrace{A^{\eta}(x'')\partial^{\rho}A^{\sigma}(x''')}$$

$$+ \underbrace{A^{\epsilon}(x')\partial^{\rho}A^{\sigma}(x''')}\ \ \underbrace{A^{\eta}(x'')\partial^{\mu}A^{\nu}(x''')}$$

$$+ \underbrace{A^{\epsilon}(x')\partial^{\nu}A^{\mu}(x''')}\ \ \underbrace{A^{\eta}(x'')\partial^{\sigma}A^{\rho}(x''')}$$

$$+ \underbrace{A^{\epsilon}(x')\partial^{\sigma}A^{\rho}(x''')}\ \ \underbrace{A^{\eta}(x'')\partial^{\nu}A^{\mu}(x''')}\Big\}$$

$$\langle\ \ |a(\vec{p}_1{}')a(\vec{p}_2{}')a(\vec{p}_3{}')N\Big[\Big(\Sigma^{\dagger}\partial_{\epsilon}\Sigma\Big)_{x'}\Big(\Theta^{\dagger}\partial_{\eta}\Theta\Big)_{x''}\Phi(x''')\Big]a^{\dagger}(\vec{p}_1)a^{\dagger}(\vec{p}_2)|\ \ \rangle. \tag{7.179}$$

Exploiting the antisymmetry of $\epsilon_{\mu\nu\rho\sigma}$, once more, simplifies this expression to:

$$\langle 1',2',3'|S|1,2\rangle^{III} = -ig_{s1}\ g_{s2}\ g_{ps} \int\limits_{-\infty}^{\infty} d^4x' \int\limits_{-\infty}^{\infty} d^4x'' \int\limits_{-\infty}^{\infty} d^4x'''\ \ \epsilon_{\mu\nu\rho\sigma}$$

$$4\Big\{ \underbrace{A^{\epsilon}(x')\partial^{\mu}A^{\nu}(x''')}\ \ \underbrace{A^{\eta}(x'')\partial^{\rho}A^{\sigma}(x''')}$$

$$+ \underbrace{A^{\epsilon}(x')\partial^{\rho}A^{\sigma}(x''')}\ \ \underbrace{A^{\eta}(x'')\partial^{\mu}A^{\nu}(x''')}\Big\}$$

$$\langle\ \ |a(\vec{p}_1{}')a(\vec{p}_2{}')a(\vec{p}_3{}')N\Big[\Big(\Sigma^{\dagger}\partial_{\epsilon}\Sigma\Big)_{x'}\Big(\Theta^{\dagger}\partial_{\eta}\Theta\Big)_{x''}\Phi(x''')\Big]a^{\dagger}(\vec{p}_1)a^{\dagger}(\vec{p}_2)|\ \ \rangle. \tag{7.180}$$

This is the formal gauge independent expression for the scattering amplitude for the two photon mediated production of a pseudoscalar neutral particle, such as $\pi^0$, in the scattering of two (different) spin zero nuclei. At this stage of our calculations all particles are considered pointlike.

### 7.8 Evaluation of $\langle \;|a(\vec{p}_1{}')a(\vec{p}_2{}')a(\vec{p}_3{}')N\left[\left(\Sigma^\dagger\partial_\epsilon\Sigma\right)_{x'}\left(\Theta^\dagger\partial_\eta\Theta\right)_{x''}\Phi(x''')\right]a^\dagger(\vec{p}_1)a^\dagger(\vec{p}_2)|\; \rangle$ which Appears in $r.h.s$ of (7.180)

Having obtained the expression for $\langle 1',2',3'|S|1,2\rangle^{III}$, we now evaluate the indicated matrix element. Towards this end note that the only surviving contribution to this matrix element comes from terms which contain the following combination of creation and annihilation operators $a^\dagger(\vec{p}_1)a^\dagger(\vec{p}_2)a^\dagger(\vec{p}_3)a(\vec{p}_1)a(\vec{p}_2)$. To pick these terms we write the *argument* of $N[\cdots]$ explicitly. It reads

$$\begin{aligned}
&\int \frac{d^3p\,''_1}{(2\pi)^3 2\omega_{p''_1}}\left[a^\dagger(\vec{p}\,''_1)e^{ip''_1\cdot x'} + a(\vec{p}\,''_1)e^{-ip''_1\cdot x'}\right] \\
&\int \frac{d^3p'''_1}{(2\pi)^3 2\omega_{p'''_1}}\left[a(\vec{p}\,'''_1)(-ip'''_{1\epsilon})e^{-ip'''_1\cdot x'} + a^\dagger(\vec{p}\,'''_1)(ip'''_{1\epsilon})e^{ip'''_1\cdot x'}\right] \\
&\int \frac{d^3p''_2}{(2\pi)^3 2\omega_{p''_2}}\left[a^\dagger(\vec{p}\,''_2)e^{ip''_2\cdot x''} + a(\vec{p}\,''_2)e^{-ip''_2\cdot x''}\right] \\
&\int \frac{d^3p'''_2}{(2\pi)^3 2\omega_{p'''_2}}\left[a(\vec{p}\,'''_2)(-ip'''_{2\eta})e^{-ip'''_2\cdot x''} + a^\dagger(\vec{p}\,'''_2)(ip'''_{2\eta})e^{ip'''_2\cdot x''}\right] \\
&\int \frac{d^3p''_3}{(2\pi)^3 2\omega_{p''_3}}\left[a(\vec{p}\,''_3)e^{-ip''_3\cdot x'''} + a^\dagger(\vec{p}\,''_3)e^{ip''_3\cdot x'''}\right].
\end{aligned} \tag{7.181}$$

By inspection, we can now write down the contribution from $N[\cdots]$ which will survive. It is

$$\begin{aligned}
&\int \frac{d^3p''_1}{(2\pi)^3 2\omega_{p''_1}}\int \frac{d^3p'''_1}{(2\pi)^3 2\omega_{p'''_1}}\int \frac{d^3p''_2}{(2\pi)^3 2\omega_{p''_2}}\int \frac{d^3p'''_2}{(2\pi)^3 2\omega_{p'''_2}}\int \frac{d^3p''_3}{(2\pi)^3 2\omega_{p''_3}} \\
&\Big\{a^\dagger(\vec{p}\,''_1)a(\vec{p}\,'''_1)a^\dagger(\vec{p}\,''_2)a(\vec{p}\,'''_2)a^\dagger(\vec{p}\,''_3) \\
&\qquad\left(e^{ip''_1\cdot x'}(-ip'''_{1\epsilon})e^{-ip'''_1\cdot x'}e^{ip''_2\cdot x''}(-ip'''_{2\eta})e^{-ip'''_2\cdot x''}e^{ip''_3\cdot x'''}\right) \\
&+ a^\dagger(\vec{p}\,'''_1)a(\vec{p}\,''_1)a^\dagger(\vec{p}\,'''_2)a(\vec{p}\,''_2)a^\dagger(\vec{p}\,''_3) \\
&\qquad\left(e^{-ip''_1\cdot x'}(ip'''_{1\epsilon})e^{ip'''_1\cdot x'}e^{-ip''_2\cdot x''}(ip'''_{2\eta})e^{ip'''_2\cdot x''}e^{ip''_3\cdot x'''}\right)\Big\}.
\end{aligned} \tag{7.182}$$

As a result, we have

$$\langle \; |a(\vec{p}_1{}')a(\vec{p}_2{}')a(\vec{p}_3{}')N\left[\left(\Sigma^\dagger\partial_\epsilon\Sigma\right)_{x'}\left(\Theta^\dagger\partial_\eta\Theta\right)_{x''}\Phi(x''')\right]a^\dagger(\vec{p}_1)a^\dagger(\vec{p}_2)| \; \rangle =$$
$$\int\frac{d^3p_1''}{(2\pi)^3 2\omega_{p_1''}}\int\frac{d^3p_1'''}{(2\pi)^3 2\omega_{p_1'''}}\int\frac{d^3p_2''}{(2\pi)^3 2\omega_{p_2''}}\int\frac{d^3p_2'''}{(2\pi)^3 2\omega_{p_2'''}}\int\frac{d^3p_3''}{(2\pi)^3 2\omega_{p_3''}}$$
$$\Big\{\langle\vec{p}\,'_1,\vec{p}\,'_2,\vec{p}\,'_3,\vec{p}\,'''_1,\vec{p}\,'''_2|\vec{p}_1,\vec{p}_2,\vec{p}\,''_1,\vec{p}\,''_2,\vec{p}\,''_3\rangle \tag{7.183}$$
$$\left(e^{ip_1''\cdot x'}(-ip_{1\epsilon}''')e^{-ip_1'''\cdot x'}e^{ip_2''\cdot x''}(-ip_{2\eta}''')e^{-ip_2'''\cdot x''}e^{ip_3''\cdot x'''}\right)$$
$$+\langle\vec{p}\,'_1,\vec{p}\,'_2,\vec{p}\,'_3,\vec{p}\,''_1,\vec{p}\,''_2|\vec{p}_1,\vec{p}_2,\vec{p}\,'''_1,\vec{p}\,'''_2,\vec{p}\,''_3\rangle$$
$$\left(e^{-ip_1''\cdot x'}(ip_{1\epsilon}''')e^{ip_1'''\cdot x'}e^{-ip_2''\cdot x''}(ip_{2\eta}''')e^{ip_2'''\cdot x''}e^{ip_3''\cdot x'''}\right)\Big\}.$$

The integrations are easily performed by noting that

$$\langle\vec{p}\,'_1,\vec{p}\,'_2,\vec{p}\,'_3,\vec{p}\,'''_1,\vec{p}\,'''_2|\vec{p}_1,\vec{p}_2,\vec{p}\,''_1,\vec{p}\,''_2,\vec{p}\,''_3\rangle =$$
$$(2\pi)^{3\times 5}\;(2)^5\omega_{p_1'}\;\omega_{p_2}{}'\;\omega_{p_3}{}'\omega_{p_1}\omega_{p_2}\delta^3(\vec{p}\,'_1-\vec{p}\,''_1)\delta^3(\vec{p}\,'_2-\vec{p}\,''_2)\delta^3(\vec{p}\,'_3-\vec{p}\,''_3) \tag{7.184}$$
$$\delta^3(\vec{p}\,'''_1-\vec{p}_1)\delta^3(\vec{p}\,'''_2-\vec{p}_2), etc.,$$

yielding the result

$$\langle \; |a(\vec{p}_1{}')a(\vec{p}_2{}')a(\vec{p}_3{}')N\left[\left(\Sigma^\dagger\partial_\epsilon\Sigma\right)_{x'}\left(\Theta^\dagger\partial_\eta\Theta\right)_{x''}\Phi(x''')\right]a^\dagger(\vec{p}_1)a^\dagger(\vec{p}_2)| \; \rangle$$
$$= e^{ip_1'\cdot x'}(-ip_{1\epsilon})e^{-ip_1\cdot x'}e^{ip_2'\cdot x''}(-ip_{2\eta})e^{-ip_2\cdot x''}e^{ip_3'\cdot x'''} \tag{7.185}$$
$$+ e^{-ip_1'\cdot x'}(ip_{1\epsilon})e^{ip_1\cdot x'}e^{-ip_2'\cdot x''}(ip_{2\eta})e^{ip_2\cdot x''}e^{ip_3'\cdot x'''}.$$

## 7.9 Expression for $\langle 1', 2', 3'|S|1, 2\rangle^{III}$ in Covariant Lorentz Gauge

Inserting (7.185) in (7.180) gives

$$\langle 1', 2', 3'|S|1, 2\rangle^{III} = 4ig_{s1}\, g_{s2}\, g_{ps} \int_{-\infty}^{\infty} d^4x' \int_{-\infty}^{\infty} d^4x'' \int_{-\infty}^{\infty} d^4x'''\ \ p_{1\epsilon} p_{2\eta}\ \epsilon_{\mu\nu\rho\sigma}$$
$$\left( e^{i(p_1'-p_1)\cdot x'} e^{i(p_2'-p_2)\cdot x''} e^{ip_3'\cdot x'''} + e^{-i(p_1'-p_1)\cdot x'} e^{-i(p_2'-p_2)\cdot x''} e^{ip_3'\cdot x'''} \right)$$
$$\Big\{ \underbrace{A^{\epsilon}(x')\partial^{\mu}A^{\nu}(x''')}\ \ \underbrace{A^{\eta}(x'')\partial^{\rho}A^{\sigma}(x''')}$$
$$+ \underbrace{A^{\epsilon}(x')\partial^{\rho}A^{\sigma}(x''')}\ \ \underbrace{A^{\eta}(x'')\partial^{\mu}A^{\nu}(x''')} \Big\} \tag{7.186}$$

In the covariant Lorentz gauge the contractions appearing in (7.186) can be evaluated using (see e.g. Ref. [53])

$$\begin{aligned}\underbrace{A^{\mu}(x)A^{\nu}(x')} &= \langle\ \ |T\big(A^{\mu}(x)A^{\nu}(x')\big)|\ \ \rangle \\ &= -i\int \frac{d^4k}{(2\pi)^4}\frac{\eta^{\mu\nu}}{k^2+i\epsilon}e^{-ik\cdot(x-x')}.\end{aligned} \tag{7.187}$$

As a consequence

$$\underbrace{A^{\mu}(x)\ \partial^{\sigma}A^{\nu}(x')} = \int \frac{d^4k}{(2\pi)^4}\frac{\eta^{\mu\nu}\ k^{\sigma}}{k^2+i\epsilon}e^{-ik\cdot(x-x')}. \tag{7.188}$$

Using (7.188) in (7.186) leads to the following expression for the desired S–matrix element

$$\langle 1', 2', 3'|S|1, 2\rangle^{III} = 4i\ g_{s1}\ g_{s2}\ g_{ps} \int_{-\infty}^{\infty} d^4x' \int_{-\infty}^{\infty} d^4x'' \int_{-\infty}^{\infty} d^4x'''\ p_{1\epsilon} p_{2\eta}\ \epsilon_{\mu\nu\rho\sigma}$$
$$\left( e^{i(p_1'-p_1)\cdot x'} e^{i(p_2'-p_2)\cdot x''} e^{ip_3'\cdot x'''} + e^{-i(p_1'-p_1)\cdot x'} e^{-i(p_2'-p_2)\cdot x''} e^{ip_3'\cdot x'''} \right)$$
$$\Bigg[ \left( \int \frac{d^4k}{(2\pi)^4}\frac{\eta^{\epsilon\nu}\ k^{\mu}}{k^2+i\epsilon}e^{-ik\cdot(x'-x''')} \right) \left( \int \frac{d^4k'}{(2\pi)^4}\frac{\eta^{\eta\sigma}\ k'^{\rho}}{k'^2+i\epsilon}e^{-ik'\cdot(x''-x''')} \right)$$
$$+ \left( \int \frac{d^4k''}{(2\pi)^4}\frac{\eta^{\epsilon\sigma}\ k''^{\rho}}{k''^2+i\epsilon}e^{-ik''\cdot(x'-x''')} \right) \left( \int \frac{d^4k'''}{(2\pi)^4}\frac{\eta^{\eta\nu}\ k'''^{\mu}}{k'''^2+i\epsilon}e^{-ik'''\cdot(x''-x''')} \right) \Bigg] \tag{7.189}$$

This expression can be further simplified and brought to a more physically transparent form by performing the integrations over $x', x''$ and $x'''$. The result of this

manipulation is:

$$\langle 1', 2', 3'|S|1, 2\rangle^{III} = 4i\ g_{s1}\ g_{s2}\ g_{ps}\ \ \epsilon_{\mu\nu\rho\sigma}\ \ p_{1\epsilon}\ p_{2\eta} \int \frac{d^4k}{(2\pi)^4} \int \frac{d^4k'}{(2\pi)^4}$$
$$\frac{\eta^{\epsilon\nu}k^{\mu}}{k^2+i\epsilon}\frac{\eta^{\eta\sigma}k'^{\rho}}{k'^2+i\epsilon}\Big[(2\pi)^{4\times 3}\delta^4(p_1'-p_1-k)\delta^4(p_2'-p_2-k')\delta^4(p_3'+k+k')$$
$$+ (2\pi)^{4\times 3}\delta^4(-p_1'+p_1-k)\delta^4(-p_2'+p_2-k')\delta^4(p_3'+k+k')\Big]$$
$$+ 4i\ g_{s1}\ g_{s2}\ g_{ps}\ \ \epsilon_{\mu\nu\rho\sigma}\ \ p_{1\epsilon}\ p_{2\eta} \int \frac{d^4k''}{(2\pi)^4} \int \frac{d^4k'''}{(2\pi)^4}$$
$$\frac{\eta^{\epsilon\sigma}k''^{\rho}}{k''^2+i\epsilon}\frac{\eta^{\eta\nu}k'''^{\mu}}{k'''^2+i\epsilon}\Big[(2\pi)^{4\times 3}\delta^4(p_1'-p_1-k'')\delta^4(p_2'-p_2-k''')\delta^4(p_3'+k''+k''')$$
$$+ (2\pi)^{4\times 3}\delta^4(-p_1'+p_1-k'')\delta^4(-p_2'+p_2-k''')\delta^4(p_3'+k''+k''')\Big] \tag{7.190}$$

## 7.10 Two Photon Mediated Production of a Electrically Neutral Scalar Particle in the High Energy Scattering of Two Charged Particles

To study the two photon mediated production of electrically neutral scalar particles such as $f_0(975)$, $a_0(980)$ and $f_0(1400)$ we choose

$\chi_1 =$ A scalar particle of mass $m_1$ and charge $Q_1$,

$\chi_2 =$ A scalar particle of mass $m_2$ and charge $Q_2$,

$\chi_3 =$ A *neutral scalar* particle of mass $m_3$.

Such a process approximates the photoproduction of a neutral scalar meson in the scattering of two spin zero nuclei. The matter field operators associated with the three particles involved are:

$$\chi_1 :\quad \Sigma(x) = \int \frac{d^3p_1''}{(2\pi)^3 2\omega_{p_1''}}\left[a(\vec{p}_1{}'')e^{-ip_1''\cdot x} + a^{\dagger}(\vec{p}_1{}'')e^{ip_1''\cdot x}\right] \tag{7.191}$$

$$\chi_2 :\quad \Theta(x) = \int \frac{d^3p_2''}{(2\pi)^3 2\omega_{p_2''}}\left[a(\vec{p}_2{}'')e^{-ip_2''\cdot x} + a^{\dagger}(\vec{p}_2{}'')e^{ip_2''\cdot x}\right] \tag{7.192}$$

$$\chi_3: \quad \Xi(x) = \int \frac{d^3p''_3}{(2\pi)^3 2\omega_{p_3{}''}} \Big[ a(\vec{p_3}'') e^{-ip_3{}''\cdot x} + a^\dagger(\vec{p_3}'') e^{ip_3{}''\cdot x} \Big] \tag{7.193}$$

with

$$\omega_p = (m^2 + \vec{p}^{\ 2})^{1/2}. \tag{7.194}$$

The interaction Lagrangian density operator for the two photon mediated production of scalar particles is

$$\begin{aligned} \mathcal{L}_{int.}(x) = g_{s1} \ \ \Sigma(x)^\dagger \ \partial_\mu \Sigma(x) \ A^\mu(x) + g_{s2} \ \ \Theta(x)^\dagger \ \partial_\mu \Theta(x) \ A^\mu(x) \\ + g_s \ \ F_{\mu\nu}(x) F^{\mu\nu}(x) \ \Xi(x), \end{aligned} \tag{7.195}$$

where we have suppressed the *Normal Ordering.* To calculate the transition amplitude

$$|i\rangle = |m_1, \vec{p}_1; \ m_2, \vec{p}_2\rangle \rightarrow |f\rangle = |m_1, \vec{p}_1{}'; \ m_2, \vec{p}_2{}'; m_3, \vec{p}_3{}'\rangle \tag{7.196}$$

all arguments of Sec. 7.7 *up to* equation (7.177) remain unmodified provided the following substitutions are made:

$$\begin{aligned} &\mathrm{g_{ps}} \rightarrow \mathrm{g_s}, \quad \epsilon_{\mu\nu\rho\sigma} \rightarrow 1, \quad \text{Superscript } \mu \rightarrow \text{Subscript } \mu, \\ &\text{Superscript } \nu \rightarrow \text{Subscript } \nu; \quad \rho \rightarrow \mu, \quad \sigma \rightarrow \nu, \quad \Phi(\mathrm{x}) \rightarrow \Xi(\mathrm{x}). \end{aligned} \tag{7.197}$$

The counterpart of (7.177) then reads

$$\langle 1', 2', 3'|S|1, 2\rangle^{III} = -ig_{s1}\ g_{s2}\ g_s \int\limits_{-\infty}^{\infty} d^4x' \int\limits_{-\infty}^{\infty} d^4x'' \int\limits_{-\infty}^{\infty} d^4x'''$$

$$\Big\{ \underbrace{A^{\epsilon}(x')\partial_{\mu}A_{\nu}(x''')}\ \ \underbrace{A^{\eta}(x'')\partial^{\mu}A^{\nu}(x''')}$$

$$+ \underbrace{A^{\epsilon}(x')\partial^{\mu}A^{\nu}(x''')}\ \ \underbrace{A^{\eta}(x'')\partial_{\mu}A_{\nu}(x''')}$$

$$- \underbrace{A^{\epsilon}(x')\partial_{\mu}A_{\nu}(x''')}\ \ \underbrace{A^{\eta}(x'')\partial^{\nu}A^{\mu}(x''')}$$

$$- \underbrace{A^{\epsilon}(x')\partial^{\nu}A^{\mu}(x''')}\ \ \underbrace{A^{\eta}(x'')\partial_{\mu}A_{\nu}(x''')}$$

$$- \underbrace{A^{\epsilon}(x')\partial_{\nu}A_{\mu}(x''')}\ \ \underbrace{A^{\eta}(x'')\partial^{\mu}A^{\nu}(x''')}$$

$$- \underbrace{A^{\epsilon}(x')\partial^{\mu}A^{\nu}(x''')}\ \ \underbrace{A^{\eta}(x'')\partial_{\nu}A_{\mu}(x''')}$$

$$+ \underbrace{A^{\epsilon}(x')\partial_{\nu}A_{\mu}(x''')}\ \ \underbrace{A^{\eta}(x'')\partial^{\nu}A^{\mu}(x''')}$$

$$+ \underbrace{A^{\epsilon}(x')\partial^{\nu}A^{\mu}(x''')}\ \ \underbrace{A^{\eta}(x'')\partial_{\nu}A_{\mu}(x''')}\Big\}$$

$$\langle\ \ |a(\vec{p}_1{}')a(\vec{p}_2{}')a(\vec{p}_3{}')N\Big[\Big(\Sigma^{\dagger}\partial_{\epsilon}\Sigma\Big)_{x'}\Big(\Theta^{\dagger}\partial_{\eta}\Theta\Big)_{x''}\Xi(x''')\Big]a^{\dagger}(\vec{p}_1)a^{\dagger}(\vec{p}_2)|\ \ \rangle. \tag{7.198}$$

This expression immediately simplifies to

$$\langle 1', 2', 3'|S|1, 2\rangle^{III} = -ig_{s1}\ g_{s2}\ g_s \int\limits_{-\infty}^{\infty} d^4x' \int\limits_{-\infty}^{\infty} d^4x'' \int\limits_{-\infty}^{\infty} d^4x'''$$

$$4\Big\{ \underbrace{A^{\epsilon}(x')\partial_{\mu}A_{\nu}(x''')}\ \ \underbrace{A^{\eta}(x'')\partial^{\mu}A^{\nu}(x''')}$$

$$- \underbrace{A^{\epsilon}(x')\partial_{\mu}A_{\nu}(x''')}\ \ \underbrace{A^{\eta}(x'')\partial^{\nu}A^{\mu}(x''')}\Big\}$$

$$\langle\ \ |a(\vec{p}_1{}')a(\vec{p}_2{}')a(\vec{p}_3{}')N\Big[\Big(\Sigma^{\dagger}\partial_{\epsilon}\Sigma\Big)_{x'}\Big(\Theta^{\dagger}\partial_{\eta}\Theta\Big)_{x''}\Xi(x''')\Big]a^{\dagger}(\vec{p}_1)a^{\dagger}(\vec{p}_2)|\ \ \rangle. \tag{7.199}$$

Evaluating

$\langle\ \ |a(\vec{p}_1{}')a(\vec{p}_2{}')a(\vec{p}_3{}')N\Big[\Big(\Sigma^{\dagger}\partial_{\epsilon}\Sigma\Big)_{x'}\Big(\Theta^{\dagger}\partial_{\eta}\Theta\Big)_{x''}\Xi(x''')\Big]a^{\dagger}(\vec{p}_1)a^{\dagger}(\vec{p}_2)|\ \ \rangle$ as in Sec. 7.8 ($\Phi(x) \to \Xi(x)$, the rest is identical) and substituting the result in the above

expression we get

$$\langle 1', 2', 3'|S|1, 2\rangle^{III} = 4ig_{s1}\ g_{s2}\ g_s \int\limits_{-\infty}^{\infty} d^4x' \int\limits_{-\infty}^{\infty} d^4x'' \int\limits_{-\infty}^{\infty} d^4x'''\ \ p_{1\epsilon}\ p_{2\eta}$$
$$\left( e^{i(p_1'-p_1)\cdot x'} e^{i(p_2'-p_2)\cdot x''} e^{ip_3'\cdot x'''} + e^{-i(p_1'-p_1)\cdot x'} e^{-i(p_2'-p_2)\cdot x''} e^{ip_3'\cdot x'''} \right)$$
$$\Big\{ \underbrace{A^\epsilon(x')\partial_\mu A_\nu(x''')}\ \ \underbrace{A^\eta(x'')\partial^\mu A^\nu(x''')}$$
$$- \underbrace{A^\epsilon(x')\partial_\mu A_\nu(x''')}\ \ \underbrace{A^\eta(x'')\partial^\nu A^\mu(x''')} \Big\} \tag{7.200}$$

To continue further, note

$$\underbrace{A^\epsilon(x)\ \partial_\mu A_\nu(x')} = \eta_{\mu\rho}\ \eta_{\nu\sigma}\ \ \underbrace{A^\epsilon(x)\ \partial^\rho A^\sigma(x)}. \tag{7.201}$$

In the covariant Lorentz gauge we have (see (7.188))

$$\underbrace{A^\epsilon(x)\ \partial^\rho A^\sigma(x')} = \int \frac{d^4k}{(2\pi)^4} \frac{\eta^{\epsilon\sigma}\ k^\rho}{k^2 + i\epsilon} e^{-ik\cdot(x-x')}. \tag{7.202}$$

As such

$$\underbrace{A^\epsilon(x)\ \partial_\mu A_\nu(x')} = \eta_{\mu\rho}\ \eta_{\nu\sigma}\ \int \frac{d^4k}{(2\pi)^4} \frac{\eta^{\epsilon\sigma}\ k^\rho}{k^2 + i\epsilon} e^{-ik\cdot(x-x')}. \tag{7.203}$$

Substituting this result in equation (7.200) gives

$$\langle 1', 2', 3'|S|1, 2\rangle^{III} = 4i\ g_{s1}\ g_{s2}\ g_s \int_{-\infty}^{\infty} d^4x' \int_{-\infty}^{\infty} d^4x'' \int_{-\infty}^{\infty} d^4x'''\ p_{1\epsilon}\ p_{2\eta}$$
$$\left(e^{i(p'_1-p_1)\cdot x'} e^{i(p'_2-p_2)\cdot x''} e^{ip'_3\cdot x'''} + e^{-i(p'_1-p_1)\cdot x'} e^{-i(p'_2-p_2)\cdot x''} e^{ip'_3\cdot x'''}\right)$$
$$\left[\left(\eta_{\mu\rho}\eta_{\nu\sigma}\int \frac{d^4k}{(2\pi)^4}\frac{\eta^{\epsilon\sigma}\ k^\rho}{k^2+i\epsilon}e^{-ik\cdot(x'-x''')}\right)\left(\int \frac{d^4k'}{(2\pi)^4}\frac{\eta^{\eta\nu}\ k'^\mu}{k'^2+i\epsilon}e^{-ik'\cdot(x''-x''')}\right)-\right.$$
$$\left.\left(\eta_{\mu\rho}\eta_{\nu\sigma}\int \frac{d^4k''}{(2\pi)^4}\frac{\eta^{\epsilon\sigma}\ k''^\rho}{k''^2+i\epsilon}e^{-ik''\cdot(x'-x''')}\right)\left(\int \frac{d^4k'''}{(2\pi)^4}\frac{\eta^{\eta\mu}\ k'''^\nu}{k'''^2+i\epsilon}e^{-ik'''\cdot(x''-x''')}\right)\right] \tag{7.204}$$

Performing the $x', x''$ and $x'''$ integrations then yields the counterpart of (7.190)

$$\langle 1', 2', 3'|S|1, 2\rangle^{III} = 4i\ g_{s1}\ g_{s2}\ g_s\ \ p_{1\epsilon}\ p_{2\eta}\ \eta_{\mu\rho}\ \eta_{\nu\sigma}\int \frac{d^4k}{(2\pi)^4}\int \frac{d^4k'}{(2\pi)^4}$$
$$\frac{\eta^{\epsilon\sigma}k^\rho}{k^2+i\epsilon}\frac{\eta^{\eta\nu}k'^\mu}{k'^2+i\epsilon}\Big[(2\pi)^{4\times 3}\delta^4(p'_1-p_1-k)\delta^4(p'_2-p_2-k')\delta^4(p'_3+k+k')$$
$$+ (2\pi)^{4\times 3}\delta^4(-p'_1+p_1-k)\delta^4(-p'_2+p_2-k')\delta^4(p'_3+k+k')\Big]$$
$$- 4i\ g_{s1}\ g_{s2}\ g_s\ \ p_{1\epsilon}\ p_{2\eta}\ \eta_{\mu\rho}\ \eta_{\nu\sigma}\int \frac{d^4k''}{(2\pi)^4}\int \frac{d^4k'''}{(2\pi)^4}$$
$$\frac{\eta^{\epsilon\sigma}k''^\rho}{k''^2+i\epsilon}\frac{\eta^{\eta\mu}k'''^\nu}{k'''^2+i\epsilon}\Big[(2\pi)^{4\times 3}\delta^4(p'_1-p_1-k'')\delta^4(p'_2-p_2-k''')\delta^4(p'_3+k''+k''')$$
$$+ (2\pi)^{4\times 3}\delta^4(-p'_1+p_1-k'')\delta^4(-p'_2+p_2-k''')\delta^4(p'_3+k''+k''')\Big] \tag{7.205}$$

## 7.11 Conclusions

In this work we have provided a general procedure to construct $(j,0)\oplus(0,j)$ covariant spinors for any spin. These covariant spinors are then used to construct [23] arbitrary–spin causal propagators. While at present we do not have acuaslity–free relativistic wave equations, which the $(j,0)\oplus(0,j)$ covariant spinors satisfy, we have established that the Weinberg Equations suffer from kinematical acausality.

# APPENDIX

# ELEMENTS OF CANONICAL QUANTUM FIELD THEORY

In this appendix we provide essential elements of the canonical quantum field theory. We will establish how our work contained in the main text of this work interfaces at various levels with canonical quantum field theory.

## A1 Casimir Operators and Pauli–Lubánski Pseudovector

Let $|\psi\rangle$ be the state of a system as observed by an inertial observer $\mathcal{O}$. If $|\psi\rangle'$ represents the state of the *same* system as observed by *another* inertial observer $\mathcal{O}'$, then

$$|\psi\rangle' = U(\{\Lambda, a\})|\psi\rangle, \tag{A1}$$

where $\{\Lambda, a\}$ characterises the transformation which relates $\mathcal{O}$ with $\mathcal{O}'$:

$$x'^{\mu} = \Lambda^{\mu}{}_{\nu} x^{\nu} + a^{\mu}. \tag{A2}$$

$U(\{\Lambda, a\})$ is an operator satisfying:

$$U(\{\overline{\Lambda}, \overline{a}\})U(\{\Lambda, a\}) = U(\{\overline{\Lambda}\Lambda, \overline{\Lambda}a + \overline{a}\}). \tag{A3}$$

However, because of the equivalence of *all* inertial observers for the description of a system it, follows that together with $|\psi\rangle$, $|\psi\rangle'$ is also a possible state as viewed by the *original* inertial observer $\mathcal{O}$. Thus the representation space on which unitary operators $U(\{\Lambda, a\})$ act contains with every $|\psi\rangle$, *all* transforms $U(\{\Lambda, a\})|\psi\rangle$, with $\{\Lambda, a\}$ as *any* Poincaré transformation.

To each solution $U(\{\Lambda, a\})$ of (A3) corresponds a representation space. The question now naturally arises: What are the quantum numbers which distinguish one representation space from another? Casimir invariants are considered the most suitable candidates for these quantum numbers.

Poincaré group has two Casimir operators:

$$C_1 = P_\mu P^\mu, \tag{A4}$$

$$C_2 = W_\mu W^\mu, \tag{A5}$$

where $W_\mu$ is defined as

$$W_\mu = \alpha \epsilon_{\mu\nu\rho\sigma} J^{\nu\rho} P^\sigma, \tag{A6}$$

with $\alpha$ a c–number constant. $W_\mu$ is called the *Pauli–Lubánski pseudovector.* It was first introduced by Lubánski [48] in 1942 with acknowledgements to Pauli.

It is readily seen that the Pauli–Lubánski operator has the following properties:

$$W_\mu P^\mu = 0. \tag{A7}$$

$$[W^\mu, P^\lambda] = 0. \tag{A8}$$

Equation (A7) follows from the vanishing of the commutator $[P^\sigma, P^\mu]$ because it makes $P^\sigma P^\mu$, which appears in $W_\mu P^\mu = \alpha \epsilon_{\mu\nu\rho\sigma} J^{\nu\rho} P^\sigma P^\mu$ symmetric in the indices $\sigma, \mu$. On the other hand $\epsilon_{\mu\nu\rho\sigma}$ is antisymmetric in the same indices. These two observations immediately yield the result (A7). The proof of (A8) is as follows

$$\begin{aligned} [W^\mu, P^\lambda] &= \alpha[\epsilon^{\mu\nu\rho\sigma} J_{\nu\rho} P_\sigma, P^\lambda] \\ &= \alpha J_{\nu\rho}[\epsilon^{\mu\nu\rho\sigma} P_\sigma, P^\lambda] + \alpha[\epsilon^{\mu\nu\rho\sigma} J_{\nu\rho}, P^\lambda] P_\sigma \\ &= \alpha \epsilon^{\mu\nu\rho\sigma} \left\{ -i(\eta^\lambda{}_\nu P_\rho - \eta^\lambda{}_\rho P_\nu) \right\} P_\sigma. \end{aligned} \tag{A9}$$

Now note that $\eta^\lambda{}_\nu = \eta_{\nu\epsilon}\eta^{\lambda\epsilon} = \delta^\lambda{}_\nu$. This yields

$$[W^\mu, P^\lambda] = i\alpha \epsilon^{\mu\nu\rho\sigma} \left\{ \delta^\lambda{}_\rho P_\nu - \delta^\lambda{}_\nu P_\rho \right\} P_\sigma = i\alpha \left\{ \epsilon^{\mu\lambda\sigma\nu} P_\nu - \epsilon^{\mu\lambda\sigma\nu} P_\nu \right\} = 0, \tag{A10}$$

where we used the complete antisymmetry of $\epsilon^{\mu\nu\rho\sigma}$ and then renamed appropriate indices. More properties of the Pauli–Lubánski operator can be found on page 195 of Tung [55].

As a consequence of the constraint (7), $W_\mu$ *has a maximum of three independent components.* Further it is translationally invariant. This implies that $C_2 \equiv W_\mu W^\mu$ is also translationally invariant. Since $W_\mu$ is a four vector, $W_\mu W^\mu$ is also invariant under pure Lorentz transformations. This establishes $C_2$ to be a Casimir operator for the Poincaré group. That $C_1$ is a Casimir operator is obvious. $P_\mu$ is translationally invariant, which makes $P_\mu P^\mu$ translationally invariant. In addition because $P_\mu$ is a four vector, $P_\mu P^\mu$ is invariant under pure Lorentz transformations. As a result $C_1$ is a Casimir operator.

Restricting ourselves to timelike and lightlike momenta $p^\mu$ we are led to two physically distinct classes of representation spaces.

## A2 States with $m \neq 0$: Timelike $p^\mu$, Spin and Little Group

Since the Pauli–Lubánski operator commutes with the energy–momentum four vector the two Casimir operators of the Poincaré group commute. So let $|\psi\rangle$ be a simultaneous eigenstate of $C_1$ and $C_2$. The Casimir invariants are readily found by considering a *standard vector* $p^\mu = (m, 0, 0, 0)$. Then we find that

$$C_1|\psi\rangle = P_\mu P^\mu|\psi\rangle = p_\mu p^\mu|\psi\rangle = m^2|\psi\rangle. \tag{A11}$$

As such the first Casimir invariant is identified as [by definition] the square of the *Poincaré invariant mass* associated with each of the states in the representation space to which $|\psi\{p^\mu = (m, 0, 0, 0)\}\rangle$ belongs.

Thus one of the quantum numbers by which a given representation space can be labelled is the Poincaré invariant mass, $m$. *All* physical states in the same representation space carry the *same* mass. This is a quantity which is to be determined *experimentally.* A theory in which this number itself could be *theoretically* calculated or related to other incalculable numbers of the present theory, such as the electronic charge $e$, the Planck constant $\hbar$, the speed of light $c$, the gravitational constant $G$[32] is not known at present.

To learn about the physical nature of the second Casimir invariant let's consider the action of $C_2$ on $|\psi\rangle$. Again, we choose the standard vector $p^\mu = (m, 0, 0, 0)$ throughout the calculations which follow. Since $p^\mu = (m, 0, 0, 0)$, the orbital angular momentum vanishes and we should replace $\vec{J}$, the total angular momentum, by $\vec{S}$, the spin angular momentum.

$$C_2\ |\psi\rangle = \alpha^2 \epsilon_{\mu\nu\rho\sigma} S^{\nu\rho} P^\sigma \epsilon^{\mu\alpha\beta\theta} S_{\alpha\beta} P_\theta\ |\psi\rangle = \alpha^2 m^2 \epsilon_{\mu\nu\rho 0} \epsilon^{\mu\alpha\beta 0} S^{\nu\rho} S_{\alpha\beta}\ |\psi\rangle \tag{A12}$$

The complete antisymmetry of $\epsilon^{\mu\nu\rho\sigma}$ implies that only $\mu \neq 0,\ \nu \neq 0,\ \rho \neq 0,\ \alpha \neq 0$ and $\beta \neq 0$ terms can survive in the *r.h.s* of the above expression. As such we rename the indices as follows: $\mu \rightarrow k,\ \rho \rightarrow l,\ \nu \rightarrow q,\ \alpha \rightarrow i,$ and, $\beta \rightarrow j$ where each of the new indices runs over $1, 2, 3$. With these substitutions equation (A12) reads

$$C_2\ |\psi\rangle = m^2 \alpha^2 \epsilon_{kql0} \epsilon^{kij0} S^{ql} S_{ij}\ |\psi\rangle. \tag{A13}$$

With the convention

$$\epsilon^{0123} = +1, \tag{A14}$$

we have

$$\epsilon_{kql0} \epsilon^{kij0} = (\delta^i_q \delta^j_l - \delta^i_l \delta^j_q). \tag{A15}$$

Therefore

$$\begin{aligned} C_2\ |\psi\rangle &= m^2 \alpha^2 [S^{ij} S_{ij} - S^{ji} S_{ij}]\ |\psi\rangle \\ &= m^2 \alpha^2 2 [\epsilon^{ijk} \epsilon_i j l S_k S^l]\ |\psi\rangle \\ &= m^2 \alpha^2 2 [2 \delta^k_l S_k S^l]\ |\psi\rangle. \end{aligned} \tag{A16}$$

That is eigenvalues of $C_2$ are proportional to those of $\vec{S}^{\ 2}$,

$$C_2\ |\psi\rangle = m^2 \alpha^2 2 \cdot 2 S_k S^k\ |\psi\rangle \tag{A17}$$

---

32 A constant related to a phenomenon which, it should be noted, extends Poincaré covariance to General covariance.

As a result if we choose

$$\alpha = \frac{1}{2}, \tag{A18}$$

and note

$$S_k S^k \ |\psi\rangle \equiv -\vec{S}^2 \ |\psi\rangle \ = \ -s(s+1) \ |\psi\rangle \tag{A19}$$

for $|\psi\rangle = |\psi\{p^\mu = (m, 0, 0, 0)\}\rangle$ we get the second Casimir invariant

$$c_2 = -m^2 s(s+1). \tag{A20}$$

Therefore the quantum numbers $m$ and $s$ distinguish one representation space from another. A physical state in the representation space of timelike momenta can thus be written as

$$|\psi\rangle_{p^\mu p_\mu > 0} : \quad |p^\mu; \ m, s, \sigma\rangle. \tag{A21}$$

with

$$P^\mu \ |p^\mu; \ m, s, \sigma\rangle = p^\mu |p^\mu; \ m, s, \sigma\rangle, \tag{A22}$$

$$\vec{S}^{\ 2} \ |p^\mu; \ m, s, \sigma\rangle = s(s+1) \ |p^\mu; \ m, s, \sigma\rangle, \tag{A23}$$

$$S_z \ |p^\mu; \ m, s, \sigma\rangle = \sigma |p^\mu; \ m, s, \sigma\rangle. \tag{A24}$$

Before we undertake the study of representation spaces for lightlike $p^\mu$ we look at the Pauli–Lubánski pseudovector in a little more detail. For this we let $p^\mu$ be *any* $p^\mu$ satisfying $p_\mu p^\mu > 0$, instead of the standard timelike momenta $p^\mu = (m, 0, 0, 0)$. The definition of the Pauli–Lubánski pseudovector (A6) yields the $0th$ [or time]

component

$$
\begin{aligned}
W^0 &= \frac{1}{2}\epsilon^{0\nu\rho\sigma} J_{\nu\rho} P_\sigma \\
&= \frac{1}{2}(\epsilon^{0123} J_{12} P_3 + \epsilon^{0132} J_{13} P_2 + \epsilon^{0213} J_{21} P_3 + \cdots + \epsilon^{0312} J_{31} P_2 + \epsilon^{0321} J_{32} P_1) \\
&= (\epsilon^{0123} J_{12} P_3 + \epsilon^{0132} J_{13} P_2 + \epsilon^{0231} J_{23} P_1) \\
&= [(+1) J_3 P_3 + (-1)(-J_2) P_2 + (+1) J_1 P_1] \\
&= J_i P_i.
\end{aligned}
\tag{A25}
$$

Since $P_\mu = (P_0, -\vec{P})$

$$W^0 = -\vec{J} \cdot \vec{P}. \tag{A26}$$

Now we obtain the spacial part of $W^\mu$:

$$
\begin{aligned}
W^1 &= \\
&\frac{1}{2}(\epsilon^{1023} J_{02} P_3 + \epsilon^{1032} J_{03} P_2 + \epsilon^{1203} J_{20} P_3 + \cdots + \epsilon^{1302} J_{30} P_2 + \epsilon^{1320} J_{32} P_0) \\
&= (\epsilon^{1023} J_{02} P_3 + \epsilon^{1032} J_{03} P_2 + \epsilon^{1230} J_{23} P_0) \\
&= [(-1) K_2 P_3 + (+1) K_3 P_2 + (-1) J_1 P_0] \\
&= [-J_1 P_0 + (-K_2 P_3 + K_3 P_2)] \\
&= [-J_1 P_0 + (K_2 P^3 - K_3 P^2)] \\
&= [-J_1 P_0 + (\vec{K} \times \vec{P})_1],
\end{aligned}
\tag{A27}
$$

where

$$(\vec{K} \times \vec{P})_1 \equiv K_2 P^3 - K_3 P^2 = -K_2 P_3 + K_3 P_2. \tag{A28}$$

Similarly one can obtain the $y$– and $z$–components of $W^\mu$ to get

$$\vec{W} = (-\vec{J} P_0 + \vec{K} \times \vec{P}). \tag{A29}$$

In the above equations we have defined

$$J_{12} = J_3 = -J_{21}, \quad J_{31} = J_2 = -J_{13}, \quad J_{23} = J_1 = -J_{32}, \tag{A30}$$

$$J_{i0} = -K_i = -J_{0i}, \qquad (i = 1, 2, 3). \tag{A31}$$

The operator

$$-\frac{W^0}{|\vec{p}|} = \frac{\vec{J} \cdot \vec{P}}{|\vec{p}|}, \tag{A32}$$

[see (A26)] can be interpreted as a *generalised helicity operator* for it measures the projection of the total (rather than spin) angular momentum on the direction of motion. The meaning of $\vec{W}$ is not as transparent. A few remarks on it may shed some light. First, the vector $\vec{K} \times \vec{P}$ is an operator which lies in a plane orthogonal to $\vec{P}$. Second, since $W_\mu P^\mu$ vanishes, there are only three independent components in $W^\mu$. Equation (A32) provides one independent component. The other two may be chosen along any two mutually orthogonal directions in the plane defined by $\vec{K} \times \vec{P}$. If one wished, these operators could be chosen as $\vec{J} \cdot \hat{a}$ and $\vec{J} \cdot \hat{b}$, with $\hat{a}$ and $\hat{b}$ as two dimensionless unit vectors in the plane defined by $\vec{K} \times \vec{P}$. Whether these operators will find any use in physical problems is not obvious. Bargmann and Wigner [9] obtained operators similar to (A26) and (A29) without attempting any physical interpretation.

For the standard vector $p^\mu$ the generalised helicity operator becomes undefined because of the null isotropy of $\vec{p} = \vec{0}$. As such let's, for sake of completeness, study the effect of $W^\mu$ on $|p^\mu = (m, 0, 0, 0);\ m, s, \sigma\rangle$

$$\begin{aligned} W^0\ |p^\mu = (m, 0, 0, 0);\ m, s, \sigma\rangle &= -\vec{J} \cdot \vec{P}\ |p^\mu = (m, 0, 0, 0);\ m, s, \sigma\rangle \\ &= -\vec{J} \cdot \vec{p}\ |p^\mu = (m, 0, 0, 0);\ m, s, \sigma\rangle = 0, \end{aligned} \tag{A33}$$

since $\vec{p} = \vec{0}$ for $|p^\mu = (m, 0, 0, 0);\ m, s, \sigma\rangle$

$$\begin{aligned} W^1\ &|p^\mu = (m, 0, 0, 0);\ m, s, \sigma\rangle \\ &= [-J_1 P_0 + (\vec{K} \times \vec{P})_1]\ |p^\mu = (m, 0, 0, 0);\ m, s, \sigma\rangle \\ &= -m J_i\ |p^\mu = (m, 0, 0, 0);\ m, s, \sigma\rangle. \end{aligned} \tag{A34}$$

For later use we define the *little group* as a set of transformations which leave $p^\mu$ unchanged. Referring to (A33) and (A34) we once again explicitly verify that $W^\mu$ has three independent components, for any a *timelike* standard vector $p^\mu = (m, 0, 0, 0)$, which are proportional to the generators $\{J_i\}$ of the *little group* $SO(3)$. *Note from* (A8) *that the Pauli–Lubánski operator commutes with* $P^\mu$ *and hence it is the generator of the Little group.*

## A3 States with $m = 0$: Lightlike $p^\mu$, Helicity and Little Group

The physical content of the Pauli–Lubánski operator depends on whether $p^\mu$ is timelike or lightlike. Both Casimir invariants obtained above vanish for $m \to 0$. In the limit $m = 0$, if the primitive arguments of continuity are to hold, we should have for a lightlike standard vector $k^\mu = (\kappa, 0, 0, \kappa)$

$$P_\mu P^\mu \, |k^\mu = (\kappa, 0, 0, \kappa);\; m = 0, \lambda\rangle = 0, \tag{A35}$$

$$W_\mu W^\mu \, |k^\mu = (\kappa, 0, 0, \kappa);\; m = 0, \lambda\rangle = 0, \tag{A36}$$

$$W_\mu P^\mu \, |k^\mu = (\kappa, 0, 0, \kappa);\; m = 0, \lambda\rangle = 0. \tag{A37}$$

So acting on the representation space to which the standard state vector $|k^\mu = (\kappa, 0, 0, \kappa);\; m = 0, \lambda\rangle$ belongs, we have the following *operator equations*

$$P_\mu P^\mu = 0, \quad W_\mu W^\mu, \quad W_\mu P^\mu = 0, \tag{A38}$$

where $\lambda$ represents a yet unidentified quantum number. This quantum number must be related, in yet unspecified fashion, in the $m \to 0$ limit to the quantum number $s$. Without loss of generality we take $P^\mu = (P^0, 0, 0, P^3)$, with the understanding that both $P^0$ and $P^3$ acting on the standard state vector $|k^\mu = (\kappa, 0, 0, \kappa);\; m = 0, \lambda\rangle$ yield $\kappa|k^\mu = (\kappa, 0, 0, \kappa);\; m = 0, \lambda\rangle$. Equations (A35) to (A37) then read

$$P_0 P^0 - P_3 P^3 = 0, \tag{A39}$$

$$W_0 W^0 - W_1 W^1 - W_2 W^2 - W_3 W^3 = 0, \tag{A40}$$

$$W_0 P^0 - W_3 P^3 = 0. \tag{A41}$$

Since both $P^0$ and $P^3$ acting on the standard state vector $|m = 0, \lambda;\; k^\mu =$

$(\kappa, 0, 0, \kappa))\rangle$ yield $\kappa|k^\mu = (\kappa, 0, 0, \kappa);\ m = 0, \lambda\rangle$ we obtain from (A41)

$$W_0 = W_3. \tag{A42}$$

Substitution of the above result in (A40) gives us the condition thus imposed on the $1st$ and $2nd$ component of the Pauli–Lubánski operator

$$W_1 W^1 = -W_2 W^2 \quad \Rightarrow \quad (W^1)^2 = -(W^2)^2. \tag{A43}$$

Now if the Pauli–Lubánsi operator is to be an observable, the square of its eigenvalues (for *each* component) should be a real and a positive number. The condition (A43) then means that $W^1$ and $W^2$ are null operators

$$W^1 = W^2 = 0. \tag{A44}$$

Thus we conclude $W^\mu$ must be proportional to $P^\mu$. Identifying this proportionality constant with $\lambda$, introduced above, we have

$$\left(W^\mu - \lambda P^\mu\right)|k^\mu = (\kappa, 0, 0, \kappa);\ m = 0, \lambda\rangle = 0. \tag{A45}$$

This proportionality constant $\lambda$ has the dimension of angular momentum, and is called *helicity.* We now undertake a more rigorous study of the representation spaces associated with timelike momenta. One of the results we will obtain is that $\lambda = \pm j$, if the operation of *parity* $\vec{x} \rightarrow -\vec{x}$ is included. Otherwise $\lambda$ is either $+j$ *or* $-j$.

We first study the form of $W^\mu$ for the standard lightlike vector $k^\mu = (\kappa, 0, 0, \kappa)$. With $\alpha = 1/2$, as before, the zeroth, or the temporal, component of $W^\mu$ is by definition (A6)

$$\begin{aligned} W^0 &= \frac{1}{2}\epsilon^{0\nu\rho\sigma} J_{\nu\rho} P_\sigma \\ &= \frac{1}{2}(\epsilon^{0123} J_{12} P_3 + \epsilon^{0213} J_{21} P_3) \\ &= J_3 P_3. \end{aligned} \tag{A46}$$

Similarly the *spacial* components of $W^\mu$ are

$$\begin{aligned} W^1 &= \frac{1}{2}\epsilon^{1\nu\rho\sigma} J_{\nu\rho} P_\sigma \\ &= \frac{1}{2}(\epsilon^{1\nu\rho 0} J_{\nu\rho} P_0 + \epsilon^{1\nu\rho 3} J_{\nu\rho} P_3) \\ &= -(J_1 P_0 + K_2 P_3), \end{aligned} \tag{A47}$$

$$W^2 = -(J_2 P_0 - K_1 P_3), \tag{A48}$$

$$W^3 = -J_3 P_0. \tag{A49}$$

Therefore, while acting on the standard state vector $|k^\mu = (\kappa, 0, 0, \kappa);\ m = 0, \lambda\rangle$, the Pauli–Lubánski operator can be written as

$$W^\mu = -\kappa(-J_3,\ J_1 + K_2,\ J_2 - K_1,\ J_3). \tag{A50}$$

Introducing

$$T_1 = J_1 + K_2, \tag{A51}$$

$$T_2 = J_2 - K_1, \tag{A52}$$

we find the Lie algebra satisfied by the generators of the Little group. It reads

$$[T_1, T_2] = 0, \quad [T_1, J_3] = -iT_2, \quad [T_2, J_3] = iT_1. \tag{A53}$$

To gain physical insight into this algebra we note from Table II that the generators of rotations and translations in *a plane* [say $x$–$y$], of the ordinary spacetime, has associated with it the following Lie algebra

$$[P_x, P_y] = 0, \quad [P_x, J_3] = -iP_y, \quad [P_y, J_3] = iP_x. \tag{A54}$$

As such the Lie algebra of the Little group for the lightlike momenta is *isomorphic* to translations and rotations in a plane. To understand the possible origin of this isomorphism, consider a set of events

$$(t^1, x^1, y^1, z^1), \quad (t^2, x^2, y^2, z^2), \cdots\cdots, (t^n, x^n, y^n, z^n) \tag{A55}$$

as described by an observer $\mathcal{O}$. If the *same* set of events are observed by another observer $\mathcal{O}_c$ whose relative velocity with respect to observer $\mathcal{O}$ is $c\hat{z}$, $c$ being the speed of light, then for $\mathcal{O}_c$ the separations $t^i - t^j$ and $z^i - z^j$ *all vanish.* That is, two out of four dimensions seem to essentially disappear. Consequently, one may be tempted not to distinguish between the events which differ only in their *t or z* values and only refer to the projection of events onto the $x$–$y$ plane. However, as one may satisfy oneself by considering a few elementary examples, the physics in $\mathcal{O}_c$ is not completely identical if *all* events were *initially* in the $x$–$y$ plane of $\mathcal{O}$. Further consider two coincident worldlines $\Gamma^1$ and $\Gamma^2$ in the $x$–$y$ plane. Let $\Gamma^1$ be associated with a photon, and $\Gamma^2$ with a neutrino. Even though $\Gamma^1$ and $\Gamma^2$ are coincident, the dynamics associated with the internal helicity degrees of freedom is *not.* This, by way of an example, shows why one should (at most) expect only an isomorphism, and not an identity, between the generators of the little group for the massless particles and the group formed by the generators of the two translations and the rotations in a plane.

A finite Little group transformation [see Ref. (34)]

$$\mathcal{R}^\mu{}_\nu k^\nu = k^\mu, \quad k^\mu = (\kappa, 0, 0, \kappa), \tag{A56}$$

like all Lorentz transformations satisfies the condition

$$\mathcal{R}^\mu{}_\rho \mathcal{R}^\nu{}_\sigma \eta_{\mu\nu} = \eta_{\rho\sigma}, \tag{A57}$$

and can be factored as

$$\mathcal{R}(\Theta, X_1, X_2) = \mathcal{R}(\Theta, 0, 0)\mathcal{R}(0, X_1, X_2). \tag{A58}$$

Where for infinitesimal transformations

$$(\Theta, X_1, X_2) \to (\theta, \chi_1, \chi_2). \tag{A59}$$

[Note: Ref. (34)–$\chi_1 = \chi_2$ here, and Ref. (34)–$\chi_2 = \chi_1$ here. Beware of other notational differences too !]

The physical states $|k^\mu = (\kappa, 0, 0, \kappa);\ m = 0, \lambda\rangle$ under the *infinitesimal* transformation of the little group, then transform as

$$
\begin{aligned}
|k^\mu = (\kappa, 0, 0, \kappa);\ m = 0, \lambda\rangle' &= U[\mathcal{R}]\ |k^\mu = (\kappa, 0, 0, \kappa);\ m = 0, \lambda\rangle \\
&= (1 - i\theta J_3 + i\chi_2 T_2 + i\chi_1 T_1)\, |k^\mu = (\kappa, 0, 0, \kappa);\ m = 0, \lambda\rangle \\
&= \Big[1 - i\theta J_3 + i\chi_2(J_2 - K_1) + i\chi_1(J_1 + K_2)\Big] |k^\mu = (\kappa, 0, 0, \kappa);\ m = 0, \lambda\rangle.
\end{aligned}
\tag{A60}
$$

Just as the generators of translations, $P^\mu$, in the ordinary spacetime span an invariant Abelian subalgebra, the generators of the Little group 'translations' $T_1$ and $T_2$ also span an invariant Abelian subalgebra .

Group Theory Break: Ref. (46), Definition: "An *invariant subalgebra* is some set of generators which goes into itself (or zero) under commutation with any element of the algebra". That is if $T$ is any generator *in the invariant subalgebra* and X is any generator *in the whole* algebra, the commutator $[T, X]$ is a generator in the invariant subalgebra (or it is zero). To quote Ref. (46) again, Abelian invariant subalgebras are "particularly annoying", because the generators in an Abelian invariant subalgebra commute with every generator in the subalgebra. The structure constants, as a consequence, vanish. Definition: If $X_a$ is a generator of a group, then

$$[X_a, X_b] = i f_{abc} X_c. \tag{A61}$$

The constants $f_{abc}$ are called *structure constants* of the group. The generators satisfy the Jacobi identity

$$[X_a, [X_b, X_c]] + \text{ cyclic permutations} = 0. \tag{A62}$$

In terms of structure constants the Jacobi identity reads

$$f_{bcd} f_{ade} + f_{abd} f_{cde} + f_{cad} f_{bde} = 0. \tag{A63}$$

One of the representations of the algebra can be found by introducing a set of

*matrices* $T_a$

$$(T_a)_{bc} \equiv -i f_{abc}, \tag{A64}$$

$$[T_a, T_b] = i f_{abc} T_c. \tag{A65}$$

Thus the structure constants themselves generate a representation of the algebra. The representation generated by the structure functions is called the *adjoint representation.* For more details the reader is directed to our source itself, Georgi [46].

We now have a choice. Either to have infinite dimensional representation (in the parameter $\lambda$), or have a one dimensional representation. Which of the representations is physically realised in nature is an interesting question, to which we (the author) have no honest theoretical answer. However the primitive arguments of physical continuity suggest we explore the possibility of the one dimensional representation. By physical continuity in this context we mean a smooth conceptual and algebraic transition from the $m \to 0$ limit to the $m = 0$ case. This representation is obtained by setting

$$T_1 \; |k^\mu = (\kappa, 0, 0, \kappa);\; m = 0, \lambda\rangle\rangle = 0, \tag{A66}$$

$$T_2 \; |k^\mu = (\kappa, 0, 0, \kappa);\; m = 0, \lambda\rangle = 0. \tag{A67}$$

Using (A66) and (A67) and identifying, as will be justified soon, the states $|k^\mu = (\kappa, 0, 0, \kappa);\; m = 0, \lambda\rangle$ as eigenstates with a definite helicity $\lambda$

$$J_3 \; |k^\mu = (\kappa, 0, 0, \kappa);\; m = 0, \lambda\rangle = \lambda \; |k^\mu = (\kappa, 0, 0, \kappa);\; m = 0, \lambda\rangle, \tag{A68}$$

yields for a *finite* Little group transformation (see (A60))

$$U[\mathcal{R}]|k^\mu = (\kappa, 0, 0, \kappa);\; m = 0, \lambda\rangle = \exp(-i\lambda\Theta[\mathcal{R}]) \; |k^\mu = (\kappa, 0, 0, \kappa);\; m = 0, \lambda\rangle. \tag{A69}$$

Introduce the following simpler notation

$$|k^{\mu} = (\kappa, 0, 0, \kappa);\ m = 0, \lambda\rangle = |\kappa, \lambda\rangle, \tag{A70}$$

In this notation Eqs. (A68) and (A69) read

$$J_3|\kappa, \lambda\rangle = \lambda\ |\kappa, \lambda\rangle, \tag{A71}$$

$$U[\mathcal{R}]\ |\kappa, \lambda\rangle = \exp(-i\lambda\Theta[\mathcal{R}])|\kappa, \lambda\rangle, \tag{A72}$$

Using (A60) and (A72) for the *infinitesimal* Little group transformations yields

$$(1 - i\theta J_3 + i\chi_2(J_2 - \mathcal{K}_1) + i\chi_1(J_1 + \mathcal{K}_2))\ |\kappa, \lambda\rangle = (1 - i\theta J_3)\ |\kappa, \lambda\rangle. \tag{A73}$$

The constraints imposed on matter fields $\phi(\kappa, \lambda)$ because of conditions (A66) and (A67) thus read

$$\left[1 - i\theta J_3 + i\chi_2(J_2 - K_1) + i\chi_1(J_1 + K_2)\right]\ \phi(\kappa, \lambda) = (1 - i\theta J_3)\ \phi(\kappa, \lambda). \tag{A74}$$

where $\{\vec{J}, \vec{K}\}$ are now the finite dimensional representations of the Lorentz algebra. To explore the physical consequences of this we begin with the observation

$$\begin{aligned} & i\chi_2(J_2 - K_1) + i\chi_1(J_1 + K_2) \\ &\quad = i\chi_2\left[(\vec{S}_R)_2 + (\vec{S}_L)_2 + i(\vec{S}_R)_1 - i(\vec{S}_L)_1\right] \\ &\qquad\qquad + i\chi_1\left[(\vec{S}_R)_1 + (\vec{S}_L)_1 - i(\vec{S}_R)_2 + i(\vec{S}_L)_2\right] \\ &\quad = (i\chi_1 + \chi_2)\left[(\vec{S}_L)_1 + i(\vec{S}_L)_2\right] + (i\chi_1 - \chi_2)\left[(\vec{S}_R)_1 - i(\vec{S}_R)_2\right]. \end{aligned} \tag{A75}$$

Then using (A71) and replacing $J_3$, in accordance with definitions introduced in Chapter 3, by $(\vec{S}_R)_3 + (\vec{S}_L)_3$ the constraint (A74) becomes three independent

conditions, which read

$$\left[(\vec{S}_R)_3 + (\vec{S}_L)_3\right] \ \phi(\kappa, \lambda) = \lambda \ \phi(\kappa, \lambda), \tag{A76}$$

$$\left[(\vec{S}_L)_1 + i(\vec{S}_L)_2\right] \ \phi(\kappa, \lambda) = 0, \tag{A77}$$

$$\left[(\vec{S}_R)_1 - i(\vec{S}_R)_2\right] \ \phi(\kappa, \lambda) = 0. \tag{A78}$$

In (A76) we used the fact that $\phi(\kappa, \lambda)$ is identified with the eigenstate of $J_3$. Since

$$\left[(\vec{S}_L)_1 + i(\vec{S}_L)_2\right] \tag{A79}$$

is a *raising* operator, which raises the $\sigma_l$ value by unity, and

$$\left[(\vec{S}_R)_1 - i(\vec{S}_R)_2\right] \tag{A80}$$

is a *lowering* operator which lowers the $\sigma_r$ value by unity, conditions (A77) and (A78) imply that $\phi(\kappa, \lambda)$ must simultaneously be eigenstates of $\vec{S}_L$ with (the maximum) eigenvalue $\sigma_l = j_l$ and of $\vec{S}_R$ with (the minimum) eigenvalue $\sigma_r = -j_r$. That is

$$(\vec{S}_R)_3 \ \phi(\kappa, \lambda) = -j_r \ \phi(\kappa, \lambda), \tag{A81}$$

$$(\vec{S}_L)_3 \ \phi(\kappa, \lambda) = +j_l \ \phi(\kappa, \lambda). \tag{A82}$$

These two equations coupled with Eq. (A76) yield a simple and remarkable result which severly restricts the type of representations allowed for a lightlike momenta, by requiring

$$\lambda \ = \ j_l - j_r. \tag{A83}$$

It must be noted that the *assumed* identification of $\phi(\kappa, \lambda)$ with one of the eigenstates is forced by conditions (A81) and (A82). To see this add (A81) and

(A82) to obtain

$$\left[(\vec{S}_R)_3 + (\vec{S}_L)_3\right]\ \phi(\kappa,\lambda) = (j_l - j_r)\ \phi(\kappa,\lambda), \tag{A84}$$

One may satisfy oneself that constraints (A81) and (A82) are a direct result of conditions (A66) and (A67), and need not to be connected to any of the intervening mathematical steps. In the limit $m \to 0$, the only degrees of freedom left out of $\sigma = -j, -j+1, \cdots\cdots, +j-1, +j$ are $\pm j$. However, *not* all representations may have massless realisations physically. For $\{\lambda = -j,\ j > 0\}$ only those fields may be physically realised which satisfy the condition

$$[\lambda = -j,\ j > 0]: \quad j_l - j_r = -j. \tag{A85}$$

Consequently physically realisable fields are

$$[\lambda = -j]: \quad (j_l + j, j_l)\ \rightarrow\ (j,0), (j+1/2, 1/2), (j+1,1), \cdots\ . \tag{A86}$$

For $\lambda = -1/2,\ \lambda = -1$ and $\lambda = -3/2$, physically realisable representations are

$$[\lambda = -1/2]: \quad (1/2,0), (1,1/2), (3/2,1), \cdots\ . \tag{A87}$$

$$[\lambda = -1]: \quad (1,0), (3/2,1/2), (2,1), \cdots\ . \tag{A88}$$

$$[\lambda = -3/2]: \quad (3/2,0), (2,1/2), (5/2,1), \cdots\ . \tag{A89}$$

Similarly for $\{\lambda = +j,\ j > 0\}$ we have the constraint

$$[\lambda = +j,\ j > 0]: \quad j_l - j_r = +j. \tag{A90}$$

The fields which may be physically realised are

$$[\lambda = +j]: \quad (j_r, j + j_r)\ \rightarrow\ (0,j), (1/2, j+1/2), (1, j+1), \cdots\ . \tag{A91}$$

For $\lambda = +1/2,\ \lambda = +1,$ and $\lambda = 3/2$, physically realisable representations are

$$[\lambda = +1/2]: \quad (0,1/2),(1/2,1),(1,3/2),\cdots \ . \tag{A92}$$

$$[\lambda = +1]: \quad (0,1),(1/2,3/2),(1,2),\cdots \ . \tag{A93}$$

$$[\lambda = +3/2]: \quad (0,3/2),(1/2,2),(1,5/2),\cdots \ . \tag{A94}$$

As a result a massless matter field is constrained to be any one the possibilities indicated by (A86) and (A91) , depending on the sign of the helicity. Even though the (1/2,1/2) representation can be shown to transform as a four–vector, it violates the constraint (A83), and is therefore physically unrealisable. Weinberg [33] adds to this line "at least until we broaden our notion of what we mean by a Lorentz transformation".

So far we have considered only *continuous* $ds^2$ preserving transformations. However, the *discrete* transformations of *parity*

$$P : \vec{x} \rightarrow -\vec{x}, \tag{A95}$$

and *time reversal*

$$T : t \rightarrow -t. \tag{A96}$$

also preserve $ds^2$. Under parity the momentum changes sign, as a result

$$P : \vec{p} \rightarrow -\vec{p}, \quad \vec{K} \rightarrow -\vec{K}, \quad \vec{J} \rightarrow \vec{J}. \tag{A97}$$

It then follows from an inspection of Lorentz transformation properties of the $(j,0)$ and $(0,j)$ matter fields that

$$P : (j,0) \leftrightarrow (0,j). \tag{A98}$$

Thus in order that we are not thrown out of the linear representation space of the matter fields under consideration by the Parity operation we *must* consider both helicity states $\lambda = \pm j$, and the associated matter fields must transform, for example, as the $(j,0) \oplus (0,j)$ representations.

We have followed in some detail the origin of $(2j+1)$ spin degrees freedom for timelike momenta, and its seemingly abrupt reduction to just one [or two if parity is incorporated] helicity degree of freedom for the lightlike momenta. Further physical insights into the nature of this change in the number of degrees of freedom may be gained by referring to Ref. (39) where Wigner poses the question why particles with non–zero mass may have more than two spin–degrees of freedom. In this context Wigner notes that it is *only* for the lightlike momenta that, the parallelness [or anti–parallelness] of spin and momentum is a Lorentz invariant concept.

A state with lightlike momenta $\vec{p}$ and helicity $\lambda$ is obtained by a Lorentz transformation

$$|\vec{p},\lambda\rangle = U[\mathcal{L}(\vec{p})]|\kappa,\lambda\rangle. \tag{A99}$$

where $\mathcal{L}(\vec{p})$ is a Lorentz transformation which takes $k^\mu \equiv (\kappa,0,0,\kappa) \rightarrow p^\mu \equiv (|\vec{p}|,\vec{p})$:

$$\mathcal{L}(\vec{p}): \quad k^\mu \equiv (\kappa,0,0,\kappa) \rightarrow p^\mu \equiv (|\vec{p}|,\vec{p}) \tag{A100}$$

$$\mathcal{L}^\mu{}_\nu k^\nu = p^\mu. \tag{A101}$$

The Lorentz transformation $\mathcal{L}(\vec{p})$ can be factorised into a pure boost and a rotation as follows

$$\mathcal{L}(\vec{p}) = R(\hat{p})\ \mathcal{B}(|\vec{p}|). \tag{A102}$$

Here $\mathcal{B}(\vec{p})$ is the boost that takes $k^\mu = (\kappa,0,0,\kappa) \rightarrow (|\vec{p}|,0,0,|\vec{p}|)$. The rotation $R(\hat{p})$ takes $(|\vec{p}|,0,0,|\vec{p}|) \rightarrow p^\mu \equiv (|\vec{p}|,\vec{p})$. The boost has the form

$$[\mathcal{B}^\mu{}_\nu] = \begin{pmatrix} \cosh(\varphi_z) & 0 & 0 & \sinh(\varphi_z) \\ 0 & 1 & 0 & 0 \\ 0 & 0 & 1 & 0 \\ \sinh(\varphi_z) & 0 & 0 & \cosh(\varphi_z) \end{pmatrix}, \quad \varphi_z \equiv \ln(\frac{|\vec{p}|}{\kappa}). \tag{A103}$$

## A4 Vacuum State and Single Particle States

To develop a formalism for *directly observable single particle states* $|\vec{p}, \sigma\rangle$ or $|\vec{p}, \lambda\rangle$ it seems necessary to have a Poincaré invariant state called the *vacuum state* $|\ \rangle$

$$|\ \rangle = U[\{\Lambda, a\}]\ |\ \rangle. \tag{A104}$$

A global phase factor by which $|\ \rangle$ and $U[\{\Lambda, a\}]\ |\ \rangle$ may differ are of no physical significance, and hence are ignored. States $|\vec{p}, \sigma\rangle$ and $|\vec{p}, \lambda\rangle$ are directly observable states and correspond to timelike and lightlike momenta respectively.

Directly observable single particle states $|\vec{p}, \sigma\rangle$ or $|\vec{p}, \lambda\rangle$ are obtained from the vacuum state through the action of creation operators $\{a^\dagger(\vec{p}, \sigma)\}$ and $\{a^\dagger(\vec{p}, \lambda)\}$

$$|\vec{p}, \sigma\rangle = a^\dagger(\vec{p}, \sigma)|\ \rangle, \tag{A105}$$

$$|\vec{p}, \lambda\rangle = a^\dagger(\vec{p}, \lambda)|\ \rangle. \tag{A106}$$

The creation operators $\{a^\dagger(\vec{p}, \sigma)\}$, and the annihilation operators $\{a(\vec{p}, \sigma)\}$ satisfy

$$[a(\vec{p}, \sigma),\ a^\dagger(\vec{p}\,', \sigma')]_\pm \equiv \delta^3(\vec{p} - \vec{p}\,'^{)}\ \delta_{\sigma\sigma'}$$

$$[a(\vec{p}, \sigma),\ a(\vec{p}\,', \sigma')]_\pm \equiv 0, \tag{A107}$$

and

$$[a(\vec{p}, \lambda),\ a^\dagger(\vec{p}\,', \lambda')]_\pm \equiv \delta^3(\vec{p} - p\ ')\ \delta_{\lambda\lambda'}$$

$$[a(\vec{p}, \lambda),\ a^\dagger(\vec{p}\,', \lambda')]_\pm \equiv 0. \tag{A108}$$

If $L(\vec{p})$ is a boost which takes a particle of mass $m \neq 0$ at rest to momentum

$\vec{p}$ then, according to our earlier discussions,

$$|\vec{p},\sigma\rangle = U[L(\vec{p})]\ |\vec{0},\sigma\rangle. \tag{A109}$$

Where, as for the $m=0$ case (see equation (A70)), we have introduced a simpler notation for $m\neq 0$ states

$$|p^\mu;\ m,s,\sigma\rangle = |\vec{p},\sigma\rangle. \tag{A110}$$

Similarly, in the terms of already defined notation for $m=0$, we have

$$|\vec{p},\lambda\rangle = U[\mathcal{L}(\vec{p})]\ |\kappa,\lambda\rangle. \tag{A111}$$

The single particle states are normalised as follows[33]

$$\langle\vec{p},\sigma|\vec{p}\,',\sigma'\rangle = \delta^3(\vec{p}-\vec{p}\,')\ \delta_{\sigma\sigma'}, \tag{A112}$$

$$\langle\vec{p},\lambda|\vec{p}\,',\lambda'\rangle = \delta^3(\vec{p}-\vec{p}\,')\ \delta_{\lambda\lambda'}. \tag{A113}$$

## A5 A Remark on Single Particle States

Due to the non-commutativity of position and momentum, these observables cannot simultaneously be measured to an arbitrary precision, for a state. As such one must assume that microscopically infinite time $T_\infty$

$$T_\infty \gg \frac{\hbar}{|\vec{p}\,|c}, \tag{A114}$$

is allowed for the *preparation* of these states with the well defined momentum $\vec{p}$. This condition is invariably satisfied in the usual scattering experiments. [We are

33 The normalisations introduced above are convenient ones for these considerations and may differ from normalisations used elsewhere in this work.

reintroducing $\hbar$ and $c$ explicitly for this discussion.] However for any microscopically finite time $T_0$ [cf. Ref. (41), section 2.13]

$$T_0 \sim \frac{\hbar}{|\vec{p}|c}, \tag{A115}$$

the momentum of the particle has a non-zero probability of being either timelike, lightlike, or spacelike. This uncertainty arises when the spacetime region, to which the measurements are confined, reaches the quantum–mechanically placed lower bound

$$\delta(ds^2) \sim \frac{\hbar^2}{\vec{p}.\vec{p}} \ . \tag{A116}$$

If one considers *gedanken microscopic observers* confined to regions with $ds^2 \sim \hbar^2/(\vec{p}.\vec{p})$, then the transformations between these observers cease to be Poincaré because of the inherent uncertainty involved with measurements of energy momentum and spacetime separations. The unpredictable, and uncontrollable, accelerations associated with these gedanken observers are locally equivalent to the existence of a gravitational field. The vacuum state, as observed by these observers, is therefore no longer the Poincaré vacuum $|\ \rangle$ introduced in Eq. (A104) above, but is replaced by the *Rindler vacuum* $|\ \rangle\rangle$ as Gerlach [47] has argued.

The Rindler vacuum has the property

$$\langle\ |\ \rangle\rangle = 0. \tag{A117}$$

Ref. [(47), p 1037]: "Rindler ... vacuum of an accelerated frame determines a Hilbert space of quantum states which is distinct from Hilbert space determined by the Minkowski [Poincaré, in our language] vacuum. There is no unitary transformation which connects elements in these two spaces." As a result, the physical states accessible to the *macroscopic inertial observers* are only a subset of *all* physical states accessible to a general observer unrestricted by Poincaré covariance. In this regard we should parenthetically, but explicitly, note that the gedanken microscopic *inertial* observers are physically ruled out by the non–commutativity of the position and momentum associated with a particle. Consequently, in the absence of gedanken microscopic inertial observers, microscopic Poincaré covariance is meaningless.

## A6 Lorentz Transformation of Single Particle States

The effect of an arbitrary Lorentz transformation $\Lambda$ on the single particle states $|\vec{p},\sigma\rangle$ is given by

$$U[\Lambda]\ |\vec{p},\sigma\rangle = U[\Lambda]U[L(\vec{p})]\ |\vec{0},\sigma\rangle, \quad U[\Lambda] \equiv U[\{\Lambda, 0\}]. \tag{A118}$$

In obtaining the *rhs* of the above equation we substituted for $|\vec{p},\sigma\rangle$ from Eq. (A109). Exploiting,

$$1 = U[L(\Lambda\vec{p})]\ U^{-1}[L(\Lambda\vec{p})], \quad U^{-1}[\Lambda] = U[\Lambda^{-1}], \tag{A119}$$

we obtain

$$\begin{aligned} U[\Lambda]\ |\vec{p},\sigma\rangle &= U[L(\Lambda\vec{p})]U^{-1}[L(\Lambda\vec{p})]U[\Lambda]U[L(\vec{p})]\ |\vec{0},\sigma\rangle \\ &= U[L(\Lambda\vec{p})]U[L^{-1}(\Lambda\vec{p})\Lambda L(\vec{p})]\ |0,\sigma\rangle \\ &= U[L(\Lambda\vec{p})]\sum_{\sigma'}\int d^3p' |\vec{p}\,',\sigma'\rangle\langle\vec{p}\,',\sigma'|U[R_W]|0,\sigma\rangle. \end{aligned} \tag{A120}$$

In the last step above we have used the *completeness relation*

$$1 = \sum_{\sigma'}\int d^3p' |\vec{p}\,',\sigma'\rangle\langle\vec{p}\,',\sigma'|, \tag{A121}$$

and identified a *pure rotation:*

$$R_W = L^{-1}(\Lambda\vec{p})\Lambda L(\vec{p}), \tag{A122}$$

called the *Wigner Rotation.* The $p'$–integration can be performed using the orthonormality condition (A112), and recalling that a pure rotation does not alter the momentum,

$$U[\Lambda]\ |\vec{p},\sigma\rangle = U[L(\Lambda\vec{p})]\ \sum_{\sigma'} |\vec{0},\sigma'\rangle\langle\vec{0},\sigma'|U[R_W]|\vec{0},\sigma\rangle. \tag{A123}$$

We now define a matrix, dependent on the suppressed spin quantum number $s$ (see (A110)), whose matrix elements are given by

$$D^{(s)}_{\sigma'\sigma}(R_W) \equiv \langle 0,\sigma'|U[R_W]|0,\sigma\rangle. \tag{A124}$$

With this definition we arrive at the remarkable result

$$U[\Lambda]\ |\vec{p},\sigma\rangle = \sum_{\sigma'} D^{(s)}_{\sigma'\sigma}(R_W)|\Lambda\vec{p},\sigma'\rangle. \tag{A125}$$

The surprising feature of this result lies in the fact that under a Lorentz transformation $\Lambda$, the transformation of the physical states with timelike momenta is completely determined by the generators of rotation $\vec{J}$, because $D^{(s)}_{\sigma'\sigma}$ which determines the transformation property of the single particle states associated with timelike momenta through (A125) can be written as:

$$D^{(s)}_{\sigma'\sigma}(R_W) \equiv \langle\vec{0},\sigma'|\exp[-\frac{i}{2}\lambda^{ij}(R_W)J_{ij}]|\vec{0},\sigma\rangle. \tag{A126}$$

Or equivalently

$$D^{(s)}_{\sigma'\sigma}(R_W) \equiv \langle\vec{0},\sigma'|\exp[-\frac{i}{2}\lambda^{ij}(R_W)]\epsilon^{ijk}J_k]|\vec{0},\sigma\rangle. \tag{A127}$$

Here, as usual, the indices $i,j,k$ run over $1,2,3$. The notation $\lambda^{ij}(R_W)$ means the transformation parameters $\lambda^{ij}$ are functions of the Wigner rotation matrix, $R_W$, for the timelike states. We would have expected the generators of the boosts to play an important role. But that turns out *not* to be the case.

Similarly, for the $m = 0$ case we obtain

$$U[\Lambda]\ |\vec{p}, \lambda\rangle = \exp[-i\lambda\Theta(\mathcal{R}_W)]\ |\kappa, \lambda\rangle. \tag{A128}$$

In the above expression we have introduced the Wigner rotation for the states associated with lightlike momenta

$$\mathcal{R}_W \equiv \mathcal{L}^{-1}(\Lambda\vec{p})\Lambda\mathcal{L}(\vec{p}). \tag{A129}$$

$\Theta(\mathcal{R})$ means that the angle $\Theta$ is a function of the rotation matrix $\mathcal{R}$.

## A7 Lorentz Transformation of Creation and Annihilation Operators

The Lorentz transformation properties of the single particle states, given by (A125) and (A128), arise out of the fundamental assumption regarding the existence of a Poincaré invariant vacuum state. Obviously the single particle states and the vacuum state transform differently under a Lorentz transformation. For this reason, *with the exception of spinless particles,* the creation and annihilation operators *cannot* transform as

$$[NotPossible] \qquad U[\Lambda]a^\dagger(\vec{p}, \sigma)U^{-1}[\Lambda] = a^\dagger(\Lambda\vec{p}, \sigma), \tag{A130}$$

$$[NotPossible] \qquad U[\Lambda]a^\dagger(\vec{p}, \lambda)U^{-1}[\Lambda] = a^\dagger(\Lambda\vec{p}, \lambda). \tag{A131}$$

The actual transformation property of $a^\dagger(\vec{p}, \sigma)$ as implied by (A125) is

$$U[\Lambda]a^\dagger(\vec{p}, \sigma)U^{-1}[\Lambda] = \sum_{\sigma'} D^{(s)}_{\sigma'\sigma}[L^{-1}(\Lambda\vec{p})\Lambda L(\vec{p})]a^\dagger(\Lambda\vec{p}, \sigma'). \tag{A132}$$

To obtain the transformation property of the annihilation operators we make the

following observations:

$$U^{\dagger}[\Lambda] = U^{-1}[\Lambda], \tag{A133}$$

$$\{U^{-1}[\Lambda]\}^{\dagger} = \{U^{\dagger}[\Lambda]\}^{\dagger} = U[\Lambda]. \tag{A134}$$

Further even though $D[\Lambda]$ is *not* unitary, $D^{(s)}[R]$ is. That is:

$$\left\{D^{(s)}[R_W]\right\}^{\dagger} = \left\{D^{(s)}[R_W]\right\}^{-1} = \left\{D^{(s)}[R_W^{-1}]\right\}. \tag{A135}$$

But since $D^{(s)}[R_W]$ is a real matrix

$$\left\{D^{(s)}[R_W]\right\}^{\dagger} = \left\{D^{(s)}[R_W]\right\}^{T}. \tag{A136}$$

Therefore

$$\left\{D^{(s)}[R_W]\right\}^{\dagger} = \left\{D^{(s)}[R_W]\right\}^{T} = D^{(s)}[R_W^{-1}]. \tag{A137}$$

Which implies

$$D^{(s)}_{\sigma'\sigma}[R_W] = D^{(s)}_{\sigma\sigma'}[R_W^{-1}]. \tag{A138}$$

Taking the adjoint of (A132) and exploiting these observations, the annihilation operators can be shown to transform, under a Lorentz transformation,

$$U[\Lambda]a(\vec{p},\sigma)U^{-1}[\Lambda] = \sum_{\sigma'} D^{(s)}_{\sigma\sigma'}[L^{-1}(\vec{p})\Lambda^{-1}L(\Lambda\vec{p})]a(\Lambda\vec{p},\sigma'). \tag{A139}$$

Similarly the transformation properties of the creation and annihilation operators for the states associated with the lightlike momenta are found to be

$$U[\Lambda]a^{\dagger}(\vec{p},\lambda)U^{-1}[\Lambda] = \exp\left(-i\lambda\Theta[\mathcal{L}^{-1}(\Lambda\vec{p})\Lambda\mathcal{L}(\vec{p})]\right) a^{\dagger}(\Lambda\vec{p},\lambda), \tag{A140}$$

$$U[\Lambda]a(\vec{p},\lambda)U^{-1}[\Lambda] = \exp\left(-i\lambda\Theta[\mathcal{L}^{-1}(\vec{p})\Lambda^{-1}\mathcal{L}(\Lambda\vec{p})]\right) a(\Lambda\vec{p},\lambda), \tag{A141}$$

where we used

$$-\Theta[\mathcal{R}_W] = \Theta[\mathcal{R}_W^{-1}]. \tag{A142}$$

## A8 Matter Field Operators for $(2j+1)$ Component Matter Fields

The Lorentz transformation properties, for the $m \neq 0$ case, of the creation and annihilation operators collected together are (see Eqs. (A132) and (A139))

$$U[\Lambda]a^\dagger(\vec{p},\sigma)U^{-1}[\Lambda] = \sum_{\sigma'} D^{(s)}_{\sigma'\sigma}[L^{-1}(\Lambda\vec{p})\Lambda L(\vec{p})]a^\dagger(\Lambda\vec{p},\sigma'), \tag{A143}$$

$$U[\Lambda]a(\vec{p},\sigma)U^{-1}[\Lambda] = \sum_{\sigma'} D^{(s)}_{\sigma\sigma'}[L^{-1}(\vec{p})\Lambda^{-1}L(\Lambda\vec{p})]a(\Lambda\vec{p},\sigma'). \tag{A144}$$

It must be recalled, and emphasised, that the result expressed by Eqs. (A143) and (A144) depend crucially on the postulated existence of a Poincaré invariant non-degenerate state called the vacuum $|\ \ \rangle$. These are precisely the transformation properties which provide us an opportunity to exploit the finite dimensional representations of the Lorentz group by introducing the *multicomponent matter field operators*

$$\Phi_n(x) \equiv \Phi_n^{(+)}(x) + \Phi_n^{(-)}(x), \tag{A145}$$

with $\Phi_n^{(\pm)}(x)$ transforming as

$$U[\Lambda]\Phi_n^{(\pm)}(x)U^{-1}[\Lambda] = \sum_m D_{nm}[\Lambda^{-1}]\Phi_m^{(\pm)}(\Lambda x). \tag{A146}$$

The $D[\Lambda]$ appearing in the *rhs* is one of the finite dimensional representations of the Lorentz group.

The *principle of the linear superposition* of the physical states suggests that the Matter field operators $\Phi(x)$ be constructed by taking linear combinations of the creation and annihilation operators. Further in order to preserve the *translational invariance,* $\Phi(x)$ must be of the form

$$\Phi_n(x) = \left(\frac{1}{2\pi}\right)^{3/2} \int \frac{d^3(p)}{\sqrt{2\omega(\vec{p})}} \sum_\sigma \Big[u_n(\vec{p},\sigma)a(\vec{p},\sigma)\exp(ip\cdot x) + (-1)^{j-\sigma}v_n(\vec{p},\sigma)b^\dagger(\vec{p},-\sigma)\exp(-ip\cdot x)\Big]. \tag{A147}$$

As is usual we have defined

$$p \cdot x \equiv p_\mu x^\mu. \tag{A148}$$

The right hand side of the defining Eq. (A147) for the field operators $\Phi_n(x)$ contains $a(\vec{p},\sigma)$ the *particle* annihilation operator, and $b^\dagger(\vec{p},-\sigma)$ the *antiparticle* creation operator. The combination $(-1)^{j-\sigma} b^\dagger(\vec{p},-\sigma)$ appears in the antiparticle creation term because $(-1)^{j-\sigma} a^\dagger(\vec{p},-\sigma)$ transforms as $a(\vec{p},\sigma)$. Like the master drummer: 'I remember that when someone had started to teach me about creation and annihilation operators, that this operator creates an electron, I said "How do you create an electron? It disagrees with the conservation of charge" ' [R. P. Feynman, Nobel Lecture], any beginning student must ask the same question. In order that our theory have appropriate particle interpretation and the Lagrangians and the hamiltonians yield the same wave equations as imposed upon us by the Poincaré covariance, the Lagrangians $L(\Phi(x), \partial_\mu \Phi(x))$ and the hamiltonians $H(\Phi(x), \partial_\mu \Phi(x))$ must be at least bilinear in $\Phi(x)$ That is $L$ or $H \sim \Phi^\dagger(x)\Phi(x)$ or $\sim \Phi^\dagger(x)\partial^{\leftrightarrow}{}_\mu \Phi(x)$. As a result $L$ or $H \sim (a^\dagger + b)(a + b^\dagger) = a^\dagger a + a^\dagger b^\dagger + ba + bb^\dagger$. These acting on the vacuum $|\ \rangle$ produce no net conserved charge. The $a^\dagger a$, for instance, acting on the vacuum state $|\ \rangle$ yields zero. The $a^\dagger b^\dagger$ creates a antiparticle–particle pair, thus producing net conserved charge of zero. The $ba$ acting on the vacuum is identically zero. And $bb^\dagger$ creates a antiparticle, and then destroys it at the same instant. This contributes zero to the net charge. Therefore, overall one has produced a particle–antiparticle pair of net conserved charge zero, and created and destroyed an antiparticle at the same instant. Thus the total conserved charge, associated with the particles involved, of the universe remains unaltered. Of course, one cannot but note that the phrase *same instant* used in the above discussion is observationally as well as theoretically of limited validity. It must, roughly speaking, be replaced by a time interval $\Delta t \sim \hbar/2mc^2$ [Where we restored $\hbar$ and $c$]. So are we to conclude that over time periods of the order of $\Delta t \sim \hbar/2mc^2$, the associated 'conserved charge' of the universe is uncertain by the amount $q = ne$, $n$ being an integer?

We now wish to know the physical interpretation of $u_n(\vec{p},\sigma)$ and $v_n(\vec{p},\sigma)$.

Towards this end consider

$$\Phi_n^{(+)}(x) = \left(\frac{1}{2\pi}\right)^{3/2} \int \frac{d^3p}{\sqrt{2\omega(\vec{p})}} \sum_{\sigma} u_n(\vec{p},\sigma) a(\vec{p},\sigma) \exp(ip\cdot x), \tag{A149}$$

and multiply on the left by $U[\Lambda]$ and on the right by $U^{-1}[\Lambda]$. Using the Lorentz transformation property of the annihilation operators given by (A139) we immediately obtain:

$$\begin{aligned} &U[\Lambda]\Phi_n^{(+)}(x)U^{-1}[\Lambda] = \\ &\quad \left(\frac{1}{2\pi}\right)^{3/2} \int \frac{d^3(p)}{\sqrt{2\omega(\vec{p})}} \sum_{\sigma} \\ &\quad u_n(\vec{p},\sigma) \sum_{\sigma'} D^{(s)}_{\sigma\sigma'}[L^{-1}(\vec{p})\Lambda^{-1}L(\Lambda\vec{p})] a(\Lambda\vec{p},\sigma') \exp(ip\cdot x). \end{aligned} \tag{A150}$$

Now implement a change of variables: $p_\mu \to (\Lambda^{-1})_\mu{}^\nu p_\nu$, so that $\vec{p} \to \Lambda^{-1}\vec{p}$ and $\Lambda^{-1}p{\cdot}x = p{\cdot}\Lambda x$. If confused, note: $\Lambda^{-1}p{\cdot}x = p{\cdot}\Lambda x \Rightarrow \Lambda\Lambda^{-1}p{\cdot}x = \Lambda p{\cdot}\Lambda x, \Leftrightarrow p{\cdot}x = \Lambda p\cdot\Lambda x$. This immediately translates into the more familiar form: $p_\mu x^\mu = p'_\mu x'^\mu$. The change of variables thus gives

$$\begin{aligned} &U[\Lambda]\Phi_n^{(+)}(x)U^{-1}[\Lambda] = \\ &\quad \left(\frac{1}{2\pi}\right)^{3/2} \int \frac{d^3(p)}{\sqrt{2\omega(\vec{p})}} \sum_{\sigma} \\ &\quad \sum_{\sigma'} u_n(\Lambda^{-1}\vec{p},\sigma) D^{(s)}_{\sigma\sigma'}[L^{-1}(\Lambda^{-1}\vec{p})\Lambda^{-1}L(\vec{p})] a(\vec{p},\sigma') \exp(ip\cdot\Lambda x). \end{aligned} \tag{A151}$$

Substituting (A149) in (A146) gives

$$\begin{aligned} &U[\Lambda]\Phi_n^{(+)}(x)U^{-1}[\Lambda] = \\ &\quad \left(\frac{1}{2\pi}\right)^{3/2} \int \frac{d^3p}{\sqrt{2\omega(\vec{p})}} \sum_{\sigma}\sum_{m} D_{nm}[\Lambda^{-1}] u_m(\vec{p},\sigma) a(\vec{p},\sigma) \exp(ip\cdot\Lambda x). \end{aligned} \tag{A152}$$

Comparison of the right hand sides of (A151) and (A152) gives the equation satisfied by the Fourier coefficients $u_n(\vec{p}, \sigma)$

$$\sum_m D_{nm}[\Lambda^{-1}] u_m(\vec{p}, \sigma) a(\vec{p}, \sigma) = \sum_{\sigma'} u_n(\Lambda^{-1}\vec{p}, \sigma) D^{(s)}_{\sigma\sigma'}[L^{-1}(\Lambda^{-1}\vec{p})\Lambda^{-1}L(\vec{p})] a(\vec{p}, \sigma'). \tag{A153}$$

Next set $\Lambda = L(\vec{p})$, and recall that $L(\vec{p})$ takes a particle from rest to momentum $\vec{p}$. $L^{-1}(\vec{p})$ takes a particle with momentum $\vec{p}$ to rest. Then the argument of the $D^{(s)}$ on the *rhs* of the above expression equals

$$L^{-1}(L^{-1}(\vec{)}p)L^{-1}(\vec{p})L(\vec{p}) = L^{-1}(\vec{0}). \tag{A154}$$

$L^{-1}(\vec{0})$, by definition, is a boost which takes $\vec{p} = \vec{0} \rightarrow \vec{p} = \vec{0}$. Hence it is an identity transformation $I$. Consequently $D^{(s)}[I]$ is also an identity matrix

$$D^{(s)}_{\sigma\sigma'}[L^{-1}(\Lambda^{-1}\vec{p})\Lambda^{-1}L(\vec{p})]\Bigg|_{\Lambda = L(\vec{p})} = \delta_{\sigma\sigma'}. \tag{A155}$$

This reduces the equation satisfied by the Fourier coefficients $u_n(\vec{p}, \sigma)$ to

$$\sum_m D_{nm}[L^{-1}(\vec{p})] u_m(\vec{p}, \sigma) = u_n(\vec{0}, \sigma). \tag{A156}$$

Exploiting the group property satisfied be the finite dimensional representations of the Lorentz group

$$\sum_{\sigma''} D_{\sigma'\sigma''}[\Lambda_1] D_{\sigma''\sigma}[\Lambda_2] = D_{\sigma'\sigma}[\Lambda_1\Lambda_2], \tag{A157}$$

we finally obtain the equation satisfied by the the Fourier coefficients $u_n(\vec{p}, \sigma)$

$$u_n(\vec{p}, \sigma) = \sum_m D_{nm}[L(\vec{p})] u_m(\vec{0}, \sigma). \tag{A158}$$

Assembling the (2j+1) $u_n(\vec{p}, \sigma)$'s in a $(2j+1)$–dimensional column vector $u(\vec{p}, \sigma)$,

the above equation for the Fourier coefficients reads

$$u(\vec{p},\sigma) = D[L(\vec{p})]u(\vec{0},\sigma). \tag{A159}$$

Similarly, we obtain

$$v(\vec{p},\sigma) = D[L(\vec{p})]v(\vec{0},\sigma). \tag{A160}$$

Matter Field Operators for the $(j,0)$ representations are obtained by the identification

$$D[L(\vec{p})] = D^{(j,0)}[L(\vec{p})] = \exp(\vec{J}\cdot\vec{\phi}). \tag{A161}$$

With this identification the Fourier coefficients $u(\vec{p},\sigma)$ [and $v(\vec{p},\sigma)$] have the interpretation of matter fields corresponding to the $(j,0)$ representation of the Lorentz group. That is

$$u(\vec{p},\sigma) \Leftrightarrow \phi^{R}_{j_r,\sigma_r}(\vec{p}). \tag{A162}$$

Or, in terms of the compact notation introduced earlier this identification reads

$$u(\vec{p},\sigma) \Leftrightarrow \phi^{R}(\vec{p}). \tag{A163}$$

Matter Field Operators for the $(0,j)$ representations are obtained by the identification

$$D[L(\vec{p})] = D^{(0,j)}[L(\vec{p})] = \exp(-\vec{J}\cdot\vec{\phi}). \tag{A164}$$

Then the Fourier coefficients are the $(0,j)$ matter fields

$$u(\vec{p},\sigma) \Leftrightarrow \phi^{L}_{j_l,\sigma_l}(\vec{p}). \tag{A165}$$

Or more compactly

$$u(\vec{p},\sigma) \Leftrightarrow \phi^{L}(\vec{p}). \tag{A166}$$

## A9 $2(2j+1)$ Component Matter Fields, Causality and Spin Statistics

As already pointed out under the operation of parity the $(j,0)$ and the $(0,j)$ representations get interchanged. It is therefore necessary to introduce a single $2(2j+1)$–component matter field

$$\psi(x) = \begin{pmatrix} \phi^{R}(x) \\ \\ \phi^{L}(x) \end{pmatrix}. \tag{A167}$$

which transforms as the $(j,0)\oplus(0,j)$ representation of the Lorentz group. For a spin–1/2 particle, it obeys the Dirac equation. For spin–1 it obeys the Spin–1 Weinberg equation, and so on. The associated matter field operators transform as

$$U[\Lambda]\Psi_\alpha(x)U^{-1}[\Lambda] = \sum_\beta \mathcal{D}^{(j)}_{\alpha\beta}[\Lambda^{-1}]\Psi_\beta(\Lambda x), \tag{A168}$$

where

$$\mathcal{D}^{(j)}[\Lambda] = \begin{pmatrix} D^{(j,0)}[\Lambda] & 0 \\ \\ 0 & D^{(0,j)}[\Lambda] \end{pmatrix}. \tag{A169}$$

Now note that

$$\begin{aligned} \{D^{(j,0)}[\Lambda]\}^\dagger &= \exp[i\vec{J}\cdot\vec{\phi} + \vec{J}\cdot\vec{\phi}] \\ 2 &= \exp[-\{i\vec{J}\cdot(-\vec{\phi}) + \vec{J}\cdot(-\vec{\phi})\}] \\ &= D^{(0,j)}[\Lambda^{-1}]. \end{aligned} \tag{A170}$$

Introducing the $2(2j+1)$ dimensional matrix (in chiral representation)

$$\beta = \begin{pmatrix} 0 & 1 \\ 1 & 0 \end{pmatrix}; \quad \beta^2 = 1, \tag{A171}$$

one easily verifies that

$$\{\mathcal{D}^{(j)}[\Lambda]\}^\dagger = \beta\ \mathcal{D}^{(j)}[\Lambda^{-1}]\ \beta. \tag{A172}$$

As a consequence, taking the hermitian conjugate of (A168) and using (A172) one finds

$$U[\Lambda]\ \overline{\Psi}_\alpha(x)\ U^{-1}[\Lambda] = \sum_\beta \overline{\Psi}(\Lambda x)\ \mathcal{D}^{(j)}_{\beta\alpha}[\Lambda], \tag{A173}$$

where we have introduced the *covariant adjoint*

$$\overline{\Psi}(x) \equiv \Psi^\dagger(x)\beta. \tag{A174}$$

Explicitly this is seen as follows. Hermitian conjugate of (A168)yields

$$\begin{aligned} U[\Lambda]\ \Psi^\dagger_\alpha(x)\ U^{-1}[\Lambda] &= \{U^{-1}[\Lambda]\}^\dagger\ \Psi^\dagger_\alpha(x)\ \{U[\Lambda]\}^\dagger \\ &= \sum_\beta \Psi^\dagger_\beta(\Lambda x)\ \{\mathcal{D}^{(j)}[\Lambda^{-1}]\}^\dagger_{\beta\alpha} \\ &[Using\ (\text{A172})] \\ &= \sum_{\beta,\lambda,\rho} \Psi^\dagger_\beta(\Lambda x)\ \beta_{\beta\lambda}\ \{\mathcal{D}^{(j)}[\Lambda]\}_{\lambda\rho}\ \beta_{\rho\alpha}. \end{aligned} \tag{A175}$$

In matrix notation, this can be written as

$$U[\Lambda]\ \Psi^\dagger(x)\ U^{-1}[\Lambda] = \Psi^\dagger(\Lambda x)\ \beta\ \mathcal{D}^{(j)}[\Lambda]\ \beta. \tag{A176}$$

Multiplying both sides by $\beta$ from the right and remembering that $U[\Lambda]$ and $\beta$ belong to different spaces [ $U[\Lambda]$ is an infinite dimensional unitary operator, while $\beta$ is a $2(2j+1)-$ dimensional :matrix in the wave function space.], we obtain the result claimed earlier at Eq. (A173):

$$U[\Lambda]\ \overline{\Psi}(x)\ U^{-1}[\Lambda] = \overline{\Psi}(\Lambda x)\ \mathcal{D}^{(j)}[\Lambda]. \tag{A177}$$

The multicomponent field operator for a general $2(2j+1)$ component $(j,0)\oplus(0,j)$ matter field can be written as

$$\Psi(x) = \left(\frac{1}{2\pi}\right)^{3/2} \int \frac{d^3p}{\sqrt{2\omega(\vec{p})}} \sum_{\sigma} \Big[\xi\ u(\vec{p},\sigma)\ a(\vec{p},\sigma)\ \exp(ip\cdot x) + \eta\ v(\vec{p},\sigma)\ b^{\dagger}(\vec{p},\sigma)\ \exp(-ip\cdot x)\Big]. \tag{A178}$$

where $\xi$ and $\eta$ are complex numbers, $\xi_0 \exp(i\theta)$ and $\eta_0 \exp(i\varphi)$ respectively, to be fixed by imposing the *causality condition*

$$\left[\Psi_\alpha(x), \Psi_\beta(x')\right]_{\pm} = 0, \qquad \text{for}\ \ \eta_{\mu\nu}(x^\mu - x'^\mu)(x^\nu - x'^\mu) < 0. \tag{A179}$$

The $[\ ,\ ]_\pm$ is a commutator for the $+$ sign and anticommutator for the $-$ sign. The $u(\vec{p},\sigma)$ and $v(\vec{p},\sigma)$ are the particle and antiparticle wave functions satisfying the transformation property

[Chiral Representation]

$$\left\{\begin{matrix} u(\vec{p},\sigma) \\ or \\ v(\vec{p},\sigma)\end{matrix}\right\}' = \begin{pmatrix} \exp(\vec{J}\cdot\vec{\phi}) & 0 \\ & \\ 0 & \exp(-\vec{J}\cdot\vec{\phi})\end{pmatrix} \left\{\begin{matrix} u(\vec{p},\sigma) \\ or \\ v(\vec{p},\sigma)\end{matrix}\right\}. \tag{A180}$$

The particle interpretation requires that the operators appearing in the Fourier transform on the *rhs* of (A178), satisfy the following properties:

$$\Big[a(\vec{p},\sigma), a^{\dagger}(\vec{p}\,',\sigma')\Big]_{\pm} = \delta(\vec{p}-\vec{p}\,')\ \delta_{\sigma\sigma'}, \tag{A181}$$

$$\Big[b(\vec{p},\sigma), b^{\dagger}(\vec{p}\,',\sigma')\Big]_{\pm} = \delta(\vec{p}-\vec{p}\,')\ \delta_{\sigma\sigma'}, \tag{A182}$$

$$\left[a(\vec{p},\sigma), b^{\dagger}(\vec{p}\,',\sigma')\right]_{\pm} = 0, \tag{A183}$$

$$\left[a(\vec{p},\sigma), b(\vec{p}\,',\sigma')\right]_{\pm} = 0, \tag{A184}$$

The Spin-1/2 spinors in the *canonical representation* are related to the ones in the chiral representation [defined by (A167)]

$$u_{CA}(\vec{p}) = \frac{1}{\sqrt{2}} \begin{pmatrix} 1 & 1 \\ 1 & -1 \end{pmatrix} \begin{pmatrix} \phi^{R}(\vec{p}) \\ \phi^{L}(\vec{p}) \end{pmatrix} = \frac{1}{\sqrt{2}} \begin{pmatrix} \phi^{R}(\vec{p}) + \phi^{L}(\vec{p}) \\ \phi^{R}(\vec{p}) - \phi^{L}(\vec{p}) \end{pmatrix}, \tag{A185}$$

and are readily verified to be (See Ref. [53, Sec. 2.5]. Also note that the $\phi^{R}(\vec{p})$ and $\phi^{L}(\vec{p})$ in the above expression correspond to $j = 1/2$)

[Canonical Representation]

$$u_{+1/2}(\vec{p}) = \left(\frac{E+m}{2m}\right)^{(1/2)} \begin{pmatrix} 1 \\ 0 \\ \frac{p_z}{E+m} \\ \frac{p_+}{E+m} \end{pmatrix}, \quad u_{-1/2}(\vec{p}) = \left(\frac{E+m}{2m}\right)^{(1/2)} \begin{pmatrix} 0 \\ 1 \\ \frac{p_-}{E+m} \\ \frac{-p_z}{E+m} \end{pmatrix},$$

$$v_{+1/2}(\vec{p}) = \left(\frac{E+m}{2m}\right)^{(1/2)} \begin{pmatrix} \frac{p_z}{E+m} \\ \frac{p_+}{E+m} \\ 1 \\ 0 \end{pmatrix}, \quad v_{-1/2}(\vec{p}) = \left(\frac{E+m}{2m}\right)^{(1/2)} \begin{pmatrix} \frac{p_-}{E+m} \\ \frac{-p_z}{E+m} \\ 0 \\ 1 \end{pmatrix}. \tag{A186}$$

They satisfy the following properties

$$\overline{u}_{(\sigma)}(\vec{p})\ u_{(\sigma')}(\vec{p}) = \delta_{\sigma\sigma'}, \tag{A187}$$

$$\overline{v}_{(\sigma)}(\vec{p})\ v_{(\sigma')}(\vec{p}) = -\delta_{\sigma\sigma'}, \tag{A188}$$

$$\overline{u}_{(\sigma)}(\vec{p})\ v_{(\sigma')}(\vec{p}) = 0, \tag{A189}$$

$$u^{\dagger}_{(\sigma)}(\vec{p})\ u_{(\sigma')} = v^{\dagger}_{(\sigma)}(\vec{p}) v_{(\sigma')} = \frac{E}{m}\delta_{\sigma\sigma'}, \tag{A190}$$

where $\sigma = +\frac{1}{2}, -\frac{1}{2}$. In addition the reader can verify the following identities

$$\sum_{\sigma} u^{\alpha}_{(\sigma)}(\vec{p})\ \overline{u}^{\beta}_{(\sigma)}(\vec{p}) = \left(\frac{\gamma\cdot p + m}{2m}\right)_{\alpha\beta}, \tag{A191}$$

$$\sum_{\sigma} v^{\alpha}_{(\sigma)}(\vec{p})\ \overline{v}^{\beta}_{(\sigma)}(\vec{p}) = \left(\frac{\gamma\cdot p - m}{2m}\right)_{\alpha\beta}, \tag{A192}$$

Here $\alpha, \beta$ are the 4-spinor indices which refer to *components* of a spinor; $(\sigma)$ runs over the eigenvalues of $\frac{1}{2}\sigma_z : \pm\frac{1}{2}$, and refers to *a* spinor (rather than components of *a* spinor).

<u>The $\xi$ and $\eta$ as determined by the Causality Condition (A179)</u>

<u>for the Spin–1/2 Particles</u>: We now wish to calculate the anticommutator

$$[\Psi_{\alpha}(x), \Psi^{\dagger}_{\beta}(x')]_{+} \equiv \{\Psi_{\alpha}(x), \Psi^{\dagger}_{\beta}(x')\} \tag{A193}$$

explicitly for the spin–$\frac{1}{2}$ case, and determine the constraints imposed on $\xi$ and $\eta$ (which appear on the *rhs* of (A178) ). The $\alpha_{th}$ component of the 4-component spin–$\frac{1}{2}$ matter field operator is

$$\begin{aligned}\Psi_{\alpha}(x) = \left(\frac{1}{2\pi}\right)^{3/2} \int \frac{d^3p}{\sqrt{2\omega(\vec{p})}} \sum_{\sigma} \\ \Big[\xi\ u_{\alpha}(\vec{p},\sigma)\ a(\vec{p},\sigma)\ \exp(ip\cdot x) + \eta\ v_{\alpha}(\vec{p},\sigma)\ b^{\dagger}(\vec{p},\sigma)\ \exp(-ip\cdot x)\Big].\end{aligned} \tag{A194}$$

with $u_{\alpha}(\vec{p},\sigma)$ and $v_{\alpha}(\vec{p},\sigma)$ components of $u(\vec{p},\sigma) \equiv u_{(\sigma)}(\vec{p})$ and $v(\vec{p},\sigma) \equiv v_{(\sigma)}(\vec{p})$ respectively (as given by (A186), in the canonical representation). Taking the hermitian conjugate of (A194) we get

$$\Psi^\dagger_\alpha(x) = \left(\frac{1}{2\pi}\right)^{3/2} \int \frac{d^3p}{\sqrt{2\omega(\vec{p})}} \sum_\sigma \Big[\xi^* \; u^\dagger_\alpha(\vec{p},\sigma) \; a(\vec{p},\sigma) \; \exp(-ip\cdot x) + \eta^* \; v^\dagger_\alpha(\vec{p},\sigma) \; b^\dagger(\vec{p},\sigma) \; \exp(ip\cdot x)\Big]. \tag{A195}$$

Remember $u$ and $u^\dagger$ live in the spinorial space and $a, a^\dagger, b, b^\dagger$ are the Fock space operators and hence can be moved across $u's$ and $v's$. The component $u_\alpha(\vec{p},\sigma)$ of the spinor $u(\vec{p},\sigma)$ are of course $c$–numbers; $u^\dagger_\alpha$ is the $\alpha_{th}$ component of $u^\dagger$ (and equals $u^*_\alpha$). To calculate the anticommutator $\{\Psi_\alpha(x), \Psi^\dagger_\beta(x')\}$ we first calculate

$$\begin{aligned}\Psi_\alpha(x) \; \Psi^\dagger_\beta(x') = \left(\frac{1}{2\pi}\right)^{3} \int \frac{d^3p \; d^3p'}{\sqrt{2\omega(\vec{p}) \; 2\omega(\vec{p}\,')}} \sum_{\sigma,\sigma'} & \Big[\xi\xi^* \; u_\alpha(\vec{p},\sigma) \; u^\dagger_\beta(\vec{p}\,',\sigma') \; a(\vec{p},\sigma) \; a^\dagger(\vec{p}\,',\sigma') \; \exp(ip\cdot x - ip'\cdot x') \\ &+ \xi\eta^* \; u_\alpha(\vec{p},\sigma) \; v^\dagger_\beta(\vec{p}\,',\sigma') \; a(\vec{p},\sigma) \; b^\dagger(\vec{p}\,',\sigma') \; \exp(ip\cdot x + ip'\cdot x') \\ &+ \eta\xi^* \; v_\alpha(\vec{p},\sigma) \; u^\dagger_\beta(\vec{p}\,',\sigma') \; b^\dagger(\vec{p},\sigma) \; a^\dagger(\vec{p}\,',\sigma') \; \exp(-ip\cdot x - ip'\cdot x') \\ &+ \eta\eta^* \; v_\alpha(\vec{p},\sigma) \; v^\dagger_\beta(\vec{p}\,',\sigma') \; b^\dagger(\vec{p},\sigma) \; b(\vec{p}\,',\sigma') \; \exp(-ip\cdot x + ip'\cdot x')\Big].\end{aligned} \tag{A196}$$

This immediately yields the anticommutator

$$\{\Psi_\alpha(x),\ \Psi^\dagger_\beta(x')\} = \left(\frac{1}{2\pi}\right)^3 \int \frac{d^3p\ d^3p'}{\sqrt{2\omega(\vec{p})\ 2\omega(\vec{p}\ ')}} \sum_{\sigma,\sigma'}$$

$$\Big[\xi\xi^*\ u_\alpha(\vec{p},\sigma)\ u^\dagger_\beta(\vec{p}\ ',\sigma')\ \{a(\vec{p},\sigma),\ a^\dagger(\vec{p}\ ',\sigma')\}\ \exp(ip\cdot x - ip'\cdot x')$$

$$+\ \xi\eta^*\ u_\alpha(\vec{p},\sigma)\ v^\dagger_\beta(\vec{p}\ ',\sigma')\ \{a(\vec{p},\sigma),\ b^\dagger(\vec{p}\ ',\sigma')\}\ \exp(ip\cdot x + ip'\cdot x')$$

$$+\ \eta\xi^*\ v_\alpha(\vec{p},\sigma)\ u^\dagger_\beta(\vec{p}\ ',\sigma')\ \{b^\dagger(\vec{p},\sigma),\ a^\dagger(\vec{p}\ ',\sigma')\}\ \exp(-ip\cdot x - ip'\cdot x')$$

$$+\ \eta\eta^*\ v_\alpha(\vec{p},\sigma)\ v^\dagger_\beta(\vec{p}\ ',\sigma')\ \{b^\dagger(\vec{p},\sigma),\ b(\vec{p}\ ',\sigma')\}\ \exp(-ip\cdot x + ip'\cdot x')\Big]$$

$$= \left(\frac{1}{2\pi}\right)^3 \int \frac{d^3p\ d^3p'}{\sqrt{2\omega(\vec{p})\ 2\omega(\vec{p}\ ')}} \sum_{\sigma,\sigma'}$$

$$\Big[\xi\xi^*\ u_\alpha(\vec{p},\sigma)\ u^\dagger_\beta(\vec{p}\ ',\sigma')\ \ \delta(\vec{p}-\vec{p}')\ \delta_{\sigma\sigma'}\ \exp(ip\cdot x - ip'\cdot x')$$

$$+\ \eta\eta^*\ v_\alpha(\vec{p},\sigma)\ v^\dagger_\beta(\vec{p}\ ',\sigma')\ \ \delta(\vec{p}-\vec{p}')\ \delta_{\sigma\sigma'}\ \exp(-ip\cdot x + ip'\cdot x')\Big]$$

$$= \left(\frac{1}{2\pi}\right)^3 \int \frac{d^3p}{2\omega(\vec{p})} \sum_{\sigma}$$

$$\Big[\xi\xi^*\ u_\alpha(\vec{p},\sigma)\ u^\dagger_\beta(\vec{p},\sigma)\ \exp[ip\cdot(x-x')]$$

$$+\ \eta\eta^*\ v_\alpha(\vec{p},\sigma)\ v^\dagger_\beta(\vec{p},\sigma)\ \exp[-ip\cdot(x-x')]\Big]. \tag{A197}$$

To evaluate the *rhs* of the above expression we now use (A191) and note that $(\gamma^0)^2 = 1$, or equivalently, $\{(\gamma^0)^2\}_{\alpha\beta} = \delta_{\alpha\beta}$,

$$\begin{aligned}
\sum_\sigma u_\alpha(\vec{p},\sigma)\ u^\dagger_\beta(\vec{p},\sigma) &= \sum_\sigma\ \sum_\rho u_\alpha(\vec{p},\sigma)\ u^\dagger_\rho(\vec{p},\sigma)\ \delta_{\rho\beta} \\
&= \sum_\sigma\ \sum_\rho u_\alpha(\vec{p},\sigma)\ u^\dagger_\rho(\vec{p},\sigma)\ \{(\gamma^0)^2\}_{\rho\beta} \\
&= \sum_\sigma\ \sum_{\rho,\lambda} u_\alpha(\vec{p},\sigma)\ u^\dagger_\rho(\vec{p},\sigma)\ \gamma^0_{\rho\lambda}\ \gamma^0_{\lambda\beta} \\
&= \sum_\sigma\ \sum_\lambda u_\alpha(\vec{p},\sigma)\ \overline{u}_\lambda(\vec{p},\sigma)\ \gamma^0_{\lambda\beta} \\
&= \sum_\lambda\ \left(\frac{\gamma\cdot p + m}{2m}\right)_{\alpha\lambda}\ \gamma^0_{\lambda\beta},
\end{aligned} \tag{A198}$$

where we used the definition $\overline{u}_\lambda \equiv \sum_\rho u^\dagger_\rho(\vec{p},\sigma)\ \gamma^0_{\rho\lambda}$. Similarly

$$\sum_\sigma v_\alpha(\vec{p},\sigma)\ v^\dagger_\beta(\vec{p},\sigma) = \sum_\lambda \left(\frac{\gamma\cdot p - m}{2m}\right)_{\alpha\lambda} \gamma^0_{\lambda\beta}. \tag{A199}$$

As a result we have

$$\begin{aligned}\{\Psi_\alpha(x),\, \Psi^\dagger_\beta(x')\} = \left(\frac{1}{2\pi}\right)^3 \int \frac{d^3p}{2\omega(\vec{p})} \sum_\lambda \\ \Big[\xi\xi^*\ \left(\frac{\gamma\cdot p + m}{2m}\right)_{\alpha\lambda} \gamma^0_{\lambda\beta}\ \exp[ip\cdot(x-x')] \\ +\,\eta\eta^*\ \left(\frac{\gamma\cdot p - m}{2m}\right)_{\alpha\lambda} \gamma^0_{\lambda\beta} \exp[-ip\cdot(x-x')]\Big].\end{aligned} \tag{A200}$$

We would now confine our considerations to the simplest spacelike separations. However the results which follow hold true for *all* spacelike separations (See Ref. [21]). The *simplest* spacelike separation is obtained by setting $x^0 = x'^0 = t$. For these spacelike separations (A200) takes the form

$$\begin{aligned}\{\Psi_\alpha(t,\vec{x}),\, \Psi^\dagger_\beta(t,\vec{x}\,')\} = \left(\frac{1}{2m}\right)\left(\frac{1}{2\pi}\right)^3 \int \frac{d^3p}{2\omega(\vec{p})} \sum_\lambda \\ \Big[\xi\xi^*\ (\gamma\cdot p + m)_{\alpha\lambda}\ \gamma^0_{\lambda\beta}\ \exp[i\vec{p}\cdot(\vec{x}-\vec{x}\,')] \\ +\,\eta\eta^*\ (\gamma\cdot p - m)_{\alpha\lambda}\ \gamma^0_{\lambda\beta} \exp[-i\vec{p}\cdot(\vec{x}-\vec{x}\,')]\Big].\end{aligned} \tag{A201}$$

Without loss of further generality , we now let $\vec{p}\to -\vec{p}$ in the term associated with the $\eta\eta^*$ term on the *rhs* of the above expression, to obtain

$$\begin{aligned}\{\Psi_\alpha(t,\vec{x}),\, \Psi^\dagger_\beta(t,\vec{x}\,')\} = \left(\frac{1}{2m}\right)\left(\frac{1}{2\pi}\right)^3 \int \frac{d^3p}{2\omega(\vec{p})} \sum_\lambda \\ \Big[\xi\xi^*\ \left(\gamma^0 p_0 - \vec{\gamma}\cdot\vec{p} + m\right)_{\alpha\lambda}\ \gamma^0_{\lambda\beta}\ \exp[i\vec{p}\cdot(\vec{x}-\vec{x}\,')] \\ +\,\eta\eta^*\ \left(\gamma^0 p_0 + \vec{\gamma}\cdot\vec{p} - m\right)_{\alpha\lambda}\ \gamma^0_{\lambda\beta}\ \exp[i\vec{p}\cdot(\vec{x}-\vec{x}\,')]\Big].\end{aligned} \tag{A202}$$

In order that the anticommutator of the spin-$\frac{1}{2}$ matter field operators vanish for spacelike separations we require

$$\xi\xi^* = \eta\eta^*. \tag{A203}$$

With this requirement, and taking note of the relation $\sum_\lambda \ \gamma^0_{\alpha\lambda}\,\gamma^0_{\lambda\beta} = \delta_{\alpha\beta}$, we get

$$\{\Psi_\alpha(t,\vec{x}),\ \Psi^\dagger_\beta(t,\vec{x}\,')\} = \left(\frac{\xi\xi^*}{2m}\right)\left(\frac{1}{2\pi}\right)^3 \int \frac{d^3p}{2\omega(\vec{p})}\, 2p_0\ \delta_{\alpha\beta}\ \exp[i\vec{p}\cdot(\vec{x}-\vec{x}\,')]. \tag{A204}$$

But by definition $p_0 = \omega(\vec{p})$,

$$\{\Psi_\alpha(t,\vec{x}),\ \Psi^\dagger_\beta(t,\vec{x}\,')\} = \left(\frac{\xi\xi^*}{2m}\right)\ \delta_{\alpha\beta}\ \left(\frac{1}{2\pi}\right)^3 \int d^3p\ \exp[i\vec{p}\cdot(\vec{x}-\vec{x}\,')]. \tag{A205}$$

Implementing the $p$–integration, we finally have the causality condition

$$\{\Psi_\alpha(t,\vec{x}),\ \Psi^\dagger_\beta(t,\vec{x}\,')\} = \left(\frac{\xi\xi^*}{2m}\right)\ \delta^3(\vec{x}-\vec{x}\,')\ \delta_{\alpha\beta}. \tag{A206}$$

We thus note that in order to arrive at the causality condition we have to impose a *constraint* upon $\xi$ and $\eta$. This constraint is given by Eq. (A203). In Weinberg's [21, p. B1323] words the constraint means that *" Every particle must have an antiparticle (perhaps itself) which enters into interactions with equal coupling strength."* This result [proved here for spin–$\frac{1}{2}$ particles, and the simplest spacelike separations] is the direct consequence of demanding causality. In a similar fashion it follows that

$$\{\Psi_\alpha(t,\vec{x}),\ \Psi_\beta(t,\vec{x}\,')\} = 0. \tag{A207}$$

In the next section we will *derive* the causality condition by considering the evolution of a quantum system from a spacelike surface to another. For the moment we note that

$$[\Psi_\alpha(t,\vec{x}),\ \Psi^\dagger_\beta(t,\vec{x}\,')] \neq \eta\ \delta^3(\vec{x}-\vec{x}\,')\ \delta_{\alpha\beta}, \tag{A208}$$

and no simple means are known to replace $\neq$ by $=$ in the above expression, except be replacing the commutator [ , ] by the anticommutator { , }. In the above equation $\eta$ is a c–number.

In order to match the widely used conventions of Bjorken and Drell [52, p.59] we choose

$\left[\text{For Spin}-\frac{1}{2}\right]$

$$\xi\xi^* = \eta\eta^* = 2m, \tag{A209}$$

so that the spin$-\frac{1}{2}$ matter field operator, given by (A178), reads

$$\begin{aligned} \Psi(x) = \sum_\sigma \int \frac{d^3p}{(2\pi)^{3/2}} \sqrt{\frac{m}{\omega(\vec{p})}} \\ \Big[u(\vec{p},\sigma)\ a(\vec{p},\sigma)\ \exp(ip\cdot x) + v(\vec{p},\sigma)\ b^\dagger(\vec{p},\sigma)\ \exp(-ip\cdot x)\Big], \end{aligned} \tag{A210}$$

and (A206) takes the slightly simpler form

$$\{\Psi_\alpha(t,\vec{x}),\ \Psi^\dagger_\beta(t,\vec{x}\ ')\} = \delta^3(\vec{x}-\vec{x}\ ')\ \delta_{\alpha\beta}. \tag{A211}$$

As a result of these considerations it is obvious that fermions and bosons behave in intrinsically different ways. The anticommutativity of the fermion creation and destruction operators does not allow any two fermions to be in the same state in a given system. This result is usually referred to as the *Pauli Exclusion Principle*. As already commented there is no known way of circumventing this anticommutativity and meeting the causality condition for fermions at the same time. It is not only at the microscopic level that this fundamental anticommutativity shows its dramatic consequences, for example much of the different chemical characteristics of elements arise from the Pauli exclusion principle, but the consequences are equally important at the macroscopic level. For instance as a result of this anticomutativity, the pressure, $P$, of an extremely degenerate electron gas (with an electrically positive background to provide overall electrical neutrality) depends on the 4/3 power of the electron density $\rho_e$ (See Ref. [49, Sec. 61]. We have restored $\hbar$ and $c$ in the formula below.)

$$P = \frac{(3\pi^2)^{1/3}}{4}\ \hbar c\ \rho_e^{4/3}. \tag{A212}$$

This pressure when balanced by the gravitational forces in astrophysical situations results in the formation of *White Dwarfs* and *Neutron Stars.* Even though the momentum-spectrum of the particles which is responsible for this pressure is constrained by the Pauli exclusion principle, these momenta have their origin in electroweak and gravitational interactions. So, given the interactions between various fermions in a system, the anticommutativity acts as an additional constraint, like the boundary conditions, in determining what states are accessible to a system. The next section may shed some light on the origin of the causality conditions for fermions and bosons. However, just as the inertial-frame-independence of the speed of light, in the context of which alone the causality conditions acquire a meaning, is a mysterious empirical fact, the same holds to some extent for the causality conditions. In regard to the last comment recall that causality conditions require us to specify spacelike separations. The concept of spacelike separations cannot be defined without reference to the constancy of the speed of light. As such we suspect that the observed constancy of the speed of light and the causality conditions are interrelated. At the least, the latter loses meaning without the former.

The above observations require a further parenthetic remark. This concerns the *finiteness* of the speed of light. For $c = \infty$, to be distinguished from $c \to \infty$, spacelike regions of spacetime disappear and the causality conditions cannot be expressed as in (A211). For comparison with spin–1 wave functions, given in Chapter 3, we rewrite the spin–$\frac{1}{2}$ spinors in the canonical representation:

[Canonical Representation]

$$u_{+\frac{1}{2}}(\vec{p}) = \begin{pmatrix} \sqrt{\frac{E+m}{2m}} \\ 0 \\ \frac{p_z}{\sqrt{2m(E+m)}} \\ \frac{p_+}{\sqrt{2m(E+m)}} \end{pmatrix}, \quad u_{-\frac{1}{2}}(\vec{p}) = \begin{pmatrix} 0 \\ \sqrt{\frac{E+m}{2m}} \\ \frac{p_-}{\sqrt{2m(E+m)}} \\ \frac{-p_z}{\sqrt{2m(E+m)}} \end{pmatrix}$$

$$v_{+\frac{1}{2}}(\vec{p}) = \begin{pmatrix} \frac{p_z}{\sqrt{2m(E+m)}} \\ \\ \frac{p_+}{\sqrt{2m(E+m)}} \\ \\ \sqrt{\frac{E+m}{2m}} \\ \\ 0 \end{pmatrix}, \quad v_{-\frac{1}{2}}(\vec{p}) = \begin{pmatrix} \frac{p_-}{\sqrt{2m(E+m)}} \\ \\ \frac{-p_z}{\sqrt{2m(E+m)}} \\ \\ 0 \\ \\ \sqrt{\frac{E+m}{2m}} \end{pmatrix}. \tag{A213}$$

This indicates that the Pauli exclusion principle, or more generally classification of particles as bosons and fermions, may be possible within the framework of finite $c$ theories. However, nonexistence of the causality conditions must not be considered as equivalent to the suspension of *causality* itself.

It is perhaps also an open question whether the distinction between fermions and bosons is unaffected by the introduction of gravitational interactions. We raise this question because it is not obvious in what way is the discussion of the next section, where we arrive at the causality condition within the framework of the Poincaré covariant structure of quantum systems, is modified by the introduction of gravitational interactions. In any case the distinction between fermions and bosons, i.e. the existence of the causality condition (A179), can always be maintained in a *local* inertial frame, even in the presence of the gravitational field.

## A10 Schwinger [24] as a Logical Continuation of Weinberg [21,33]: Action Principle, Origin of Causality Conditions

We have seen that demanding Poincaré covariance and introducing a vacuum state $|\ \rangle$, such that

$$|\vec{p},\sigma\rangle = a^\dagger(\vec{p},\sigma)|\ \rangle, \qquad a(\vec{p},\sigma)|\ \rangle = 0; \tag{A214}$$

naturally leads to the introduction of the matter field operators

$$\begin{aligned}\Psi(x) = \left(\frac{1}{2\pi}\right)^{3/2} \int \frac{d^3p}{\sqrt{2\omega(\vec{p})}} \sum_\sigma \\ \Big[\xi\ u(\vec{p},\sigma)\ a(\vec{p},\sigma)\ \exp(ip\cdot x) + \eta\ v(\vec{p},\sigma)\ b^\dagger(\vec{p},\sigma)\ \exp(-ip\cdot x)\Big],\end{aligned} \tag{A215}$$

which contain the basic degrees of freedom. As such the *physical observables,* must be constructed as functionals of $\Psi(x)$.

Definition: A *functional* $F[x(t)]$ gives a number for each function $x(t)$, see Ref. [40, Sec. 7.2].

Our discussion so far has been confined to the *free fields.* Given a free state, nothing really evolves. Existence of free states is unconfirmable without some form of interactions. It is the potentiality of a system to interact, even if the interaction is confined to infinitesimal regions of spacetime, which allows for a measurement. The *dynamical evolution* of a system arises from the introduction of *interactions.* Discovering interactions and incorporating them in some unified fashion in the quantum systems seems to be a one of the major occupations of the physicists in the modern era.

A parenthetic remark: Should *all* possible *fundamental* interactions be expected to exist in the limit of two particle limit? To clarify, even though electromagnetic interaction can, and does exist, between $n \geq 2$ particles, it *can* be detected at the 2 particle level. The questions is, is it possible that there exist interactions which have no observable consequences if only two particles are involved? That is, are there interactions which manifest themselves only at the n-particle level, with $2 < n \leq \infty$. The actual upper limit for $n$ for a system of nucleons is $n \sim 10^{80}$, the number of nucleons in the observable universe. The question which we have raised here seems relevant not only to Physics but to Philosophy as well, for one can never from a practical point of view disprove the existence of an interactions for large n.

In order to study the dynamical evolution we need to specify a boundary value problem. That is, given $\Psi(x)$ on a particular surface $\Sigma$, we study its evolution (say in the direction perpendicular to $\Sigma$ ). For this specification on $\Sigma$ to be consistent with causality, $\Psi(x)$ must be specified at *physically independent* spacetime points. That is, as noted by Schwinger [24], spacetime points which cannot be connected even by signals propagating at the speed of light. A continuous set of such points forms a spacelike surface $\Sigma$. If $A$ is a complete set of commuting hermitian operators constructed out of $\Psi(x)$, and as Schwinger puts it "attached" to $\Sigma$, then to $\Sigma$ one can associate a basis $|\Sigma, a\rangle$. Of particular interest is the transformation which takes the physical system from a spacelike surface $\Sigma_2$ to another spacelike surface $\Sigma_1$. This evolution is represented as

$$|\Sigma_1, a_1\rangle = U_{2\to 1}|\Sigma_2, a_2\rangle, \tag{A216}$$

$$A_1 = U_{2\to 1}\ A_2\ \overset{-1}{U}_{2\to 1}, \quad U^{\dagger}_{2\to 1} = \overset{-1}{U}_{2\to 1}. \tag{A217}$$

The set of eigenvalues $\{a_1\}$ and $\{a_2\}$ are identical. In the absence of any interactions we expect $U_{2\to 1}$ to be related to $U(\{\Lambda, a\})$. Notational Comments: $a$ in $U(\{\Lambda, a\}$ is the translational parameter defined in

$$x'^{\mu} = \Lambda^{\mu\nu} x^{\nu} + a^{\mu}, \tag{A218}$$

while $a$ in $\{a_1\}$ or $|\Sigma, a\rangle$ refer to a set of eigenvalues of a general complete set of commuting observables $A$. $\{a\}$ is a set in which each $a$ is in turn a set of eigenvalues specifying a state. The *transformation function* can be written completely in terms of the basis vectors on the surface $\Sigma_2$ and the unitary operator $\overset{-1}{U}_{2\to 1}$

$$\langle\Sigma_1, a_1|\Sigma_2, a_2\rangle = \langle\Sigma_2, a_2|\overset{-1}{U}_{2\to 1}|\Sigma_2, a_2\rangle. \tag{A219}$$

The operator $U_{2\to 1}$ describes the development of the system from $\Sigma_2 \to \Sigma_1$ and involves, not only the detailed dynamical characteristics of the system in this space time region , but also the choice of observables on the surfaces $\Sigma_1$ and $\Sigma_2$. Any infinitesimal change in the quantities on which the transformation function depends induces a corresponding change in $\overset{-1}{U}_{2\to 1}$

$$\delta\langle\Sigma_1, a_1|\Sigma_2, a_2\rangle = \langle\Sigma_2, a_2|\delta\overset{-1}{U}_{2\to 1}|\Sigma_2, a_2\rangle. \tag{A220}$$

The variations $\delta U_{2\to1}$ and $\delta \overset{-1}{U}_{2\to1}$:

$$\begin{aligned} U_{2\to1} &\to U_{2\to1} + \delta U_{2\to1} \\ \overset{-1}{U}_{2\to1} &\to \overset{-1}{U}_{2\to1} + \delta \overset{-1}{U}_{2\to1}. \end{aligned} \tag{A221}$$

must satisfy certain conditions in order to preserve unitarity of $U_{2\to1}$. First we must have

$$U_{2\to1}\overset{-1}{U}_{2\to1} = (U_{2\to1} + \delta U_{2\to1})(\overset{-1}{U}_{2\to1} + \delta \overset{-1}{U}_{2\to1}). \tag{A222}$$

That is

$$U_{2\to1}\ \delta \overset{-1}{U}_{2\to1} = -\ \delta U_{2\to1}\ \overset{-1}{U}_{2\to1}, \tag{A223}$$

where in deriving this result a term $\mathcal{O}(\delta U_{2\to1}\ \delta \overset{-1}{U}_{2\to1})$ has been neglected.

The second constraint is provided by the condition

$$\overset{-1}{U}_{2\to1} = U^{\dagger}_{2\to1}, \tag{A224}$$

which yields

$$\delta \overset{-1}{U}_{2\to1} = \delta U^{\dagger}_{2\to1}. \tag{A225}$$

<u>It is readily seen that</u> $U_{2\to1}\delta \overset{-1}{U}_{2\to1}$ <u>is antihermitian.</u> From (A223) we have

$U_{2\to1}\delta \overset{-1}{U}_{2\to1} = -\delta U_{2\to1}\overset{-1}{U}_{2\to1}$, but $\overset{-1}{U}_{2\to1} = U^{\dagger}_{2\to1}$

and hence, $U_{2\to1}\delta \overset{-1}{U}_{2\to1} = -\delta U_{2\to1}U^{\dagger}_{2\to1}$. Using (A225) for $\delta U_{2\to1}$ we

get $U_{2\to1}\ \delta \overset{-1}{U}_{2\to1} = -(\delta U^{\dagger}_{2\to1})^{\dagger}\ U^{\dagger}_{2\to1} = -(\delta \overset{-1}{U}_{2\to1})^{\dagger}\ U^{\dagger}_{2\to1} = -(U_{2\to1}\ \delta \overset{-1}{U}_{2\to1})^{\dagger}$. As a consequence there exists an infinitesimal hermitian operator $\delta W_{2\to1}$

$$\delta W_{2\to1} = -i\ U_{2\to1}\ \delta \overset{-1}{U}_{2\to1}, \tag{A226}$$

so that

$$\delta \overset{-1}{U}_{2\to1} = i\ \overset{-1}{U}_{2\to1}\ \delta W_{2\to1}. \tag{A227}$$

Substituting $\delta \overset{-1}{U}_{2\to 1}$ from (A227) in (A220) yields

$$\delta\langle\Sigma_1, a_1|\Sigma_2, a_2\rangle = i\langle\Sigma_2, a_2|\overset{-1}{U}_{2\to 1}\ \delta W_{2\to 1}|\Sigma_2, a_2\rangle. \tag{A228}$$

Identifying $\langle\Sigma_2, a_2|\overset{-1}{U}_{2\to 1}$ by $\langle\Sigma_1, a_1|$ (see Eqs. (A216) and (A217)) we obtain

$$\delta\langle\Sigma_1, a_1|\Sigma_2, a_2\rangle = i\langle\Sigma_1, a_1|\delta W_{2\to 1}|\Sigma_2, a_2\rangle. \tag{A229}$$

In addition to the unitarity of $U_{2\to 1}$, we now use the *completeness relation* on the spacelike surface $\Sigma_2$

$$1 = \int |\Sigma_2, a_2\rangle da_2\langle\Sigma_2, a_2|, \tag{A230}$$

to obtain an important property of $\delta W_{2\to 1}$ in order to eventually postulate its general form. Using the completeness relation (A230) we write the transformation function

$$\langle\Sigma_1, a_1|\Sigma_3, a_3\rangle = \int\langle\Sigma_1, a_1|\Sigma_2, a_2\rangle da_2\langle\Sigma_2, a_2|\Sigma_3, a_3\rangle. \tag{A231}$$

As a result of variations given by (A221) and variations

$$\begin{aligned} U_{3\to 1} &\to U_{3\to 1} + \delta U_{3\to 1} \\ \overset{-1}{U}_{3\to 1} &\to \overset{-1}{U}_{3\to 1} + \delta\overset{-1}{U}_{3\to 1}, \end{aligned} \tag{A232}$$

the change in the transformation function $\langle\Sigma_1, a_1|\Sigma_3, a_3\rangle$ given by (A231) can be written as

$$\begin{aligned} \delta\langle\Sigma_1, a_1|\Sigma_3, a_3\rangle = &\int \{\delta\langle\Sigma_1, a_1|\Sigma_2, a_2\rangle\}\, da_2\langle\Sigma_2, a_2|\Sigma_3, a_3\rangle \\ &+ \int\langle\Sigma_1, a_1|\Sigma_2, a_2\rangle da_2\,\{\delta\langle\Sigma_2, a_2|\Sigma_3, a_3\rangle\}\,. \end{aligned} \tag{A233}$$

Using (A229)for the general variations $\delta\langle\Sigma_i, a_i|\Sigma_j, a_j\rangle$ in the above expression and replacing $\int|\Sigma_2, a_2\rangle da_2\langle\Sigma_2, a_2|$ by the unit operator 1 we get

$$\langle\Sigma_1, a_1|\delta W_{3\to 1}|\Sigma_3, a_3\rangle = \langle\Sigma_1, a_1|\delta W_{2\to 1}|\Sigma_3, a_3\rangle + \langle\Sigma_1, a_1|\delta W_{3\to 2}|\Sigma_3, a_3\rangle. \tag{A234}$$

Since the spacelike surfaces $\Sigma_1, \Sigma_2, \Sigma_3$ are completely arbitrary, we have the general property which the generators $\delta W_{j\to i}$ of the infinitesimal transformations must

satisfy

$$\delta W_{3\to 1} = \delta W_{3\to 2} + \delta W_{2\to 1}. \tag{A235}$$

We thus see that the unitarity and the completeness yield an additive law for the composition of the infinitesimal generators of evolution, from a spacelike surface $\Sigma_j$ another spacelike surface $\Sigma_i$, $\delta W_{j\to i}$. The additive requirement (A235)suggests that a finite evolution, the generators of the evolution have the form

$$W_{2\to 1} \equiv \int_{\Sigma_2}^{\Sigma_1} d^4x \; \Omega[x]. \tag{A236}$$

Individual systems are described by stating that $\Omega[x]$ is a Poincaré covariant hermitian function of the fields and their derivatives. For the moment we assert

$$\Omega[x] \equiv \Omega\left(\Psi(x), \partial_\mu \Psi(x)\right), \tag{A237}$$

in order to discover some of its general properties and physical significance. The covariance of $\Omega[x]$, and therefore of $W_{2\to 1}$, guarantees that our fundamental dynamical principle

$$\begin{aligned} &\delta\langle\Sigma_1, a_1|\Sigma_2, a_2\rangle \\ &\qquad = i\langle\Sigma_1, a_1|\delta W_{2\to 1}|\Sigma_2, a_2\rangle \\ &\qquad = i\langle\Sigma_1, a_1|\delta \int_{\Sigma_2}^{\Sigma_1} d^4x \; \Omega[x] \; |\Sigma_2, a_2\rangle, \end{aligned} \tag{A238}$$

is unaltered in form by Poincaré transformations or change in coordinate systems.

It should be parenthetically noted, for the moment, that Schwinger [24] points out that an exception must be made for discrete transformations, such as time reversal. He argues that the requirement of the invariance under time reversal imposes a general restriction upon the algebra of the field operators – the connection between the spin and the statistics.

From (A237) and (A238) we note that the evaluation of $\delta\langle\Sigma_1, a_1|\Sigma_2, a_2\rangle$ requires the knowledge of

$$\delta W_{2\to 1} = \delta \int_{\Sigma_2}^{\Sigma_1} d^4x \; \Omega(\Psi(x), \partial_\mu \Psi(x)). \tag{A239}$$

The evaluation of $\delta W_{2\to 1}$ involves adding

i) The independent effects of changing the matter field operators at each point by $\delta\Psi(x)$, $\Psi(x) \to \Psi'(x) = \Psi(x) + \delta\Psi(x)$, and

ii) of altering the region of integration by a displacement $\delta x^\mu$ of the points on the boundary surfaces $\Sigma_1$ and $\Sigma_2$. On $\Sigma_1$ and $\Sigma_2$: $x^\mu \to x'^\mu = x^\mu + \delta x^\mu$.

Thus

$$\delta W_{2\to 1} = \int_{\Sigma_2}^{\Sigma_1} d^4x \; \delta\Omega + \left( \int_{\Sigma_1} - \int_{\Sigma_2} \right) d\Sigma_\mu \delta x^\mu \Omega, \tag{A240}$$

where:

$$\begin{aligned} \delta\Omega &\equiv \left(\frac{\partial \Omega}{\partial \Psi}\right) \delta\Psi + \frac{\partial \Omega}{\partial(\partial_\mu \Psi)} \; \delta\partial_\mu \Psi \\ &= \left(\frac{\partial \Omega}{\partial \Psi}\right) \delta\Psi + \frac{\partial \Omega}{\partial(\partial_\mu \Psi)} \; \partial_\mu \delta\Psi. \end{aligned} \tag{A241}$$

By adding and subtracting

$$\partial_\mu \left( \frac{\partial \Omega}{\partial(\partial_\mu \Psi)} \right) \delta\Psi \tag{A242}$$

to the *rhs* of the above expression we obtain

$$d\Omega = \left[ \frac{\partial \Omega}{\partial \Psi} - \partial_\mu \left( \frac{\partial \Omega}{\partial(\partial_\mu \Psi)} \right) \right] \delta\Psi + \partial_\mu \left[ \frac{\partial \Omega}{\partial(\partial_\mu \Psi)} \; \delta\Psi \right]. \tag{A243}$$

Substituting (A243) in (A240) then yields the needed expression for the variation

in $W_{2\to1}$

$$\delta W_{2\to 1} = \int\limits_{\Sigma_2}^{\Sigma_1} d^4x \ \left[\frac{\partial\Omega}{\partial\Psi} - \partial_\mu\left(\frac{\partial\Omega}{\partial(\partial_\mu\Psi)}\right)\right]\delta\Psi + \int\limits_{\Sigma_2}^{\Sigma_1} d^4x \ \partial_\mu \left[\frac{\partial\Omega}{\partial(\partial_\mu\Psi)} \ \delta\Psi\right] + \left(\int\limits_{\Sigma_1} - \int\limits_{\Sigma_2}\right) d\Sigma_\mu \delta x^\mu \Omega. \tag{A244}$$

Using the divergence theorem, the second term on the *rhs* of the above expression

$$\int\limits_{\Sigma_2}^{\Sigma_1} d^4x \ \partial_\mu \left[\frac{\partial\Omega}{\partial(\partial_\mu\Psi)} \ \delta\Psi\right] \tag{A245}$$

can be written as

$$\left(\int\limits_{\Sigma_1} - \int\limits_{\Sigma_2}\right) d\Sigma_\mu \ \frac{\partial\Omega}{\partial(\partial_\mu\Psi)} \ \delta\Psi + \int\limits_{\Sigma_{21}} \ d\Sigma_\mu \ \frac{\partial\Omega}{\partial(\partial_\mu\Psi)} \ \delta\Psi, \tag{A246}$$

where $\Sigma_{21}$ is a surface joining the spacelike surfaces $\Sigma_1$ and $\Sigma_2$ at their boundaries at infinity. Under the usual assumption that the matter field operators [that is, their expectation value for the physical states under consideration] vanish on $\Sigma_{21}$, we get

$$\delta W_{2\to 1} = \int\limits_{\Sigma_2}^{\Sigma_1} d^4x \ \left[\frac{\partial\Omega}{\partial\Psi} - \partial_\mu\left(\frac{\partial\Omega}{\partial(\partial_\mu\Psi)}\right)\right]\delta\Psi + \left(\int\limits_{\Sigma_1} - \int\limits_{\Sigma_2}\right) d\Sigma_\mu \ \left[\frac{\partial\Omega}{\partial(\partial_\mu\Psi)} \ \delta\Psi + \Omega \ \delta x^\mu.\right] \tag{A247}$$

Before continuing with our calculations let us explicitly note that the *positional order* of operators in $\Omega(\Psi(x), \partial_\mu\Psi(x))$ must not be altered in the course of implementing the variations. Accordingly, the algebraic [commutators or anti-commtators] properties of $\delta\Psi(x)$ are involved in obtaining $\delta W_{2\to1}$. For simplicity, we have, following Schwinger, introduced the explicit assumption that the commutation properties of $\delta\Psi(x)$ and the structure of $\Omega$ must be so related that identical contributions are produced by the terms that differ fundamentally only in the position of $\delta\Psi(x)$ .

In order that works of Schwinger [24] and Weinberg [21, 33] yield the same Poincaré covariant equations of motion, $\Omega(\Psi(x), \partial_\mu\Psi(x))$ must be interpreted as the *Lagrangian density* $\mathcal{L}(\Psi(x), \partial_\mu\Psi(x))$

$$\Omega(\Psi(x), \partial_\mu\Psi(x)) = -\mathcal{L}(\Psi(x), \partial_\mu\Psi(x)) \tag{A248}$$

so that $\delta W_{2\to1}$ can be identified with the *action operator.* The *principle of stationary action operator* is then demanded rather than postulated, and yields the following Euler Lagrange equations of motion

$$\frac{\partial\mathcal{L}}{\partial\Psi} - \partial_\mu\left(\frac{\partial\mathcal{L}}{\partial(\partial_\mu\Psi)}\right) = 0 \tag{A249}$$

and the generator of evolution attached to a spacelike surface is

$$F(\Sigma) = +\int\limits_\Sigma d\Sigma_\mu\left[\frac{\partial\mathcal{L}}{\partial(\partial_\mu\Psi)}\ \delta\Psi + \mathcal{L}\ \delta x^\mu\right]. \tag{A250}$$

The reason for choosing the the + sign on the *rhs* of the above expression is associated with fact that we wish to have the conventional signs in the commutation relations and other definitions. This should become obvious by the end of this section.

The physical interpretation of the generators of evolution $F(\Sigma)$ becomes clear from the following simple considerations. The change in the transformation function resulting from variations defined above can be written in terms of $F(\Sigma)$ as

$$\delta\langle\Sigma_1, a_1|\Sigma_2, a_2\rangle \equiv \langle{\Sigma_1}', a_1'|{\Sigma_2}', a_2'\rangle - \langle\Sigma_1, a_1|\Sigma_2, a_2\rangle. \tag{A251}$$

For the infinitesimal transformations we can write

$$\begin{aligned} |\Sigma', a'\rangle &= U(\Sigma)\ |\Sigma, a\rangle \\ &= [1 + iF(\Sigma)]\ |\Sigma, a\rangle. \end{aligned} \tag{A252}$$

Neglecting terms $\mathcal{O}(F(\Sigma_1)\ F(\Sigma_2))$ the change in the transformation can now be written as

$$\delta\langle\Sigma_1, a_1|\Sigma_2, a_2\rangle = i\langle\Sigma_1, a_1|[F(\Sigma_2) - F(\Sigma_1)]|\Sigma_2, a_2\rangle. \tag{A253}$$

Comparison with (A238) then yields

$$\delta W_{2\to 1} = F(\Sigma_2) - F(\Sigma_1). \tag{A254}$$

The physical significance of this result reads: Action integral operator

$$W_{2\to 1} = -\int_{\Sigma_2}^{\Sigma_1} d^4x\ \mathcal{L}(\Psi(x), \partial_\mu\Psi(x)) \tag{A255}$$

is *unaltered* by the infinitesimal variations *in the interior* of the region bounded by $\Sigma_2$ and $\Sigma_1$. It depends only on the operator $F(\Sigma)$ attached to the boundary surfaces involved.

The equations of motion follow by demanding that $\delta W_{2\to 1}$ vanish and making the identification for $F(\Sigma)$ provided by (A250). The form of the Lagrangian density $\mathcal{L}(\Psi(x), \partial_\mu\Psi(x))$, for the *free matter fields,* is determined by requiring that the Euler Lagrange equations of motion (A249) be identical with the equations of motion obtained for a given representation of the $SU_R(2)\otimes SU_L(2)$ as formulated by Weinberg.

As a general procedure let

$$\Gamma\ \Psi(x) = 0, \tag{A256}$$

with $\Gamma$ an appropriate differential operator, be the equation of motion corresponding to some representation of the $SU_R(2)\otimes SU_L(2)$. Then the Lagrangian density

is

$$\mathcal{L}(\Psi(x), \partial_\mu \Psi(x)) = \overline{\Psi}(x)\ \Gamma\ \Psi(x) + \text{A total divergence.} \tag{A257}$$

The total divergence is so chosen as to keep the Lagrangian density hermitian. The Lagrangian Density Operator for the Dirac Field: As an example for the

$$(\frac{1}{2}, 0) \oplus (0, \frac{1}{2}) \tag{A258}$$

Dirac field we have

$$\Gamma = (i\gamma^\mu \partial_\mu - m). \tag{A259}$$

Introduce

$$\mathcal{L}_0 \equiv \overline{\Psi}(x)\ (i\gamma^\mu \partial_\mu - m)\ \Psi(x), \tag{A260}$$

and note

$$\begin{aligned} \mathcal{L}_0^\dagger &= \Psi^\dagger(x) \left[-i(\gamma^\mu)^\dagger \overleftarrow{\partial}_\mu - m\right] (\overline{\Psi}(x))^\dagger \\ &= \Psi^\dagger(x) \left[-i(\gamma^0)^\dagger \overleftarrow{\partial}_0 - i(\gamma^i)^\dagger \overleftarrow{\partial}_i - m\right] (\Psi^\dagger(x)\beta)^\dagger. \end{aligned} \tag{A261}$$

Since the expressions in the covariant form should be representation independent we work in the chiral representation. Then $\beta = \gamma^0$, $(\gamma^0)^\dagger = \gamma^0$, $(\gamma^i)^\dagger = -\gamma^i$ and $\{\gamma^\mu, \gamma^\nu\} = 2\eta^{\mu\nu}$. These observations, yield

$$\begin{aligned} \mathcal{L}_0^\dagger &= \Psi^\dagger(x) \left[-i\gamma^0 \overleftarrow{\partial}_0 + i\gamma^i \overleftarrow{\partial}_i - m\right] (\gamma^0 \Psi(x)) \\ &= \overline{\Psi}(x) \left[-i\gamma^0 \overleftarrow{\partial}_0 - i\gamma^i \overleftarrow{\partial}_i - m\right] \Psi(x) \\ &= \overline{\Psi}(x) \left[-i\gamma^\mu \overleftarrow{\partial}_\mu - m\right] \Psi(x). \end{aligned} \tag{A262}$$

That $\mathcal{L}_0$ is not hermitian is immediately observed by noting

$$\begin{aligned} \mathcal{L} - \mathcal{L}^\dagger &= i\overline{\Psi}(x)\ \gamma^\mu\ (\partial_\mu \Psi(x)) + i(\partial_\mu \Psi(x))\ \gamma^\mu\ \Psi(x) \\ &= i\partial_\mu \left(\overline{\Psi}(x)\ \gamma^\mu\ \Psi(x)\right). \end{aligned} \tag{A263}$$

Since $(\mathcal{L}_0 - \mathcal{L}_0^\dagger)^\dagger = -(\mathcal{L}_0 - \mathcal{L}_0^\dagger)$ we immediately conclude that $i\partial_\mu \left(\overline{\Psi}(x)\gamma^\mu\Psi(x)\right)$

is antihermitian. That is:

$$\{i\partial_\mu \left(\overline{\Psi}(x)\ \gamma^\mu\ \Psi(x)\right)\}^\dagger = -i\partial_\mu \left(\overline{\Psi}(x)\ \gamma^\mu\ \Psi(x)\right). \qquad \text{(A264)}$$

Now if we introduce

$$\mathcal{L} \equiv \mathcal{L}_0 - \frac{i}{2}\partial_\mu \left(\overline{\Psi}\ \gamma^\mu\ \Psi\right), \qquad \text{(A265)}$$

then using (A264) we get

$$\mathcal{L}^\dagger = \mathcal{L}_0^\dagger + \frac{i}{2}\partial_\mu \left(\overline{\Psi}\ \gamma^\mu\ \Psi\right). \qquad \text{(A266)}$$

Substituting for $\mathcal{L}^\dagger$ from (A263) immediately yields

$$\mathcal{L}^\dagger = \mathcal{L}_0 - \frac{i}{2}\partial_\mu \left(\overline{\Psi}\ \gamma^\mu\ \Psi\right) = \mathcal{L}. \qquad \text{(A267)}$$

Thus we establish that $\mathcal{L}$ defined by (A265), which is of the general form (A257), is the required Poincaré covariant hermitian Lagrangian density operator for the Dirac field. Thus the *hermitian* Dirac Lagrangian (A265) can be written as

$$\begin{aligned}\mathcal{L}_{DIRAC} &= \frac{i}{2}\left[\overline{\Psi}\ \gamma^\mu(\partial_\mu\Psi) - (\partial_\mu\overline{\Psi})\ \gamma^\mu\ \Psi\right] - m\overline{\Psi}\Psi \\ &= \frac{i}{2}\overline{\Psi}\gamma^\mu \overleftrightarrow{\partial}_\mu\Psi - m\overline{\Psi}\Psi.\end{aligned} \qquad \text{(A268)}$$

We end this example of constructing the Lagrangian density operator from a known wave equations by asking: Is

$$\mathcal{L} = \mathcal{L}_0 - \frac{i}{2}(\mathcal{L}_0 - \mathcal{L}_0^\dagger), \qquad \text{(A269)}$$

with

$$\mathcal{L}_0 \equiv \overline{\Psi}(x)\ \Gamma\ \Psi(x), \qquad \text{(A270)}$$

the *general* expression for the hermitian Lagrangian density operator?

*Functional variation* (also called *local variation* by Roman [51] ) in $\Psi(x)$ is defined by

$$\Delta\Psi(x) \equiv \Psi'(x') - \Psi(x). \tag{A271}$$

The points $x'^{\mu}$ and $x^{\mu}$ refer to the *same* geometrical point in different frames. The *total* $\Psi(x)$ is defined as :

$$\delta\Psi(x) \equiv \Psi'(x) - \Psi(x). \tag{A272}$$

A slightly rewritten passage from Ref. [51] makes the definitions more clear: "$\Psi'(x)$ differs from $\Psi(x)$ because of two reasons. First the field is described in a new frame so that $x$ in the argument of $\Psi'(x)$ is *not* the same geometrical point as the argument in of $\Psi(x)$, but is rather the point that in the *primed frame* has the same numerical values for the coordinates as had the point of definition of $\Psi(x)$ in the unprimed frame. Secondly, we also envisage a relabeling of the field components, usually a linear mixing of $\psi_\alpha(x)$ among themselves." With the help of a Taylor expansion we get

$$\begin{aligned} \Delta\Psi(x) &= \Psi'(x) + \partial_\nu \Psi'(x)\delta x^\nu - \Psi(x) \\ &= \delta\Psi(x) + \partial_\nu \Psi'(x)\delta x^\nu. \end{aligned} \tag{A273}$$

In terms of the variations $\Delta\Psi(x)$ and $\delta x^\mu$ the generator of evolution, given by (A250), becomes

$$\begin{aligned} F(\Sigma) &= \int\limits_\Sigma d\Sigma_\mu \left[ \frac{\partial \mathcal{L}}{\partial(\partial_\mu \Psi)} \left[\Delta\Psi(x) - (\partial_\nu \Psi(x))\delta x^\nu\right] + \mathcal{L}\delta x^\mu \right] \\ &= \int\limits_\Sigma d\Sigma_\mu \left[ \frac{\partial \mathcal{L}}{\partial(\partial_\mu \Psi)} \Delta\Psi(x) - \left( \frac{\partial \mathcal{L}}{\partial(\partial_\mu \Psi)} (\partial_\nu \Psi) - \delta^\mu{}_\nu \mathcal{L} \right) \delta x^\nu \right]. \end{aligned} \tag{A274}$$

On introducing the four vector

$$\Pi^\mu(x) = \frac{\partial \mathcal{L}}{\partial(\partial_\mu \Psi)}, \tag{A275}$$

the generator of evolution can be written as

$$F(\Sigma) = \int_{\Sigma} d\Sigma_\mu \Big[\Pi^\mu(x)\Delta\Psi(x) - \big[\Pi^\mu \partial_\nu \Psi(x) - \delta^\mu{}_\nu \mathcal{L}\big]\delta x^\nu\Big]. \tag{A276}$$

The infinitesimal variations in general operators are generated by the unitary operator

$$U = 1 + iF, \quad U^{-1} = 1 - iF. \tag{A277}$$

So if $A$ is an operator under consideration, the change induced by the transformation (A277) is

$$\begin{aligned} \delta A &\equiv UAU^1 - A \\ &= (1 + iF)A(1 - iF) - A. \end{aligned} \tag{A278}$$

To order $\mathcal{O}(F)$

$$\delta A = i[F, A]. \tag{A279}$$

The Rules of Quantisation or the Causality Conditions are obtained by considering transformations for which $\delta x^\nu = 0$. Then $\delta\Psi(x) = \Delta\Psi(x)$, and the generator of evolution given by (A276) takes the form

$$F(\Sigma) = \int_{\Sigma} d\Sigma_\mu \; \Pi^\mu_\alpha \; \delta\Psi_\alpha(x), \tag{A280}$$

where we have explicitly introduced the multicomponent indices, $\alpha$, for $\Psi(x)$. These are the four spinorial indices if we are considering the Dirac field, for example. Equation (A279) then yields

$$\delta\Psi_\beta(x) = i \left[\int_{\Sigma} d\Sigma_\mu(x') \; \Pi^\mu_\alpha(x') \; \delta\Psi_\alpha(x'), \; \Psi_\beta(x)\right] \tag{A281}$$

$$\delta\Pi^\nu_\beta(x) = i \left[\int_{\Sigma} d\Sigma_\mu(x') \; \Pi^\mu_\alpha(x') \; \delta\Psi_\alpha(x'), \; \Pi^\nu_\beta(x)\right] \tag{A282}$$

for all $x \in \Sigma$. The simplest spacelike surface $\Sigma$ is a constant-time surface $x'^0 =$

$x^0 = t$. For the constant– time surfaces the above equations read

$$\delta\Psi_\beta(t,\vec{x}) = i\int\Big[\Pi^0_\alpha(t,\vec{x}\ ')\ \delta\Psi_\alpha(t,\vec{x}\ '),\ \Psi_\beta(t,\vec{x})\Big]\ d^3\vec{x}\ ' \tag{A283}$$

$$\delta\Pi^\nu_\beta(t,\vec{x}) = i\int\Big[\Pi^0_\alpha(t,\vec{x}\ ')\ \delta\Psi_\alpha(t,\vec{x}\ '),\ \Pi^\nu_\beta(t,\vec{x})\Big]\ d^3\vec{x}\ '. \tag{A284}$$

We will now see that the *simplest* solution of these integral equations are the equal–time causality conditions. Towards this end let's note the following identities

$$\Big[AB,C\Big] = \Big[A,C\Big]B + A\Big[B,C\Big] \tag{A285}$$

$$\Big[AB,C\Big] = A\Big\{C,B\Big\} - \Big\{C,A\Big\}B. \tag{A286}$$

<u>Solution 1, The Bosonic Solution:</u> Using (A285) in (A283) we have

$$\begin{aligned}\delta\Psi_\beta(t,\vec{x}) = i\int\Bigg(&\Big[\Pi^0_\alpha(t,\vec{x}\ '),\ \Psi_\beta(t,\vec{x})\Big]\delta\Psi_\alpha(t,\vec{x}\ ')\\ &+\Pi^0_\alpha(t,\vec{x}\ ')\Big[\delta\Psi_\alpha(t,\vec{x}\ '),\ \Psi_\beta(t,\vec{x})\Big]\Bigg)d^3\vec{x}\ '.\end{aligned} \tag{A287}$$

The *simplest* solution is obtained by setting

$$\Big[\Pi^0_\alpha(t,\vec{x}\ '),\ \Psi_\beta(t,\vec{x})\Big] = -i\ \delta_{\alpha\beta}\ \delta(\vec{x}-\vec{x}\ ') \tag{A288}$$

$$\Big[\delta\Psi_\alpha(t,\vec{x}\ '),\ \Psi_\beta(t,\vec{x})\Big] = 0. \tag{A289}$$

But since $\alpha$ and $\beta$ do not refer to any specific components of the multicomponent field operators, (A289) implies

$$\Big[\delta\Psi_\alpha(t,\vec{x}\,'),\ \Psi_\beta(t,\vec{x})\Big] + \Big[\Psi_\alpha(t,\vec{x}\,'),\ \delta\Psi_\beta(t,\vec{x})\Big] = 0, \tag{A290}$$

or equivalently

$$\delta\Big[\Psi_\alpha(t,\vec{x}\,'),\ \Psi_\beta(t,\vec{x})\Big] = 0. \tag{A291}$$

Since this result holds for arbitrary variations, we conclude

$$\Big[\Psi_\alpha(t,\vec{x}\,'),\ \Psi_\beta(t,\vec{x})\Big] = 0. \tag{A292}$$

In order to obtain the commutator $[\Pi^0_\alpha(t,\vec{x}\,'),\ \Pi^0_\beta(t,\vec{x})]$ from (A284) we note that adding a total divergence $\partial_\nu f^\nu(\Psi(x), \partial_\mu\Psi(x))$ to the Lagrangian density operator leaves the equations of motion unaltered while changes the generator of evolution $F(\Sigma)$ to $F(\Sigma) + \int_\Sigma d\Sigma_\mu\ \delta f^\mu$. Choosing

$$f^\mu(\Psi(x), \partial_\mu\Psi(x)) = -\Pi^\mu_\alpha(x)\ \Psi_\alpha(x), \tag{A293}$$

equation (A282) [with $F \to F + f^\mu$] can be rewritten as

$$\begin{aligned}\delta\Pi^\nu_\beta(x) &= i\Big[\int_\Sigma d\Sigma_\mu(x')\ \Big(\Pi^\mu_\alpha(x')\ \delta\Psi_\alpha(x') - \delta\Pi^\mu_\alpha(x')\ \Psi_\alpha(x') \\ &\qquad\qquad - \Pi^\mu_\alpha(x')\ \delta\Psi_\alpha(x')\Big),\ \Pi^\nu_\beta(x)\Big] \\ &= -i\Big[\int_\Sigma d\Sigma_\mu(x')\ \delta\Pi^\mu_\alpha(x')\ \Psi_\alpha(x'),\ \Pi^\nu_\beta(x)\Big].\end{aligned} \tag{A294}$$

As before we choose $\Sigma$ to be a constant time surface $x^{0\prime} = x^0 = t$. With this choice of $\Sigma$ we have

$$\delta\Pi^\nu_\beta(x) = -i\int\Big[\delta\Pi^0_\alpha(t,\vec{x}\,')\ \Psi_\alpha(t,\vec{x}\,'),\ \Pi^\nu_\beta(t,\vec{x})\Big]\ d^3\vec{x}\,'. \tag{A295}$$

In this fashion we have brought (A284) to a more useful form. Setting $\nu = 0$ we

have

$$\delta\Pi^0_\beta(x) = -i\int\bigg(\Big[\delta\Pi^0_\alpha(t,\vec{x}\,'),\ \Pi^0_\beta(t,\vec{x})\Big]\Psi_\alpha(t,\vec{x}\,') + \delta\Pi^0_\alpha(t,\vec{x}\,')\Big[\Psi_\alpha(t,\vec{x}\,'),\ \Pi^0_\beta(t,\vec{x})\Big]\bigg)\ d^3\vec{x}\,'. \qquad \text{(A296)}$$

Using (A288) in the second term, and repeating steps similar to the ones used in obtaining (A291) , we get the additional commutation relation

$$\Big[\Pi^0_\alpha(t,\vec{x}\,'),\ \Pi^0_\beta(t,\vec{x})\Big] = 0. \qquad \text{(A297)}$$

Therefore for constant time spacelike surfaces, $\Sigma$, the *simplest* set of commutation relation which solve the integral equations (A283) and (A284) is

$$\Big[\Psi_\beta(t,\vec{x}),\ \Pi^0_\alpha(t,\vec{x}\,')\Big] = i\ \delta_{\alpha\beta}\ \delta(\vec{x}-\vec{x}\,') \qquad \text{(A298)}$$

$$\Big[\Psi_\alpha(t,\vec{x}\,'),\ \Psi_\beta(t,\vec{x})\Big] = 0. \qquad \text{(A299)}$$

$$\Big[\Pi^0_\alpha(t,\vec{x}\,'),\ \Pi^0_\beta(t,\vec{x})\Big] = 0. \qquad \text{(A300)}$$

For our purposes we simply note that Bosons are described by matter field operators which satisfy these commutation relations. That this is the case can be verified by considering specific cases. The simplest such exercise can be carried out with a scalar field (as, for example, in Ref. [53, Sec. 4.1).

<u>Solution 2, The Fermionic Solution:</u> Using (A286) in (A283) we have

$$\delta\Psi_\beta(t,\vec{x}) = i\int\bigg(\Pi^0_\alpha(t,\vec{x}\,')\Big\{\Psi_\beta(t,\vec{x}),\ \delta\Psi_\alpha(t,\vec{x}\,')\Big\} - \Big\{\Psi_\beta(t,\vec{x}),\ \Pi^0_\alpha(t,\vec{x}\,')\Big\}\delta\Psi_\alpha(t,\vec{x}\,')\bigg)d^3\vec{x}\,'. \qquad \text{(A301)}$$

The *simplest* solution is obtained by setting

$$\Big\{\Psi_\beta(t,\vec{x}),\ \Pi^0_\alpha(t,\vec{x}\,')\Big\} = i\ \delta_{\alpha\beta}\ \delta(\vec{x}-\vec{x}\,') \qquad \text{(A302)}$$

$$\Big\{\Psi_\alpha(t,\vec{x}\,'),\ \Psi_\beta(t,\vec{x})\Big\} = 0. \tag{A303}$$

Similarly (A284) yields

$$\Big\{\Pi^0_\alpha(t,\vec{x}\,'),\ \Pi^0_\beta(t,\vec{x})\Big\} = 0. \tag{A304}$$

Without proof we note, as above, that Fermions are described by matter field operators which satisfy these anticommutation relations. That this is the case can be verified by considering specific cases. The simplest such exercise has already been carried out in the last section for the spin-$\frac{1}{2}$ fermions.

As just noted, earlier in this chapter we established that solution 2 is satisfied, in particular, by the matter field operators associated with the Dirac field: $[(1/2,0)\oplus(0,1/2)]$. The real (or complex) scalar field is a specific example of solution 1. These are specific examples of a more general theorem, called the *Spin Statistics Theorem*, which argues that particles with half integral spins, called *fermions*, are associated with the solution 2; and the integral spin particles, called *bosons*, are associated with the solution 1. However, we must explicitly note with some emphasis that these solutions, obtained here, are the *simplest* solutions of the integral equations (A283) and (A284). There seems to be no reason, a priori, to rule out the possibility of other solutions consistent with the basic interpretational scheme of Quantum Mechanics.

## A11 Conservation Laws and Time Evolution

Under the infinitesimal variations

$$x^\mu \rightarrow x'^\mu = x^\mu + \delta x^\mu \tag{A305}$$

$$\Psi(x) \rightarrow \Psi'(x') = \Psi(x) + \Delta\Psi(x), \tag{A306}$$

with $\Delta\Psi(x)$ defined by (A273), the physical states transform as

$$|\Sigma', a'\rangle = U(\Sigma)\ |\Sigma, a\rangle = [1 + iF(\Sigma)]\ |\Sigma, a\rangle. \tag{A307}$$

Using (A279) the change induced in the multicomponent matter field operators

can be written as

$$\delta\Psi_\alpha(x) = i[F(\Sigma), \Psi_\alpha(x)]. \tag{A308}$$

The generator of these changes is

$$F(\Sigma) = \int\limits_{\Sigma} d\Sigma_\mu \left[\Pi^\mu_\alpha(x)\ \Delta\Psi_\alpha(x) - \Big[\Pi^\mu_\alpha\ \partial_\nu\Psi_\alpha(x) - \delta^\mu{}_\nu\ \mathcal{L}\Big]\delta x^\nu\right]. \tag{A309}$$

The principle of the stationary action operator, introduced in the last chapter, and the observation (see (A254))

$$\delta W_{2\to 1} = F(\Sigma_2) - F(\Sigma_1), \tag{A310}$$

together imply $F(\Sigma_2) = F(\Sigma_1)$, that is we have a constant of motion. Symbolically,

$$\frac{\delta F(\Sigma)}{\delta\Sigma} = 0, \tag{A311}$$

where $x$ is an arbitrary point on the spacelike surface $\Sigma$. For $\Sigma =$ a constant–time surface the invariance of the action operator $[\delta W = 0]$ for every variation which can be expressed as (1) and (2) yields a conservation law

$$\frac{d}{dt}F(t) = 0. \tag{A312}$$

We thus see that when one considers evolution of a system from one constant–time surface to another the invariance of the action operator under (A305) and (A306) yields the conservation laws expressed by (A312). However, the conservation laws expressed by (A312) may *not* be the totality of the conservation laws associated with a particular system. These extra conservation laws appear, for example, when one demands invariance under local phase transformations. Some of these extra conservation laws are actually associated with a modification in the "free" matter field equations, and as such extend the class of variations given by (A305) and (A306) which do not modify the free matter field equations of motion.

The simplest conservation law is obtained by requiring the invariance of the action operator under the four translations

$$x^\mu \to x'^\mu = x^\mu + a^\mu \tag{A313}$$

with $a^\mu$ a *real* constant. Consequently, in accord with definition (A272):

$$\begin{aligned}\delta\Psi_\alpha(x) &= \Psi'_\alpha(x) - \Psi_\alpha(x)\\ &= \Psi_\alpha(x-a) - \Psi_\alpha(x)\\ &= -\partial_\nu\Psi_\alpha(x)\ a^\nu\end{aligned} \tag{A314}$$

In the above we have used the fact that only the change is that of having gone to a new frame without the introduction of any additional physical degrees of freedom. Substituting (A314) in (A273) we have the result:

$$\Delta\Psi_\alpha(x) = 0. \tag{A315}$$

As a result, using (A309), the generator of the infinitesimal translations defined by (A313) is found to be

$$F(\Sigma) = -a^\nu \int\limits_\Sigma d\Sigma_\mu \Big[\Pi^\mu_\alpha\ \partial_\nu\Psi_\alpha(x) - \delta^\mu{}_\nu\ \mathcal{L}\Big]. \tag{A316}$$

Introducing the *energy momentum tensor (density) operator* $\theta^\mu{}_\nu(x)$

$$\theta^\mu{}_\nu(x) = \Pi^\mu_\alpha\ \partial_\nu\Psi_\alpha(x) - \delta^\mu{}_\nu\ \mathcal{L} \tag{A317}$$

we have

$$F(\Sigma) = -a^\nu \int\limits_\Sigma d\Sigma_\mu\ \theta^\mu{}_\nu. \tag{A318}$$

Demanding the action operator to be stationary under (A313) then translates into the expression

$$0 = \frac{\delta\ F(\Sigma)}{\delta\Sigma(x)} \equiv -a^\nu \left[\lim_{\Delta\to 0}\left\{\frac{\int_{\Sigma'} d\Sigma_\mu\ \theta^\mu{}_\nu - \int_\Sigma d\Sigma_\mu\ \theta^\mu{}_\nu}{\Delta}\right\}\right] \tag{A319}$$

[Here $\Delta$ = volume between $\Sigma'$ and $\Sigma$] which on using the Gauss' theorem with the assumption that $\theta^\mu{}_\nu(x)$ vanishes at the boundary surface at infinity connecting the boundaries of $\Sigma'$ and $\Sigma$, yields

$$0 = \frac{\delta\ F(\Sigma)}{\delta\Sigma} = -a^{\nu}\ \partial_{\mu}\theta^{\mu}{}_{\nu}. \tag{A320}$$

The statement that a certain operator vanishes at the surface at infinity means that we confine ourselves only to those states for which the expectation value of the operator vanishes at infinity. If one has physical states which do not satisfy this requirement then one must review the derivation of all equations of motion and conservation laws *ab initio.*

Now since the infinitesimal translation $a^{\nu}$ is arbitrary, we have the conservation law

$$\partial_{\mu}\theta^{\mu}{}_{\nu} = 0. \tag{A321}$$

Or, equivalently

$$\partial_{\mu}\theta^{\mu\nu} = 0, \quad (\nu = 0, 1, 2, 3). \tag{A322}$$

For well known reasons (see, for example, [53, p.91]) it is customary to introduce the *canonical energy momentum tensor (density) operator* :

$$T^{\mu\nu} = \theta^{\mu\nu} + \partial_{\lambda} f^{\lambda\mu\nu} \tag{A323}$$

with $f^{\lambda\mu\nu} = -f^{\mu\lambda\nu}$, so that

$$\partial_{\mu}\partial_{\lambda} f^{\lambda\mu\nu} = 0. \tag{A324}$$

The $f^{\lambda\mu\nu}$ is so chosen as to make $T^{\mu\nu}$ *symmetric.*

*Definitions:* The 00 component of $T^{\mu\nu}$

$$T^{00} \equiv \mathcal{H}(x), \tag{A325}$$

is called the energy density operator of the matter field in the region surrounding the point $x$. The energy momentum (operator) four vector of the matter field is

defined by

$$P^{\nu} = \int_{\Sigma} d\Sigma_{\mu}(x)\ T^{\mu\nu}(x), \quad (\nu = 0, 1, 2, 3). \tag{A326}$$

If we choose $\Sigma$ to be a constant–time surface, then

$$P^{\nu} = \int T^{0\nu}\ d^3x, \tag{A327}$$

The $0th$ component of $P^{\nu}$

$$P^0 = \int T^{00}\ d^3x = \int \mathcal{H}(x)\ d^3x \equiv H. \tag{A328}$$

is the total matter field energy operator, or the *Hamiltonian* of the system under consideration. The associated conservation law reads

$$\frac{d}{dt} P^{\nu} = 0. \qquad \text{(A329))}$$

Similarly by demanding invariance of the action operator under spacial rotations yields the *angular momentum operator* [see, for example: Ref. [51, pp. 71-73], Ref. [53, pp. 91-92]

$$M^{\mu\nu} = \int (T^{0\mu} x^{\nu} - T^{0\nu} x^{\mu}) d^3x \tag{A330}$$

$$\frac{d}{dt} M^{\mu\nu} = 0. \tag{A331}$$

It should be noted [see, for example, Ref. [53, p.91] that even though the canonical energy momentum tensor (density) is not unique the energy and momentum in the field are.

We now obtain a fundamental equation for the evolution of the matter field operators. This equation when used in conjunction with the canonical commutation [or anticommutation] relations is seen to be equivalent to the Euler Lagrange equations of motion.

Exploiting the freedom given by (A323) requires that the generator of infinitesimal translations given by (A318) be replaced by

$$F(\Sigma) = -a^{\nu} \int\limits_{\Sigma} d\Sigma_{\mu}\; T^{\mu}{}_{\nu} = -a_{\nu} \int\limits_{\Sigma} d\Sigma_{\mu}\; T^{\mu\nu} = -a_{\nu} P^{\nu}. \tag{A332}$$

This generator induces the following [see (A314)] change in the multicomponent matter field operators

$$\delta \Psi_{\alpha}(x) = -\partial_{\nu} \Psi_{\alpha}(x)\; a^{\nu}. \tag{A333}$$

Equations (A332) and (A333) coupled with the general results of the last section imply that

$$\begin{aligned} -\partial_{\nu} \Psi_{\alpha}(x)\; a^{\nu} &= i[F(\Sigma), \Psi_{\alpha}(x)] \\ &= -ia^{\nu}[P_{\nu}, \Psi_{\alpha}(x)]. \end{aligned} \tag{A334}$$

Since $a^{\nu}$ is arbitrary

$$\partial_{\nu} \Psi_{\alpha}(x) = i[P_{\nu}, \Psi_{\alpha}(x)]. \tag{A335}$$

Or, for a general operator $\mathcal{O}(x)$

$$\partial_{\nu} \mathcal{O}(x) = i[P_{\nu}, \mathcal{O}(x)]. \tag{A336}$$

Setting $\nu = 0$ in (A335) and using (A328) yields the well known *Heisenberg Equation of Motion*

$$\frac{d}{dt} \Psi_{\alpha}(t, \vec{x}) = i[H, \Psi_{\alpha}(t, \vec{x})]. \tag{A337}$$

## A12 Heisenberg, Schrödinger and Dirac/Interaction Pictures

Dropping the multicomponent indices in Eq. (A337) the Heisenberg equation of motion reads

$$\frac{d}{dt}\Psi(t,\vec{x}) = i[H, \Psi(t,\vec{x})]. \tag{A338}$$

It has the solution

$$\Psi(t,\vec{x}) = e^{iHt}\ \Psi(0,\vec{x})\ e^{-iHt} \tag{A339}$$

as can be verified by direct substitution of (A339) in (A338). Under the evolution we have$|\Sigma', a'\rangle = [1 + F(\Sigma)]\ |\Sigma, a\rangle$. Since $F(\Sigma)$ is a constant of motion, the state vectors $|\Sigma, a\rangle \equiv |a\rangle$ are time independent:

$$\frac{\partial}{\partial t}|a\rangle = 0. \tag{A340}$$

This description of quantum systems is called the *Heisenberg picture.*

An equivalent description , called the *Schrödinger picture,* is defined by means of the unitary transformation

$$|a,t\rangle^{(S)} = e^{-iHt}\ |a\rangle \tag{A341}$$

$$\Psi^{(S)} = e^{-iHt}\ \Psi(t,\vec{x})\ e^{iHt} \tag{A342}$$

with the following identifications

$$\Psi^{(S)}(\vec{x}) = \Psi(0,\vec{x}) \tag{A343}$$

$$H^{(S)} = H(t=0) = H. \tag{A344}$$

As a result *all* time dependence is now contained in the state vectors $|a,t\rangle^{(S)}$. The matter field operators and the Hamiltonian are independent of time. Operating

(A341) from left by

$$i\frac{\partial}{\partial t} \tag{A345}$$

and using the fact that the Heisenberg state vectors are time independent [see (A340)] we obtain the equation of motion for the Schrödinger state vector

$$i\frac{\partial}{\partial t}|a,t\rangle^{(S)} = H|a,t\rangle^{(S)}. \tag{A346}$$

Intermediate between these two pictures is the *Dirac* or the *Interaction picture.* One starts with the decomposition of the hamiltonian

$$H^{(I)} = H_0^{(I)} + H_{int.}^{(I)}, \tag{A347}$$

where

a) $H_0^{(I)}$ is the time independent *free* field hamiltonian: $H_0^{(I)} = H_0(t=0)$, and

b) $H_{int.}^{(I)}$ is the time dependent *interaction* hamiltonian assumed to vanish at $t = \pm\infty$.

With this decomposition the time evolution is shared partly by the matter field operators (or any other operator) and partly by the state vectors through the following *definitions:*

$$|a,t\rangle^{(I)} = \exp(iH_0^{(I)}t)\ |a,t\rangle^{(S)} \tag{A348}$$

$$\Psi^{(I)}(t,\vec{x}) = \exp(iH_0^{(I)}t)\ \Psi^{(S)}(\vec{x})\ \exp(-iH_0^{(I)}t). \tag{A349}$$

Combining (A349) with (A342), and (A348)with (A341)yields the relation between the Interaction and Heisenberg pictures:

$$|a,t\rangle^{(I)} = \exp(iH_0t)\exp(-iHt)\ |a\rangle. \tag{A350}$$

$$\begin{aligned}\Psi^{(I)}(t,\vec{x}) &= \exp(iH_0^{(I)}t)\exp(-iHt)\ \Psi(t,\vec{x})\ \exp(iHt)\exp(-iH_0^{(I)}t)\\ &= \exp(iH_0^{(I)}t)\ \Psi^{(S)}(\vec{x})\ \exp(-iH_0^{(I)}t)\end{aligned} \tag{A351}$$

At $t = 0$ , the Heisenberg, Schrödinger and the Interaction pictures all coincide. Unless $H_0^{(I)}$ and $H$ both commute with the commutator $[H_0^{(I)}, H]$ the exponentials in the above expressions *cannot* be combined into one term like $\exp[i(H_0^{(I)} - H)t]$.

To see the physical motivation for the definitions of the Interaction picture (A348) and (A349) let's look at the equation of motion for the state vectors in the Interaction picture:

$$\begin{aligned}\frac{\partial}{\partial t}|a,t\rangle^{(I)} &= iH_0^{(I)}\exp(iH_0^{(I)}t)\exp(-iHt)|a\rangle - \exp(iH_0^{(I)}t)\ iH\ \exp(-iHt)|a\rangle \\ &= iH_0^{(I)}|a,t\rangle^{(I)} - \exp(iH_0^{(I)}t)\ iH\ \exp(-iH_0^{(I)}t)\exp(iH_0^{(I)}t)\ \exp(-iHt)|a\rangle \\ &= iH_0^{(I)}|a,t\rangle^{(I)} - i\exp(iH_0^{(I)}t)\ H\ \exp(-iH_0^{(I)}t)|a,t\rangle^{(I)}.\end{aligned} \tag{A352}$$

Now note that

$$H^{(I)} \equiv \exp(iH_0^{(I)}t)\ H\ \exp(-iH_0^{(I)}t), \tag{A353}$$

therefore the time evolution of a state vector in the Interaction picture is

$$\frac{\partial}{\partial t}|a,t\rangle^{(I)} = iH_0^{(I)}|a,t\rangle^{(I)} - iH^{(I)}|a,t\rangle^{(I)}. \tag{A354}$$

But since $H_{int.}^{(I)} \equiv H^{(I)} - H_0^{(I)} = H - H_0^{(I)}$ we have

$$i\frac{\partial}{\partial t}|a,t\rangle^{(I)} = H_{int.}^{(I)}|a,t\rangle^{(I)}. \qquad \text{(A355))}$$

Similarly from (A351), keeping in mind that the Schrödinger matter field operator is time independent, we obtain

$$\frac{d}{dt}\Psi^{(I)}(t,\vec{x}) = iH_0^{(I)}\ \Psi^{(I)}(t,\vec{x}) - i\Psi^{(I)}(t,\vec{x})\ H_0^{(I)} \tag{A356}$$

$$\frac{d}{dt}\Psi^{(I)}(t,\vec{x}) = i[H_0^{(I)},\Psi^{(I)}(t,\vec{x})]. \tag{A357}$$

We thus arrive at the following physical interpretation for the Interaction picture:

a) The Interaction picture state vector $|a,t\rangle^{(I)}$ is completely determined by the interaction hamiltonian $H_{int.}^{(I)}(t)$.

b) The time evolution of the field operators depends entirely on the free field hamiltonian $H_0^{(I)}$, the part of the hamiltonian which has no time dependence.

Further since the interaction–picture

$$\Psi^{(I)}(t, \vec{x}) \tag{A358}$$

and

$$\Pi^{0(I)}(t, \vec{x}) \tag{A359}$$

are related through a unitary transformation to the Heisenberg–picture $\Psi(t, \vec{x})$ and $\Pi^0(t, \vec{x})$, the interaction–picture $\Psi$ and $\Pi^0$ obey the same algebra as the Heisenberg picture $\Psi$ and $\Pi^0$. This algebra has already been presented for the Heisenberg–picture matter field operators in the previous appendix.

## A13 U Matrix

We now concentrate on the solution of the equation (A355), which provides us with the evolution of the interaction–picture state vector. According to the fundamental linear structure of quantum mechanics the interaction–picture state vector at time $t_1$ must be related to the state vector at time $t_0$ through a $H_{int.}$ dependent unitary matrix (We will re-establish this below. Also note that we are dropping the superscript $(I)$, crowning the interaction picture objects, for the rest of discussion. An exception to this simplification in notation will be made whenever a confusion is likely.)

$$|a, t_1\rangle = U(t_1, t_0)|a, t_0\rangle. \tag{A360}$$

Setting $t_1 = t_0$ yields the obvious property of the $U$ matrix

$$U(t_0, t_0) = 1. \tag{A361}$$

In order that $U(t, t_0)$ followed by $U(t_1, t)$ results in the same evolution as $U(t_1, t_0)$,

$U$ should satisfy the property

$$U(t_1, t)\ U(t, t_0) = U(t_1, t_0). \tag{A362}$$

Taking $t_1$ equal to $t_0$ we infer that

$$U(t_0, t)\ U(t, t_0) = U(t_0, t_0) = 1. \tag{A363}$$

This implies that

$$U(t, t_0)^{-1} = U(t_0, t). \tag{A364}$$

In addition to these constraints which the U matrix must satisfy we now establish that $U(t, t_0)$ is a unitary operator. Towards this end substitute for $|a, t\rangle$ from (A360) into (A355), and taking care that it is $|a, t\rangle$ and not $|a, t_0\rangle$ which depends on the running $t$, we readily obtain the equation satisfied by the U matrix:

$$i\frac{\partial}{\partial t}U(t, t_0) = H_{int.}(t)\ U(t, t_0). \tag{A365}$$

Next take the hermitian conjugate of this equation (and restricting ourselves to hermitian interaction hamiltonians $H_{int.}$)

$$-i\frac{\partial}{\partial t}U(t, t_0)^\dagger = U(t, t_0)^\dagger\ H_{int.}(t). \tag{A366}$$

and use (A365) and (A366) to evaluate

$$\frac{d}{dt}\Big[U(t, t_0)^\dagger\ U(t, t_0)\Big] = 0. \tag{A367}$$

That is $U(t, t_0)^\dagger\ U(t, t_0)$ is a *constant* matrix. Eq. (A363) implies

$$U(t_0, t_0)^\dagger = 1. \tag{A368}$$

Therefore, $U(t_0, t_0)^\dagger\ U(t_0, t_0) = 1$. We thus have the general result

$$U(t, t_0)^\dagger\ U(t, t_0) = 1, \qquad \forall\, t. \tag{A369}$$

Finally multiplying this equation from the right by $U(t, t_0)^{-1}$, we get

$$U(t, t_0)^\dagger = U(t, t_0)^{-1}, \tag{A370}$$

thus establishing the unitarity of $U(t, t_0)$.

It may also be noted parenthetically that since (See (A350) and (A360))

$$|a,t\rangle^{(I)} = \exp(iH_0^{(I)}t)\ \exp(-iHt)|a\rangle \tag{A371}$$

$$|a,t\rangle^{(I)} = U(t,t_o)|a,t_0\rangle^{(I)}, \tag{A372}$$

and $|a,t_0\rangle$ for $t_0 = 0$ coincides with the Heisenberg ket $|a\rangle$, we have

$$U(t,0) = \exp(iH_0^{(I)}t)\ \exp(-iHt). \tag{A373}$$

As result, referring to the first line in (A351), the relation between matter field operators in the Interaction picture and and the Heisenberg picture may be written as

$$\Psi^{(I)}(t,\vec{x}) = U(t,0)\ \Psi(t,\vec{x})\ U(t,0)^{-1} \tag{A374}$$

Notice that we here reintroduced appropriate superscripts designating a picture.

The differential equation

$$i\frac{\partial}{\partial t}U(t,t_0) = H_{int.}(t)\ U(t,t_0). \tag{A375}$$

with the boundary condition

$$U(t_0,t_0) = 1, \qquad \text{(A376))}$$

has the formal solution

$$U(t,t_0) = 1 - i\int_{t_0}^{t} dt_1\ H_{int.}(t_1)\ U(t_1,t_0), \quad t \geq t_0. \tag{A377}$$

Equation (A377) gives

$$U(t_1,t_0) = 1 - i\int_{t_0}^{t_1} dt_2\ H_{int.}(t_2)\ U(t_2,t_0), \quad t_1 \geq t_2 \geq t_0. \tag{A378}$$

Substituting (A378) on the $rhs$ of (A377) yields

$$U(t,t_0) = 1 - i\int_{t_0}^{t} dt_1\ H_{int.}(t_1)\left\{1 - i\int_{t_0}^{t_1} dt_2\ H_{int.}(t_2)\ U(t_2,t_0)\right\}$$
$$= 1 - i\int_{t_0}^{t} dt_1\ H_{int.}(t_1) + (-i)^2\int_{t_0}^{t} dt_1 \int_{t_0}^{t_1} dt_2\ H_{int.}(t_1)H_{int.}(t_2)U(t_2,t_0). \tag{A379}$$

Continuing this iterative procedure we obtain the perturbative expansion

$$\begin{aligned} U(t,t_0) &= 1 - i\int_{t_0}^{t} dt_1\ H_{int.}(t_1) \\ &+ (-i)^2\int_{t_0}^{t} dt_1 \int_{t_0}^{t_1} dt_2\ H_{int.}(t_1)H_{int.}(t_2) \\ &+ \ldots + (-i)^n\int_{t_0}^{t} dt_1 \int_{t_0}^{t_1} dt_2 \ldots \int_{t_0}^{t_{n-1}} dt_n\ H_{int.}(t_1)\ldots H_{int.}(t_n) + \ldots . \end{aligned} \tag{A380}$$

The next we discuss the standard trick of making all the upper limits on the integrals in (A380) identical. Before we proceed with this somewhat lengthy exercise let's note that this section is largely based on Sec. 6.1 of Ref. [4]. To begin note that relabelling: $t_1 \leftrightarrow t_2$ gives

$$\int_{t_0}^{t} dt_1 \int_{t_0}^{t_1} dt_2\ H_{int.}(t_1)H_{int.}(t_2) = \int_{t_0}^{t} dt_2 \int_{t_0}^{t_2} dt_1\ H_{int.}(t_2)H_{int.}(t_1). \tag{A381}$$

Next step is the observation

$$\int_{t_0}^{t} dt_2 \int_{t_0}^{t_2} dt_1 = \int_{t_0}^{t} dt_2 \int_{t_0}^{t} dt_1\ \theta(t_2 - t_1), \tag{A382}$$

because

$$\int_{t_0}^{t} dt_2 \int_{t_0}^{t} dt_1 \; \theta(t_2 - t_1) = \int_{t_0}^{t} dt_2 \int_{t_0}^{t_2} dt_1 \; \theta(t_2 - t_1) + \int_{t_0}^{t} dt_2 \int_{t_2}^{t} dt_1 \; \theta(t_2 - t_1). \quad \text{(A383)}$$

For the first term on the *rhs* $\theta(t_2 - t_1)$ equals unity because $t_2 > t_1$ while for the second term $\theta(t_2 - t_1)$ vanishes because $t_2 < t_1$. Now we invert the order, *not* relabel as before, of integration on the *rhs* of (A382)

$$\begin{aligned} \int_{t_0}^{t} dt_2 \int_{t_0}^{t_2} dt_1 &= \int_{t_0}^{t} dt_1 \int_{t_0}^{t} dt_2 \; \theta(t_2 - t_1) \\ &= \int_{t_0}^{t} dt_1 \left\{ \int_{t_0}^{t_1} dt_2 \; \theta(t_2 - t_1) + \int_{t_1}^{t} dt_2 \; \theta(t_2 - t_1) \right\}. \end{aligned} \quad \text{(A384)}$$

For the first term on the *rhs* $t_2 < t_1$, and hence $\theta(t_2 - t_1) = 0$. On the other hand $t_2 > t_1$ for the second term, and $\theta(t_2 - t_1) = 1$. As a result we establish:

$$\int_{t_0}^{t} dt_2 \int_{t_0}^{t_2} dt_1 = \int_{t_0}^{t} dt_1 \int_{t_1}^{t} dt_2. \quad \text{(A385)}$$

Using the result just established in (A381)then gets us a little closer to our final goal

$$\int_{t_0}^{t} dt_1 \int_{t_0}^{t_1} dt_2 \; H_{int.}(t_1) H_{int.}(t_2) = \int_{t_0}^{t} dt_1 \int_{t_1}^{t} dt_2 \; H_{int.}(t_2) H_{int.}(t_1). \quad \text{(A386)}$$

With the help of this result we write

$$\begin{aligned} &\int_{t_0}^{t} dt_1 \int_{t_0}^{t_1} dt_2 \; H_{int.}(t_1) H_{int.}(t_2) \\ &= \frac{1}{2} \int_{t_0}^{t} dt_1 \int_{t_0}^{t_1} dt_2 \; H_{int.}(t_1) H_{int.}(t_2) + \frac{1}{2} \int_{t_0}^{t} dt_1 \int_{t_0}^{t_1} dt_2 \; H_{int.}(t_1) H_{int.}(t_2) \\ &= \frac{1}{2} \int_{t_0}^{t} dt_1 \int_{t_0}^{t_1} dt_2 \; H_{int.}(t_1) H_{int.}(t_2) + \frac{1}{2} \int_{t_0}^{t} dt_1 \int_{t_1}^{t} dt_2 \; H_{int.}(t_2) H_{int.}(t_1). \end{aligned} \quad \text{(A387)}$$

Now is the right time to take a time out from the trickery and define the *time ordered product.* From Ref. [51, p. 96], *time ordered product* for a set of *arbitrary* field operators $\Psi_\alpha(x^\alpha), \Psi_\beta(x^\beta), \ldots \Psi_\epsilon(x^\epsilon)$ is defined as

$$T\Big[\Psi_\alpha(x^\alpha)\Psi_\beta(x^\beta)\ldots\Psi_\epsilon(x^\epsilon)\Big] = (-1)^f\Psi_\omega(x^\omega)\Psi_\lambda(x^\lambda)\ldots\Psi_\rho(x^\rho), \tag{A388}$$

where, on the *rhs*, the field operators are the same ones as on the left but are arranged in such a order that

$$t^\omega \geq t^\lambda \geq \ldots \geq t^\rho, \tag{A389}$$

and $f =$ the number of necessary transpositions among the *fermion* field operators that are needed to achieve the ordering.

We now return back to equation (A387) to conclude the discussion of the trick. The second *rhs* has two terms. For the first term $t_1 > t_2$, while for the second term $t_2 > t_1$. Consequently we observe that the product $H_{int.}(t_1)H_{int.}(t_2)$ in the first term and the product $H_{int.}(t_2)H_{int.}(t_1)$ in the second term are in the time ordered form for each of the respective terms. Therefore finally, we have the result

$$\begin{aligned}
\int_{t_0}^{t} dt_1 \int_{t_0}^{t_1} dt_2\ H_{int.}(t_1)H_{int.}(t_2) &= \frac{1}{2}\int_{t_0}^{t} dt_1 \int_{t_0}^{t_1} dt_2\ T\Big[H_{int.}(t_1)H_{int.}(t_2)\Big] \\
&+ \frac{1}{2}\int_{t_0}^{t} dt_1 \int_{t_1}^{t} dt_2\ T\Big[H_{int.}(t_2)H_{int.}(t_1)\Big] \qquad \text{(A390)} \\
&= \frac{1}{2}\int_{t_0}^{t} dt_1 \int_{t_0}^{t} dt_2\ T\Big[H_{int.}(t_1)H_{int.}(t_2)\Big].
\end{aligned}$$

Regarding the *rhs* of (A390) an important remark needs to be made explicitly. The time ordered product also prescribes a change of sign for each transposition of the fermion field operators. Since $H_{int.}(t)$ always contains *pairs* of fermion field operators [if any are contained] the interchanges in the positions of $H_{int.}(t)$ prescribed by time ordering always involves an even number of minus signs.

The $n_{th}$ order term in the perturbative expansion (A380) can be treated similarly, with the result

$$\begin{aligned}&\int_{t_0}^{t} dt_1 \int_{t_0}^{t_1} dt_2 \ldots \int_{t_0}^{t_{n-1}} dt_n \; H_{int.}(t_1)\ldots H_{int.}(t_n) \\ &\quad = \frac{1}{n!}\int_{t_0}^{t} dt_1 \int_{t_0}^{t} dt_2 \ldots \int_{t_0}^{t} dt_n \; T\Big[H_{int.}(t_1)\ldots H_{int.}(t_n)\Big]\end{aligned} \tag{A391}$$

Using the result (A391) the perturbative expansion (A380) takes the form

$$\begin{aligned}U(t,t_0) &= 1 + \sum_{n=1}^{\infty}\frac{(-i)^n}{n!}\int_{t_0}^{t} dt_1 \ldots \int_{t_0}^{t} dt_n \; T\Big[H_{int.}(t_1)\ldots H_{int.}(t_n)\Big] \\ &\equiv T\exp\Big[-i\int_{t_0}^{t} H_{int.}(t')dt'\Big]\end{aligned} \tag{A392}$$

This result can be cast into a manifestly Poincaré covariant form by realising that the hamiltonian operator $H_{int.}(t)$ is defined in terms of the hamiltonian density operator $\mathcal{H}_{int.}(t,\vec{x})$ as follows:

$$H_{int.}(t) \equiv \int d^3x \; \mathcal{H}_{int.}(t,\vec{x}). \tag{A393}$$

The U-Matrix then reads

$$\begin{aligned}U(\Sigma_t,\Sigma_{t_0}) &= 1 + \sum_{n=1}^{\infty}\frac{(-i)^n}{n!}\int_{\Sigma_{t_0}}^{\Sigma_t} d^4x_1 \ldots \int_{\Sigma_{t_0}}^{\Sigma_t} d^4x_n \; T\Big[\mathcal{H}_{int.}(x_1)\ldots \mathcal{H}_{int.}(x_n)\Big] \\ &\equiv T\exp\Big[-i\int_{\Sigma_{t_0}}^{\Sigma_t} d^4x' \; \mathcal{H}_{int.}(x')\Big]\end{aligned} \tag{A394}$$

In the above expression $\Sigma_{t_0}$ and $\Sigma_t$ are the constant $t_0$ and constant $t$ surfaces respectively.

## A14 S Matrix

Most of the physics done at particle accelerators deals with scattering problems. Even the table top experiments of atomic and nuclear physics are often best viewed in terms of scattering processes. In a typical scattering problem we have $i$ *free particles* at $t = -\infty$, which can be represented by an initial state

$$|i\rangle = a^\dagger(\vec{p}, \sigma) \ldots a^\dagger(\vec{p}\,', \sigma') \mid \ \rangle. \tag{A395}$$

The $rhs$ of the above expression contains $i$ creation operators. The creation operator $a^\dagger(\vec{p}, \sigma)$ creates a particle with momentum $\vec{p}$ and $\sigma$ refers to the $z$ component of $J_z$ as viewed in the rest frame of the particle. The creation operators do not necessarily refer to the same spin. This description implies that $|i\rangle$ is an eigenstate of the free hamiltonian (besides other compatible observables)

$$H_0^{(i)} \; |i\rangle = E_0^{(i)} \; |i\rangle. \tag{A396}$$

The particles are brought together to a "small" region where most of the physical evolution of interest takes place. The out product are $f$ *free particles* at $t = +\infty$

$$|f\rangle = a^\dagger(\vec{p}, \sigma) \ldots a^\dagger(\vec{p}\,'', \sigma'') \mid \ \rangle. \tag{A397}$$

The final state $|f\rangle$ is assumed to be an an eigenstate of yet another free hamiltonian (besides other compatible observables)

$$H_0^{(f)} \; |f\rangle = E_0^{(f)} \; |f\rangle. \tag{A398}$$

In these notes we assume

$$H_0^{(i)} = H_0^{(f)}. \tag{A399}$$

This is a *non trivial* assumption. As an illustrating example, this assumption excludes a process in which the ingoing particles are an electron and a proton, while the outgoing particles are a hydrogen atom and a photon. Such processes must be treated carefully. However, relaxing the assumption (A399) should pose no undue difficulties. For the specific example just cited we know how to solve the bound state problem for a hydrogen atom. Therefore we know the spectrum of the outgoing states, the final state must be represented as a superposition of the eigenstates of $H_0^{(f)}$ rather than $H_0(i)$ which has protons and electrons as free states. The eigenstates of $H_0^{(f)}$ are, in this example, the ground state of the hydrogen atom and various excited states along with energy-momentum conserving photon(s).

We note that not all the ingoing ($t = -\infty$) or outgoing ($t = +\infty$) particles may be fundamental particles. As a *working* definition of a *fundamental particle* we adopt the following criteria

a) The particle be pointlike

b) It be represented by one of the representations of $SU(2)_R \otimes SU(2)_L$ introduced in Appendix.

c) Its magnetic moment $\mu$ be such that the $g - factor$

$$\mu = g(1 + a)\mu_B \tag{A400}$$

be the same as that associated with the simplest matter field coupling with the the electromagnetic field. For the Dirac spinors this simplest coupling is, of course, the standard minimal coupling. In (A400) $\mu_B$ is the generalised Bohr magneton

$$\mu_B = \frac{e}{mc} j\hbar, \tag{A401}$$

where $j$ = spin of the particle and we have explicitly written the speed of light as $c$. The *"a"* is to be calculated perturbatively from the theory and depends upon the detailed content of the vacuum and various interactions present.

The criteria enumerated above are not necessarily independent. In the lowest order approximation we do not distinguish between a fundamental and a composite particle. For example if we are considering $\overline{p} - p$ annihilation yielding a $\pi^0$, the simplest interaction hamiltonian density operator is

$$\mathcal{H}_{int.}(x) = g\ \overline{\Psi}(x)\gamma^5\Psi(x)\Phi(x). \tag{A402}$$

if $\overline{p}, p$ and $\pi^0$ are all considered fundamental particles. The composite nature of the particles is then introduced by introducing a *phenomenological* scalar function of the various momenta and spin orientations, called a *form factor,* $F(\vec{p}, \sigma; \vec{p}\,', \sigma'; \vec{k})$ as follows

$$\mathcal{H}^{ph.}_{int.}(x) = g\ F(\vec{p}, \sigma; \vec{p}\,', \sigma'; \vec{k})\ \overline{\Psi}(x)\gamma^5\Psi(x)\Phi(x). \tag{A403}$$

It would be a great advancement if one could formulate a *practical* procedure to solve the bound state problem for composite particles. The bound state wave functions so obtained could then be used for the scattering process. This however has not been accomplished so far. As such, in nuclear physics, one resorts to the use of *form factors.*

In view of the above discussion the probability amplitude to make a transition from $|i\rangle$ at $t = -\infty$ to $|f\rangle$ at $t = +\infty$ is

$$\langle f|S|i\rangle \tag{A404}$$

where the S-matrix is defined by

$$\begin{aligned} S \equiv& U(t = +\infty, t_0 = -\infty) \\ &= 1 + \sum_{n=1}^{\infty} \frac{(-i)^n}{n!} \int_{-\infty}^{\infty} d^4x_1 \ldots \int_{-\infty}^{\infty} d^4x_n\ T\Big[\mathcal{H}_{int.}(x_1)\ldots\mathcal{H}_{int.}(x_n)\Big]. \end{aligned} \tag{A405}$$

The integration now runs over *all* space time. This formula first appeared in Ref. [37], and is now commonly called the *Dyson Formula.* It is the heart and the starting point of all perturbative calculations in the canonical perturbation theory.

## A15 Expansions of $\cosh(2\vec{J}\cdot\vec{\varphi})$ and $\sinh(2\vec{J}\cdot\vec{\varphi})$ For Arbitrary Spin

Here we provide expansions for $\cosh(2\vec{J}\cdot\vec{\varphi})$ and $\sinh(2\vec{J}\cdot\vec{\varphi})$. In the identities below we have defined $\eta = (2\vec{J}\cdot\hat{p}\,)$

INTEGER SPIN:

$$\cosh(2\vec{J}\cdot\vec{\varphi}) = 1 + \sum_{n=0}^{j-1} \frac{(\eta^2)(\eta^2-2^2)(\eta^2-4^2)\ldots(\eta^2-(2n)^2)}{(2n+2)!}\sinh^{2n+2}\varphi, \quad \text{(A406)}$$

$$\sinh(2\vec{J}\cdot\vec{\varphi}) = \eta\cosh\varphi\,\sum_{n=0}^{j-1} \frac{(\eta^2-2^2)(\eta^2-4^2)\ldots(\eta^2-(2n)^2)}{(2n+1)!}\sinh^{2n+1}\varphi. \quad \text{(A407)}$$

HALF INTEGER SPIN:

$$\cosh(2\vec{J}\cdot\vec{\varphi}) = \cosh\varphi\,\left[1 + \sum_{n=1}^{j-1/2} \frac{(\eta^2-1^2)(\eta^2-3^2)\ldots(\eta^2-(2n-1)^2)}{(2n)!}\sinh^{2n}\varphi\right], \quad \text{(A408)}$$

$$\sinh(2\vec{J}\cdot\vec{\varphi}) = \eta\sinh\varphi\,\left[1 + \sum_{n=1}^{j-1/2} \frac{(\eta^2-1^2)(\eta^2-3^2)\ldots(\eta^2-(2n-1)^2)}{(2n+1)!}\sinh^{2n}\varphi\right]. \quad \text{(A409)}$$

# VITA

Dharam Vir Ahluwalia was born on October 20, 1952 in Fateh Pur, near Kurukshetra, India. After obtaining B. Sc. (Hons., 1972) and M. Sc. (1974) in physics from University of Delhi (India), he obtained a M. A. (1982) in film-making and physics from State University of New York at Buffalo and a M. S. (1987) in physics from Texas A&M University. He was a National Science Talent Scholar of the National Council of Educational Research and Training (India) for the period 1969-74 and the youngest awardee to receive an All India Invention Talent Competition Award in 1974 from the National Council of Educational Research and Training (India). During his studies at Texas A&M University he took informal leave of absence to spend a significant time at Northeastern University and King's College (London), and worked for the Fall of 1989 at the Jet Propulsion Laboratory, Pasadena, California. For his M. A. work he made a short film entitled "Echoing Shadows," which was screened in several national and international film festivals here and broad. He can be reached through:

Department of Physics
Texas A&M University
College Station, Texas 77843-4242.